%% file: ms.tex
\documentclass[acmtog, nonacm]{acmart}
\pdfoutput=1
\include{settings}
\include{commands}

\acmJournal{TOG}
\begin{document}

% Title portion
%\title{Sparse Surface Constraints for Inverse Elasticity Problems}
\title{Sparse Surface Constraints for Combining Physics-based Elasticity Simulation and Correspondence-Free Object Reconstruction}

\author{Sebastian Weiss}
\orcid{0000-0003-4399-3180}
%\affiliation{
	%\institution{Technical University of Munich}
	%\streetaddress{Arcisstraße 21}
	%\city{Munich}
	%\country{Germany}}
\email{sebastian13.weiss@tum.de}

\author{Robert Maier}
%\affiliation{
	%\institution{Technical University of Munich}
	%\streetaddress{Arcisstraße 21}
	%\city{Munich}
	%\country{Germany}}
\email{robert.maier@in.tum.de}

\author{R\"udiger Westermann}
%\affiliation{
	%\institution{Technical University of Munich}
	%\streetaddress{Arcisstraße 21}
	%\city{Munich}
	%\country{Germany}}
\email{westermann@tum.de}

\author{Daniel Cremers}
%\affiliation{
	%\institution{Technical University of Munich}
	%\streetaddress{Arcisstraße 21}
	%\city{Munich}
	%\country{Germany}}
\email{cremers@tum.de}

\author{Nils Thuerey}
%\affiliation{
	%\institution{Technical University of Munich}
	%\streetaddress{Arcisstraße 21}
	%\city{Munich}
	%\country{Germany}}
\email{nils.thuerey@tum.de}

\affiliation{
	\institution{\newline Technical University of Munich}
	\streetaddress{Arcisstraße 21}
	\city{Munich}
	\country{Germany}}

\begin{abstract}
We address the problem to infer physical material parameters and boundary conditions from 
the observed motion of a homogeneous deformable object via the solution of an inverse problem. Parameters are estimated from potentially unreliable real-world data sources such as sparse observations without correspondences. We introduce a novel Lagrangian-Eulerian optimization formulation, including a cost function that 
%reliably and efficiently 
penalizes differences to observations during an optimization run. This formulation matches correspondence-free, sparse observations from a single-view depth sequence with a finite element simulation of deformable bodies. In conjunction with an efficient hexahedral discretization and a stable, implicit formulation of collisions, our method can be used in demanding situation to recover a variety of material parameters, ranging from Young's modulus and Poisson ratio to gravity and stiffness damping, and even external boundaries. 
In a number of tests using synthetic datasets and real-world measurements, we analyse the robustness of our approach and the convergence behavior of the numerical optimization scheme.  
\end{abstract}

%
% The code below should be generated by the tool at
% http://dl.acm.org/ccs.cfm
% Please copy and paste the code instead of the example below.
%

%\ccsdesc[300]{Computing methodologies~Interactive simulation}
%\ccsdesc[100]{Computing methodologies~Scientific visualization}
%\ccsdesc[100]{Computing methodologies~Collision detection}

%
% End generated code
%


\begin{teaserfigure}
\centering
\includegraphics[width=\linewidth]{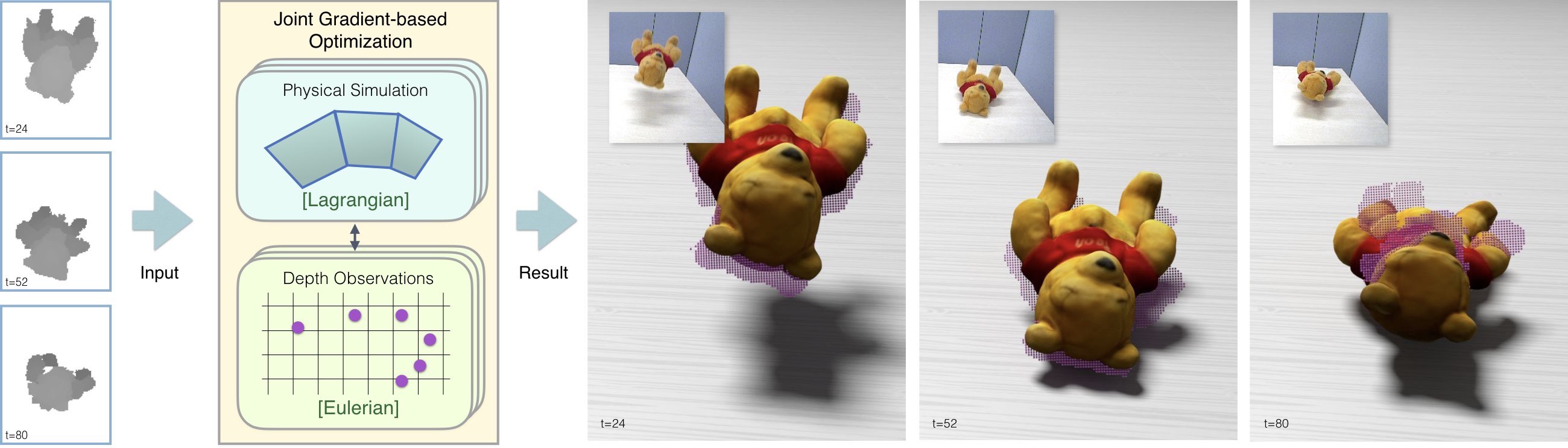}
\caption{Algorithm overview: From a sparse sequence of single-view depth observations,
    we jointly compute a physical explanation and a matching of the observations with a simulation model.
    In this way, we can perform an end-to-end gradient-based optimization to estimate a wide range of material parameters as well as collision geometries. The three images show a material and damping reconstruction for a plush toy. The ground truth images (not used in the optimization) are shown in the top left insets.} 
	\Description{A visual overview of our algorithm: starting with a rough set of depth observations, we jointly compute a physical explanation and a matching of the observations with the simulation.  In this way, we can perform gradient-based end-to-end optimization to estimate arbitrary simulation parameters from material behavior to collision geometry. The three images show a material and damping reconstruction for a plush toy. The ground truth color images (not used in the optimization) are shown in the top left insets.}
\label{fig:teaser}
\end{teaserfigure}

\emergencystretch=1em
\maketitle

%\commentSebi{Remove hypersetup\{draft\} in settings.tex when the pdfendlink-error is fixed}

%\begin{bibunit}[ACM-Reference-Format]
\input{Body}

%\putbib
%\end{bibunit}

%\vspace{10pt}
%\newpage

% Bibliography
\bibliographystyle{ACM-Reference-Format}
\bibliography{ms}
%\bibliographystyle{ACM-Reference-Format}
%\bibliography{main}{literature}{Literature}

\clearpage 
%\newrefsection

\begin{center}\LARGE{Supplemental Material - Appendix}\end{center}

% Appendix
\appendix
%\begin{bibunit}[ACM-Reference-Format]

\input{Appendix}

\end{document}

%% file: settings.tex
\usepackage[bookmarksnumbered,unicode]{hyperref}

\usepackage{booktabs} % For formal tables

\usepackage{rotating}

\usepackage{graphicx}
\graphicspath{{images/}}

\usepackage{float}
\usepackage{subcaption}
\usepackage{marginnote}
\usepackage{tikz}
\usepackage{pgfplots,pgfplotstable}
\usepackage[percent]{overpic}

\usetikzlibrary{fit,decorations.pathreplacing,calc,decorations.pathmorphing,patterns,trees,arrows,arrows.meta}
\usepgfplotslibrary{groupplots}
\usetikzlibrary{pgfplots.groupplots}

\usepackage{pgfplots}
\pgfplotsset{compat=1.3}
\usepackage{amsmath}
\usepackage{amsthm}
\usepackage{amsfonts}
\usepackage{amssymb}
\usepackage{dsfont}
\usepackage{bm}
\usepackage{xfrac}

\makeatletter
\@definecounter{equation}
\def\equation{$$\refstepcounter{equation}}
\def\endequation{\eqno \hbox{\@eqnnum}$$\@ignoretrue}
\def\@eqnnum{{\normalsize\normalfont \normalcolor (\theequation)}}
\newenvironment{smalleq}{%
   \skip@=\baselineskip
   \Small
   \baselineskip=\skip@
   \equation
}{\endequation \ignorespacesafterend}
\newenvironment{tinyeq}{%
   \skip@=\baselineskip
   \tiny
   \baselineskip=\skip@
   \equation
}{\endequation \ignorespacesafterend}
\makeatother

\usepackage{rotating}
\usepackage{tabularx}
\usepackage{csvsimple}
\usepackage{adjustbox}
\usepackage{makecell}

\usepackage{algorithmicx}
\usepackage{algpseudocode}
\usepackage{algorithm}
\makeatletter
\newcounter{algorithmicH}% New algorithmic-like hyperref counter
\let\oldalgorithmic\algorithmic
\renewcommand{\algorithmic}{%
  \stepcounter{algorithmicH}% Step counter
  \oldalgorithmic}% Do what was always done with algorithmic environment
\renewcommand{\theHALG@line}{ALG@line.\thealgorithmicH.\arabic{ALG@line}}
\makeatother
\algrenewcommand\algorithmicindent{1.0em}% Decrease intentation

\usepackage{color}
\usepackage[normalem]{ulem}
\definecolor{nilscolor}{rgb}{.6,0.2,0.0}
\definecolor{sebicolor}{rgb}{0.188, 0.192, 0.874}
\definecolor{ruedigercolor}{rgb}{0.0, 0.5, 0.0}

\definecolor{nilscolorC}{rgb}{.9,0.4,0.0}
\definecolor{sebicolorC}{rgb}{0.396, 0.4, 0.823}
\definecolor{ruedigercolorC}{rgb}{0.2, 0.6, 0.2}
\definecolor{revisionCol}{rgb}{0.188, 0.192, 0.874}
  \newcommand{\commentNils}[1]{}
  \newcommand{\commentSebi}[1]{}
  \newcommand{\commentRuediger}[1]{}
  \newcommand{\nils}[1]{#1}
  \newcommand{\sebi}[1]{#1}
  \newcommand{\ruediger}[1]{#1}

%% file: commands.tex
\newcommand{\N}{\mathbb{N}}

\newcommand{\R}{\mathbb{R}}
\newcommand{\I}{\mathbf{1}}
\DeclareMathOperator{\tr}{tr}
\DeclareMathOperator{\dx}{dx}

\DeclareMathOperator{\ds}{ds}
\DeclareMathOperator{\divergence}{div}

\newcommand{\minuseq}{\mathrel{-}=}

\newcommand{\clearemptydoublepage}{%
  \ifthenelse{\boolean{@twoside}}{\newpage{\pagestyle{empty}\cleardoublepage}}%
  {\clearpage}}

\makeatletter
\newcommand*{\rom}[1]{\expandafter\@slowromancap\romannumeral #1@}
\makeatother

\newcolumntype{R}[2]{%
    >{\adjustbox{angle=#1,lap=\width-(#2)}\bgroup}%
    l%
    <{\egroup}%
}
\newcounter{rowcntr}[table]

\newcolumntype{N}{>{\refstepcounter{rowcntr}\alph{rowcntr})}r}
\AtBeginEnvironment{tabular}{\setcounter{rowcntr}{0}}

\newcommand{\etAl}{\emph{et al.}\mbox{ }}

%% file: Body.tex
%====================================================================================
%====================================================================================
\section{Introduction}\label{sec:Introduction}

Parameterizing the deformation behavior of an object
such that it can be recovered in a numerical simulation 
is important in many real-world applications, such as object tracking, pose estimation, and 
computer animation.
% leave out the following?
%  and medical tissue analysis. The estimation of these parameters, e.g., with empirical correlation estimation or instrumented tests, is difficult and error-prone, and often infeasible due to time and cost constraints.
In this work, we address the challenging inverse problem to infer the values 
of physical material parameters from sparse observations, i.e., a single depth image per timestep, 
that can be acquired easily in real-world scenarios, 
e.g., with a single commodity depth camera.
% of object motion via non-linear optimization.
While previous work has likewise targeted reconstructing materials from measured deformations and visual observations, it typically relies either on carefully controlled lab settings \cite{bickel2009capture,miguel2016modeling,zehnder2017metasilicone}, or on dense observations \cite{deAguiar:2008:PCS:1360612.1360697,wang2015deformation,Innmann.2016}.
In contrast, we focus on real-world interactions, such as falling and colliding objects, that are recorded from a single viewpoint. Correspondingly, our goal is not a highly accurate and generic material model for fabrication or medical applications, but rather a plausible explanation that enables to analyze, understand and simulate the behaviour of objects in everyday environments.
% Our approach relates to previous methods for establishing correspondences between tracked objects and synthetic deformable models. These methods either infer mainly geometric shape deformations rather than accurate physical behavior \cite{deAguiar:2008:PCS:1360612.1360697,Innmann.2016}, or need sufficient amounts of dense data in the form of scanned 3D point clouds \cite{wang2015deformation}.

To match sparse observations to the simulated object without an explicit feature tracking step, we propose a novel cost function that is used together with a differentiable simulation method for soft bodies.
%\nils{Our approach ''disambiguates'' sparse observations, e.g., a single depth image, with the help of a differentiable simulation method for soft bodies, and proposes a gradient-based non-linear optimization to match the depth measurements over time with a physical simulation. To achieve this, we introduce a novel cost function that can match sparse observations with the simulated object without the need to explicitly track features. Here we make the assumption that the observed data is related to the object we want to recover by a homogeneous linear elastic model, expressed by a system of differential equations.}
%solve the open, and fundamental problem of matching a sparse set of constraints with an elasticity simulation.
%
%
We seek to optimize for a wide range of parameters controlling the motion of an elastic body, such as Young's modulus, Poisson ratio, mass, initial linear and angular velocity, and Rayleigh damping, as well as collision plane position and orientation. To enable
the robust optimization we leverage a differentiable solver in combination with the adjoint method.
The adjoint method provides analytic gradients for all matched surface points. As we will demonstrate 
below, this is significantly more stable than, e.g., finite differences, which have to rely on changes 
of the final cost function values. In addition, the adjoint method effectively allows us to optimize 
for multiple parameters with a negligible increase in computations, as only a single backward pass is
required to simultaneously compute gradients for all parameters.
%efficient joint optimization over multiple parameters, we formulate the optimization algorithm based on the adjoint method. 
%We show improved numerical stability of the adjoint method over the finite difference method for the proposed cost function, due to the regularization properties of the adjoint optimization pass.
%Furthermore, the adjoint method requires only one single backward pass to simultaneously compute gradients for all parameters, as opposed to e.g. the finite difference method which requires a forward pass for every parameter.}
%\nils{\ruediger{To enable the efficient optimization for many parameters at once,} we formulate our optimization algorithm based on the adjoint method.}
%\ruediger{Therefore, we} 
%\nils{propose a novel cost function
%that can incorporate sparse surface constraints without requiring any tracked features. % "explicitly" removed

Our approach further supports the inclusion of collisions via an implicit formulation. This is 
important for real-world scenarios, as the collision response of materials typically contains 
a wide range of information for the inference of a suitable parameterization.
%can be included implicitly in the time integration of the body dynamics.
%

We take inspiration from techniques for fluid simulation, in particular, 
the Eulerian formulation of advective motion,
and present a novel hybrid Lagrangian-Eulerian optimization algorithm.
We leverage a hexahedral discretization with implicitly embedded boundaries to arrive at a
simulation algorithm that is fully differentiable, and provides robust gradients for inverse problems.
This makes it amenable to optimizations like the adjoint method, and a flexible tool
for a variety of future algorithmic combinations, e.g., to incorporate soft body physics 
into deep learning methods.

%The central contributions of our work are:
The central contributions of our work target inverse elasticity problems. 
In this context, we propose:
\begin{itemize}
	\item A novel formulation for sparse and correspondence-free
		surface constraints, e.g. measurements from RGB-D cameras. % in inverse problems.
	\item A hybrid Lagrangian-Eulerian approach that yields robust gradients for
	    solving inverse elasticity problems.
		%with surface constraints that provide robust gradients.
	\item An efficient discretization scheme with implicit object boundaries
	    and collision constraints based on a hexahedral simulation grid.
	%\item The introduction of a discretization scheme for soft body 
		%FEM simulations based on a regular grid with implicit object boundaries.
	%\item The efficient embedding of collision constraints and their use for 
	    %material parameter prediction.
\end{itemize}
In combination, these contributions make it possible to estimate 
material parameters and boundary conditions in complex situations using simple hardware setups. We demonstrate this capability and analyze the plausibility of inferred results in a number of experiments using computer simulations with ground truth behavior and observed deformations of real-world objects.

%====================================================================================
%==================================================================================== 
\section{Related Work}\label{sec:RelatedWork}

We build upon a number of techniques in physics-based elastic body simulation: we employ linear elasticity, and 
%discretize the governing system of partial differential equations using 
a discretization based on hexahedral finite elements with a corotational formulation of strain \cite{muller2002stable,MichaelHauth.2003,georgii2008CorotatedFE,Dick:2011:CUDAFEM}.
%, in combination with a 
%The efficient numerical solutions with fast solvers such as 
%multigrid solver \cite{Dick:2011:CUDAFEM,mcadams2010parallel}.
%\comment Sebi{No Multigrid is used!}
Discretization variants and details are thoroughly discussed in previous work \cite{sifakis2015finite,Wu:2015}.
\begin{figure*}[tp]
    \centering
       \includegraphics[width=0.9\textwidth]{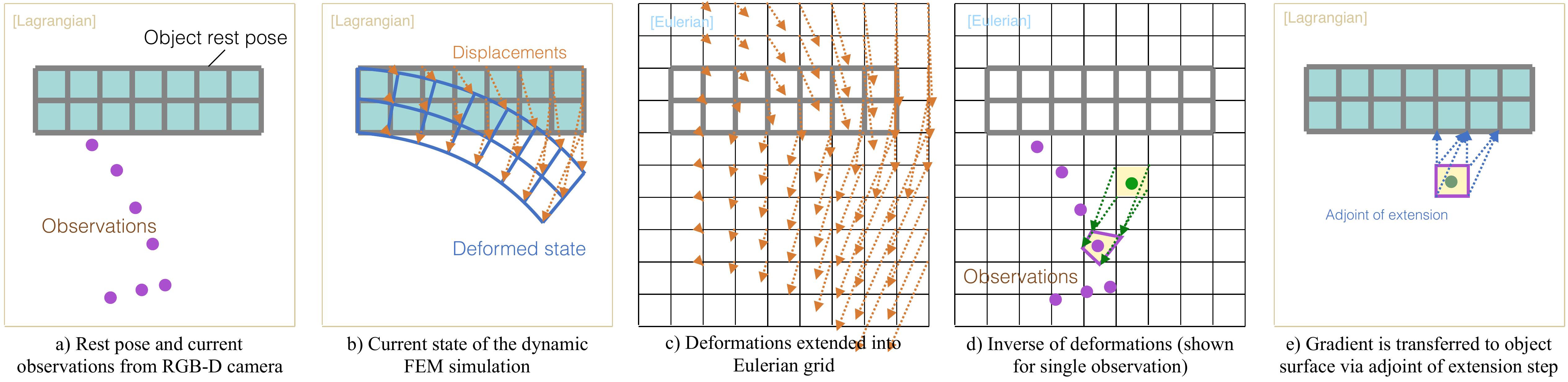}
    \caption{Algorithm overview 
    %of the Sparse Surface Constraints algorithm
    \ruediger{: Given the object in rest pose and sparse point observations, elastic displacements are first computed in an Lagrangian framework and then resampled to an Eulerian grid. For each observation, the inverse displacement is computed, and displaced observations are matched to grid cells. 
    \sebi{A cost function penalizes the distance of these observations to the object's surface.}
    Finally, the gradient \sebi{of the cost function} is transferred to the object surface via the adjoint of the extension step.}}
		\Description{Overview of the Sparse Surface Constraints algorithm: Given are the object in rest pose and sparse point observations. The algorithm first computes the elastic displacements in an Lagrangian framework. Then, these displacements are extended into an Eulerian grid. For each observation, the inverse displacement is solved for, and the observations are matched into grid cells. Finally, the gradient is transferred to the object surface via the adjoint of the extension step.}
		\label{fig:SSCOverview}
\end{figure*}
%
%To accurately represent complex object boundaries, hexahedral simulation grids at high resolution would be required. To overcome this limitation, the boundary can be 
To reduce the need for very fine computational meshes, the boundary can be incorporated into the basis functions
via enrichment functions \cite{Belytschko:1999:XFEM}. In computer graphics, this has been used to accurately simulate cuts \cite{XFEM10,KBT17}. While the construction of the enrichments depends on the type of discontinuity and the interface location, the Cut Finite Element Method (\textit{cutFEM}) makes it possible to use geometry-independent and non-boundary-fitted meshes \cite{Hansbo.2014}, which is a similar concept to \textit{Immersed Boundary Methods} for fluids \cite{Mittal.2005,Ferstl:2014:TVCG}.
%The integration domain is then restricted with subgrid accuracy to the part of the domain that is covered by the material.
%In fluid simulations, similar approaches falling into the category of \textit{Immersed Boundary Methods} have been used to accurately resolve the fluid-air interface \cite{Mittal.2005,Ferstl:2014:TVCG}.

%\nils{more regular simulation, TODO integrate?
%teran \cite{teran2005robust} %  title={Robust quasistatic finite elements and flesh simulation},
%teran multigrid \cite{mcadams2010parallel} %  title={A parallel multigrid Poisson solver for fluids simulation on large grids},
% done: bernhard \cite{zehnder2017metasilicone}
%
%melina, inverse shells with contact , \cite{ly2018inverse}
%shin sueda , muscle tendons, hands \cite{sueda2008musculotendon} %  title={Musculotendon simulation for hand animation},
%shin,  meta material library , fabrication \cite{chen2015data} %  title={Data-driven finite elements for geometry and material design},
%}

Besides the "classical" forward simulation of elastic material, a substantial amount of research has been devoted to inverse elasticity simulation. In the computer graphic community, inverse elasticity simulations are often used for artistic control.
Previous works can be split into approaches that compute external forces that make the deformable body follow a prescribed animation \cite{BarbicJernejdaSilvaMarcoPopovicJovan.2009, Barbic.2012,Schulz.2014}, and approaches that let an elastic body deform into its rest shape via internal forces \cite{Coros.2012, Chen.2014}. Further applications of inverse methods in computer graphics include skeletal-based character animation \cite{Kim.2017}, and thin shell deformations \cite{Bergou.2007}.
Inverse solvers were additionally proposed for shells with friction and contact effects \cite{ly2018inverse}, or for reconstructing rigid body collisions \cite{monszpart2016smash}.

%While these approaches take the deformations, i.e., the key-poses, as input, 
Wang~\etAl \shortcite{wang2015deformation} instead capture motion trajectories and use virtual forces derived from these trajectories to compute an alignment between an elastic FE-mesh and the captured point cloud.
The alignment procedure is split into the computation of a reference shape that best matches the observed shape, and the estimation of the deformation parameters of the captured shape. Our approach for determining material parameters from recorded depth images is inspired particularly by this work, yet instead of using a gradient-free downhill simplex method we formulate the inverse problem as a constrained minimization problem that is solved using the adjoint method \cite{McNamara.2004}.
Therefore, we demonstrate the efficient calculation of the gradient of the cost function with respect to the optimization parameters at every iteration.
The reconstruction from only one single depth camera as well as the integration of collisions further distinguishes our approach.
%\sebi{Furthermore, Wang~\etAl \shortcite{wang2015deformation} did not include collisions in their simulation and reconstruction and required three synchronized cameras as observations}. \comment Sebi{Anmerkung von Robert, besser von Wang abgrenzen?}
We do not employ a rest pose estimation, but instead assume that it can, e.g., be obtained from an initial scan.
In recent years, also first attempts have been made to replace the elasticity model itself by neural networks \cite{wang2019neuralmat,luo2018deepwarp}, an avenue that would yield interesting benefits in conjunction with algorithms for object reconstruction.

Especially for the handling of non-linear and large deformations,
reduced order models have shown impressive results. Specific adaptations of linear modal bases using derivatives and modal warping have been described by Barbi\u{c} and James \cite{barbivc2005real} and Choi and Ko \cite{choi2005modal}, respectively. The rotation-strain space model by Pan~\etAl \shortcite{pan2015subspace} preserves the key characteristics of deformable bodies with a lower-dimensional configuration space representation. Yang~\etAl \shortcite{yang2013boundary,yang2015expediting}
present a linear inertia mode technique, Xu~\etAl \shortcite{xu2015interactive} demonstrate the use of reduced models for designing material properties of three-dimensional elastic objects, and Brandt~\etAl\shortcite{brandt2018hyper} enable real-time simulation of high-resolution meshes in specifically designed subspaces using projective dynamics and hyper-reduction. In some of these works, contact handling has been combined with reduced model simulation \cite{xu2014implicit,brandt2018hyper}.

Reconstructing deforming objects geometrically is a long-stand\-ing topic in research. E.g., methods were proposed to capture characters via locally rigid parts \cite{pekelny2008articulated}, for
template based capturing of freely deforming objects \cite{li2009robust}, 
and non-rigid reconstructions based on depth videos \cite{Innmann.2016,slavcheva2017killingfusion}, to name just a few.
While these works share our goal to capture deforming objects, they focus on constructing a 
geometric representation, while our method focuses on the construction of a physical explanation for the object.
%TODO new geometry / vision section ,  computer vision, depth data
% ? \cite{kerl2013robust} % title={Robust odometry estimation for RGB-D cameras},
% defo obj tracking
%\cite{slavcheva2017killingfusion} %  title={Killingfusion: Non-rigid 3d reconstruction without correspondences},
%
% from r4
% joint surface reconstruction and registration , point clouds, less relevant?
%\cite{huang2007bayesian} %  title={Bayesian surface reconstruction via iterative scan alignment to an optimized prototype},
%
%relevant: identify and track rigid components from single video \cite{pekelny2008articulated} %  title={Articulated object reconstruction and markerless motion capture from depth video},
%
%relevant: two scale, refinement, template based; similar to non-rigid ICP , deforming clown etc. \cite{li2009robust} % title={Robust single-view geometry and motion reconstruction},

Inverse elasticity simulation has also been used in medical imaging to estimate the material properties.
In the medical setting, only single physical parameters like stiffness are typically estimated.
For instance, the study by Gokhale~\etAl \shortcite{Gokhale.2008} use dense constraints in 2D images to estimate stiffness parameters.
Kroon and Holzapfel  \shortcite{Kroon2008EstimationOT} use a similar approach including an element partitioning strategy to estimate elastic properties.
Posterior distributions over linear elastic material parameters are estimated via Monte-Carlo Markov-Chain by Risholm~\etAl \shortcite{PetterRisholmJamesRossGeorgeR.WashkoWilliamM.Wells.2011}, while inhomogeneous elasticity parameters of shell-like surface structures are
targeted by Zhao~\etAl \shortcite{Zhao.2017}.
Similar approaches have been used in computer graphics to capture complex materials, e.g., to design deformable objects for fabrication
Bickel~\etAl \shortcite{bickel2009capture,bickel2010design}, and to obtain accurate parameterizations of cloth behavior Wang~\etAl \cite{wang2011data}. 
More recently, researchers have also investigated building libraries of material behavior \cite{chen2015data}
and measuring complex and non-uniform materials \cite{miguel2016modeling,zehnder2017metasilicone}.
Although the works above share our overall goals and achieve impressive material estimates,
they inherently rely on full correspondences between observations and the geometric models, and thus are difficult to apply outside of lab environments.

%====================================================================================
%====================================================================================
\section{Overview and Physical Model}\label{sec:OverviewPhysicalModel}

We assume that a scanned representation of the observed object in rest pose exists.
At the core of our method, we iteratively optimize the physical parameters that govern the object's deformation behavior.
In a forward step, a Lagrangian elasticity simulation is performed with the current parameter estimates.
We use a finite element discretization of the displacement field with tri-linear shape functions, and employ a rotational
invariant formulation of the strain tensor using the corotated strain formulation.
The computed deformations are then matched and compared to a set of observations, e.g. given by depth images from a commodity RGB-D sensor.
This step utilizes a Cartesian grid as Eulerian representation,
which is facilitated by the hexahedral Lagrangian discretization used in the forward simulation.
The matching step, constrained by the physical deformation properties, provides a similarity measure 
and enables the calculation of gradients of the cost function that is to be minimized.
These gradients are back-traced through the simulation using the adjoint method,
in order to obtain updates for the physical parameters via a gradient-based optimizer.
This process is illustrated in Fig.~\ref{fig:SSCOverview}.
\sebi{The parameters that can be reconstructed are gravity, Young's modulus, Poisson ratio, mass and stiffness damping and the collision plane position. For example, optimizing for the collision plane allows us to reconstruct collisions even if the actual point of impact is obstructed from the camera.}
Due to the inherent non-linearity, we evaluate a batch of optimization runs using perturbed initial values, 
and choose the best match as final result.

%====================================================================================
%====================================================================================

In the following, we briefly summarize the used elasticity model. Details of
our particular discretization scheme are provided in Sec.~\ref{sec:FEM}.
\sebi{
The reference configuration of the observed, scanned object is given as a signed distance function (SDF) $\phi_0 : \R^3\rightarrow\R$ with the object occupying the space where $\phi$ is negative: $\Omega_r:=\{\mathbf{x}\in\R^3:\phi_0(\mathbf{x})<0\}$.
For a displacement of each point of the object at time $t\geq 0$ ($t\in\R_0^{+}$) given
by $u:\Omega^r \times t \rightarrow \R^3$,}
the linear \textit{Green strain tensor} 
$E(u) :=$ $\frac{1}{2} ( \nabla u + (\nabla u)^T )$
is used to compute the second order \textit{Piola-Kirchoff stress tensor} as
$P(u) := 2 \mu E(u) + \lambda \tr(E(u)) \I$,
% \begin{equation}
% 	P(u) := 2 \mu E(u) + \lambda \tr(E(u)) \I\,
% \label{eq:StressTensor}
% \end{equation}
with the \textit{Lam\'e coefficients} $\mu$ and $\lambda$ derived from the Young's modulus $k$ and the Poisson ratio $\rho$.
The dynamic behavior of a deformable object is then governed by the system of partial differential equations
\begin{subequations}
\label{eq:StrongPDEdynamic}
\begin{alignat}{4}
	m\ddot{u}-\divergence P(u) &=& f_B &\text{\ \ in  } \Omega^r \times \R_0^{+} \\
	u &=& u_D &\text{\ \ on  } \Gamma^r_D \times \R_0^{+}  \\
	P(u) \cdot n &=& f_S &\text{\ \ in  } \Gamma^r_N \times \R_0^{+}.
%	u &=& u^0 &\text{\ \ in  } \Omega^r \times \{0\} \\
%	\dot{u} &=& \dot{u}^0 &\text{\ \ in  } \Omega^r \times \{0\} \ ,
\end{alignat}
\end{subequations}
%with mass $m$, Dirichlet boundaries $\Gamma^r_D$ and Neumann boundaries $\Gamma^r_N$.
\sebi{The body forces are denoted by $f_B$ and include e.g. the gravity.} The boundary conditions on the boundary $\Gamma^r_D$ consist of Dirichlet boundary conditions, which prescribe the displacement $u_D$ on $\Gamma^r_D$, and Neumann boundary conditions, which prescribe external surface forces $f_S$ on $\Gamma^r_N$. $\mathbf{n}$ is the unit outer
normal on $\Gamma^r_N$, and $m$ the body's mass.
For the dynamic case, this leads to an initial value problem, using the initial shape $\phi_0$ and material parameters. %the elasticity solver computes
In our work, problem (\ref{eq:StrongPDEdynamic}) is discretized in space using a hexahedral simulation grid with implicitly embedded boundaries, and in time by applying a Newmark time integration scheme.

%====================================================================================
%====================================================================================
\section{Sparse Surface Constraints}\label{sec:Cost}

The goal of our work is to estimate unknown material parameters from observed object deformations with an optimization algorithm.
One fundamental problem is to find a cost function $J$ that can reliably and efficiently penalize differences to observations. We first review existing cost
functions in order to motivate our proposed cost function.

\subsection{Existing Cost Functions}\label{sec:Cost:Existing}

\begin{figure*}%
\centering
\begin{subfigure}{0.24\textwidth}
	\includegraphics[width=\textwidth,trim=0 0 0 0,clip]{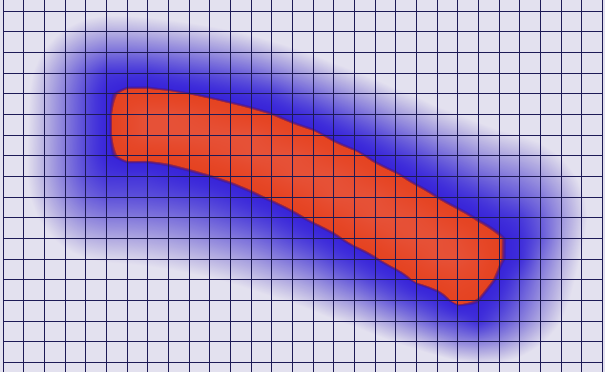}%
	\caption{\footnotesize Simulated deformation. }
	\label{fig:CostFunctionComparison:pic}
\end{subfigure}
~
\begin{subfigure}{0.24\textwidth}
	\includegraphics[width=\textwidth]{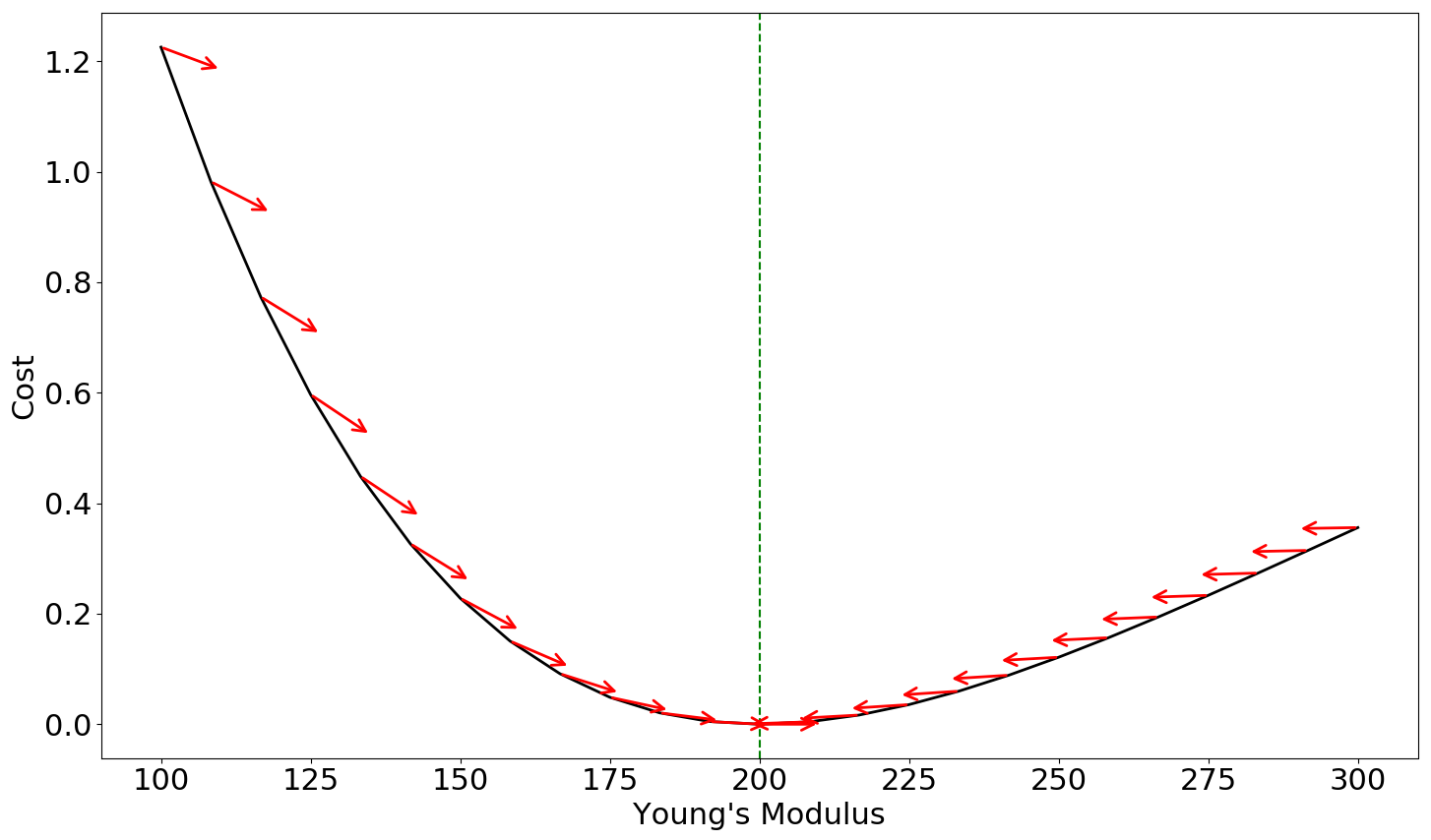}%
	\caption{\footnotesize $J_{\text{DISP}}$: Exact vertex displacements.}
	\label{fig:CostFunctionComparison:ref}
\end{subfigure}
~
\begin{subfigure}{0.24\textwidth}
	\includegraphics[width=\textwidth,trim=0 0 0 25,clip]{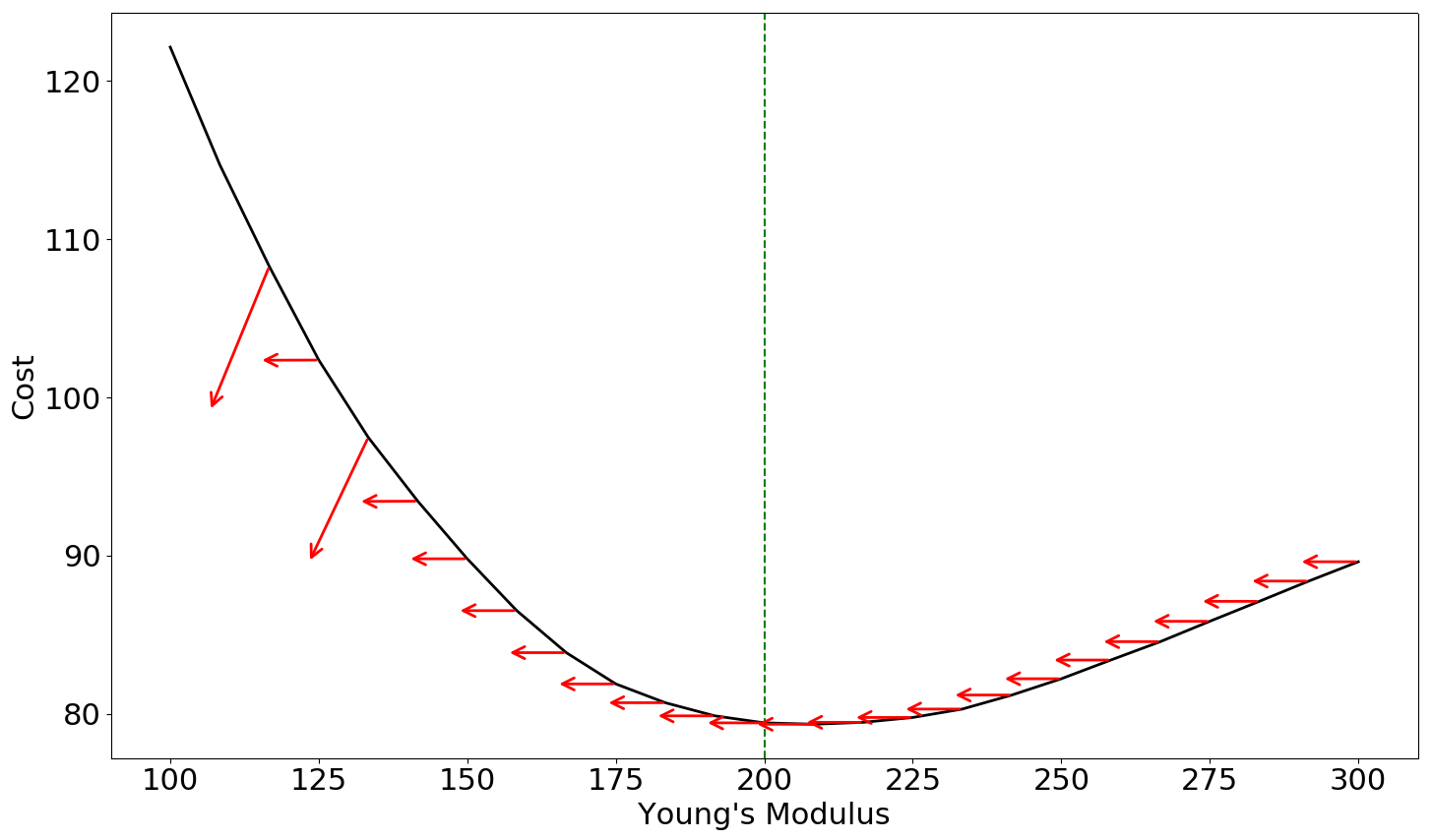}%
	\caption{\footnotesize $J_{\text{SDF}}$: Per-cell SDF differences.}
	\label{fig:CostFunctionComparison:sdf} 
\end{subfigure}
~
\begin{subfigure}{0.24\textwidth}
	\includegraphics[width=\textwidth,trim=0 0 0 25,clip]{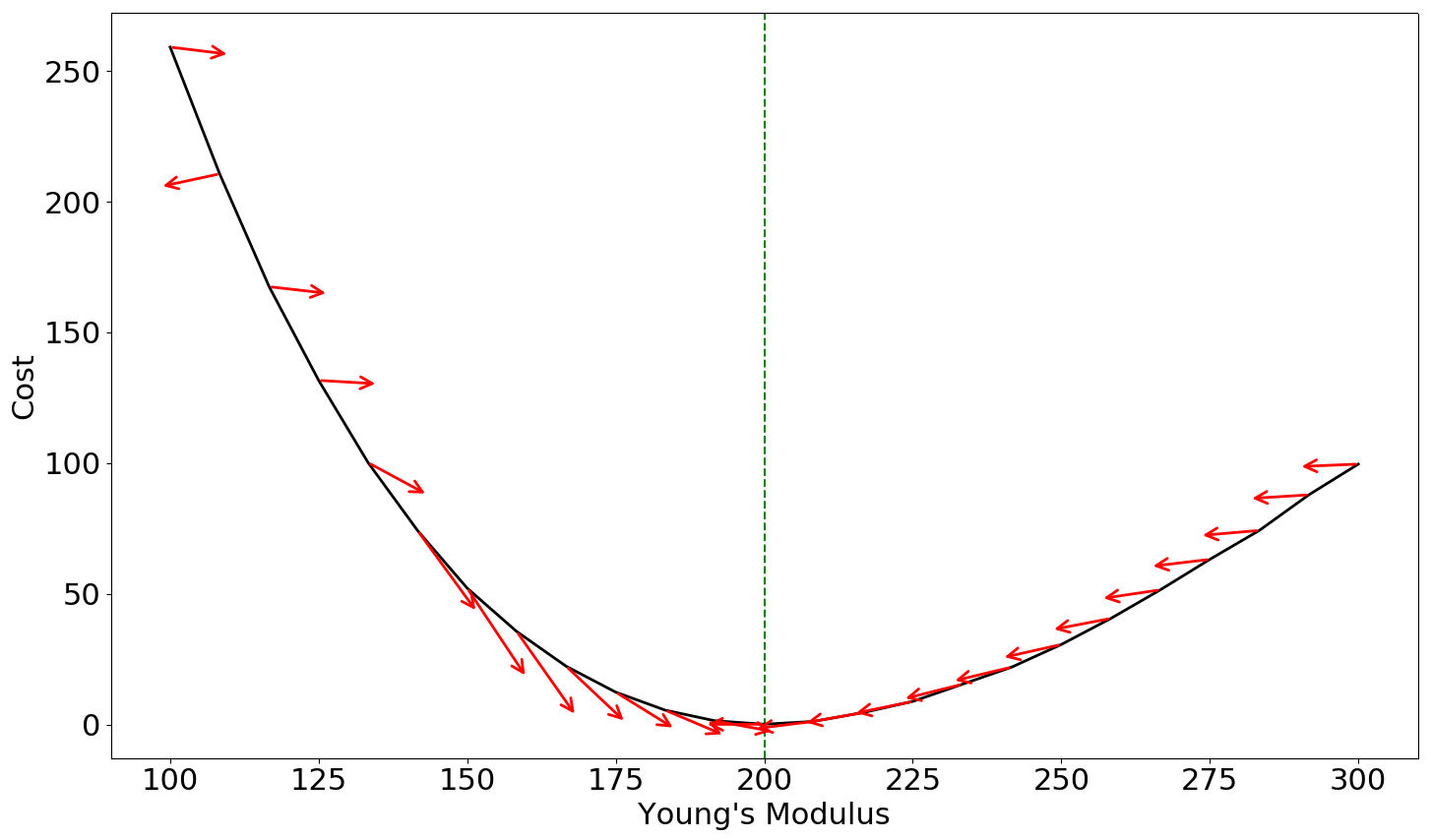}%
	\caption{\footnotesize $J_{\text{SSC}}$: Point-wise SDF inversion.}
	\label{fig:CostFunctionComparison:ssc}
\end{subfigure}
\caption{Comparison of cost functions for the elastic simulation in (a). % from Fig.~\ref{fig:PartialObservations}.
	(b) Known vertex correspondences enable accurate gradient estimation.
    % life scenarios typically cannot rely on a full set of correspondences.
    (c) Per-cell SDF differences give mostly wrong gradient estimates due to the highly non-linear 
	advection step: For a Young's modulus less than 200 the gradients all point into the wrong direction. 
	(d) By using the proposed cost function $J_{SSC}$, almost everywhere can the gradients be recovered accurately.
    }%
	\Description{Comparison of the three cost functions for a 2D bar example (a).  While the gradients are accurate and reliable for known correspondences in (b), real life scenarios typically cannot rely on a full set of correspondences. The SDF version in (c) fails due to the induced, highly non-linear advection step: the gradients all point in the wrong direction for a Young's modulus less than 200.  In contrast, our sparse formulation in (d) can recover robust gradients most of which (all apart from one in this example) point in the correct direction.}%
\label{fig:CostFunctionComparison}%
\end{figure*}

A natural choice for a cost function is one that considers 
squared differences of per-vertex displacements and
derivatives over time $t$, as illustrated in Fig.~\ref{fig:CostFunctionComparison:ref}
and proposed in previous works \cite{Coros.2012,PanManocha.2017}:
\begin{equation}
	J_{\text{DISP}}(u) := \sum_{t=1}^T{\frac{1}{2} w_t \left\| u^{(t)} - u^{(t)}_{\text{obs}} \right\|^2_{\Omega^r} + \frac{1}{2} v_t \left\| \dot{u}^{(t)} - \dot{u}^{(t)}_{\text{obs}} \right\|^2_{\Omega^r}}.
\label{eq:CostFunction1}
\end{equation}
Here, $u_{\text{obs}}$ denotes observed positions, e.g., from a tracking procedure.
The weighting terms $w_{t},v_{t}$ are optional to, e.g., model the reliability of measurements.
This cost function requires ground truth vertex displacements, which are often not available.
Feature tracking methods \cite{Schulman2013,wang2015deformation} can circumvent this problem
by explicitly matching observations to vertices in the simulation mesh.
However, the matching step does not provide derivatives, and as such
cannot be used in conjunction with gradient-based optimization schemes.
In addition, explicit matching approaches typically only couple to a simulation
via external force estimates, and are thus decoupled from the actual parameter estimation step.
Errors in the computed vertex displacements will invariably propagate into the inferred physical simulation.

Another cost function variant that is especially popular for fluid control uses squared differences of the SDF values $\phi$ per domain point \cite{McNamara.2004}.
The function can be formulated as
\begin{equation}
	J_{\text{SDF}}(\phi^{(t)}) := \sum_{t=1}^T{\frac{1}{2} \left\| w_t \left( \phi^{(t)} - \phi_{\text{obs}}^{(t)} \right) \right\|^2_{\Omega} },
\label{eq:CostFunction2}
\end{equation}
with $\phi_{\text{obs}}$ denoting the observation, i.e. a target SDF in this case. % and the meaning of $w$ is as before.
The SDF-based cost function, however, requires a full SDF representation $\phi^{(t)}$,
and while guesses about the complete 3D shape of the observation could be made,
erroneous estimates can easily mislead the optimization procedure.
Furthermore, the current SDF \sebi{$\phi^{(t)}$} needs to be calculated by means of an advection step \sebi{from $\phi_0$ and $u^{(t)}$}, e.g.,
with a semi-Lagrangian method.
While this is possible in general,
it requires highly non-linear gradient evaluations for inverting the displacement field,
which often leads to diverging optimizations in practice.
This is demonstrated in Fig.~\ref{fig:CostFunctionComparison:sdf}, where even for the simple 2D setup the
gradients increasingly deteriorate for lower values of the Young's modulus.

\subsection{Proposed Cost Function for Sparse Surface Constraints}\label{sec:Cost:SSC}

%We propose a novel cost function variant that is inspired by the 
We instead propose a SDF-based formulation that
is able to incorporate sparse constraints 
without requiring any explicit feature matching.
We assume that in time step $t$, $N^{(t)}$ points---with world space positions $\mathbf{x}_{t,i}\in\R^3$---are observed via depth images. % from a Kinect sensor.
Here, we denote single vectors in bold face, e.g., $\mathbf{x}$, while vector fields are denoted with non-bold letters as before.
Since observed points are located at the object boundary where $\phi=0$,
our new {\em sparse surface constraint} (SSC) cost function aims to minimize the SDF values at these locations via
\begin{equation}
	J_{\text{SSC}}(\phi) := \sum_{t=1}^T{ \sum_{i=0}^{N^{(t)}}{ w_{t,i} \frac{1}{2} \left( \phi^{(t)}(\mathbf{x}_{t,i}) \right)^2}}.
\label{eq:CostFunction3}
\end{equation}

To avoid advecting the full SDF, we introduce a new method that intrinsically encapsulates our underlying sparseness assumption by solving for a point-wise inversion of the body motion.
This method builds upon the assumption that the simulated displacements $u^{(t)}$ do not destroy the signed distance
property of $\phi$. Then we can approximate $\phi^{(t)}$---the deformed SDF---
by evaluating the initial SDF at the images of displaced locations, i.e., by going along the inverse displacement field, without the need to reinitialize the SDF:
\begin{equation}
	\phi^{(t)}(\mathbf{x}) \approx \phi^{(0)}\left(\mathbf{x} + u^{(t)}(\mathbf{x})^{-1}\right) ,
\label{eq:PartialObservations-PhiT}
\end{equation}
\sebi{where $u(\mathbf{x})^{-1}$ indicates the inverse function, i.e. $\mathbf{y}=u(\mathbf{x})^{-1} \rightarrow \mathbf{x}=u(\mathbf{y})$.}
However, this requires the inverse displacement field, which cannot simply be computed by back-tracing as in divergence-free fluid flows.
\sebi{Since every point $\mathbf{x}$ of the deformed object can be associated with a matching point $\mathbf{x'}$ of the undeformed object, the problem can be reformulated in the following way:}
Since we know the point \sebi{$\mathbf{x}$} with $\mathbf{x}=\mathbf{x}'+u^{(t)}(\mathbf{x}')$ (and hence $\mathbf{x}'=\mathbf{x}+u^{(t)}(\mathbf{x})^{-1}$),
we can compute the (yet unknown) index $(i,j,k)$ of the hexahedral simulation cell containing $\mathbf{x}'$.
With $\mathcal{N}=\{i+[0,1],j+[0,1],k+[0,1]\}$ denoting the eight corners of the cell $(i,j,k)$,
and $\mathbf{x}'_{l \in \mathcal{N}}$
their reference locations, let
$\mathbf{x}_{l \in \mathcal{N}}$
be the displaced locations of these corners, i.e., $\mathbf{x}_{l}=\mathbf{x}'_{l}+\mathbf{u}^{(t)}_{l}$.
Then, the location of point $\mathbf{x}'$ can be computed by trilinear interpolation of the eight
reference corner locations, with the cell-wise interpolation weights $\alpha, \beta, \gamma$.

By further assuming that the interpolation weights don't change during the advection, i.e.,
$\mathbf{x}'$ is interpolated from
$\mathbf{x}'_{l}$ with the same weights as $\mathbf{x}$ is interpolated from $\mathbf{x}_{l}$
(see Fig.~\ref{fig:PartialObservations-Interpolation}),
the same weights for interpolating positions can be used to interpolate the SDF values.
This allows us to
formulate Alg.~\ref{alg:PartialObservations-GeneralAlgorithm} for computing $\phi^{(t)}(\mathbf{x})$:
\vspace{-0.4cm}
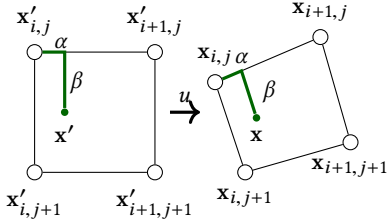
\begin{figure}[htb]%
	\centering%
	\begin{tikzpicture}[scale=0.4]
		\node[draw, circle, inner sep=2pt] (R1) at (0,0) [label=below:{$\mathbf{x}'_{i,j+1}$}] {};
		\node[draw, circle, inner sep=2pt] (R2) at (4,0) [label=below:{$\mathbf{x}'_{i+1,j+1}$}] {};
		\node[draw, circle, inner sep=2pt] (R3) at (0,4) [label={$\mathbf{x}'_{i,j}$}] {};
		\node[draw, circle, inner sep=2pt] (R4) at (4,4) [label={$\mathbf{x}'_{i+1,j}$}] {};
		\draw (R1) -- (R2) -- (R4) -- (R3) -- (R1);
		\node[fill, circle, green!40!black, inner sep=1pt] (RX) at (1,2) [label=below:{$\mathbf{x}'$}] {};
		\draw[very thick, green!40!black] (R3) -- (1,4) -- (RX);
		\node at (0.9,4.3) {$\alpha$};
		\node at (1.4,3) {$\beta$};
		
		\draw[very thick,->] (4.5,2) -- (5,2) node[above] {$u$} -- (5.5,2);
		
		\node[draw, circle, inner sep=2pt] (D1) at (7,0) [label=below:{$\mathbf{x}_{i,j+1}$}] {};
		\node[draw, circle, inner sep=2pt] (D2) at (10.5,1) [label=below:{$\mathbf{x}_{i+1,j+1}$}] {};
		\node[draw, circle, inner sep=2pt] (D3) at (6,3) [label={$\mathbf{x}_{i,j}$}] {};
		\node[draw, circle, inner sep=2pt] (D4) at (9.5,4.5) [label={$\mathbf{x}_{i+1,j}$}] {};
		\draw (D1) -- (D2) -- (D4) -- (D3) -- (D1);
		\node[fill, circle, green!40!black, inner sep=1pt] (DX) at (7.375,1.8125) [label=below:{$\mathbf{x}$}] {};
		\draw[very thick, green!40!black] (D3) -- (6.875,3.375) -- (DX);
		\node at (6.9,3.8) {$\alpha$};
		\node at (7.8,2.7) {$\beta$};
	\end{tikzpicture}
\caption{Interpolation for reference and deformed configurations.}%
\Description{An interpolated point from the reference configuration is displaced in the deformed configurations by interpolating the new deformed corner positions with the same interpolation weights.}%
\label{fig:PartialObservations-Interpolation}%
\end{figure}

\begin{algorithm}[H]
	\caption{Compute $\phi^{(t)}(\mathbf{x})$ based on $\phi^{(0)}$ and $u^{(t)}$}
	\begin{algorithmic}[1]
		\Statex \textbf{Input:} The observed point $\mathbf{x}$
		\For{\textbf{each} cell $i,j,k$}
			\State Compute $\alpha,\beta,\gamma$ with Newton solve of \\
				\hspace{2cm} $\mathbf{x} = \text{interpolate}(\mathbf{x}_{l \in \mathcal{N}},\alpha,\beta,\gamma)$
			\If{$(\alpha,\beta,\gamma) \in [0,1]^3$}
			    \State \textbf{return} $\phi^{(t)}(\mathbf{x}) = \text{interpolate}(\bm{\phi}^{(0)}_{l \in \mathcal{N}},\alpha,\beta,\gamma)$
		    \EndIf
		\EndFor
	\end{algorithmic}
	\label{alg:PartialObservations-GeneralAlgorithm}
\end{algorithm}

The key step here is solving for the unknown trilinear interpolation weights.
This requires finding a solution within the cell space $[0,1]^3$ of a non-linear system of equation in three variables. For this, we employ \sebi{a} Newton iteration that typically converges within a few iterations.
Recall that the standard trilinear interpolation is given as
\begin{align}
       f(\alpha,\beta,\gamma) &= (1-\alpha)(1-\beta)(1-\gamma)\mathbf{x}_1 + \alpha(1-\beta)(1-\gamma)\mathbf{x}_2  \nonumber \\
                                                      & \ \ \ + \cdots + \alpha \beta \gamma \mathbf{x}_8 \nonumber \\
                              &= \mathbf{z}_1 + \alpha \mathbf{z}_2 + \beta \mathbf{z}_3 + \gamma \mathbf{z}_4 + \alpha \beta \mathbf{z}_5 + \alpha \gamma \mathbf{z}_6 + \beta \gamma \mathbf{z}_7 \nonumber \\
                                                     & \ \ \ + \alpha \beta \gamma \mathbf{z}_8 .
\label{eq:trilinearInterpolation}
\end{align}
Then, a Newton iteration is computed as $\alpha\beta\gamma^{t+1}=\alpha\beta\gamma-J^{-1} f(\alpha\beta\gamma^t)$,
with
\begin{equation}
J =
\begin{pmatrix}
		 | & \\
		\mathbf{z}_2 + \beta \mathbf{z}_5 + \gamma \mathbf{z}_6 + \beta\gamma \mathbf{z}_8 & \cdots \\
		 | &
	\end{pmatrix}.
\end{equation}
being the Jacobian.

In case of a self-intersection, it can happen that a single observed point $\mathbf{x}$ is located in two different cells. 
%This is e.g. the case if self-intersections occur.
Upon detecting this case we match the cell with the lowest interpolated SDF-value $\phi^{(t)}(\mathbf{x})$.

\subsection{Extension of Displacements}\label{sec:Cost:Diffusion}

To ensure that $J_{SSC}$ can be evaluated
for all cells that possibly contain an observed point,
the displacements, which are provided by the FE solver only at locations covered by the object,
need to be extended into the ambient space around the object. 
%We detail this extension step below.
For that purpose, we extend the displacements around the rest pose on the Eulerian grid via a 
%
% keep diffusion here, refer to "extension" otherwise. I think that's more intuitive
solving a Poisson problem for a diffusion process.
% Poisson-based diffusion process.} 
%
%For performance reasons, the extrapolation is only performed on a narrow band around the surface, which is a common operation and seen e.g. also in fluid simulation.
As is commonly done, e.g., for level-set methods \cite{osher2006level}, the extension is only performed in a narrow band around the surface.
The width of the narrow band $\phi_{\text{max}}$ naturally defines an upper bound per point for our cost function. The width specifies the maximum allowed distance of a matched point to the surface. Thus, points that are further away from the surface can be ignored and induce a constant cost of $\frac{1}{2}\phi_{\text{max}}^2$.
%
% ---NEW---
%
All cells that receive displacements are implicitly matched with observations via the SSC. Once observations and displacements
are brought together, the adjoint method ensures that the information travels back to the relevant nodes on the FE mesh. 
This process is visualized in Fig.~\ref{fig:SSCOverview}c) to e). Note that this two-step process -- extension first, 
then matching via deformation -- differs from standard procedure in Eulerian solvers \cite{bridson2015fluid} and in 
previous work \cite{wang2015deformation}. 
There, the object itself is typically deformed, and correspondences established with the deformed state. We found
that the latter variant is not suitable for gradient-based optimizations, the primary reason being the inherently divergent displacement
field of the deformable object, which precludes the use of commonly used methods for advection. This variant corresponds
to the aforementioned $J_{\text{SDF}}$ formulation, and an example of the sub-par gradients it yields can be seen 
in Fig.~\ref{fig:CostFunctionComparison:sdf}).
Hence, our reconstructions will focus on  $J_{\text{SSC}}$ in the following.
% end ---NEW---

While for convex objects $\phi_{\text{max}}$ can be chosen arbitrarily, for concave objects like the dragon in \autoref{fig:result:dragonSm} the narrow band can self-penetrate leading to points that cannot be matched uniquely to the surface via the adjoint method. Therefore, in all our experiments we heuristically limit $\phi_{\text{max}}$ to approximately half the 
size of the smallest cavity. A more detailed quantitative evaluation of how this parameter influences the accuracy of the 
solutions can be found in Appendix~\ref{app:Stability:maxSdf}.
%
%In practice, we found that the size of the narrow band can influence the obtained solution.
%Unless extremely small values of less than a voxel for $\phi_{\text{max}}$ are chosen, this limitation does not degrade the quality of the optimization. 
%Especially for later time steps when the current simulation is far away from the observations, 
If a small value for $\phi_{\text{max}}$ was chosen, it can happen that no points can be matched, especially for later time steps when the current simulation is far away from the observations.
This does not pose a problem for our optimization as long as early time steps tie the 
simulation to the observations.
As the optimizer converges towards a tighter match between observations and cells, more and more points from later timesteps are included in the cost function and improve the results.
%

%The width of narrow band is chosen to be at least $\phi_{\text{max}}$ voxels so that all points that could be matched from outside the object are inside this narrow band.
%}

To summarize, with one pass over the computational grid, we can compute the inverse mappings
(as interpolation weights) for all observed points $\mathbf{x}$, so that $J_{\text{SSC}}$
can be evaluated. All steps in the evaluation of $J_{\text{SSC}}$, as well as the extension step, can be
efficiently differentiated and incorporated into an inverse elasticity solver for optimizing the material parameters.
Fig.~\ref{fig:CostFunctionComparison:ssc} demonstrates
the capability of $J_{\text{SSC}}$ to accurately estimate gradients.

%====================================================================================
%====================================================================================
\section{Inverse Elasticity Solver}\label{sec:Adjoint}

By using a forward solver for Eq.~\eqref{eq:StrongPDEdynamic} in combination with
our proposed cost function $J_{\text{SSC}}$,
we propose an optimization framework for the unknown material parameters using the adjoint method \cite{McNamara.2004}.
Let $\mathbf{u} \in \R^{U}$ be the $U\in\N$ states of the system, i.e., the output variables
such as the computed displacements $u^{(t)}$ and velocities $\dot{u}^{(t)}$ for each timestep.
Let $\mathbf{p} \in \R^{P}$ be the $P$ control parameters of the system, i.e., the estimated material parameters that are used as input variables in the forward pass.
\sebi{
The general optimization problem is then defined as 
\begin{subequations}
    \begin{align}
    \text{minimize } & J(\mathbf{u}, \mathbf{p})  \ , \ J : \R^U \times \R^P \rightarrow \R \label{eq:GeneralCost} \\
    \text{subject to } & \mathbf{E}(\mathbf{u}, \mathbf{p}) = \mathbf{0} \ , \ \mathbf{E} : \R^U \times \R^P \rightarrow \R^{U} \label{eq:GeneralProblem}
    \end{align}
\label{eq:GeneralProblemAndCost}%
\end{subequations}%
with a problem-specific function~$\mathbf{E}(\mathbf{u}, \mathbf{p})$ that relates the control parameters to the state variables and a cost function $J(\mathbf{u}, \mathbf{p})$.
}%
% The general problem is then defined as
% \begin{equation}
% 	\mathbf{E}(\mathbf{u}, \mathbf{p}) = \mathbf{0} \ , \ \mathbf{E} : \R^U \times \R^P \rightarrow \R^{U},
% \label{eq:GeneralProblem}
% \end{equation}
% with a problem-specific function~$\mathbf{E}$ that relates the control parameters to the state variables.
% In our scenario, $\mathbf{E}$ is the forward solver for problem (Eq.~\eqref{eq:StrongPDEdynamic}).
% The goal is to optimize the control parameters with respect our cost function
% \begin{equation}
% 	J_{\text{SSC}}(\mathbf{u}, \mathbf{p})  \ , \ J : \R^U \times \R^P \rightarrow \R .
% \label{eq:GeneralCost}
% \end{equation}
The gradient $\frac{d J}{d\mathit{p}}$, which is needed in the optimization, is computed by first solving
\begin{equation}
 \frac{\partial\mathbf{E}}{\partial\mathbf{u}} y = \frac{\partial J}{\partial\mathbf{u}} ,
	\label{eq:GeneralGradient:a}
\end{equation}
for $y$, and then computing
\begin{equation}
	\frac{d J}{d\mathbf{p}} = -y^T \frac{\partial\mathbf{E}}{\partial\mathbf{p}} + \frac{\partial J}{\partial\mathbf{p}}.
	\label{eq:GeneralGradient:b}
\end{equation}
The advantage of the adjoint method is that just a single linear system (Eq.~\eqref{eq:GeneralGradient:a}) has to be solved initially.
Afterwards, arbitrary control parameters can be added to the final vector-matrix multiplication (Eq.~\eqref{eq:GeneralGradient:b}).
Therefore, the computational cost of a single gradient evaluation is mostly independent of
the number of control parameters.

\begin{figure}[h]%
\includegraphics[width=\columnwidth]{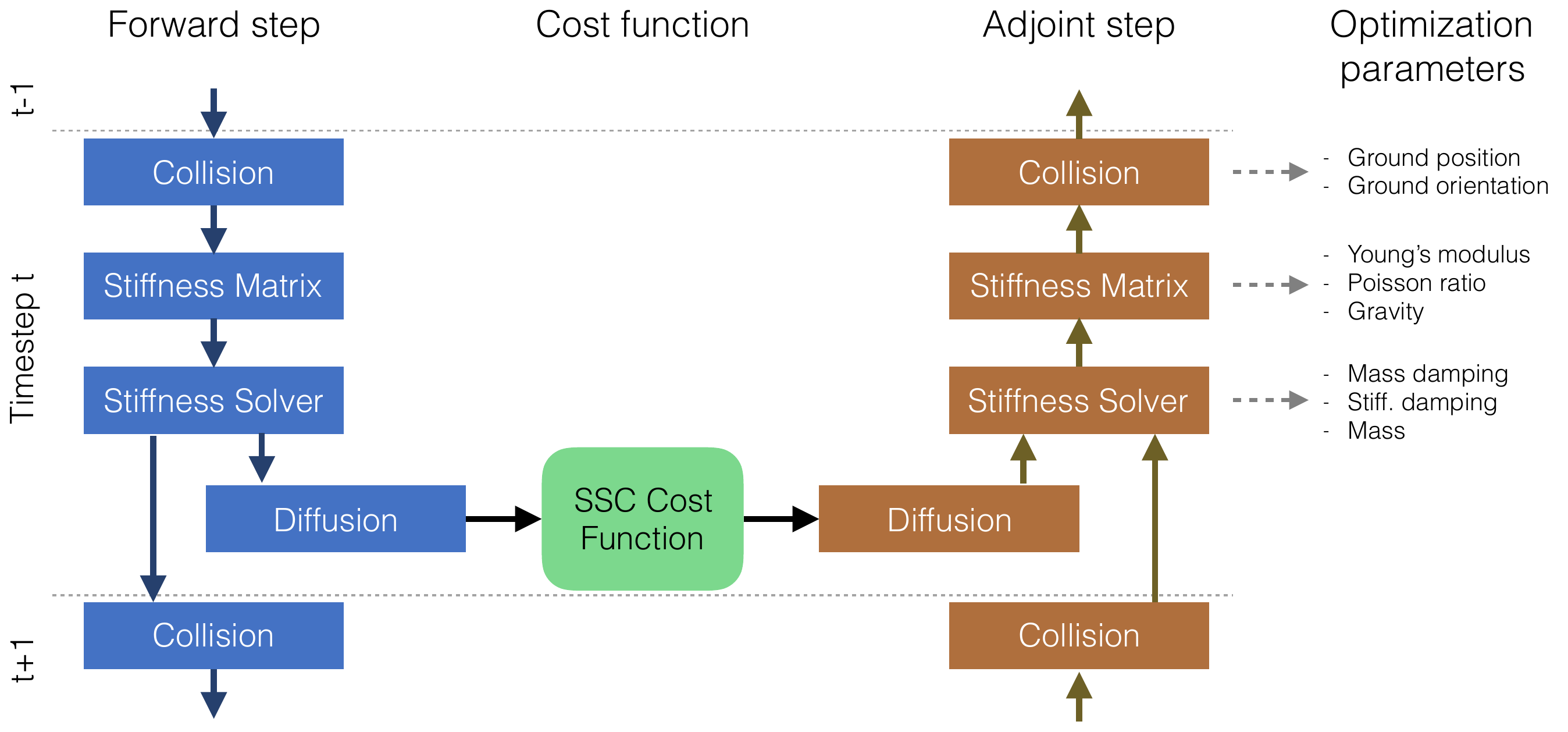}%
	\caption{Outline of the forward and adjoint steps.}%
	\Description{Outline of the forward and adjoint steps: In each time step, the forward step computes the collision forces, the stiffness matrix, solves for the displacements and extends the displacements into the grid. The cost function takes these extended displacements and computes the cost value and gradients.  In the adjoint simulation, the steps are reversed. First, the adjoint of the extension, then adjoint solver (producing gradients of the damping parameters and mass), the stiffness matrix assembly (gradients of Young's modulus, Poisson ratio and gravity) and finally the adjoint of the collision (gradients of the ground position and orientation).}%
	\label{fig:Adjoint}%
\end{figure}

\sebi{
For a problem that can be formulated as a sequence of function calls
\begin{subequations}
\begin{align}
    \mathbf{x}_1 &\leftarrow f_1(\mathbf{x_0}) \\
    \mathbf{x}_2 &\leftarrow f_2(\mathbf{x_1}) \\
    & \hdots \nonumber
\end{align}
\label{eq:FunctionGradient:a}
\end{subequations}
the general problem matrix $\mathbf{E}$ becomes triagonal and the adjoint step, i.e., Eq.~\eqref{eq:GeneralGradient:a} and~\eqref{eq:GeneralGradient:b}, simplifies to
\sebi{\cite{McNamara.2004}}
\begin{subequations}
\begin{align}
    & \hdots \nonumber \\
    \mathbf{\hat{x}}_1 &\leftarrow \mathbf{\hat{x}}_1 + \left(\frac{\partial f_2(\mathbf{x_1})}{\partial \mathbf{x}_1}\right)^T \mathbf{\hat{x}}_2 \\
    \mathbf{\hat{x}}_0 &\leftarrow \mathbf{\hat{x}}_0 + \left(\frac{\partial f_1(\mathbf{x_0})}{\partial \mathbf{x}_0}\right)^T \mathbf{\hat{x}}_1.
\end{align}
\label{eq:FunctionGradient:b}
\end{subequations}
Here, $x$ denotes the adjoint of a variable, which stores the accumulated gradient $\hat{x}$. Taking the adjoint of an operation $f$ means applying the transposed derivative, as shown in Eq.~\eqref{eq:FunctionGradient:b}.}
An overview of the different steps that are considered in the adjoint method is depicted in Fig.~\ref{fig:Adjoint}.

In each time step, the system matrix $E$ in Eq.~\eqref{eq:GeneralProblem} captures collision handling (\autoref{sec:Collision}), stiffness matrix assembly with corotation (App.~\ref{app:PhysicalModel:FEM}), 
%(Sec.~\ref{sec:DiscreteEquations} and \ref{app:PhysicalModel:Corotation}), 
stiffness solve (\autoref{sec:TimeIntegration})
and displacement extension (\autoref{sec:Adjoint:Cost}).
In the adjoint pass, the order of operations is reversed. Starting from the last frame, the adjoint variables of the displacements and velocities are computed with the derivatives of the operations $\frac{\partial\mathbf{E}}{\partial\mathbf{u}}$. This gives the adjoint state $y$ in Eq.~\eqref{eq:GeneralGradient:a}.
Instead of computing the gradients of the control parameters afterwards as in the general adjoint method Eq.~\eqref{eq:GeneralGradient:b}, we found it to be more efficient to assemble the gradients directly within the respective adjoint operations as indicated in Fig.~\ref{fig:Adjoint} and Eq.~\eqref{eq:FunctionGradient:b}.

%------------------------------------------------------------------------------------
\subsection{Sparse Surface Constraint: Gradient Evaluation}\label{sec:Adjoint:Cost}
Our cost function formulation allows for an efficient gradient calculation within the adjoint framework.
From the forward problem in Alg.~\ref{alg:PartialObservations-GeneralAlgorithm}, the index of the
cell that contains an observed point $\mathbf{x}$ is known.
Then, computing the adjoint of the trilinear interpolation with respect to the interpolation weights is straight forward, %trivial to derive and omitted here.
yet computing the adjoint of the inverse of the interpolation, i.e. solving for the weights, is more challenging. In principle, we could mechanically compute the adjoint of the Newton iteration,
but a much simpler solution can be derived by taking the derivative of the tri-linear interpolation as a whole.
Therefore, let us express the cell-wise interpolation as $E$ from the adjoint method:
\begin{align*}
&	\mathbf{E}(\mathbf{u}=\{\alpha,\beta,\gamma\}, p=\{\mathbf{z}_1, ...,\mathbf{z}_8\}) := \mathbf{z}_1 + \alpha \mathbf{z}_2 + \beta \mathbf{z}_3 + \gamma \mathbf{z}_4 +  \\
& \;\  \ \ \ \ \  \alpha \beta \mathbf{z}_5 + \alpha \gamma \mathbf{z}_6 + \beta \gamma \mathbf{z}_7 + \alpha \beta \gamma \mathbf{z}_8 - \mathbf{x} = \mathbf{0} ,
\end{align*}
Then, the matrix
\begin{align*}
	A = \frac{\partial \mathbf{E}}{\partial \mathbf{u}} =
		\begin{pmatrix}
			 | & \\
			\mathbf{z}_2 + \beta \mathbf{z}_5 + \gamma \mathbf{z}_6 + \beta\gamma \mathbf{z}_8 & \cdots \\
			 | &
		\end{pmatrix} \in\R^{3 \times 3}.
\end{align*}
is exactly the same matrix that is used in the Newton iteration in Alg.~\ref{alg:PartialObservations-GeneralAlgorithm}.
Next, the derivative of the trilinear interpolation with respect to the control points $p=(\mathbf{z}_1,...,\mathbf{z}_8)$ is computed as
\begin{align*}
&      F = \frac{\partial \mathbf{E}}{\partial \mathbf{p}} = -
             \begin{pmatrix}
                    | & | & | & | & | & | & | & | \\
                    1 & \alpha & \beta & \gamma & \alpha\beta & \alpha\gamma & \beta\gamma & \alpha\beta\gamma \\
                    | & | & | & | & | & | & | & |
             \end{pmatrix} \in \R^{3 \times 8} .
\end{align*}
With this matrix, the adjoint variables of $\mathbf{z}_1$ to $\mathbf{z}_8$ are computed as
$\left(\hat{\mathbf{z}_1},...,\hat{\mathbf{z}_8}\right) = (\alpha', \beta', $ $\gamma') F$.
Finally, the adjoint values $\hat{z_1},...,\hat{z_8}$ are added to the adjoint values
of the per-vertex displacements $\hat{\mathbf{u}}_{i,j,k}, ...,$ $\hat{\mathbf{u}}_{i+1,j+1,k+1}$, by taking the adjoint of the mapping from $\mathbf{x}_1,...,\mathbf{x}_8$ to $\mathbf{z}_1,...,\mathbf{z}_8$ (Eq.~\eqref{eq:trilinearInterpolation}).

Our cost function yields gradients on the whole computational grid. The gradients are
transferred back to the active nodes with the adjoint of the Eulerian extensions of the displacements.
%
% SEBASTIAN: Moved into Section 4
%To extend the displacements from the surface cells into the outer cells, we perform a Poisson diffusion on the Eulerian grid
%around the rest pose. The width of the extension could be determined based on the maximum velocities in the simulation, yet
%all of our experiments indicate that an extension about a fixed width of up to 10 cells is sufficient.

If the current state of the simulation shows large differences to the observations,
some points are matched with cells that lie outside the object. The adjoint of the extension
then "pulls back" the gradients from these points onto the object's surface, and connects
them to the Lagrangian simulation.
We refer to the Appendix for a discussion of how to compute the gradients of the damping parameters.
The lengthy description of how the remaining gradients are computed is omitted here. It will be provided together with the source code of our implementation.

%------------------------------------------------------------------------------------
%------------------------------------------------------------------------------------
%------------------------------------------------------------------------------------

\section{FEM Discretization}\label{sec:FEM}

The deformable body is discretized by means of hexahedral elements with trilinear shape functions,
which are aligned on a regular Cartesian grid. 
We further embed the object boundary into the simulation grid, and consider cells that are partly filled with material (see Appendix~\ref{app:PhysicalModel:FEM}).
Physical properties like Young's modulus and Poisson's ratio are specified globally for simplicity, but could be assigned on a per-element basis.
The hexahedral discretization yields simplified expressions of the stiffness terms, i.e., since all elements
have the same shape, the same stiffness matrix can be used for all of them (up to scaling according to the respective element's elastic modulus). Thus, we require only a single element stiffness matrix and can significantly accelerate the setup phase
for the simulation. Since the
discretization always refers to the undeformed model state, no further calculations are required even if the object geometry deforms and hexahedra become of
different shapes. 
\sebi{The whole simulation is executed on the GPU using \textit{cuMat}~\cite{cuMat} as linear algebra library.}
Furthermore, and as described in Sec.~\ref{sec:Cost}, an Eulerian discretization of the space around the object can be obtained from a hexahedral discretization in a straight forward way, and sparse
point sets can be included as observations without requiring explicit feature matching.
%Since the object can move over long distances in space during the simulation, we always compute a
%bounding box of the current shape--- including a safety margin---and discretize only the domain covered by this box.
%\comment Sebi{The bounding box is only relevant for computing the full advected SDF, which is only needed for rendering, not for the simulation itself}

%------------------------------------------------------------------------------------

%------------------------------------------------------------------------------------
\subsection{Time integration}\label{sec:TimeIntegration}
Once the stiffness matrix $K$, the mass matrix $M$ and the force vector $f$ are assembled from the per-element matrices,
we introduce Raleigh damping via the matrix $D$, and
solve the resulting linear system
\begin{equation}
	M\ddot{u}+D\dot{u}+Ku=f
\label{eq:TimeDependentLinearSystem}
\end{equation}
with a Newmark scheme that takes the following form \cite{NewmarksMethodOfDirectIntegration}:
\begin{align}
	& \left( \frac{1}{\theta \Delta t}M + D + \theta \Delta t K \right)u^{(t)} \nonumber \\
	=& \left( \frac{1}{\theta \Delta t}M + D + (1-\theta) \Delta t K \right)u^{(t-1)} + \frac{1}{\theta} M \dot{u}^{(t-1)} + \Delta t f
	\label{eq:Newmark1}
\end{align}
with $\frac{1}{2} \leq \theta < 1$ and
\begin{equation}
	\dot{u}^{(n)} = \frac{1}{\theta \Delta t} (u^{(t)} - u^{(t-1)}) - \frac{1-\theta}{\theta}\dot{u}^{(t-1)} .
\label{eq:Newmark1Time}
\end{equation}
If $f$ is time-dependent, it is given as
\begin{equation}
	f=\theta f^{(t)} + (1-\theta) f^{(t-1)} .
\label{eq:Newmark1b}
\end{equation}
Here, time-splitting of the forces in Eq.~\eqref{eq:Newmark1b} is required to make the collisions numerically stable.

The hyper-parameter $\theta$ in the Newmark time integration scheme was set to 0.6 in all of our experiments. As confirmed by a number of tests, values of $\theta$ between $0.5$ and $0.75$ do not result in any noticeable differences. 
Only for large timesteps and low Rayleigh damping does a large value of $\theta=0.99$ introduce undesirable damping. 
For a comparison of different values of $\theta$ we refer to the supplemental video.

%------------------------------------------------------------------------------------
\subsection{Collisions}\label{sec:Collision}

In order to include collisions in our inverse solver framework,
we employ the penalty method \cite{Bridson.2002,Coros.2012}.
Wherever the object penetrates another object, e.g. a ground plane, a virtual spring is attached to it that generates a repulsive force that is added as a Neumann boundary in the next timestep.
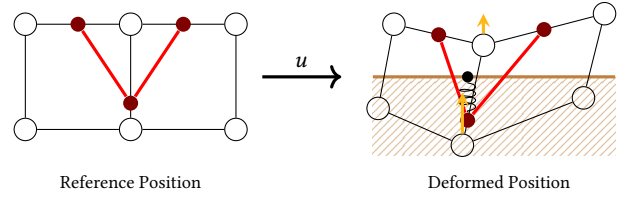
\begin{figure}[htb]%
	\centering%
	\begin{tikzpicture}[scale=0.7]
		% Reference
		\node[draw, circle, inner sep=3pt] (R1) at (0,0) {};
		\node[draw, circle, inner sep=3pt] (R2) at (2,0) {};
		\node[draw, circle, inner sep=3pt] (R3) at (4,0) {};
		\node[draw, circle, inner sep=3pt] (R4) at (0,2) {};
		\node[draw, circle, inner sep=3pt] (R5) at (2,2) {};
		\node[draw, circle, inner sep=3pt] (R6) at (4,2) {};
		\draw (R1) -- (R2);
		\draw (R2) -- (R3);
		\draw (R4) -- (R5);
		\draw (R5) -- (R6);
		\draw (R1) -- (R4);
		\draw (R2) -- (R5);
		\draw (R3) -- (R6);
		\node[fill, circle, inner sep=2pt,red!50!black] (BR1) at (1,2) {};
		\node[fill, circle, inner sep=2pt,red!50!black] (BR2) at (2,0.5) {};
		\node[fill, circle, inner sep=2pt,red!50!black] (BR3) at (3,2) {};
		\draw[very thick, red] (BR1) -- (BR2);
		\draw[very thick, red] (BR2) -- (BR3);
		\node[] at (2, -1) {\footnotesize Reference Position};
		
		% Transformation
		\draw[very thick,->] (4.5,1) -- (5.25,1) node[above] {$u$} -- (6,1);
		
		%Deformed
		\fill[pattern=north east lines, pattern color=brown!50!white] (6.6,-0.5) rectangle (11.2,1);
		\draw[brown, very thick] (6.6,1) -- (11.2,1);
		\node[draw, circle, inner sep=3pt] (D1) at (6.7,0.4) {};
		\node[draw, circle, inner sep=3pt] (D2) at (8.3,-0.3) {};
		\node[draw, circle, inner sep=3pt] (D3) at (10.6,0.6) {};
		\node[draw, circle, inner sep=3pt] (D4) at (7,2) {};
		\node[draw, circle, inner sep=3pt] (D5) at (8.7,1.6) {};
		\node[draw, circle, inner sep=3pt] (D6) at (11,2.2) {};
		\draw (D1) -- (D2);
		\draw (D2) -- (D3);
		\draw (D4) -- (D5);
		\draw (D5) -- (D6);
		\draw (D1) -- (D4);
		\draw (D2) -- (D5);
		\draw (D3) -- (D6);
		\node[fill, circle, inner sep=2pt,red!50!black] (BD1) at (7.85,1.8) {};
		\node[fill, circle, inner sep=2pt,red!50!black] (BD2) at (8.4,0.175) {};
		\node[fill, circle, inner sep=2pt,red!50!black] (BD3) at (9.85,1.9) {};
		\draw[very thick, red] (BD1) -- (BD2);
		\draw[very thick, red] (BD2) -- (BD3);
		\node[circle,fill, inner sep=1.5pt,black] (BDp2) at (8.4, 1) {};
		\draw[decoration={aspect=0.3, segment length=2.5pt, amplitude=3pt,coil},decorate] (BDp2) -- (BD2);
		\draw[very thick, yellow!70!red, ->,>=stealth] (D2) -- (8.3,0.7);
		\draw[very thick, yellow!70!red, ->,>=stealth] (D5) -- (8.7,2.2);
		\node[] at (9, -1) {\footnotesize Deformed Position};
	\end{tikzpicture}
\caption{Collision handling in the hexahedral FEM simulation.}%
\Description{Collision detection and handling for the grid simulation: When the object penetrates the surface, repulsion forces are generated and added to the grid nodes of the cell that contains the colliding surface.}
\label{fig:CollisionGrid}%
\end{figure}

Let $x=\text{dist}(\mathbf{x})$ be the penetration depth of point $\mathbf{x}$.
Then, the force of a spring is described by Hooke's Law: $\mathbf{f} = -k x \mathbf{n}$ with the stiffness factor $k$ and outer normal vector $\mathbf{n}$.
In our case, the spring must not exert an attractive force towards the surface when the objects are not penetrating. Therefore, the force has to be clamped:
\begin{equation}
	\mathbf{f}_c = -k \min(0, x) \mathbf{n}.
\label{eq:Collision1}
\end{equation}

\begin{figure}%
\centering
\begin{subfigure}{0.15\textwidth}
	\includegraphics[trim=300 0 120 0,clip,width=1.\textwidth]{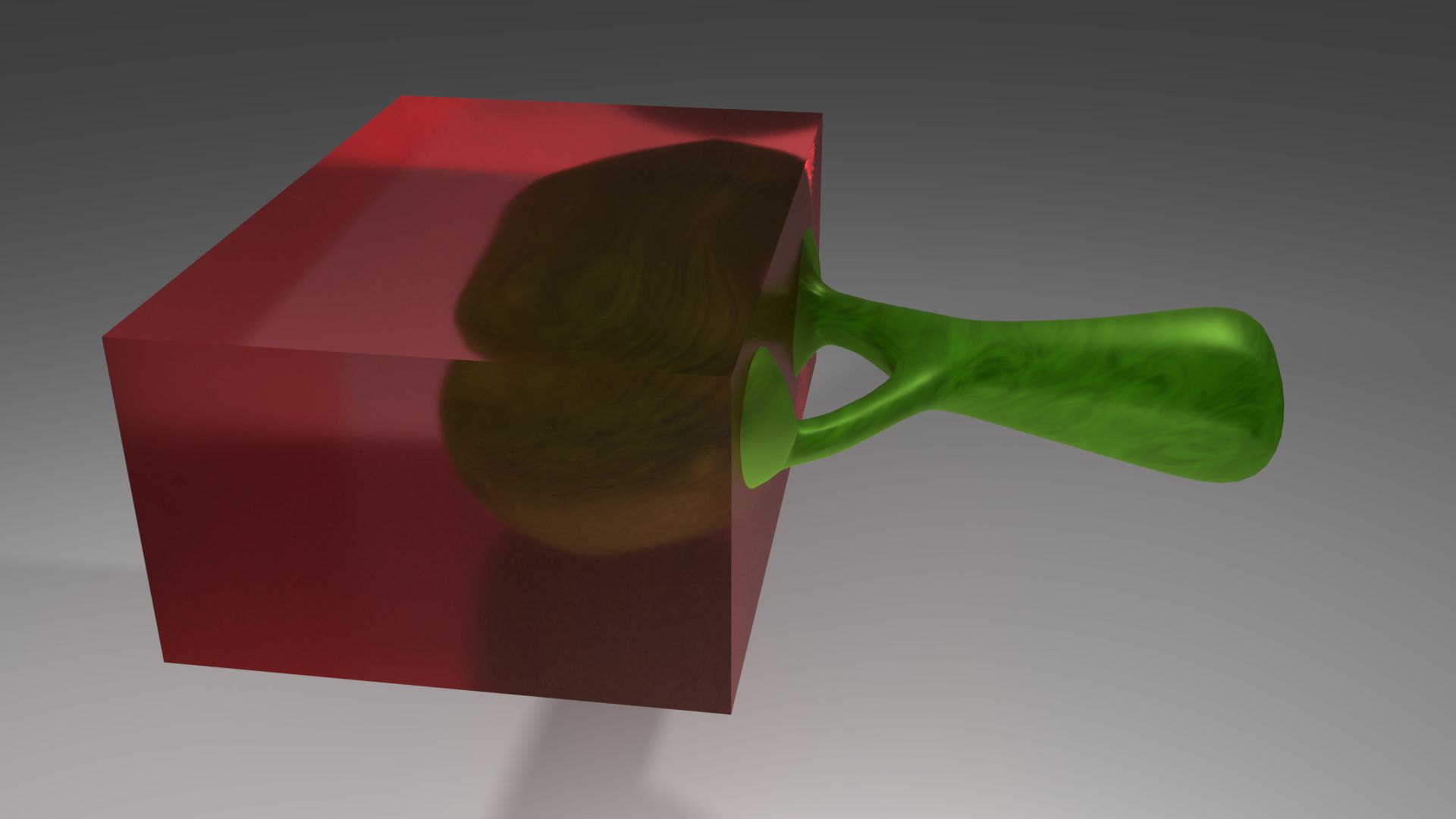}%
	\caption{\footnotesize t=0}	\label{fig:TreeGradients:t0}
\end{subfigure}
~
\begin{subfigure}{0.15\textwidth}
	\includegraphics[trim=300 0 120 0,clip,width=1.\textwidth]{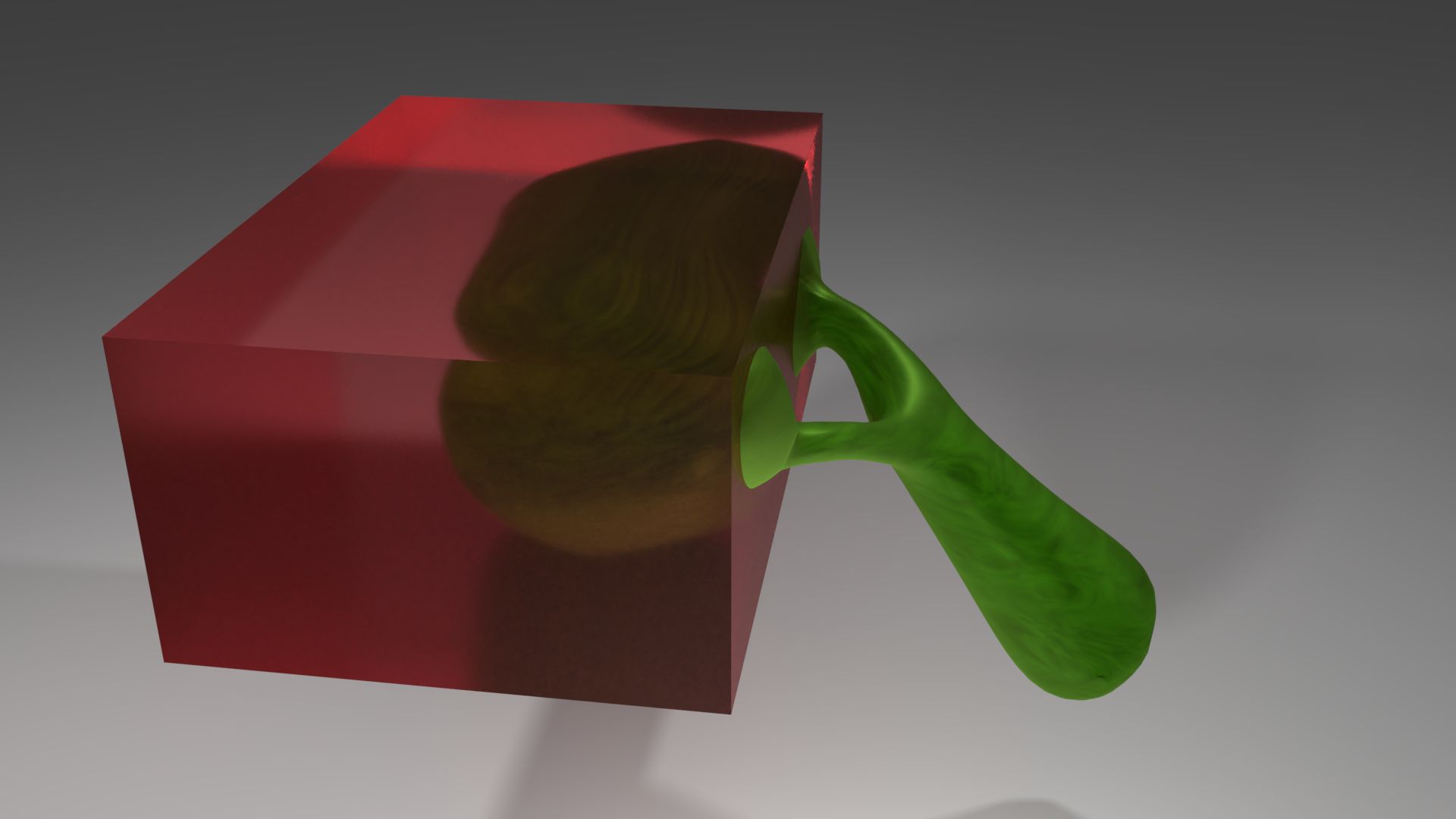}%
	\caption{\footnotesize t=20}	\label{fig:TreeGradients:t1}
\end{subfigure}
~
\begin{subfigure}{0.15\textwidth}
	\includegraphics[trim=300 0 120 0,clip,width=1.\textwidth]{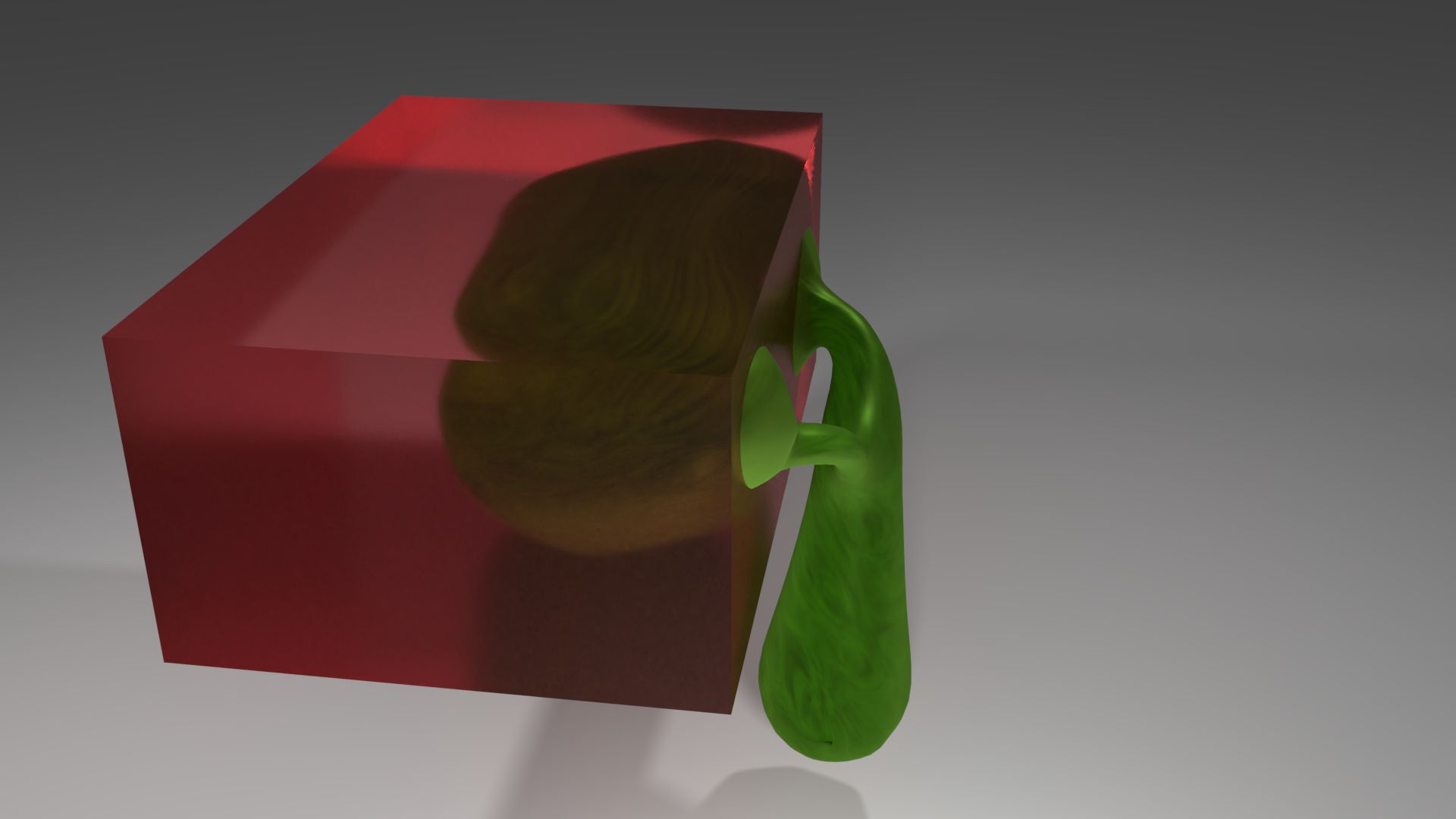}%
	\caption{\footnotesize t=40}	\label{fig:TreeGradients:t2}
\end{subfigure}
\\
\begin{subfigure}{0.22\textwidth}
	\includegraphics[width=\textwidth]{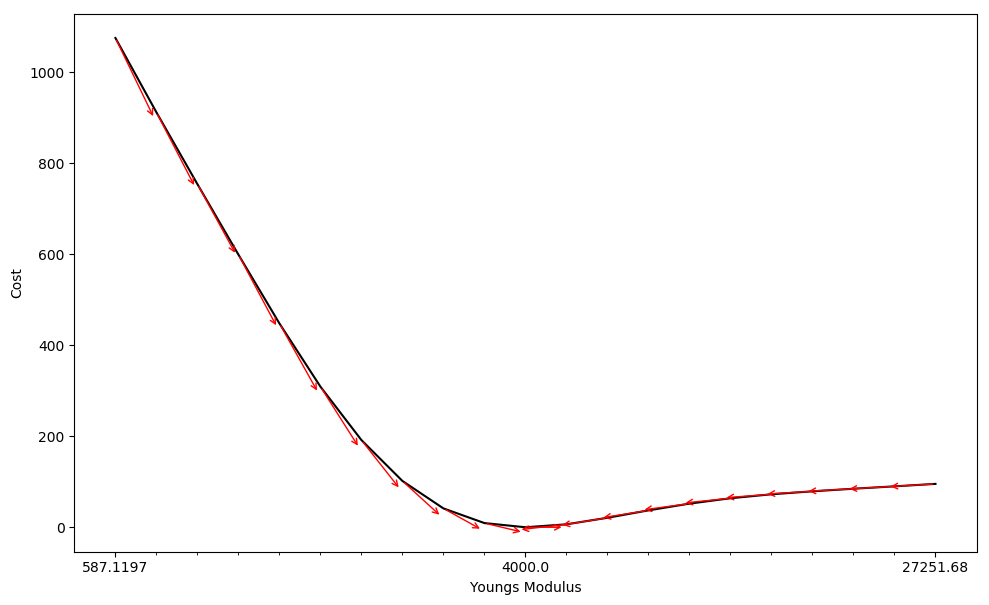}%
	\caption{\footnotesize Reference: Direct Displacements}
	\label{fig:TreeGradients:a}
\end{subfigure}
\begin{subfigure}{0.22\textwidth}
	\includegraphics[width=\textwidth]{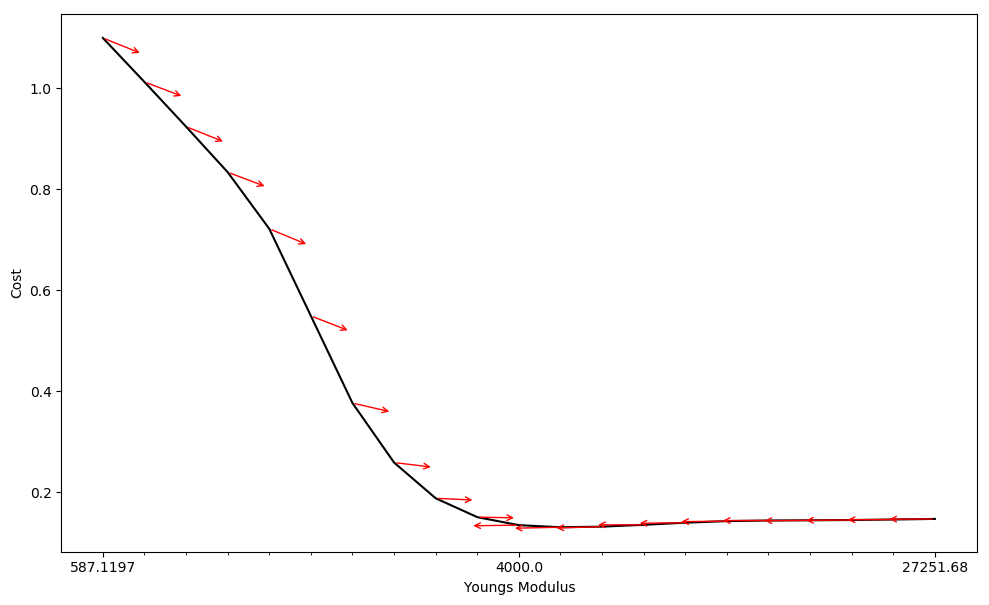}%
	\caption{\footnotesize Sparse Surface Constraints}
	\label{fig:TreeGradients:b}
\end{subfigure}
% \begin{subfigure}{0.16\textwidth}
% 	\includegraphics[width=\textwidth,trim=0 0 0 25,clip]{images/Tree_DirectDisplacements.png}%
% 	\caption{\footnotesize Reference}
% 	\label{fig:TreeGradients:a}
% \end{subfigure}
% ~\hspace{-0.2cm}
% \begin{subfigure}{0.16\textwidth}
% 	\includegraphics[width=\textwidth,trim=0 0 0 25,clip]{images/Tree_PointsAllCams.png}%
% 	\caption{\footnotesize SSC using 6 views}
% 	\label{fig:TreeGradients:b}
% \end{subfigure}
% ~\hspace{-0.2cm}
% \begin{subfigure}{0.16\textwidth}
% 	\includegraphics[width=\textwidth,trim=0 0 0 25,clip]{images/Tree_PointsOneCamAllSteps.png}%
% 	\caption{\footnotesize SSC using one view}
% 	\label{fig:TreeGradients:c}
% \end{subfigure}
% \begin{subfigure}{0.16\textwidth}
% 	\includegraphics[width=\textwidth,trim=0 0 0 25,clip]{images/Tree_PointsOneCamPartStepsNoNoise.png}%
% 	\caption{\footnotesize As (c), with every 10th timestep}
% 	\label{fig:TreeGradients:d}
% \end{subfigure}
% ~
% \begin{subfigure}{0.16\textwidth}
% 	\includegraphics[width=\textwidth,trim=0 0 0 25,clip]{images/Tree_PointsOneCamPartStepsNoise0_1.png}%
% 	\caption{\footnotesize As (d), with Gaussian noise}
% 	\label{fig:TreeGradients:e}
% \end{subfigure}
\caption{Three selected frames (a-c) of a bending tree simulation. Cost functions and gradients for optimizing for Young's modulus (ground truth of 4000) using Direct Displacements as reference (d) and cost function $J_{\text{SSC}}$ (e). 
Sparse Surface Constraints were evaluated every 10th timestep, using one single camera with Gaussian noise variance of three voxels. Gradients robustly point into the direction of the ground truth.}%
\Description{Plots showing the cost function and gradients for different simulation settings: more or less cameras, more noise, less timesteps. With increasing difficulty (less cameras, less timesteps, more noise), the cost function becomes more noisy, but the gradients still point in the correct direction.}
\label{fig:TreeGradients}%
\end{figure}

To obtain a stable simulation, we propose the following two improvements over previous approaches using repulsive forces as collision response.
First, the hard minimum in Eq.~\eqref{eq:Collision1} is replaced by a soft minimum \cite{Cook.2010,Machler.2012}
\begin{align}
	            & \mathbf{f}_c = -k \, \text{softmin}(0, x) \mathbf{n} \nonumber \\
	\text{with  } & \text{softmin}_{\alpha}(a,b) := -\ln\left(e^{-a\alpha}+e^{-b\alpha}\right) / \alpha .
\label{eq:Collision3}
\end{align}
Note that this step makes the minimum differentiable, which is necessary to consider the collision response in the adjoint method.
Second, the collision forces have to be included implicitly in the Newmark integrator, i.e., Eq.~\eqref{eq:Newmark1b}.
However, since the collision force at the next timestep $\mathbf{f}_c^{(n)}$ is not known,
it is approximated using the time derivative of Eq.~\eqref{eq:Collision3} as
\begin{equation}
	\mathbf{f}_c^{(n)} \approx \mathbf{f}_c^{(n-1)} + \Delta t \frac{\partial}{\partial t}\mathbf{f}_c^{(n-1)} .
\label{eq:Collision4}
\end{equation}

The accompanying video shows the influence of this collision formulation
for a dynamic simulation.

%====================================================================================
%====================================================================================
\section{Results and Evaluation}

In the following, we analyze the accuracy, robustness behavior and performance of our approach. All of our experiments were performed on a desktop system equipped with
an Intel Xeon W-2123 CPU, 64 GB RAM and a Nvidia RTX 2070 GPU.
We analyze both synthetic datasets, to be able to compare to ground truth material parameters, as well as live captures of a fallen teddy bear and a pillow. If not otherwise mentioned, synthetic datasets are forward simulated using the described finite-element scheme, and depth images are rendered and provided as sparse constraints. For the live captures we demonstrate how well a forward simulation with the estimated parameters recovers the recorded body dynamics.
Setup parameters for all experiments are given in Table~\ref{tab:ResultsStatistics}.
We show forward simulations by rendering a triangle mesh that is deformed with the simulated displacements.

\subsection{Gradient estimation}
First, we analyze the capability of the new cost function $J_{\text{SSC}}$
to deal with sparse observations and to provide reliable gradients.
Our key observation is that the proposed formulation is very robust against sparsity of observations and noise, and that only the absolute cost values become increasingly unstable with increasing sparsity and noise level.
As an example we use a bending tree model (see Fig.~\ref{fig:TreeGradients}a-c),
simulated over 70 timesteps, and we synthesize observations using a camera resolution of $50\times50$.
The object is fixed via Dirichlet boundaries, shown as red box,
while the green region freely moves under the influence of gravity.
In the reference forward simulation, the Young's modulus was set to $4000$.
Fig.~\ref{fig:TreeGradients} shows the cost function and resulting gradients (red arrows)
for varying Young's moduli. The gradients reliably pull back the Young's modulus to the
target value of $4000$. $J_{\text{SSC}}$ and gradient estimation performed equally well for other test cases.

Fig.~\ref{fig:TreeGradients:a} shows the behavior when $J_{\text{disp}}$ from Eq.~\eqref{eq:CostFunction1} is evaluated directly on corresponding vertices. Even though these correspondences cannot be determined in general, we use the resulting convex cost function with smooth gradients as reference.
We then simulate a real-world setting: Only one camera is used, the camera provides observations only every 10th timestep as the framerate is often limited, and the observations are affected by noise. The proposed Sparse Surface cost function can still accurately determine the gradients as shown in \autoref{fig:TreeGradients:b}.
%Fig.~\ref{fig:TreeGradients:b} shows the optimization behaviour when our novel cost function $J_{\text{SSC}}$ is used. We simulate the placement of 6 cameras around the object, and use the resulting depth images in the optimization. Now the cost function is no longer convex, as large distances of the current state to the
%constraints can lead to a reduced number of points being matched to the object surface.
%However, our method is still capable of computing gradients that consistently point into the correct direction.
%When only a single camera is used (Fig.~\ref{fig:TreeGradients:c}), the optimization behavior doesn't change significantly,
%although about 500 point observations were provided by the single depth image.

%To make the problem harder, we break up frame-to-frame coherence by using only every 10\textsuperscript{th} timestep to evaluate the cost function and gradients (Fig.~\ref{fig:TreeGradients:d}). Interestingly, while the values become more unstable, reliable gradients can still be computed. A similar behavior is observed when uniform Gaussian noise (with a $\sigma$ of 3 voxels) is added to the depth values (Fig.~\ref{fig:TreeGradients:e}). Only one gradient points into the wrong direction, whereas
%the central tendency---modulo scaling of the cost function values---is maintained.
%We find this especially remarkable, since \ruediger{it indicates that our method can cope well with very sparse and noisy observations} (a supporting visualization is shown in the accompanying video).

\begin{figure}[bt!]
    \centering
    \begin{subfigure}{0.49\linewidth}
    	\begin{overpic}[trim=300 0 300 0,clip,width=\textwidth]{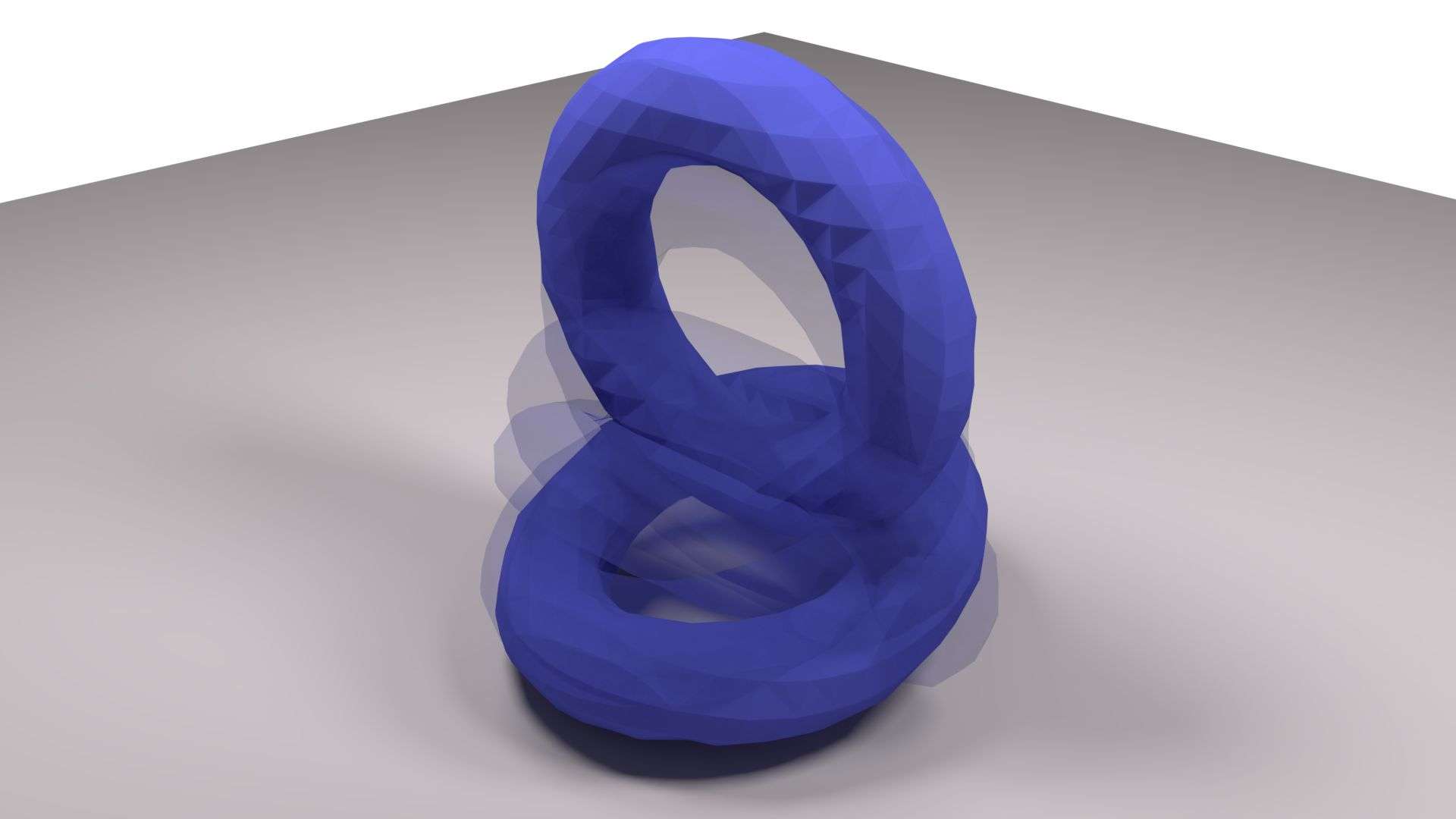}%
             \put(5,5){\textcolor{black}{(a)}}%
        \end{overpic}%
    \end{subfigure}
    \begin{subfigure}{0.49\linewidth}
    	\begin{overpic}[trim=300 0 300 0,clip,width=\textwidth]{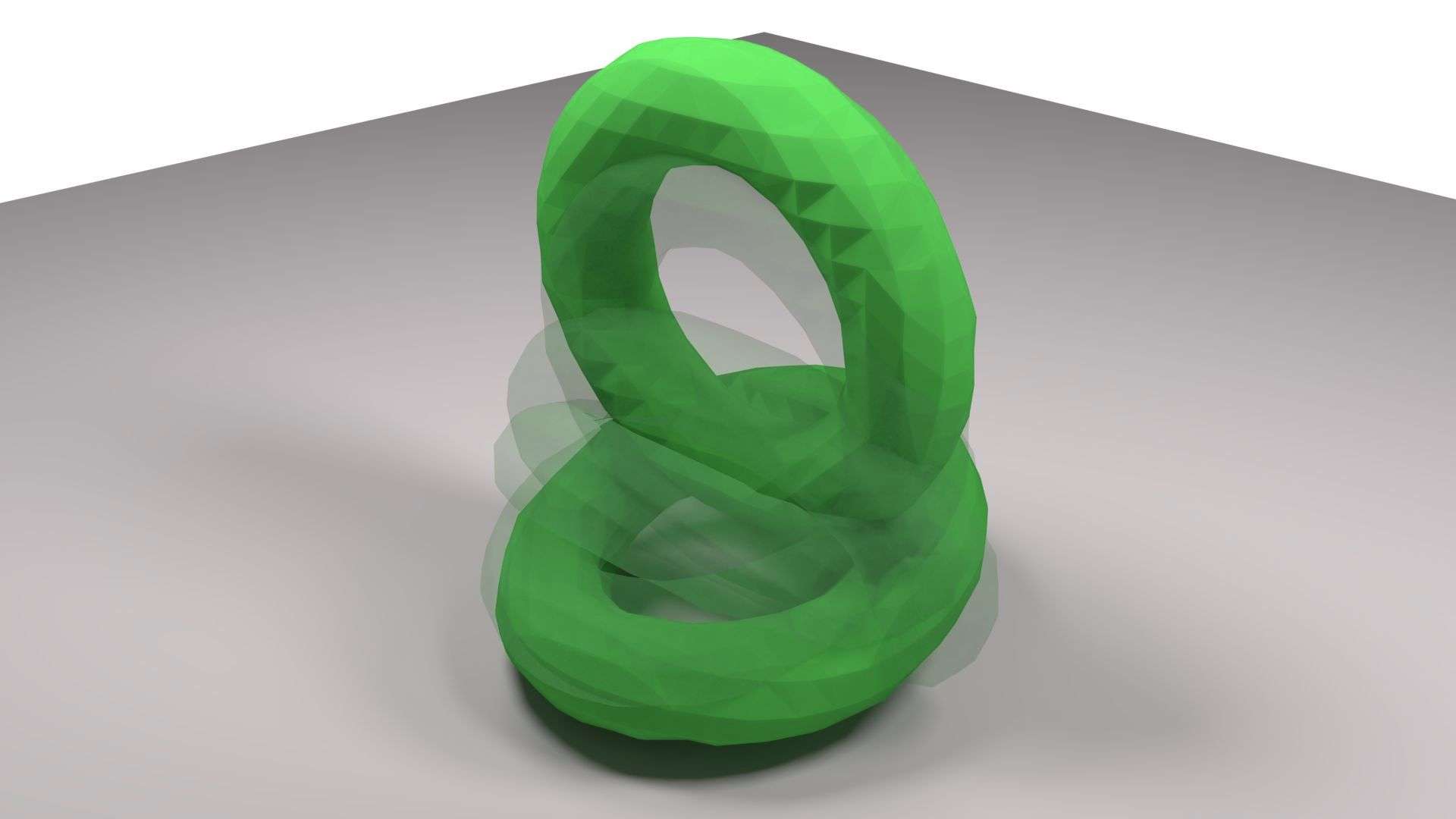}%
             \put(5,5){\textcolor{black}{(b)}}%
        \end{overpic}%
    \end{subfigure}
    \caption{We reconstruct the material properties of a synthetic elastic torus bouncing off the ground floor
	with our method. Simulation using our reconstruction in green (b) very closely 
	matches the ground truth (a). The two solid tori depict the simulation state at time $t=0$ and $t=40$, the transparent frames are in-between frames.}
    \label{fig:result:torusMain}
\end{figure}

Next, we perform tests with the torus data set (see \autoref{fig:result:torusMain}) to analyze the stability of the optimizer over an increasing number of timesteps and noise magnitudes.
A detailed analysis for this as well as the following experiments can be found in Appendix~\ref{app:Stability:TimeAndNoise} and~\ref{app:Stability:FiniteDifferences}.
Our experiments demonstrate that, although numerical errors accumulate over multiple timesteps, the gradients still reliably point towards the minimum so that the ground truth can be reconstructed. Furthermore, we show that increasing noise in the observation does not 
deteriorate the stability of our method. % and the computed gradients.

%\ruediger{
%In Appendix~\ref{app:Stability:TimeAndNoise} we further provide a more detailed stability analysis with the torus data set (see \autoref{fig:result:torus}), to demonstrate the influence of the number of timesteps and noise magnitude. The more timesteps are used, the more robust the optimizer is required to cope with the increasing noise level in the gradients, see \autoref{sec:Optimizer}. We show later that simulating more than 100 timesteps is nevertheless possible in real-world examples. Furthermore, increasing artificially the measurement noise of the simulated cameras does not degrade the quality of the results. Instead, the gradients even get slightly more stable as more noise allows to smooth out possible imprecisions in the matching step.
%}

Finally, we highlight that the adjoint method is crucial for our method by comparing it to
a finite difference scheme.
%provide a comparison between the adjoint method and the finite difference method on computing gradients in the optimization procedure.
%Besides these drawbacks, we also demonstrate considerably weaker stability of the finite difference method due to the noisy absolute values of the cost function. 
Our experiments clearly demonstrate an significantly improved robustness of the adjoint optimization compared to the finite difference method. 
Optimization runs using the finite difference scheme typically converge to local minima and fail to reconstruct the ground truth value 
(details are given in \autoref{fig:Stability:OptimFD} in the Appendix). %\commentSebi{Enough or more text?}
%In none of our experiments did an optimization run with the 
%\commentNils{TODO graphs are missing for he following statement, right? The gradients computed by finite differences caused all X out of X runs to diverge, while all X runs with the adjoint method found the correct solution.}
Intuitively, the adjoint method evaluates the gradient of each observed point locally and back-traces the gradient
via our differentiable solver, independent of the number of matched points. In this way, robust optimizations are achieved.
Besides lower robustness, the following two disadvantages of the finite difference method for computing gradients are worth mentioning: Firstly, it requires one separate forward pass for every parameter instead of one single backward pass for all parameters as with the adjoint method. Secondly, it requires an additional hyper-parameter to specify the step size for finite difference computations.
%The finite difference method, however, only uses the cost function value and therefore highly depends on the number of points matched, which slightly varies for each simulation run.
%\ruediger{
%We also provide a comparison between the adjoint method and the finite difference method on computing gradients in the optimization procedure (see Appendix~\ref{app:Stability:FiniteDifferences}). From a conceptual point of view, the finite difference method has two disadvantages: Firstly, it requires one separate forward pass for every parameter instead of one single backward pass for all parameters as with the adjoint method. Secondly, it requires an additional hyper-parameter to specify the step size for finite difference computations.
%Besides these drawbacks, we also demonstrate considerably weaker stability of the finite difference method due to the noisy absolute values of the cost function. In none of our experiments did an optimization run with the gradients computed by finite differences converge to a satisfactory solution.
%}

%====================================================================================
%====================================================================================
\subsection{Different Optimizers}\label{sec:Optimizer}

%We additionally compare three different optimizers
In combination with our proposed cost function $J_{\text{SSC}}$ the following optimizers were tested:
\begin{itemize}
	\item A simple gradient descent (GD) method with adaptive stepsize control \cite{Fletcher2015BarzilaiBorwein}.
	\item The R-Prop algorithm \cite{riedmiller1993direct} that uses the direction of the gradients ignoring their magnitudes.
	\item The standard L-BFGS algorithm \cite{Byrd1995LBFGS} as implemented in \textit{Eigen} \cite{Yixuan2016LBFGS}.
\end{itemize}
In a number of experiments given in Appendix~\ref{app:Stability:Optimizers} we show superior stability of the R-Prop algorithm. For optimizations with a single parameter, both GD and R-Prop find the global optimum, whereas L-BFGS tends to overshoot the optimum and gets stuck in a local minimum. The experiments were performed on synthetic data sets for which ground truth parameters exist (\autoref{fig:result:dragonSm} and \autoref{fig:Stability:BallMultipleVars:ImagesSm}).
For multi-parameter optimization, R-Prop outperforms all other methods and finds the local minimum with the lowest cost. 
%% SEBI: Pillow-Ramp not introduced yet
%For the Pillow-Ramp scenario, one of the real-world examples described in \autoref{sec:realworld} with seven degrees of freedom, GD even diverges and the optimization stops early because it ends up in a configuration that breaks the simulation.

%\sebi{
%The results of the experiments can be found in Appendix~\ref{app:Stability:Optimizers}.
%Based on those experiments, we chose R-Prop as the optimizer for all following examples.
%}

\subsection{Synthetic tests}\label{sec:results:torus}

\begin{figure}
    \centering
    \begin{overpic}[width=0.9\linewidth]{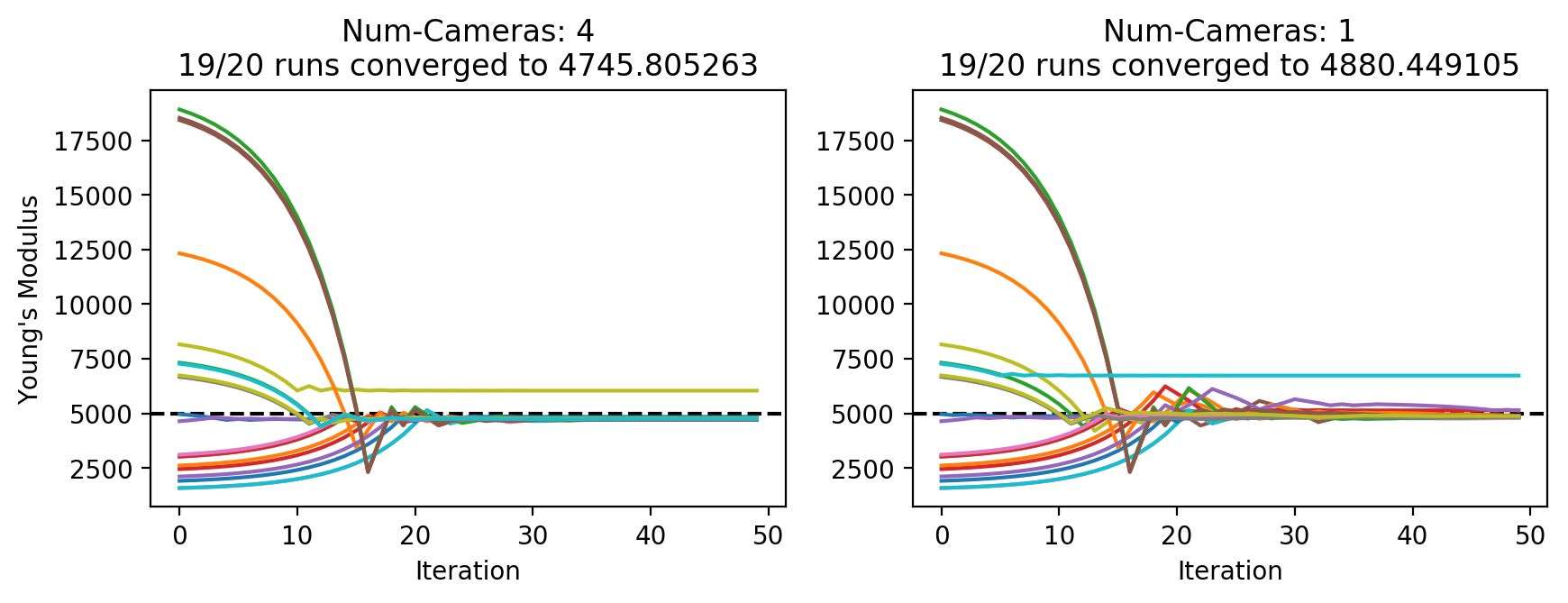}
        \put(-5,30){\textcolor{black}{(a)}}
    \end{overpic}
    \begin{overpic}[width=0.9\linewidth]{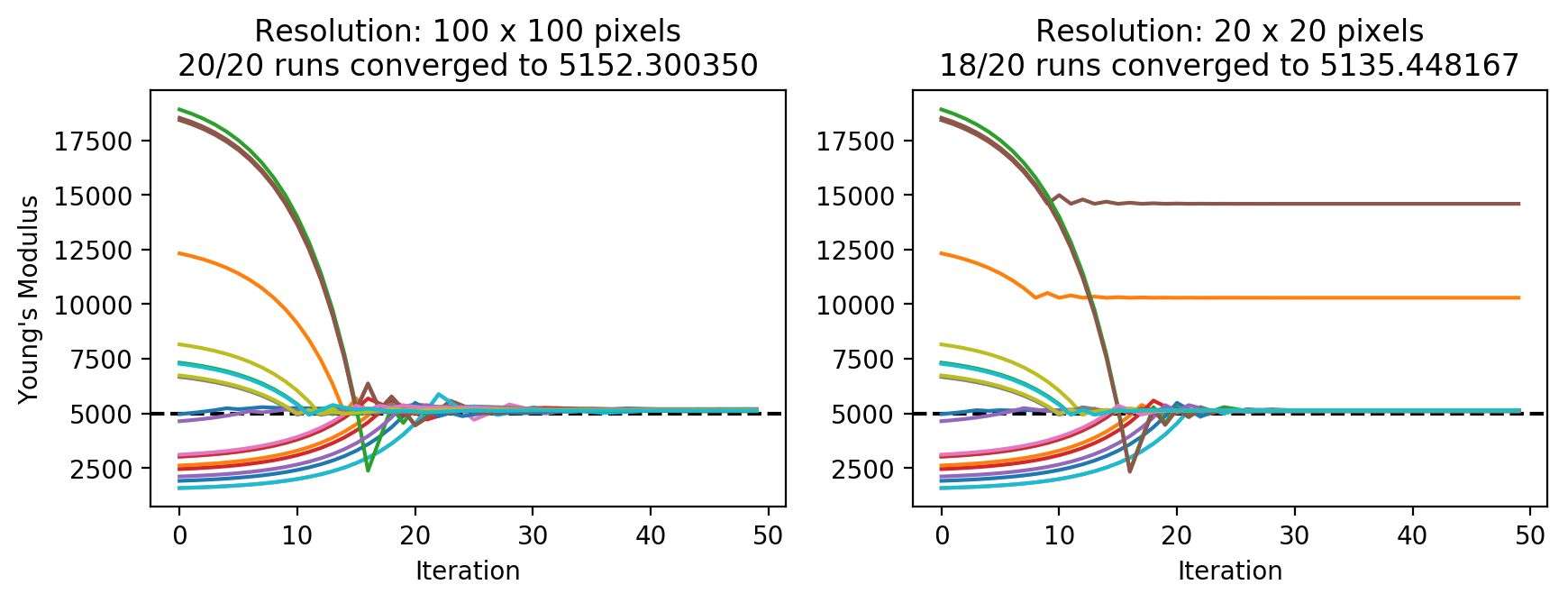}
        \put(-5,30){\textcolor{black}{(b)}}
    \end{overpic}
    \caption{Convergence behavior of our method towards a ground truth Young's modulus of 5000 
    for different numbers of cameras (a) and resolutions of observed depth images (b). In (a), depth resolution was $50x50$, in (b), a single camera was used.
    %Excerpt from the full stability test in \autoref{fig:Stability:NumCameras} in the Appendix, showing that the number of cameras has no influence on the convergence of our algorithm.
    }
    \label{fig:Stability:NumCamResSmall}
\end{figure}

%\nils{In order to demonstrate the accuracy and robustness of our method,
%we evaluated the performance of the optimization framework on three synthetic examples
%for which we can compute errors w.r.t. ground truth: a torus, a bouncing ball, and a Stanford dragon model.}
In order to demonstrate the accuracy and robustness of our method,
we evaluated the performance of the optimization framework on two synthetic examples with known ground truth values: a torus (\autoref{fig:result:torusMain}) and a bouncing ball (\autoref{fig:Stability:BallMultipleVars:ImagesSm}).
Our first set of tests (see Appendix~\ref{app:Stability:FullOptimization})
indicate that our optimization works robustly even with very few cameras
and very low resolutions, i.e. sparse observations. $20\times20$ depth values were sufficient
to reconstruct the ground truth state in 18 out of 20 runs, which increases to 20 out of 20 for a resolution of $100\times100$.
Convergence plots for varying numbers of cameras and resolutions of depth images are given in Fig. \ref{fig:Stability:NumCamResSmall}.
The plots show that our method
yields the ground truth Young's modulus in the vast majority of cases. The relative errors (averaged 
over all converged runs) is $2.39\%$ and $5.08\%$ for 4 and 1 camera, respectively, and $2.71\%$ and $3.05\%$ for $100x100$ and $20x20$ depth images, respectively.
%adding more cameras or a higher resolution of the cameras does not influence the reconstruction quality. However, since it is possible that the optimization get stuck in a local minimum, it is run several times with randomly perturbed initial values.
% cam error: 2.39%, 5.08%
% res error: 2.71%, 3.05% 

\begin{figure}
    \centering
    \begin{subfigure}{0.43\linewidth}
    	\begin{overpic}[trim=350 100 350 50,clip,width=\textwidth]{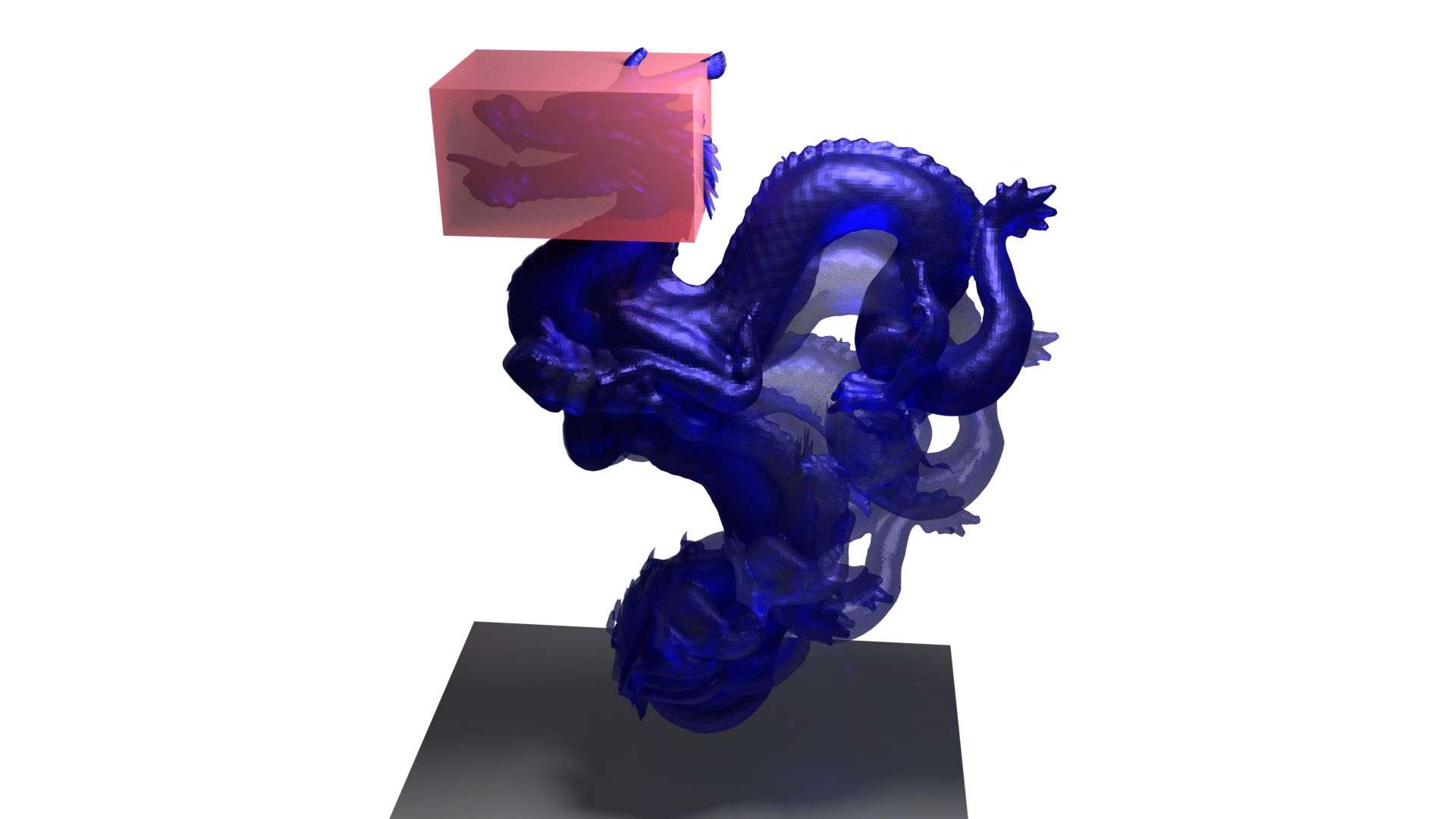}%
             \put(10,5){\textcolor{white}{(a)}}%
        \end{overpic}%
    \end{subfigure}
    \begin{subfigure}{0.43\linewidth}
    	\begin{overpic}[trim=350 100 350 50,clip,width=\textwidth]{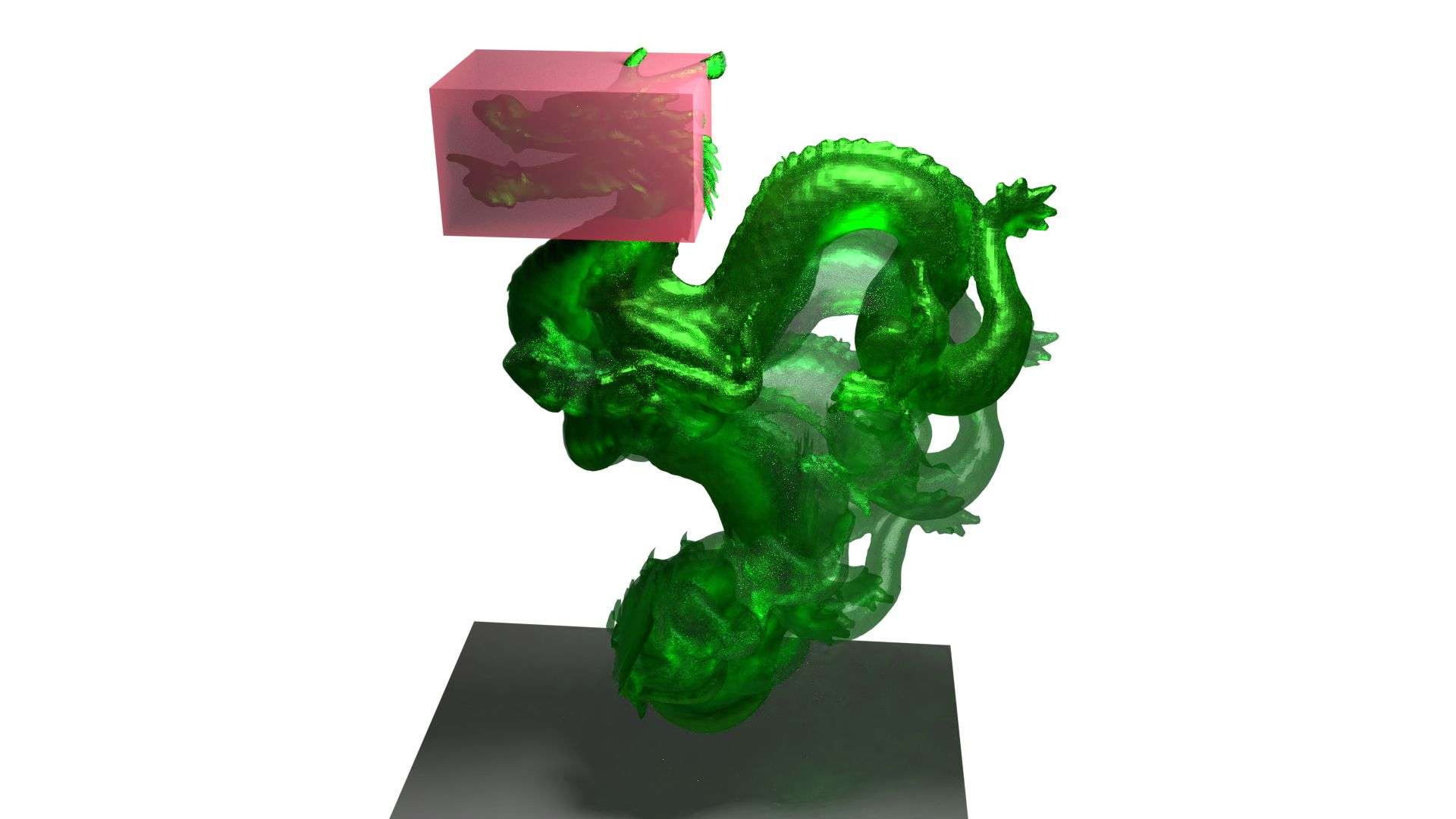}%
             \put(10,5){\textcolor{white}{(b)}}%
        \end{overpic}%
    \end{subfigure}
    \caption{
		Our proposed cost function $J_{SSC}$ is especially suitable for complex geometries. These images show a 
		ground truth simulation (a) and our reconstruction (b), which yields very small 
		reconstruction errors, leading to a visually indistinguishable result.
		}
    \label{fig:result:dragonSm}
\end{figure}

In a second set of tests (see Appendix~\ref{app:Stability:BoundaryConditions}), we demonstrate that the optimization is very robust against changes in the boundary conditions. 
We let a ball with constant material properties bounce off a ground plane. The ball's initial velocity and the height and orientation of the ground plane are randomly sampled.
For all settings, the optimization converges in 18 to 20 cases of 20 runs in total, several examples of which are shown in Fig.~\ref{fig:Stability:BallMultipleVars:ImagesSm}.
In addition, our approach can handle complex geometries. We evaluate a synthetic data set for a deformable Stanford dragon, as shown in Fig.~\ref{fig:result:dragonSm}.

% \begin{figure}
%     \centering
%     \includegraphics[width=\linewidth]{images/Stability/TorusOptimization_ResolutionTest_small.png}
%     \caption{
%     Excerpt from the full stability test in \autoref{fig:Stability:Resolution} in the Appendix, showing that the pixel resolution of the simulated camera does not change the convergence quality of our algorithm.
%     }
%     \label{fig:Stability:ResolutionSmall}
% \end{figure}

A final test (see Appendix~\ref{app:Stability:MultipleParameters}) sheds light on how stable multiple parameters can be jointly optimized by the optimization process. Again, we let a ball bounce off a ground plane, but this time optimize jointly for gravity, the Young's modulus and the stiffness damping. While gravity is estimated exactly, the Young's modulus and stiffness damping show a wide range of values. At the same time, however, all runs converge to a solution with very low cost that visually indistinguishable from the ground truth.
A closer look at the results reveals that the runs that converge to a state with a high Young's modulus reach a low value of stiffness damping, and vice versa. 
%These two parameters therefore balance each other out in this scenario.

\begin{figure}
    \centering
    \begin{subfigure}[c]{0.32\linewidth}
    	\begin{overpic}[trim=600 10 0 0,clip,width=\textwidth]{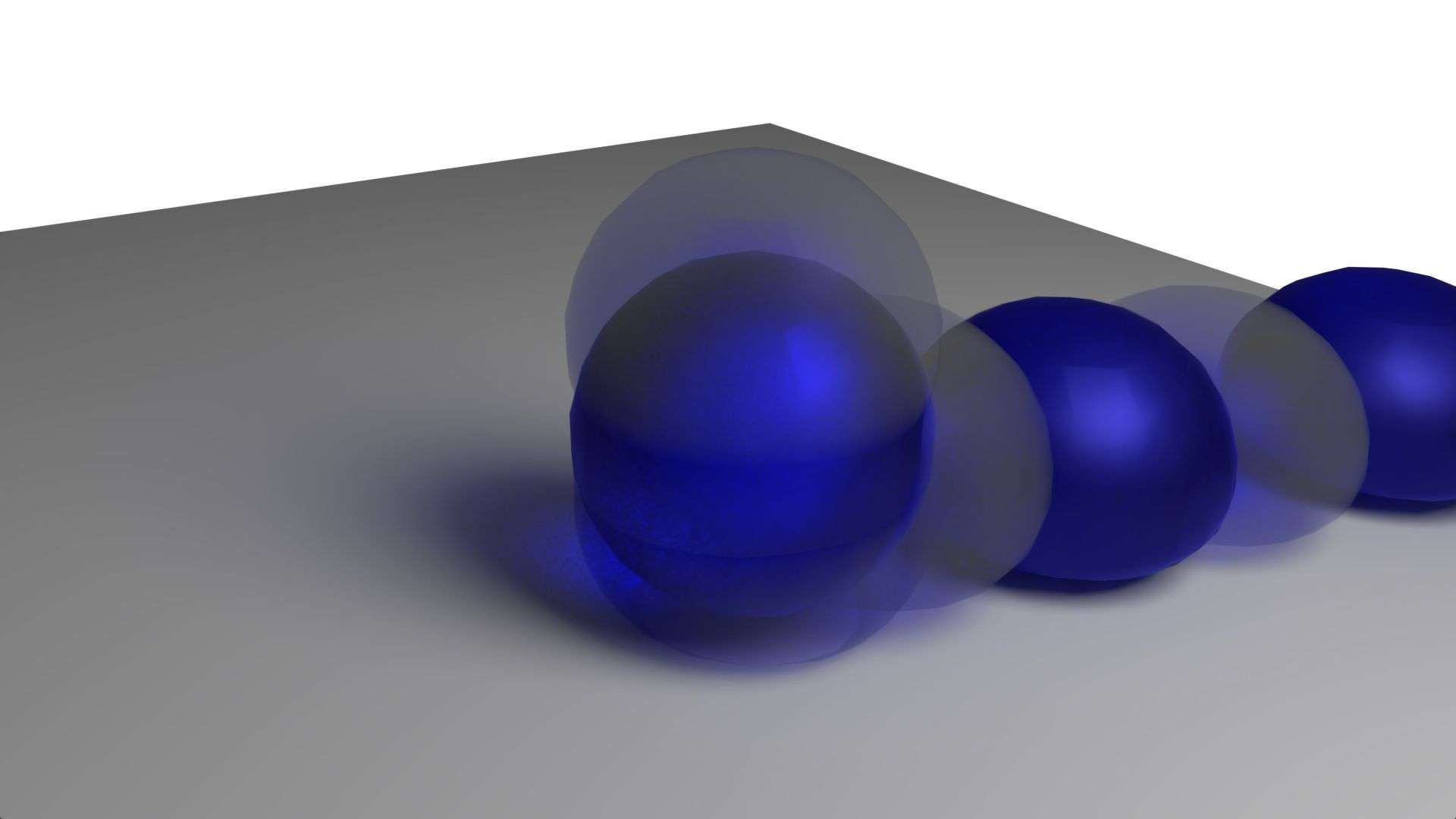}%
             \put(5,5){\textcolor{white}{(a)}}%
        \end{overpic}%
    \end{subfigure}
    \begin{subfigure}[c]{0.32\linewidth}
    	\includegraphics[trim=600 10 0 0,clip,width=\textwidth]{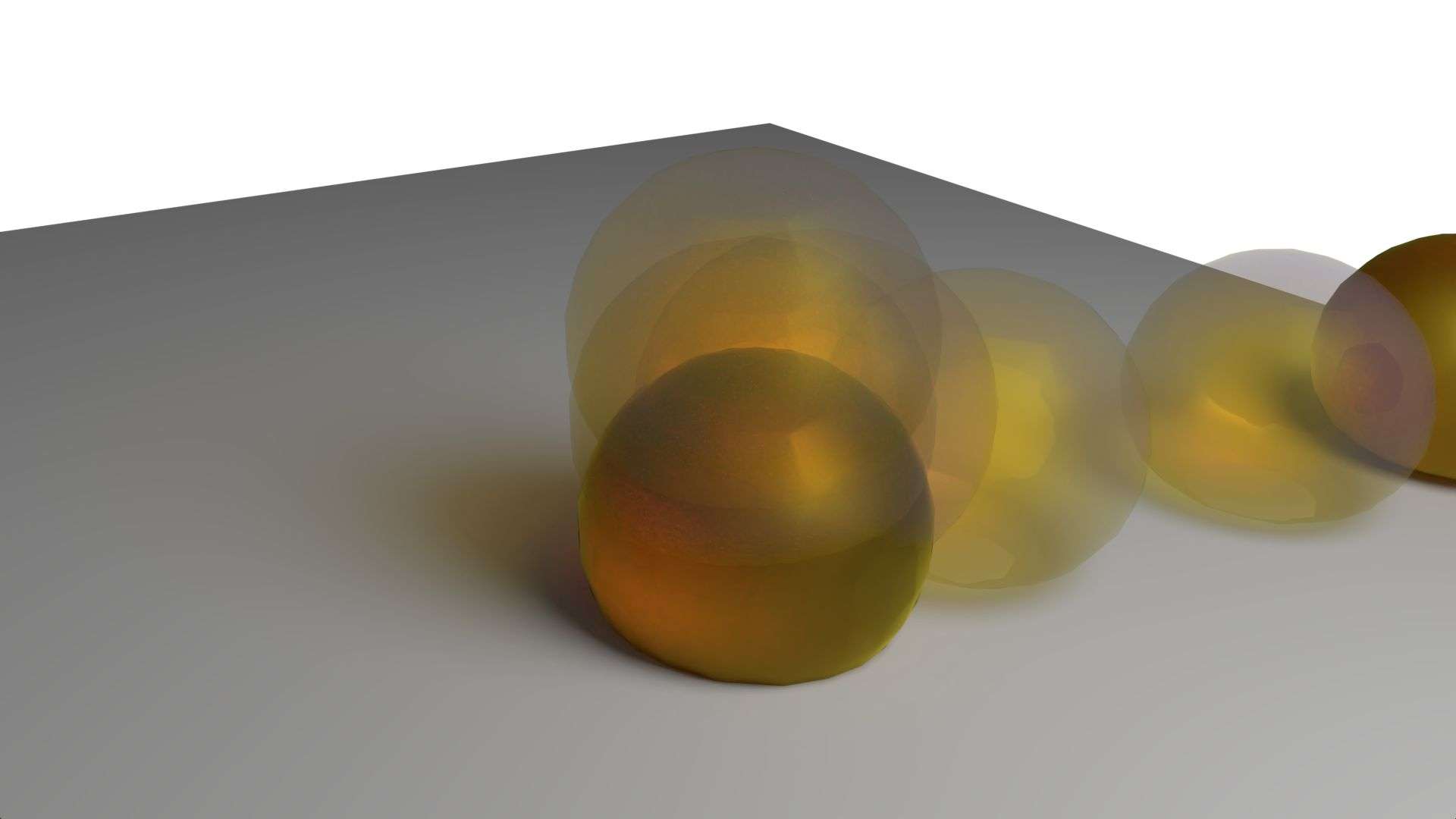}\\%
    	\begin{overpic}[trim=600 10 0 0,clip,width=\textwidth]{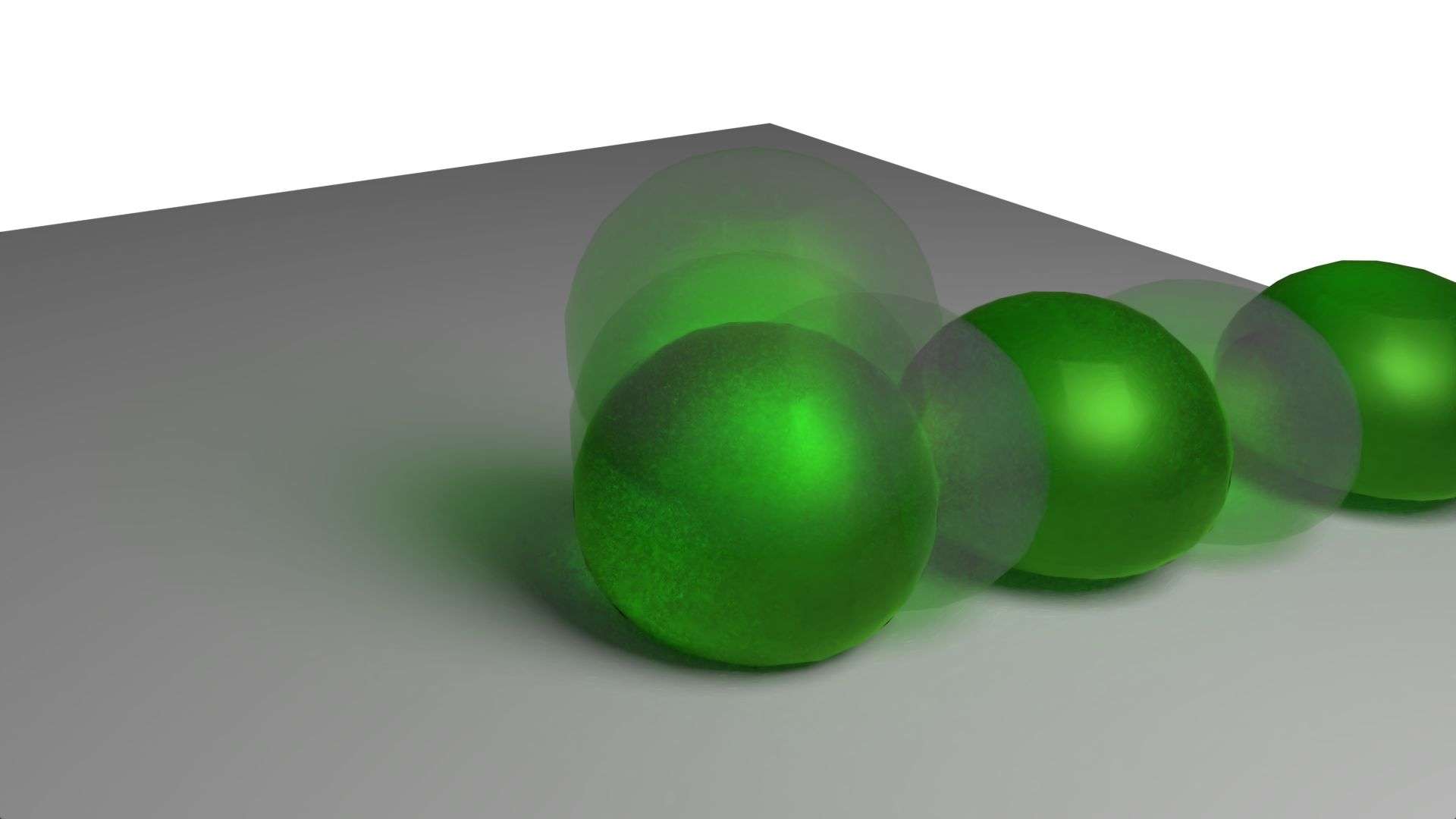}%
             \put(5,89){\textcolor{white}{(b)}} \put(5,5){\textcolor{white}{(d)} }%
        \end{overpic}%
    \end{subfigure}
    \begin{subfigure}[c]{0.32\linewidth}
    	\includegraphics[trim=600 10 0 0,clip,width=\textwidth]{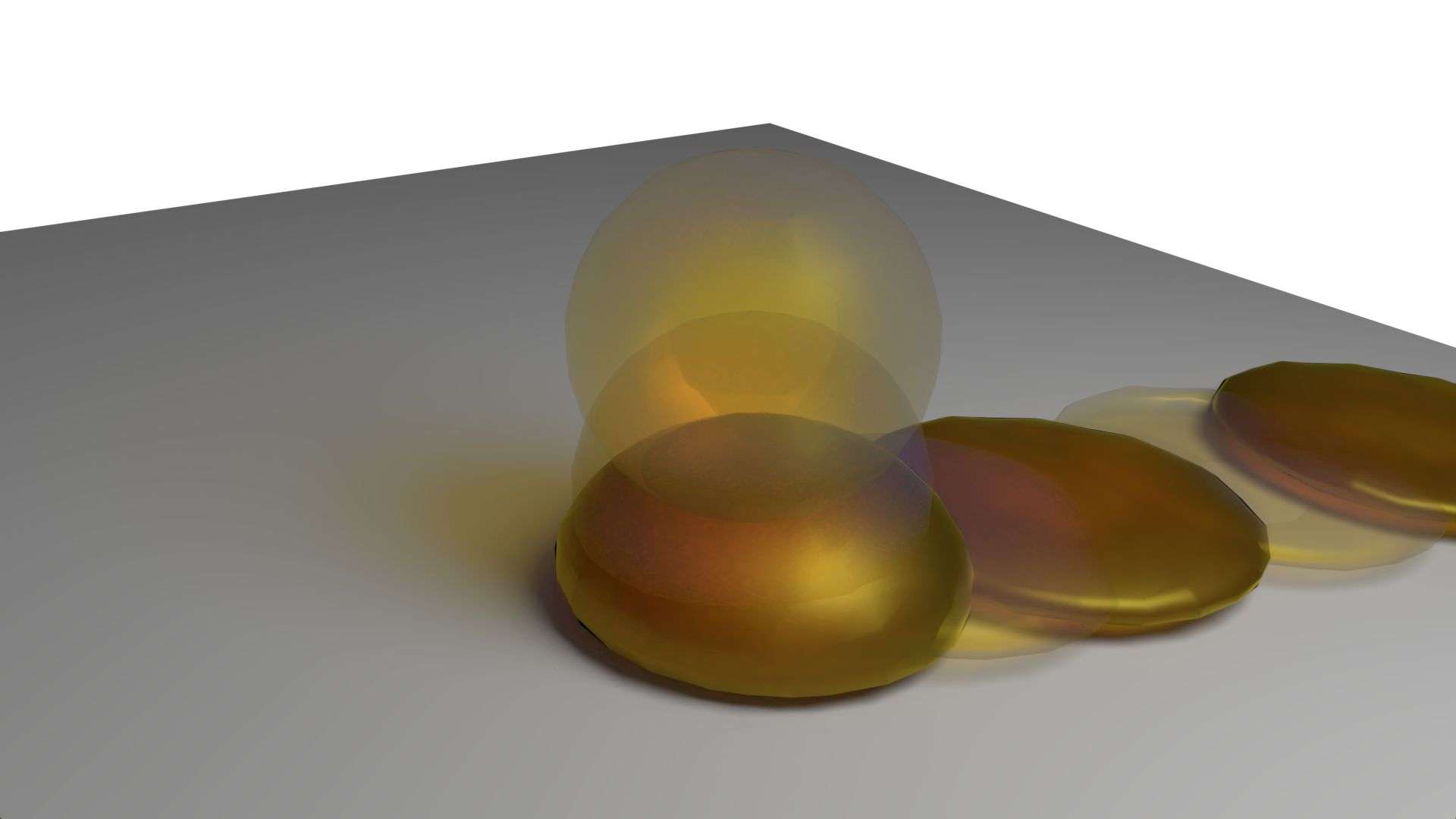}\\%
    	\begin{overpic}[trim=600 10 0 0,clip,width=\textwidth]{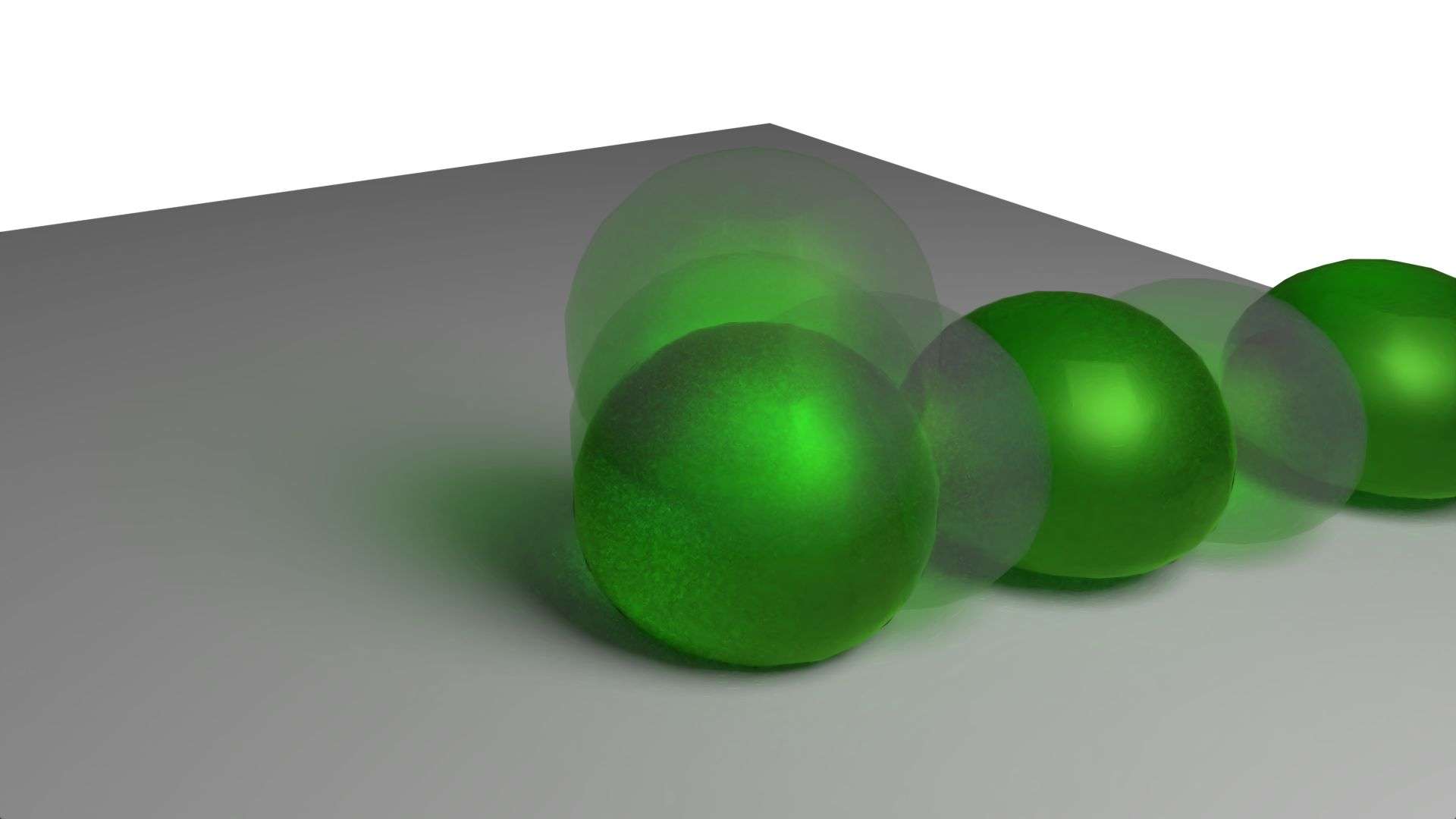}%
             \put(5,89){\textcolor{white}{(c)}} \put(5,5){\textcolor{white}{(e)}}%
        \end{overpic}%
    \end{subfigure}
    \caption{Also when optimizing for multiple parameters at once, our algorithm can robustly 
		reconstruct the ground truth motion (a), despite starting with very different initial conditions (b,c).
		For (b), the solid rendered frames indicate the different rolling behavior of the very stiff material 
		in comparison to (a), while the very soft material in (c) leads to strong deformations.
		Our sparse surface constraints allow for a robust matching, that is visually indistinguishable from the 
		ground truth, i.e. (d) reconstructed from (b), and (e) from (c). Quantified errors are given in the appendix.}
    \label{fig:Stability:BallMultipleVars:ImagesSm}
\end{figure}

These results illustrate that the optimization problem we solve is highly non-convex, and the optimizer
can arrive at different solutions yielding equally good matches with the observations. This is a property our optimization approach shares with other non-linear optimization algorithms. As shown in the previous 
experiment, our algorithm typically successfully finds one of the possible solution.
In practice, we alleviate this ambiguity by running a batch of optimizations with 
slightly perturbed initial conditions.
Our cost function lets us reliably choose the best match from the batch, which we use as the final result. 
%We will employ this strategy for the following real-world test cases.
We employ this strategy in all of the following experiments.
%, including those using real-world measurements demonstrated below.}
%\sebi{We therefore start the optimization from multiple randomly sampled initial configurations and choose the reconstruction with the lowest cost as the final result.}

\subsection{Real-World Data}\label{sec:realworld}
In addition to the synthetic test cases, we use three live captures of real-world objects to analyze the capabilities of our approach.
The first test case is a plush teddy, of which a high-quality initial pose is obtained with a 3D scanner.
An image sequence of the teddy falling onto the floor is then recorded with a commodity RGB-D
camera\footnote{An {\em Asus Xtion ProLive camera}.} (see Fig.~\ref{fig:Teddy:Color}). Of this sequence, we use only the depth (D) channel as input for the optimizer (see  Fig.~\ref{fig:Teddy:Depth}).
From the initial configuration shown in Fig.~\ref{fig:Teddy:Initial}, the optimizer estimates gravity, the Young's modulus, as well as the mass- and stiffness damping parameters.

\paragraph*{\sebi{Physical Units}}\label{sec:PhysicalUnits}
\ruediger{
%The simulation and reconstruction by itself is performed without units. By just using the depth observations from an arbitrary camera matrix, it is impossible to assign physical units. 
The material estimates are performed in a virtual, unit-less coordinate system. To convert the estimated values to physical units, the simulated gravity needs to be scaled to relate the object's size in the virtual space to its size in the real world. Furthermore, the object's mass (the Young's modulus depends on it) and the time step are required.
Since the object's mass cannot be recovered from the observations, we weight the objects beforehand. The time step is given by the known camera framerate of 60fps.}

\sebi{More specifically, we measure the size of of the object $S'$ in virtual meters $m'$ via the signed distance function of the input $\phi_0$ that defines the object in reference configuration $\Omega_r$. Given the size of the object $S$ in meters, we can compute the first scaling factor $f_\text{size}=\sfrac{S}{S'}$ in $\sfrac{m}{m'}$.
Next, let $M$ in $kg$ be the mass of the real object. The parameter $m$ in the physical model Eq.~\eqref{eq:StrongPDEdynamic} specifies the mass density. Hence the virtual mass $M'$ in $kg'$ is given by $M' = mV$ where $V$ is the object's volume computed as 
$V=f_{\text{size}}^3 \int_{\Omega_r}{1\dx}$ . 
The scaling factor for the mass is then given as $f_\text{mass}=\sfrac{M}{M'}$ in $\sfrac{kg}{kg'}$.
Last, we parametrize the simulation with the real-world time step, hence $f_\text{time}=\sfrac{1s}{1s'}$.
With these scaling factors, we can convert the value of the virtual Young's modulus $k$ (see Sec.\ref{sec:OverviewPhysicalModel}) into real-world units. The unit of the Young's modulus is Pascal ($\sfrac{kg m}{s^2}$), hence the real-world value is given as $k\cdot \sfrac{f_\text{mass} f_{\text{size}}}{f_\text{time}^2}$.
% NOTE: the gravity in the simulation has a factor 8 built-in. You have to multiply the value by 8 before conversion.
Similarly, let $g$ be the reconstructed gravitational acceleration in virtual units. The real-world gravity is then given by $g\cdot \sfrac{f_{\text{size}}}{f_\text{time}^2}$.
%The conversion of the gravity is more involved due to the Rayleigh damping that introduces an error of up to $5\%$ from the scaled gravity parameter. We therefore measure instead how fast the object falls in our virtual space without collision or Dirichlet boundaries, fit a parabola through the displacements and take that as our value for the gravity. 
}%

% \Description{
% %The conversion of the simulation result is independent of the underlying simulation method.
% It is worth noting that all measurement steps introduce errors, and our simulation introduces a potentially unphysical Rayleigh damping.
% %Furthermore, these physical quantities are subject to large rounding errors, introduced by measurement errors in the size and mass of the object and due to the non-physical Rayleigh damping. 
% %Especially the Rayleigh damping 
% Especially the latter introduces forces that weaken the gravity force similar to an outer viscous medium~\cite{wilson2002three,adhikari2004rayleigh}. Similarly, since our soft collision model repulses the object already before the contact point, it can require a stronger gravity force to compensate that effect. Hence, the material
% parameters we obtain from the scans are typically less accurate than those obtained from more specialized laboratory experiments~\cite{miguel2016modeling,zehnder2017metasilicone}. 
% However, as we discuss below, 
% the yield plausible estimates within the scope of our material model.
% }

\paragraph*{\sebi{Simulation Results}}
For our real-world cases, the accuracy of the estimated material parameters is affected by measurement noise as well as the non-physical damping distribution that is assumed in the simulation.
%Furthermore, these physical quantities are subject to large rounding errors, introduced by measurement errors in the size and mass of the object and due to the non-physical Rayleigh damping. 
%Especially the Rayleigh damping 
Especially the latter introduces forces that reduce the gravity force, similar to an outer viscous medium~\cite{wilson2002three,adhikari2004rayleigh}. Similarly, since the soft collision model repulses the object already before the contact point is reached, it can require a stronger gravity force to compensate this effect. Hence, the estimated material
parameters are typically less accurate than those obtained from more specialized laboratory experiments~\cite{miguel2016modeling,zehnder2017metasilicone}. 
%However, as we discuss below, 
%the yield plausible estimates within the scope of our material model.

\begin{figure*}%
\centering
%\begin{subfigure}{\textwidth}
     % width=0.9
    \includegraphics[width=0.9\textwidth]{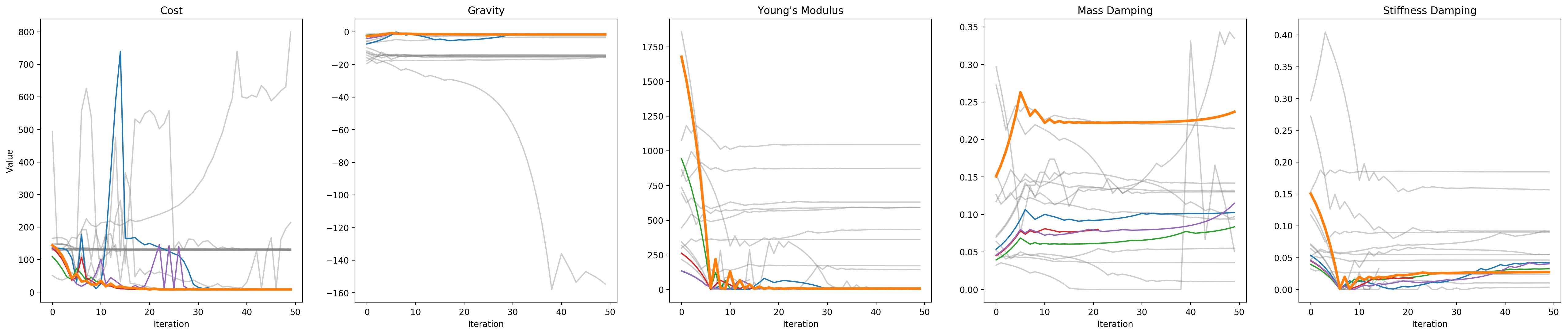}
    %\caption{\ruediger{Convergence plots for teddy, using 18 randomly selected parameter sets. Plots of best 5 runs drawn in color, plot of best run thickened.}}
    %\label{fig:Teddy:Optimization}
%\end{subfigure}
\caption{Convergence plots for teddy, using 18 randomly selected initial parameter sets. Plots of best 5 runs drawn in color, plot of best run thickened.}%
%\Description{Selected frames from the teddy test case: RGB-D observations from the real-world camera, initial configuration of the optimization and reconstructed solution. Plots of the optimization process of the teddy test case. First, the gravity and Young's modulus are matched. During this time, the mass damping is high and the stiffness damping low. After the gravity and Young's modulus have converged. The mass damping is reduced to zero and the stiffness damping increases to the final solution.}
\label{fig:Teddy:Optimization}%
\end{figure*}

%To account for non-linearities in the optimization, we run the optimization 18 times with randomly perturbed initial values.
%The teddy was simulated over 80 timesteps, with 50 iterations per optimization cycle, and 18 perturbed initial conditions. 
\sebi{The teddy was optimized for 50 iterations. Each iteration consists of a forward and adjoint simulation over 80 timesteps, and using 18 perturbed initial conditions.}
The plots of the single runs are shown in \autoref{fig:Teddy:Optimization}.
%There seem to be two local minima in the gravity parameter, one at around $-1.5$ and one at around $-15$, the best runs all converge to the first minima.
The best five runs exhibit large differences in the reconstructed values, while having an equally low cost.
In the accompanying video, these five runs are compared side-by-side and closely  match the bouncing behaviour of the teddy. The teddy, however, tilts to the side after the first bounce. We attribute this behaviour to inhomogeneities in the material composition, which cannot be reconstructed by our model.
For the Young's modulus, the initial value is several magnitudes higher than the reconstructed values of around 8, showing that the optimization is stable over a wide range of values.
Furthermore, the stiffness and mass damping seem to be antimodal, i.e. a high value of mass damping is  compensated to some extend by low stiffness damping, and vice versa.

\sebi{For the best run, \autoref{tab:TeddyPillow} shows the reconstructed values with physical units if possible, and for the 5 best runs\autoref{tab:RealWorld:Numbers:Teddy} gives the recorded parameter values of the simulation. The initial configuration and the reconstructed configuration for the best run is shown in \autoref{fig:Teddy:Initial} and \autoref{fig:Teddy:Reconstruction}, respectively.
As one can see, the gravity is slightly off from the ground truth of $9.81\frac{m}{s^2}$, possibly due to the Rayleigh damping. 
The reconstructed value of 590Pa is around five times softer than e.g. polystyrene foam of ca. 2500Pa. This is a plausible value since from the bouncing behavior, the teddy is softer than typical polystyrene foam.
}

\begin{table}[tb!]
    \centering \small
    \input{plots/RealWorld/TeddyPillowTableSmall.tex}
    \caption{\sebi{Input units and final reconstructions of real-world cases.}}
    \label{tab:TeddyPillow}
\end{table}
%Recorded observations are overlayed with purple dots.
%As one can see, the optimization accurately reconstructs gravity and can also closely match the bouncing behaviour of the teddy. The teddy, however, tilts to the side after the first bounce. We attribute this behaviour to inhomogeneities in the material composition, which cannot be reconstructed by our model.

%Our algorithm successfully matches the bouncing behavior of the first contact with the ground, as can be seen in Fig.~\ref{fig:Teddy:Reconstruction}.
%Convergence plots of the optimization process are shown in Fig.~\ref{fig:TeddyPlots}. It is interesting to note that during the first iterations of the algorithm, where the optimizer tries to find the best gravity and Young's modulus, high mass damping is favored over high stiffness damping. Once gravity and Young's modulus have converged, the optimizer sets the mass damping to zero and favors high stiffness damping instead.

%Third, a simplified version of the Pillow-Ramp test is performed, here the pillow simply bounces on the flat ground without the sloped plane.
We also reconstruct a pillow falling onto a flat floor.
We let the optimizer simultaneously reconstruct gravity, Young's modulus, mass- and stiffness damping parameters, with 15 perturbed initial configurations. 
Our method finds a plausible match for the complex deformations of the pillow,
the details for which can be found in the accompanying video and in Appendix~\ref{sec:Appendix:RealWorld}\sebi{, the reconstructed values for the best run can be found in \autoref{tab:TeddyPillow}}.

\begin{figure*}%
\centering
%\begin{subfigure}{0.9\textwidth}
    %0.9
    \includegraphics[width=0.9\textwidth]{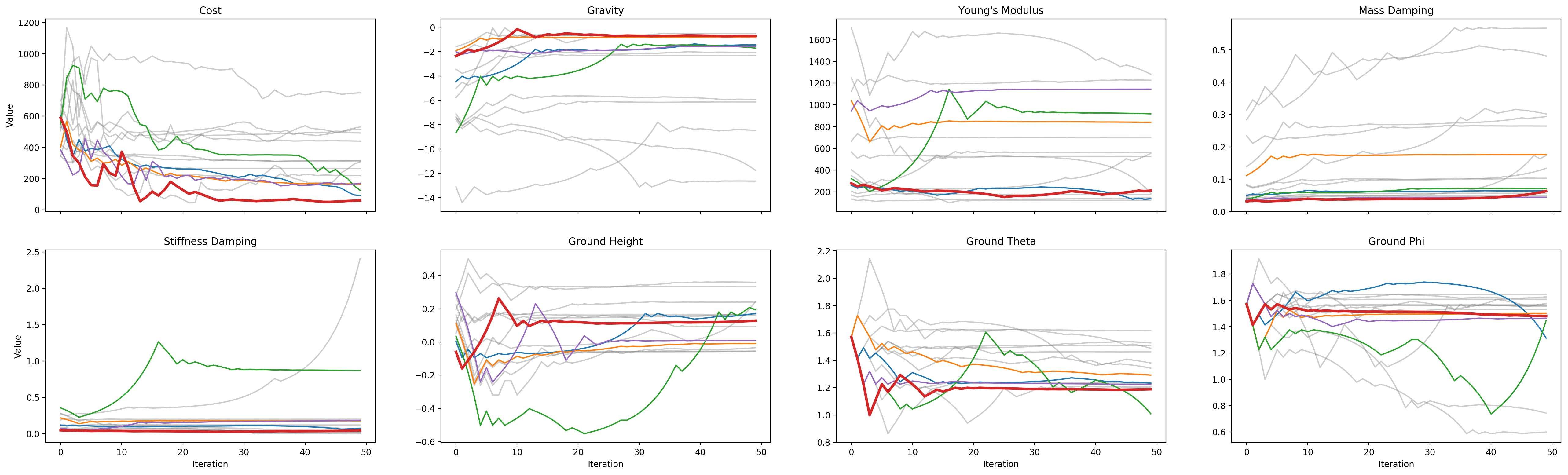}
    %\caption{\sebi{Optimization started from 15 different initial values. The best five runs are drawn in color, the best run is drawn in thick lines and displayed in the renderings above.}}
    %\label{fig:Pillow:Optimization}
%\end{subfigure}
\caption{Convergence plots for pillow, using 18 randomly selected initial parameter sets. Plots of best 5 runs drawn in color, plot of best run thickened.}%
%\Description{Selected frames from the Pillow-Ramp test case: RGB-D observations from the real-world camera, initial configuration of the optimization and reconstructed solution.Plots of the optimization process of the pillow test case. Here, ground position, angle, gravity, Young's modulus, mass- and stiffness damping are reconstructed.}
%\label{fig:PillowPlots}%
\label{fig:Pillow:Optimization}
\end{figure*}

Last, we recorded the pillow falling onto a ramp and bouncing off, 
in order to let our algorithm reconstruct the object as well as the collision geometry.
The recorded RGB-D sequence is shown in Fig.~\ref{fig:Pillow:Color} and Fig.~\ref{fig:Pillow:Depth}.
%As an initial configuration, a flat floor \sebi{with random height} and a \sebi{random} material is assumed (Fig.~\ref{fig:Pillow:Initial}). 
%\sebi{To analyse the convergence behaviour of the optimizer, we optimize using 20 different initial configurations using randomly determined floor height and material.}
We let the optimizer simultaneously reconstruct gravity, Young's modulus, mass- and stiffness damping parameters, 
the ground plane height, as well as the three degrees of freedom of the ground plane orientation, with 20 perturbed initial configurations.
%as well as the three degrees of freedom of the ground plane: orientation and position.
%The resulting simulation and convergence plots are shown in Fig.~\ref{fig:Pillow:Reconstruction} and Fig.~\ref{fig:PillowPlots}, respectively.
Our method recovers both a plausible orientation of the ramp and the object's 
material parameters purely from the sequence of depth images. 
To our knowledge, this is the first simultaneous physical reconstruction
%time that a physically plausible elastic simulation 
of a deformable object and its environment from a single depth video.
Since we compute collisions against a single plane, only the tilted ramp is reconstructed while the second collision with the flat table is ignored. Furthermore, since we do not consider friction in the underlying physical model, the simulation cannot accurately match the speed of the sliding pillow.
%The plots of the single runs are displayed in \autoref{fig:Pillow:Optimization}, and they indicate that this use case is even more challenging for the optimizer than the previous one including the teddy. 
This is a challenging test case for our method, as indicated by the plots in \autoref{fig:Pillow:Optimization}.
The parametric ambiguities, e.g. the same contact point can be obtained with higher and steep or low and flat ground planes, lead to noticeable differences in the reconstructions. 
\nils{The difficult setup of the ramp test case also leads to material estimates
in \autoref{tab:TeddyPillow} which indicate that our models tries to compensate
for the missing physical phenomena by increasing the material stiffness significantly.
Nonetheless, our final result yields a realistic reconstruction of the initial impact and still partially matches the observed sliding behavior.}
%Even the best five runs show high variations, which are caused by ambiguities in the parameters. For example, with a higher but steeper ground plane, a similar contact point can be obtained as with a lower but flatter ground.
%The reconstructed values for the best run are listed in \autoref{tab:TeddyPillow} and for the best five runs in \autoref{tab:RealWorld:Numbers:PillowRamp}, the initial configuration and the reconstructed configuration for the best run are shown in \autoref{fig:Pillow:Initial} and \autoref{fig:Pillow:Reconstruction}, respectively.

\paragraph*{\sebi{Timings}}
Performance details for our solver and optimization
are given in Table~\ref{tab:ResultsStatistics}. Overall, our forward solve is 
very fast, with less than one second per timestep for all our examples. The backward pass typically takes
a similar time, resulting in a total runtime of e.g. around 40 minutes for the teddy.
%\commentNils{insert? X} minutes for a typical optimization
%run.}
\begin{table}%
\footnotesize
\centering
\begin{tabular}{r|l|l|l|l|l|l}
					& Dragon	& Torus   & Ball	  & Teddy   & Pillow-Fl. & Pillow-R. \\ \hline
\# \sebi{active} nodes 		& 2489   	& 1138    & 770 	  & 2440    & 3836    & 4390    \\
\# elements 		& 1562		& 690	  & 516	      & 1716    & 2848    & 3284    \\
\# \sebi{diffused} nodes 		& 11587		& 4670	  & 2605	  & 14612   & 19006   & 21530   \\
\# timesteps 		& 10		& 40	  & 20		  & 100     & 450     & 175     \\
\# cameras 			& 1			& 1		  & 1		  & 1       & 1       & 1       \\
Camera res.         & 50x50     & 50x50   & 50x50     & 320x240 & 320x240 &320x240 \\
Obs. $n$'th     	& 1         & 1 	  & 1		  & 5       & 5       & 5       \\
$\phi_{\text{max}}$ & 4         & 5       & 2         & 10      & 10      & 10      \\
\# ini. cond.       & 2         & 20      & 20-60     & 18      & 10      & 15      \\ \hline
Forw. sim. (s)		& 0.318		& 0.095	  & 0.077	  & 0.803   & 0.240   & 0.264   \\
Cost eval. (s) 		& 0.172		& 0.092   & 0.199	  & 0.165   & 0.085   & 0.153   \\
Adj. sim. (s)		& 0.462		& 0.225	  & 0.058	  & 0.764   & 0.213   & 0.292   \\
\# iter. steps 		& 50		& 100	  & 30		  & 50      & 50      & 50     \\
\end{tabular}
\caption{Timing and model statistics for different test cases. Timings are per timestep. Dragon, Torus and Ball are synthetic datasets, Teddy, Pillow-Flat and Pillow-Ramp refer to real-world scans. ''Obs. $n$'th'' denotes the interval in simulation steps between observations, while ''Ini. cond.'' denotes the number of randomly perturbed initial conditions used for the optimization.}
\label{tab:ResultsStatistics}
\end{table}

\newcommand{\teddyWidth}{0.18} % 0.18 or 0.16
\begin{figure*}%
\centering
\begin{subfigure}{\textwidth}
    \centering
    \begin{overpic}[trim=40 70 40 0,clip,width=\teddyWidth\textwidth]{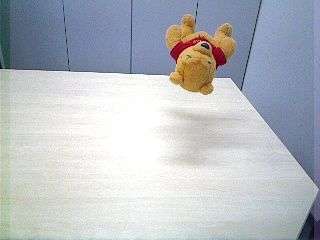}%
        \put(5,5){(a)}%
    \end{overpic}%
	\includegraphics[trim=40 70 40 0,clip,width=\teddyWidth\textwidth]{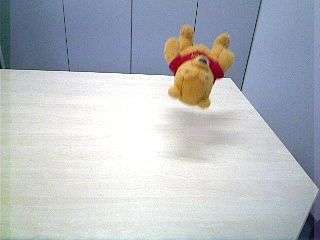}%
	\includegraphics[trim=40 70 40 0,clip,width=\teddyWidth\textwidth]{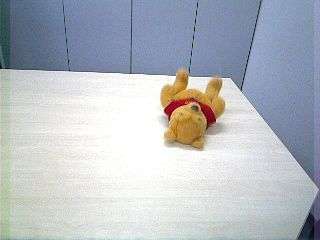}%
	\includegraphics[trim=40 70 40 0,clip,width=\teddyWidth\textwidth]{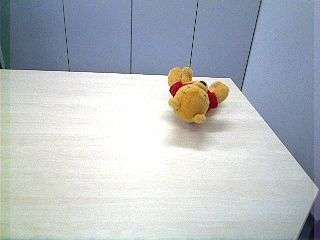}%
	\includegraphics[trim=40 70 40 0,clip,width=\teddyWidth\textwidth]{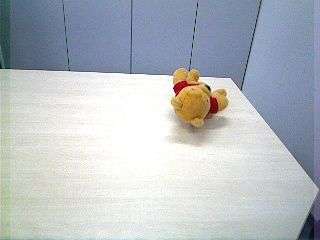}%
	%\caption{\sebi{Color Observation}}
	\phantomsubcaption\label{fig:Teddy:Color}
\end{subfigure}
\\%\vspace{-0.1cm}
\begin{subfigure}{\textwidth}
    \centering
    \begin{overpic}[trim=40 70 40 0,clip,width=\teddyWidth\textwidth]{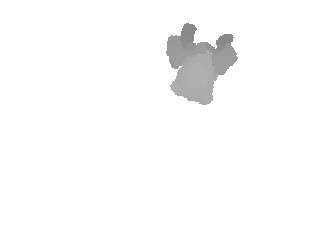}%
        \put(5,5){(b)}%
    \end{overpic}%
	\includegraphics[trim=40 70 40 0,clip,width=\teddyWidth\textwidth]{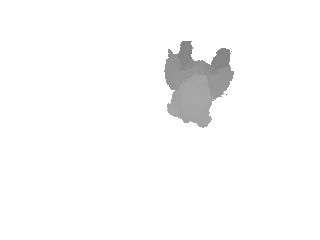}%
	\includegraphics[trim=40 70 40 0,clip,width=\teddyWidth\textwidth]{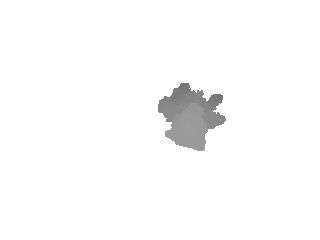}%
	\includegraphics[trim=40 70 40 0,clip,width=\teddyWidth\textwidth]{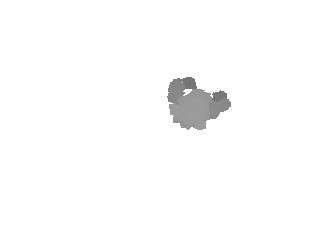}%
	\includegraphics[trim=40 70 40 0,clip,width=\teddyWidth\textwidth]{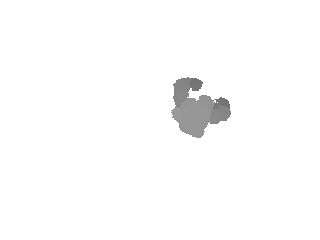}%
	%\vspace{-0.7cm}
	%\caption{\sebi{Depth Observation}}  % WARNING - shrunk space for depth
	\phantomsubcaption\label{fig:Teddy:Depth}
\end{subfigure}
\\
\begin{subfigure}{\textwidth}
    \centering
    \begin{overpic}[trim=260 160 0 0,clip,width=\teddyWidth\textwidth]{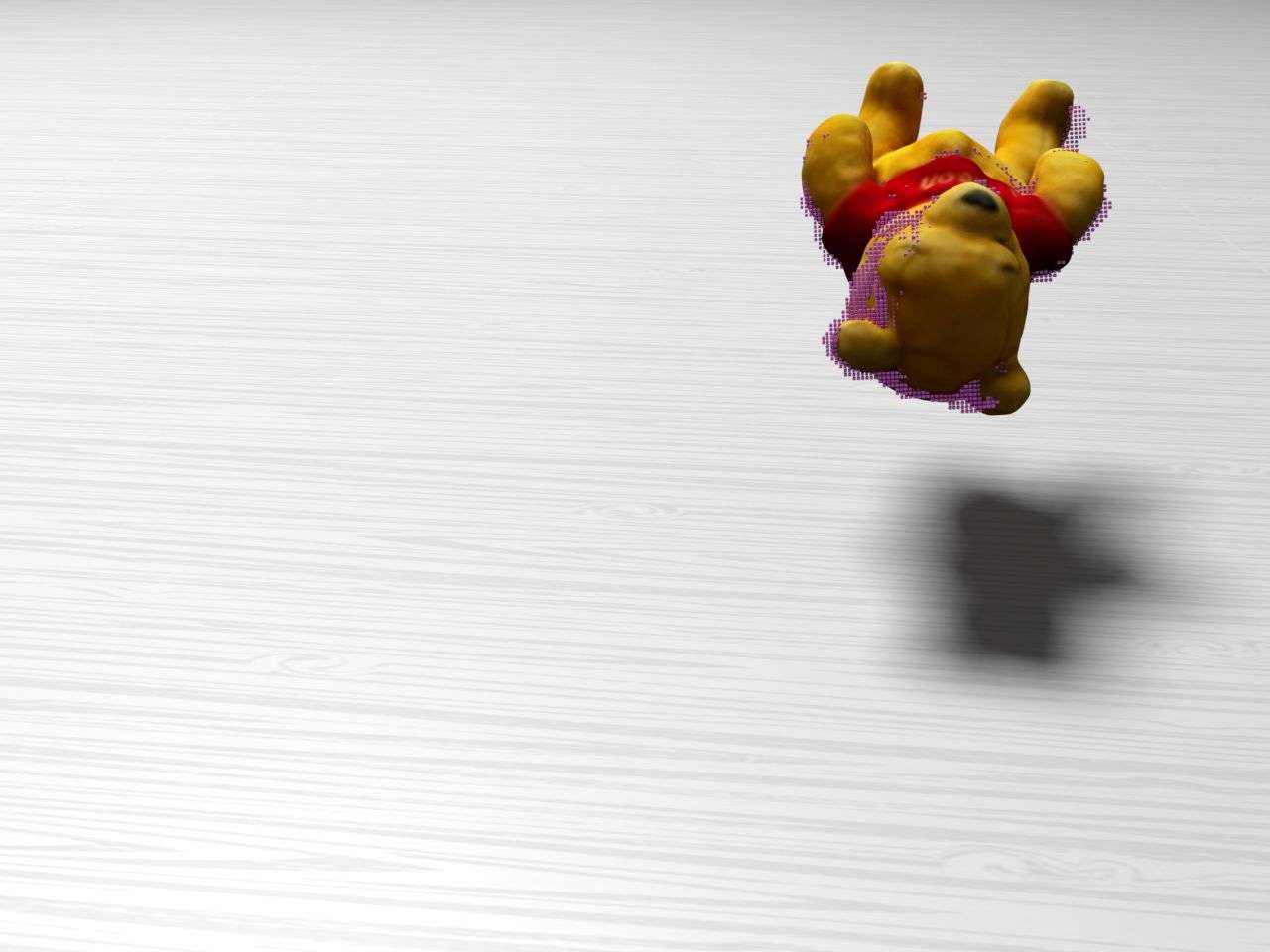}%
        \put(5,5){(c)}%
    \end{overpic}%
	\includegraphics[trim=260 160 0 0,clip,width=\teddyWidth\textwidth]{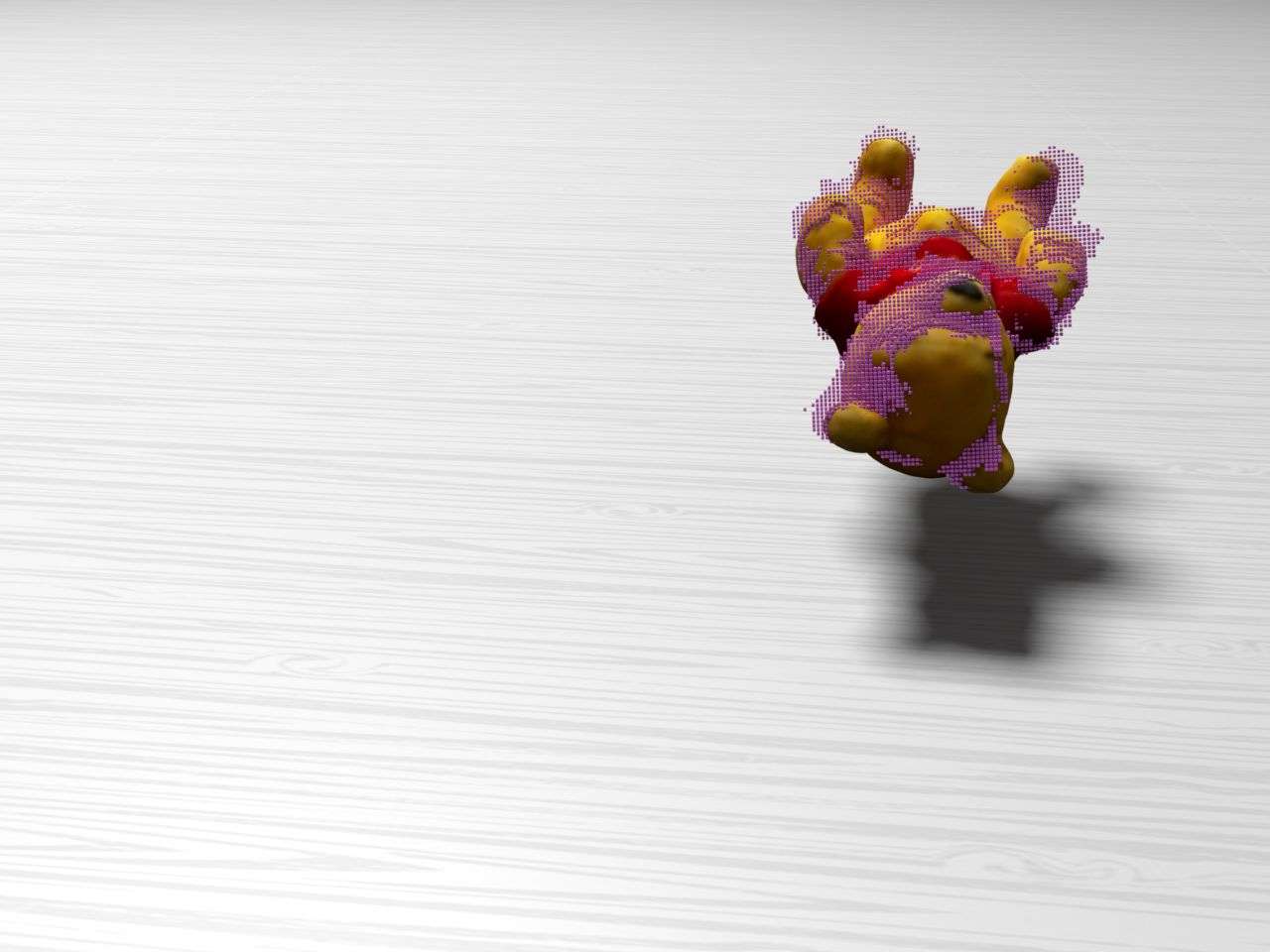}%
	\includegraphics[trim=260 160 0 0,clip,width=\teddyWidth\textwidth]{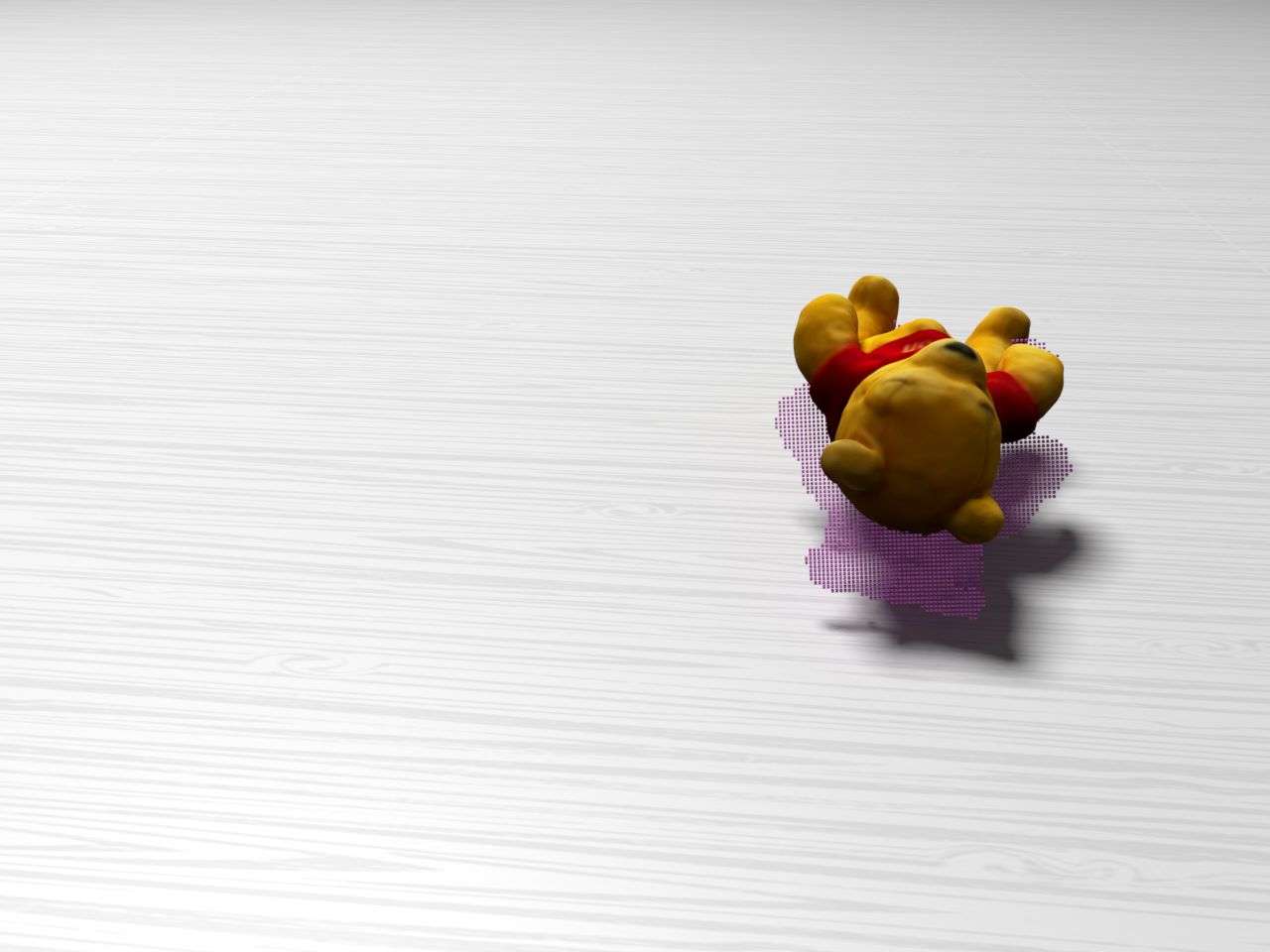}%
	\includegraphics[trim=260 160 0 0,clip,width=\teddyWidth\textwidth]{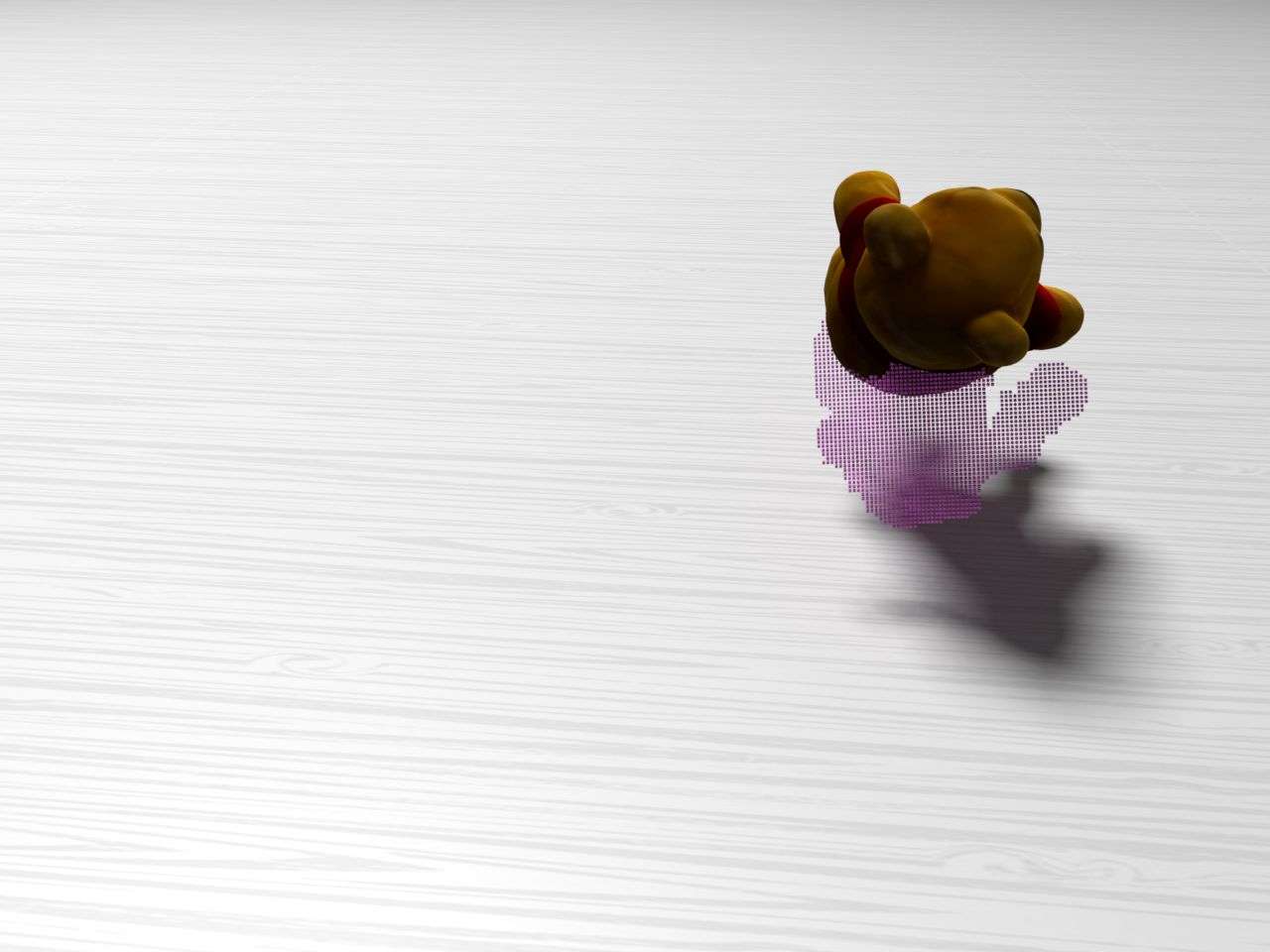}%
	\includegraphics[trim=260 160 0 0,clip,width=\teddyWidth\textwidth]{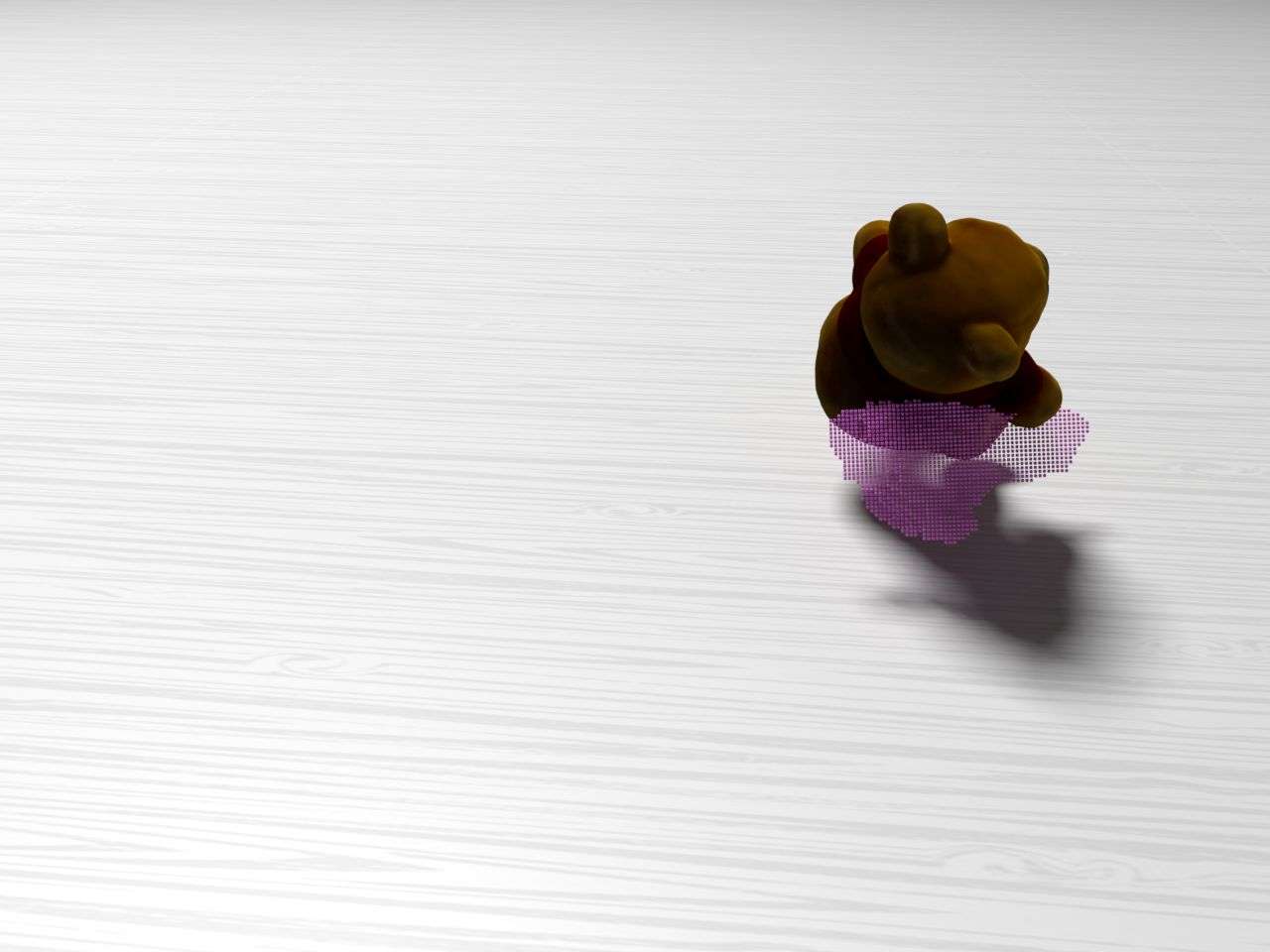}%
	%\caption{\sebi{Initial configuration for the optimization}}
	\phantomsubcaption\label{fig:Teddy:Initial}
\end{subfigure}
\\
\begin{subfigure}{\textwidth}
    \centering
    \begin{overpic}[trim=260 160 0 0,clip,width=\teddyWidth\textwidth]{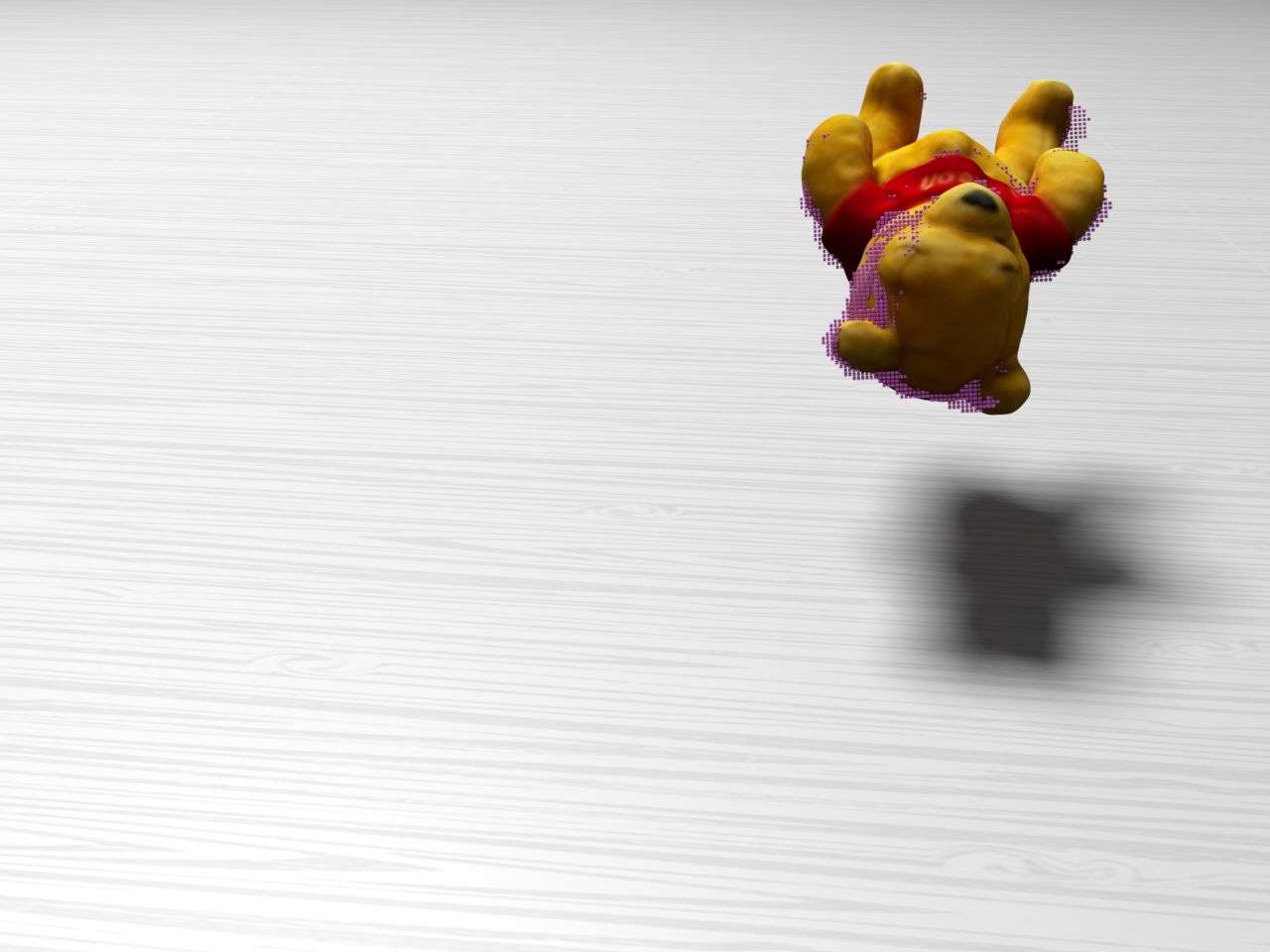}%
        \put(5,5){(d)}%
    \end{overpic}%
	\includegraphics[trim=260 160 0 0,clip,width=\teddyWidth\textwidth]{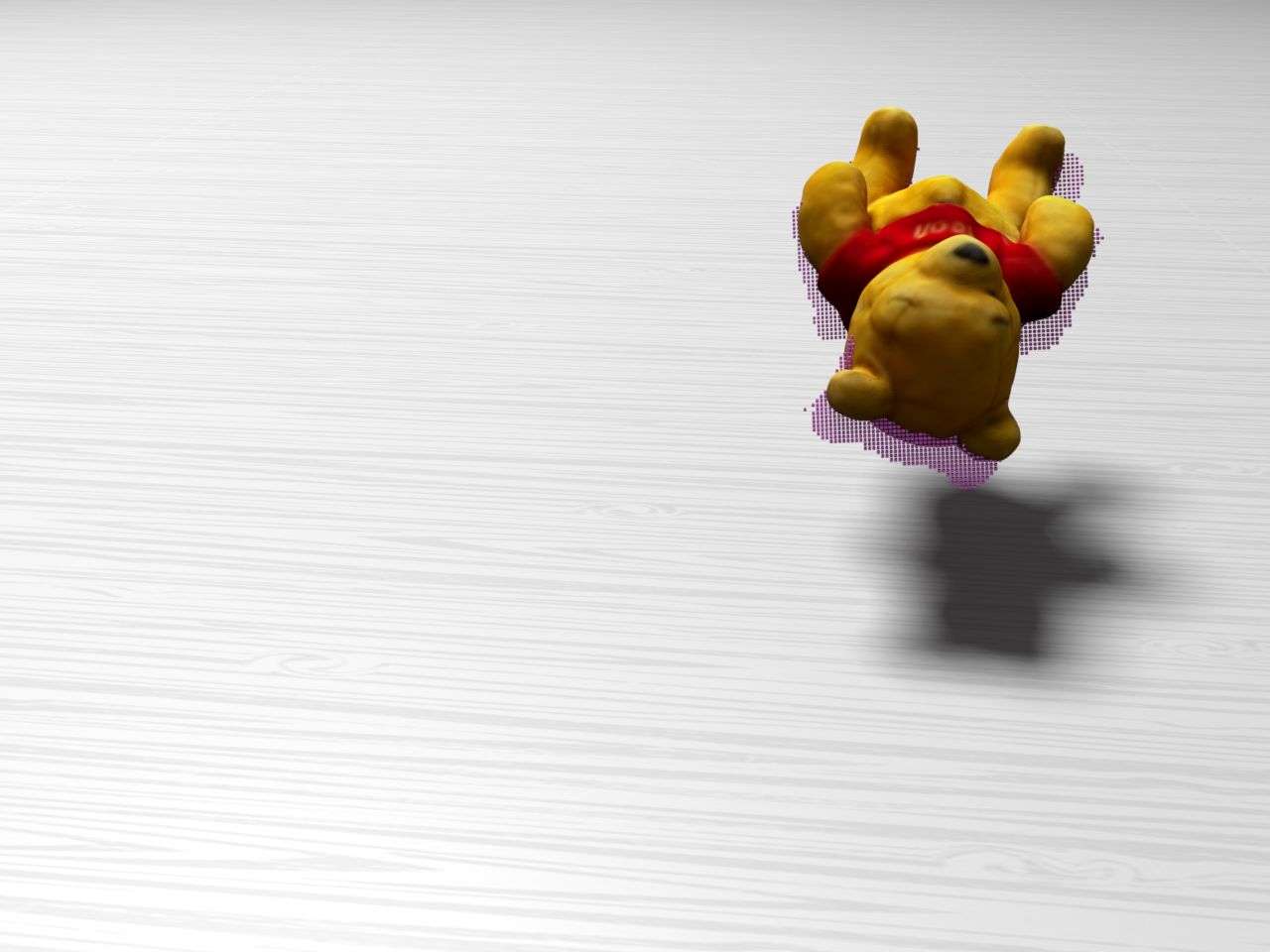}%
	\includegraphics[trim=260 160 0 0,clip,width=\teddyWidth\textwidth]{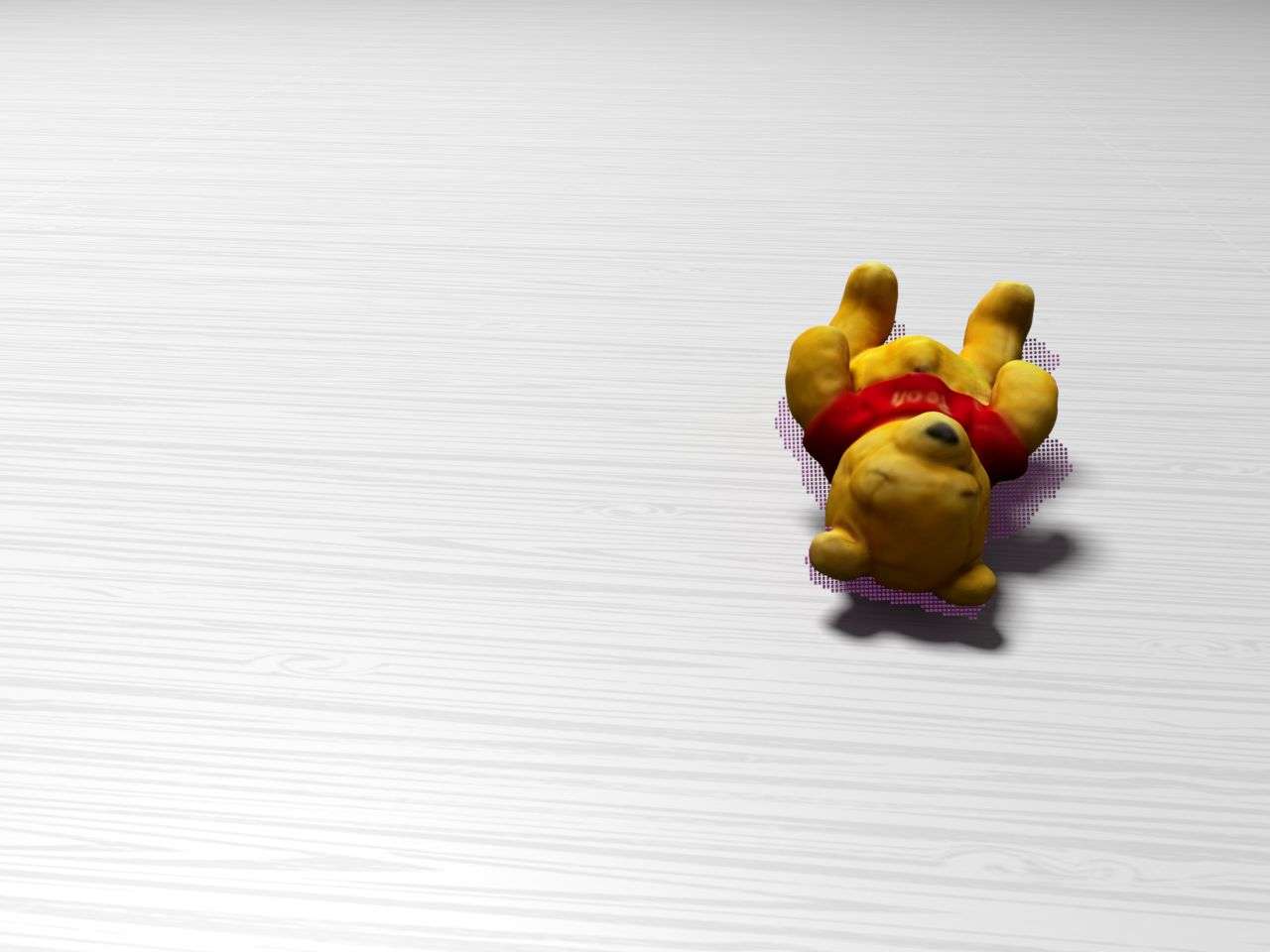}%
	\includegraphics[trim=260 160 0 0,clip,width=\teddyWidth\textwidth]{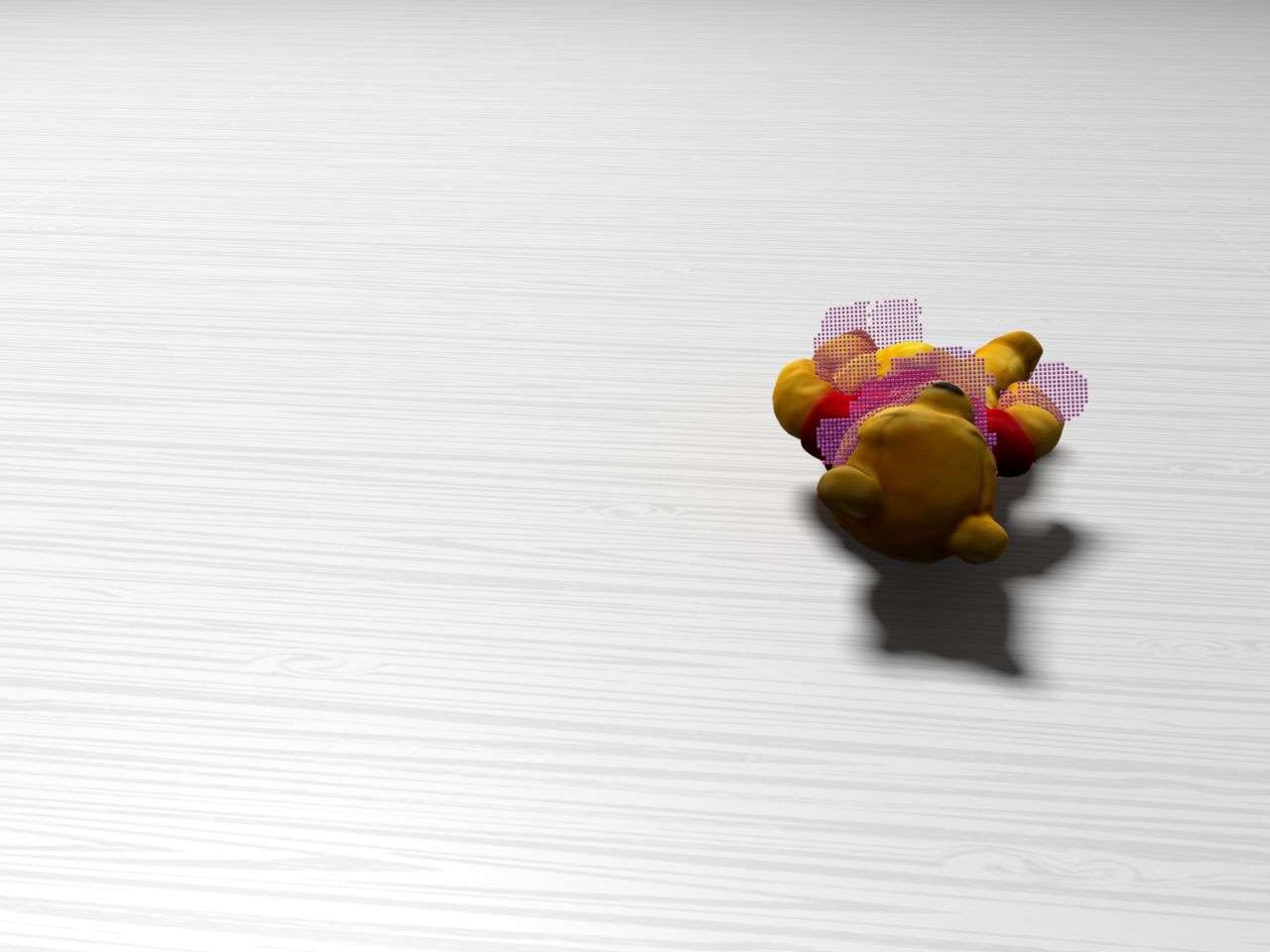}%
	\includegraphics[trim=260 160 0 0,clip,width=\teddyWidth\textwidth]{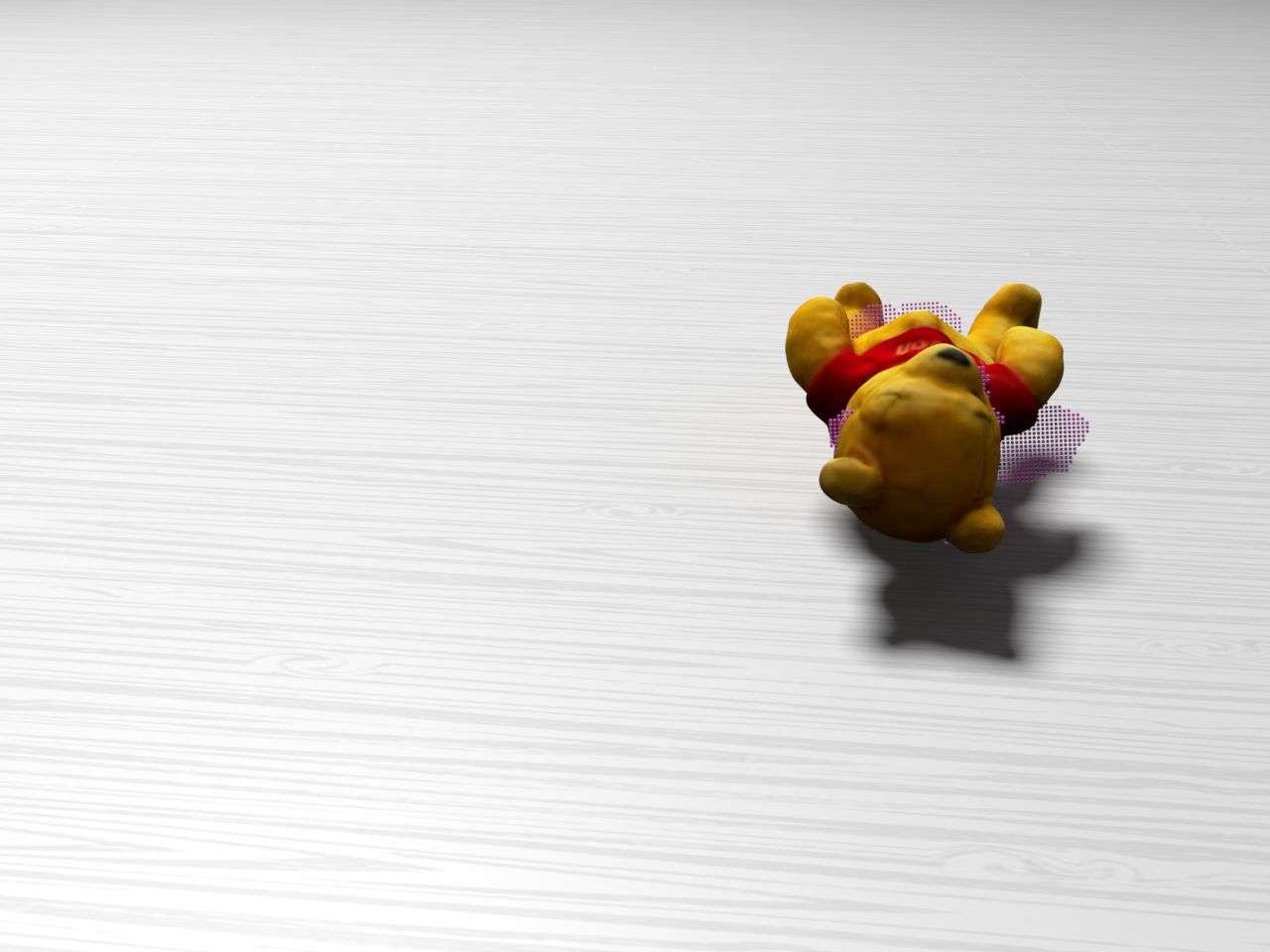}%
	%\caption{\sebi{Reconstructed solution}}
	\phantomsubcaption\label{fig:Teddy:Reconstruction}
\end{subfigure}
\caption{From top to bottom: Observed colors, observed depths, initial configuration, reconstructed model (purple dots indicate observations). Each row shows a sequence of steps over time.}
\label{fig:TeddyImages}%
\end{figure*}
%\\ %\vspace{0.2cm}

\section{Outlook and Limitations}

We see our novel formulation for gradient-based inverse parameter estimation using sparse constraints and physical priors as an important step toward more reliable object reconstruction.
There is huge potential for computer vision tasks to improve unseen or occluded motion, such as the backside of an object, via physical priors.
It will also be particularly interesting to investigate the incorporation of soft body physics into deep learning methods via our differentiable formulation. By shifting the workload to a physics-based training process it is potentially possible to train neural networks in a fully or partially unsupervised manner.

It should also be noted that our approach comes with some limitations and restrictions. In contrast to previous work \cite{wang2015deformation} we require an initial object pose. It would be highly interesting to combine our method with the rest pose estimation proposed there.
In our current approach we consider the extension region only in proximity to the domain covered by the object. Thus, the cost function cannot be evaluated at points that are observed far outside this region. To handle such cases, it will be interesting to investigate multi-scale approaches that can efficiently propagate deformations into a wider region around the object.
As our algorithm can arrive at multiple solutions with different parameter values,
we are also interested in decreasing these ambiguities by introducing additional priors 
or domain-specific knowledge about the observed materials.

In addition, it will be a very interesting and fruitful direction for future work to improve the physical models to account, e.g., for friction.
To incorporate such effects, it is necessary to include more physical accurate collision solvers using, e.g., using Linear Complementary Problems \cite{Anitescu.1997, Mordatch.2012,Foutayena.2014}. More flexible and realistic material models, such as the Neo-Hookean model \cite{smith2018stable}, would also be highly interesting additions.

%====================================================================================
%====================================================================================
\section{Conclusion}\label{sec:Conclusion}

We have proposed a framework to infer an elastic simulation of a deformable
object directly from a sparse set of depth images obtained from one or
more cameras.  To this end, we coupled a finite-element elasticity
simulation framework with the depth sensor measurements via a loss
function that imposes sparse surface constraints.
In this way we can infer plausible elastic simulation parameters,
even those for collision geometry,
from a very limited number of observations --- for
example a single stream of depth images.  All parameters of the
simulation are inferred directly by first-order optimization
algorithms applied in a hybrid Lagrangian-Eulerian formulation. We
have validated our method quantitatively and qualitatively on a variety of
simulated and real observations. In addition, we have demonstrated that our method
can robustly recover complex material behavior in real-world scenarios.

\newcommand{\pillowWidth}{0.22} %0.22 or 0.19555
\begin{figure*}%
\centering
\begin{subfigure}{\textwidth}
    \centering
	\begin{overpic}[width=\pillowWidth\textwidth]{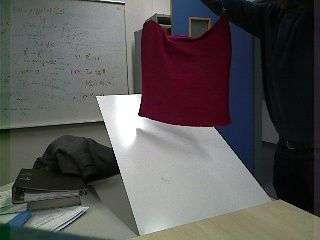}%
        \put(5,5){\textcolor{white}{(a)}}%
    \end{overpic}%
	\includegraphics[width=\pillowWidth\textwidth]{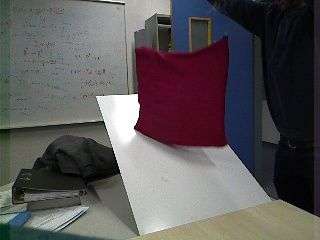}%
	\includegraphics[width=\pillowWidth\textwidth]{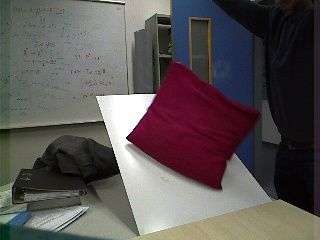}%
	\includegraphics[width=\pillowWidth\textwidth]{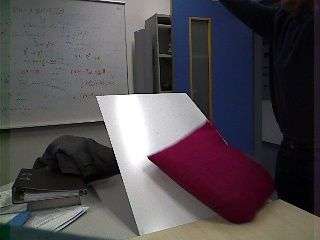}%
	%\caption{\sebi{Color Observation}}
	\phantomsubcaption\label{fig:Pillow:Color}
\end{subfigure}
\\
\begin{subfigure}{\textwidth}
    \centering
	\begin{overpic}[width=\pillowWidth\textwidth]{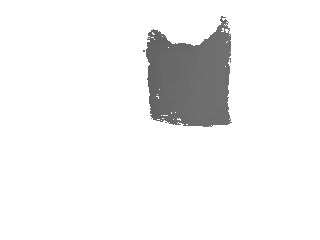}%
        \put(5,5){(b)}%
    \end{overpic}%
	\includegraphics[width=\pillowWidth\textwidth]{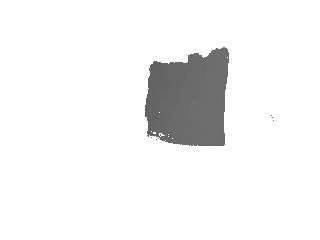}%
	\includegraphics[width=\pillowWidth\textwidth]{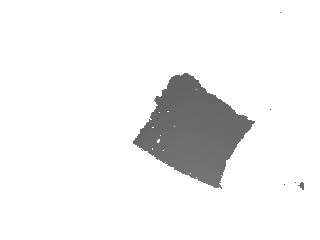}%
	\includegraphics[width=\pillowWidth\textwidth]{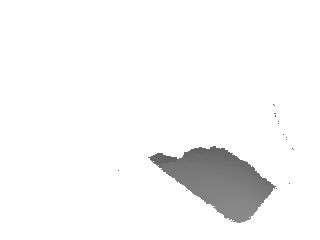}%
	%\caption{\sebi{Depth Observation}}
	\phantomsubcaption\label{fig:Pillow:Depth}
\end{subfigure}
\\
\begin{subfigure}{\textwidth}
    \centering
	\begin{overpic}[width=\pillowWidth\textwidth]{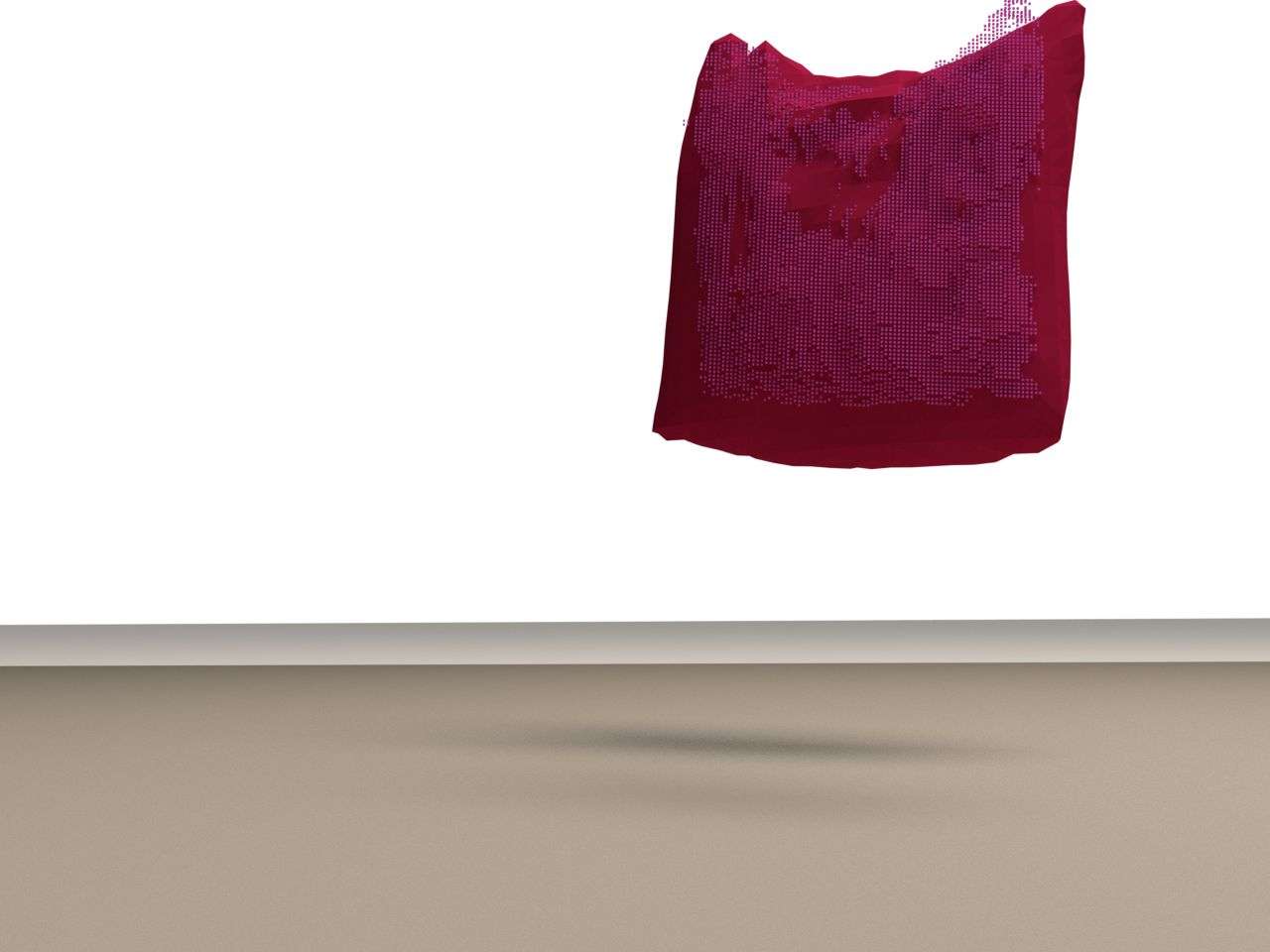}%
        \put(5,5){(c)}%
    \end{overpic}%
	\includegraphics[width=\pillowWidth\textwidth]{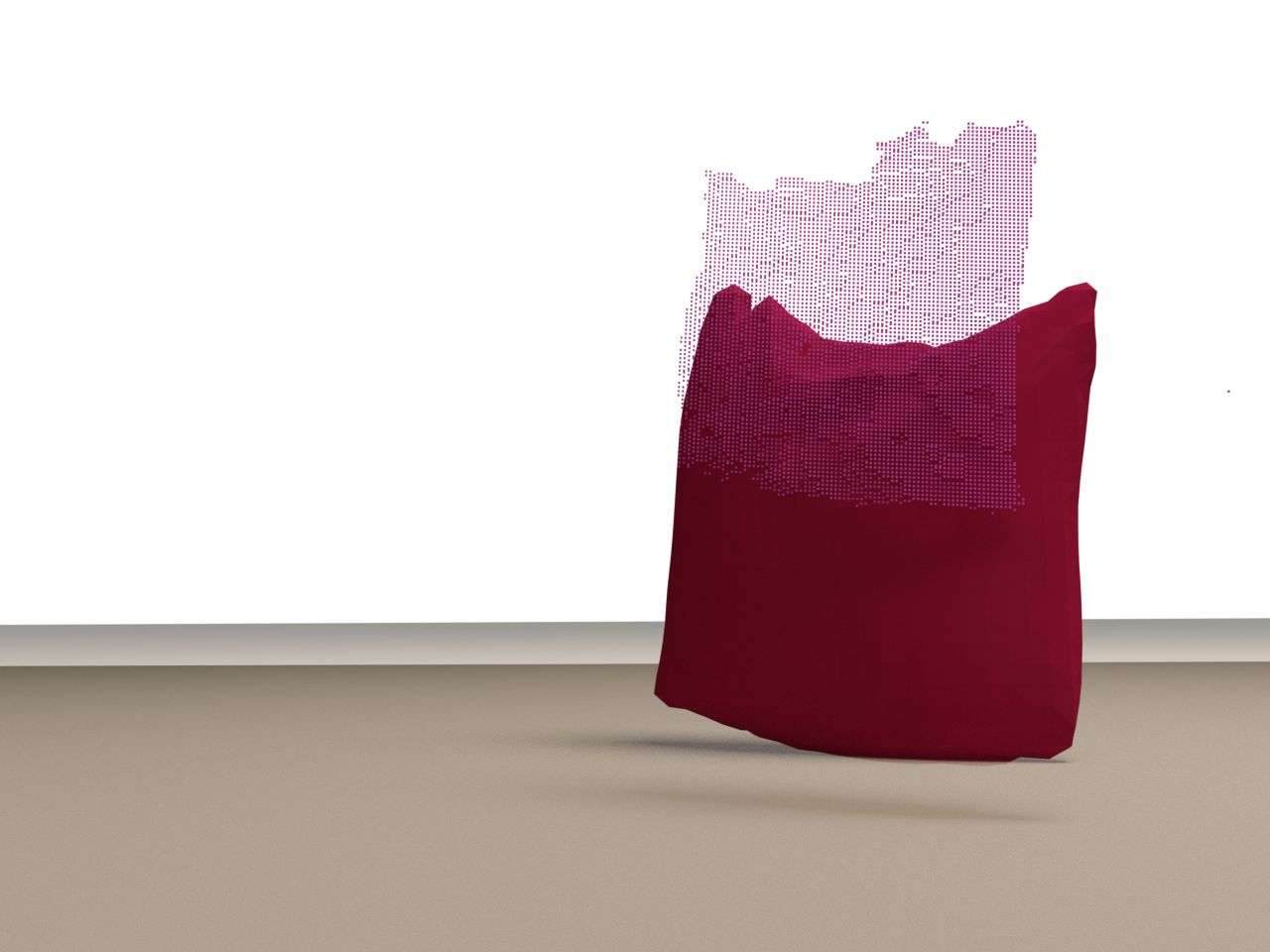}%
	\includegraphics[width=\pillowWidth\textwidth]{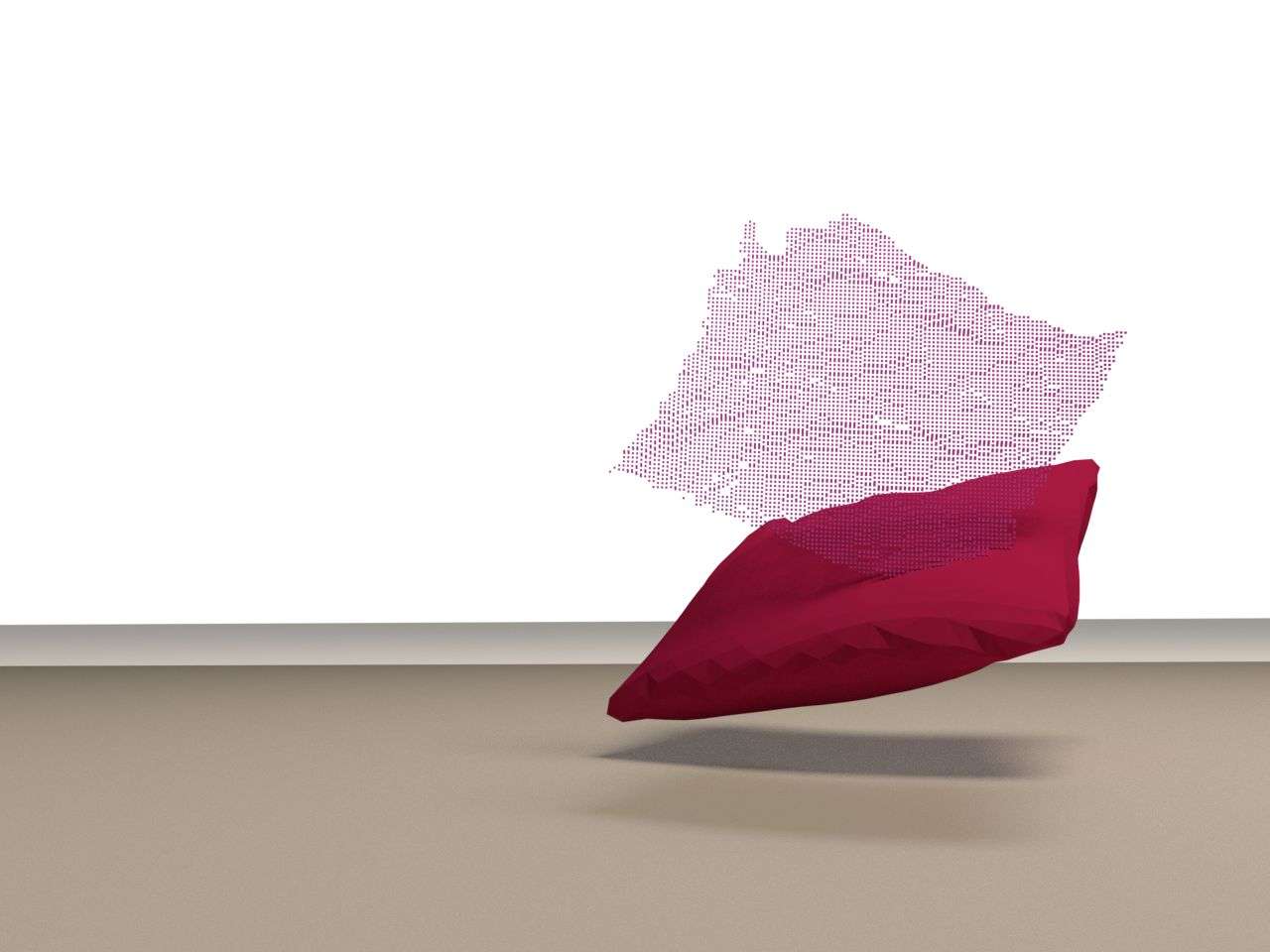}%
	\includegraphics[width=\pillowWidth\textwidth]{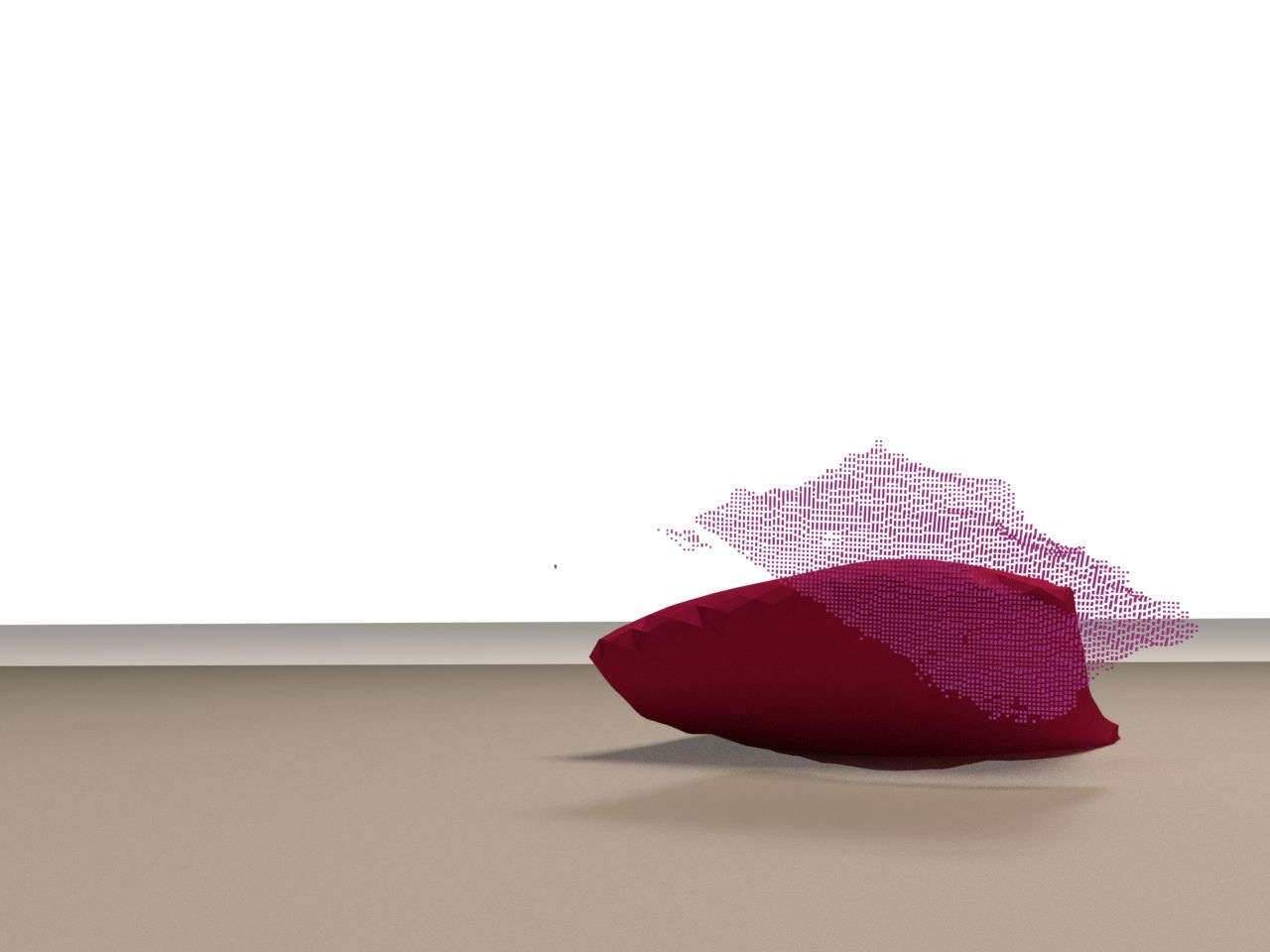}%
	%\caption{\sebi{Initial configuration for the optimization}}
	\phantomsubcaption\label{fig:Pillow:Initial}
\end{subfigure}
\\
\begin{subfigure}{\textwidth}
    \centering
	\begin{overpic}[width=\pillowWidth\textwidth]{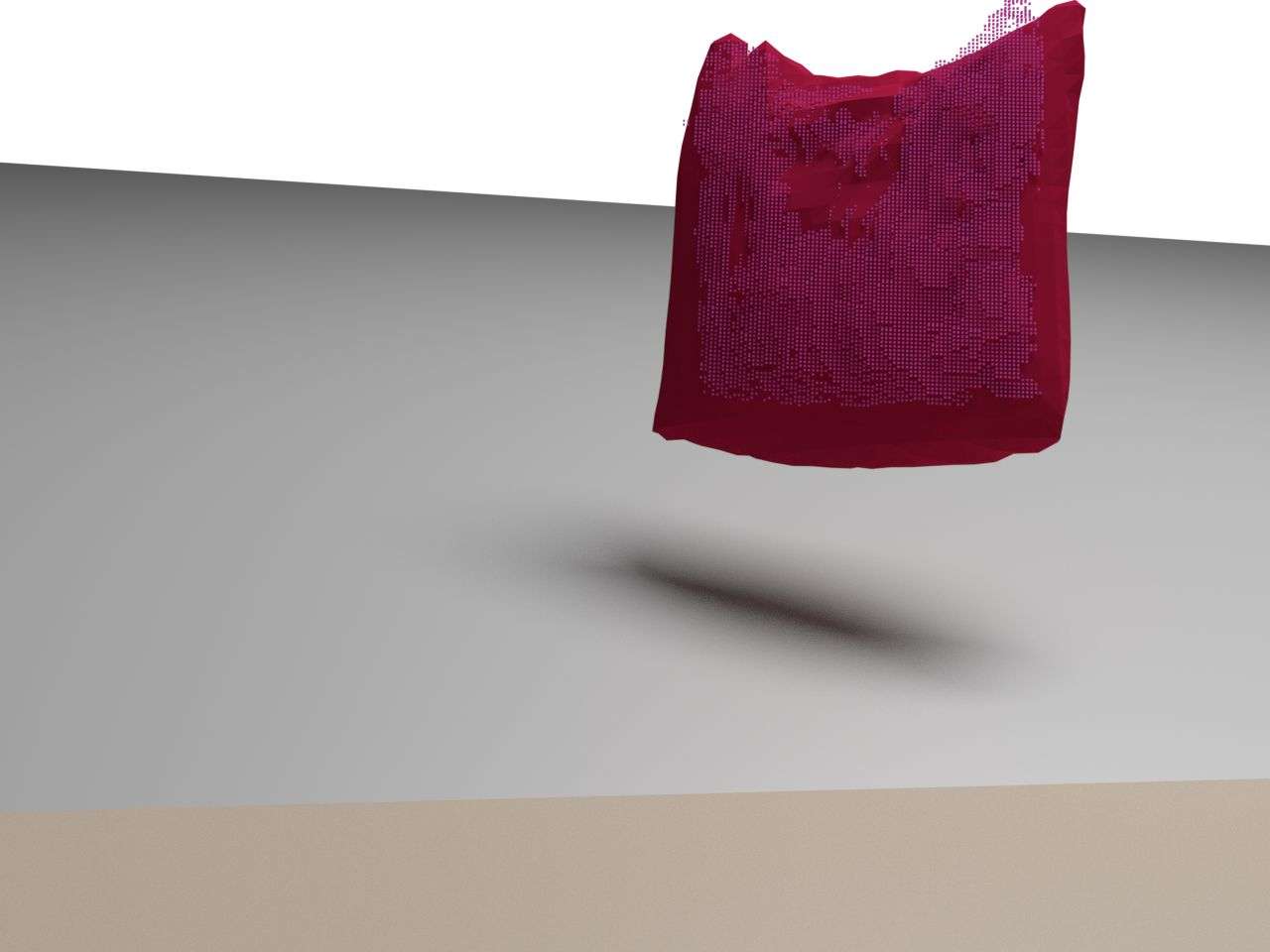}%
        \put(5,5){(d)}%
    \end{overpic}%
	\includegraphics[width=\pillowWidth\textwidth]{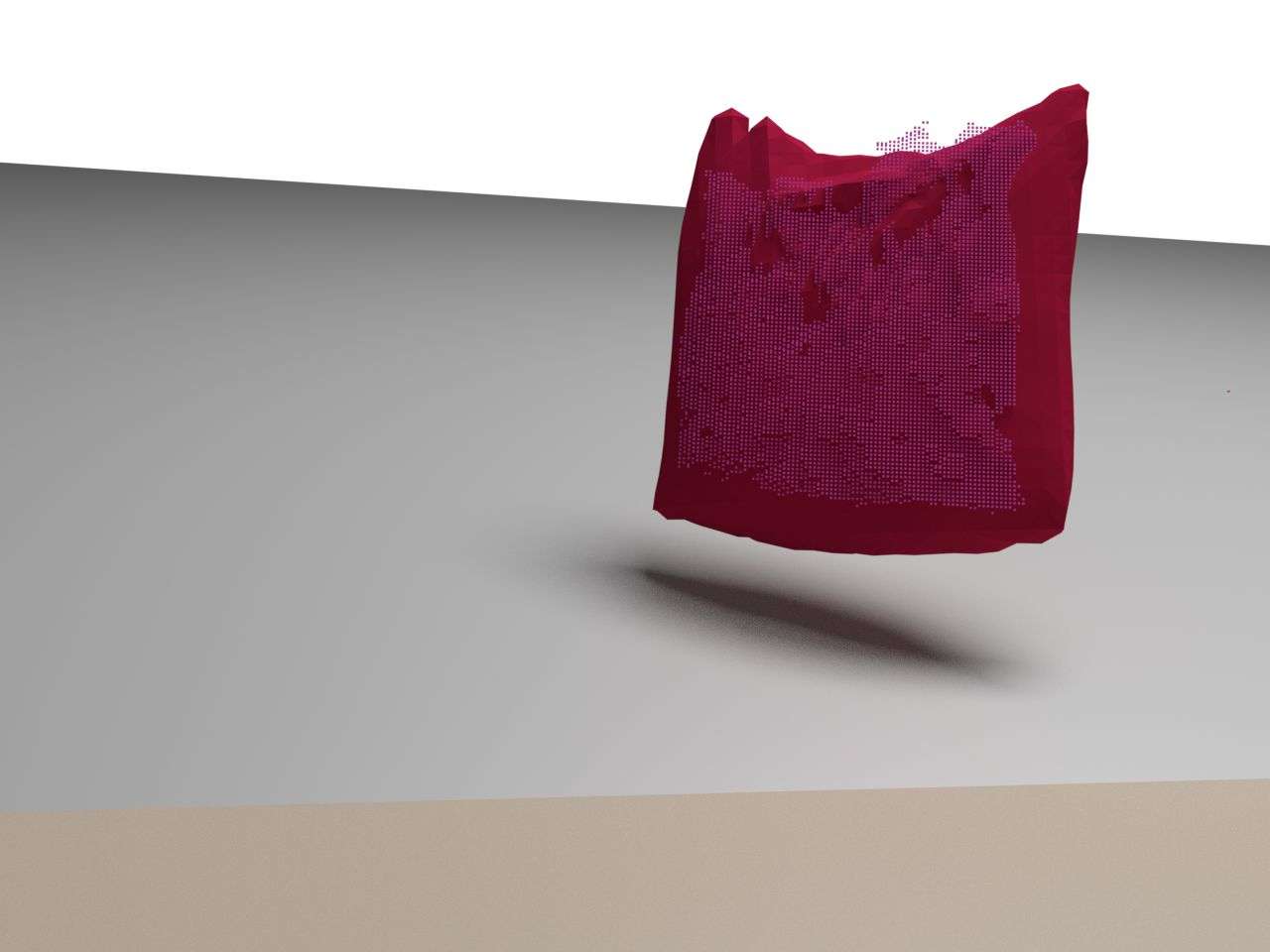}%
	\includegraphics[width=\pillowWidth\textwidth]{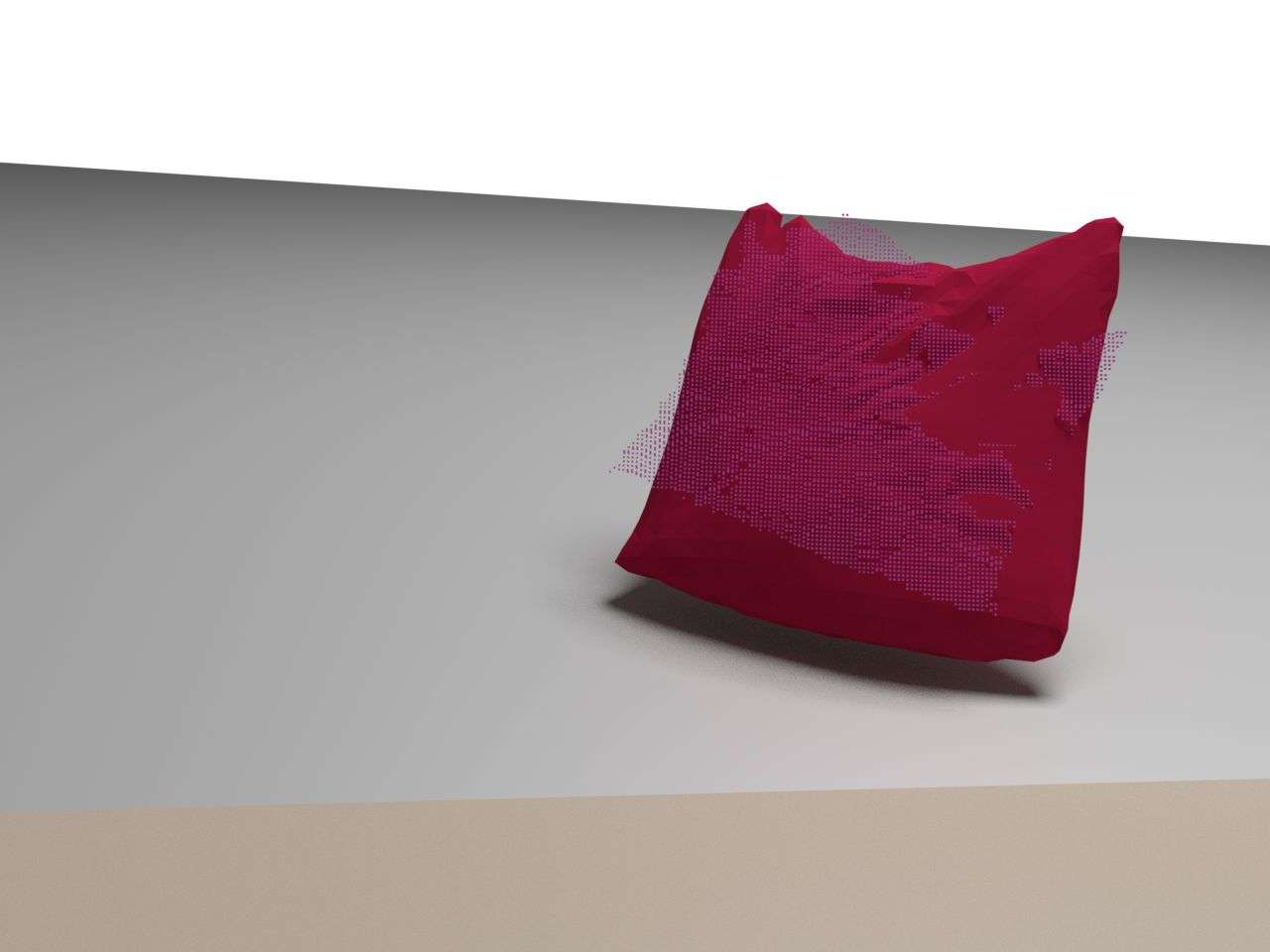}%
	\includegraphics[width=\pillowWidth\textwidth]{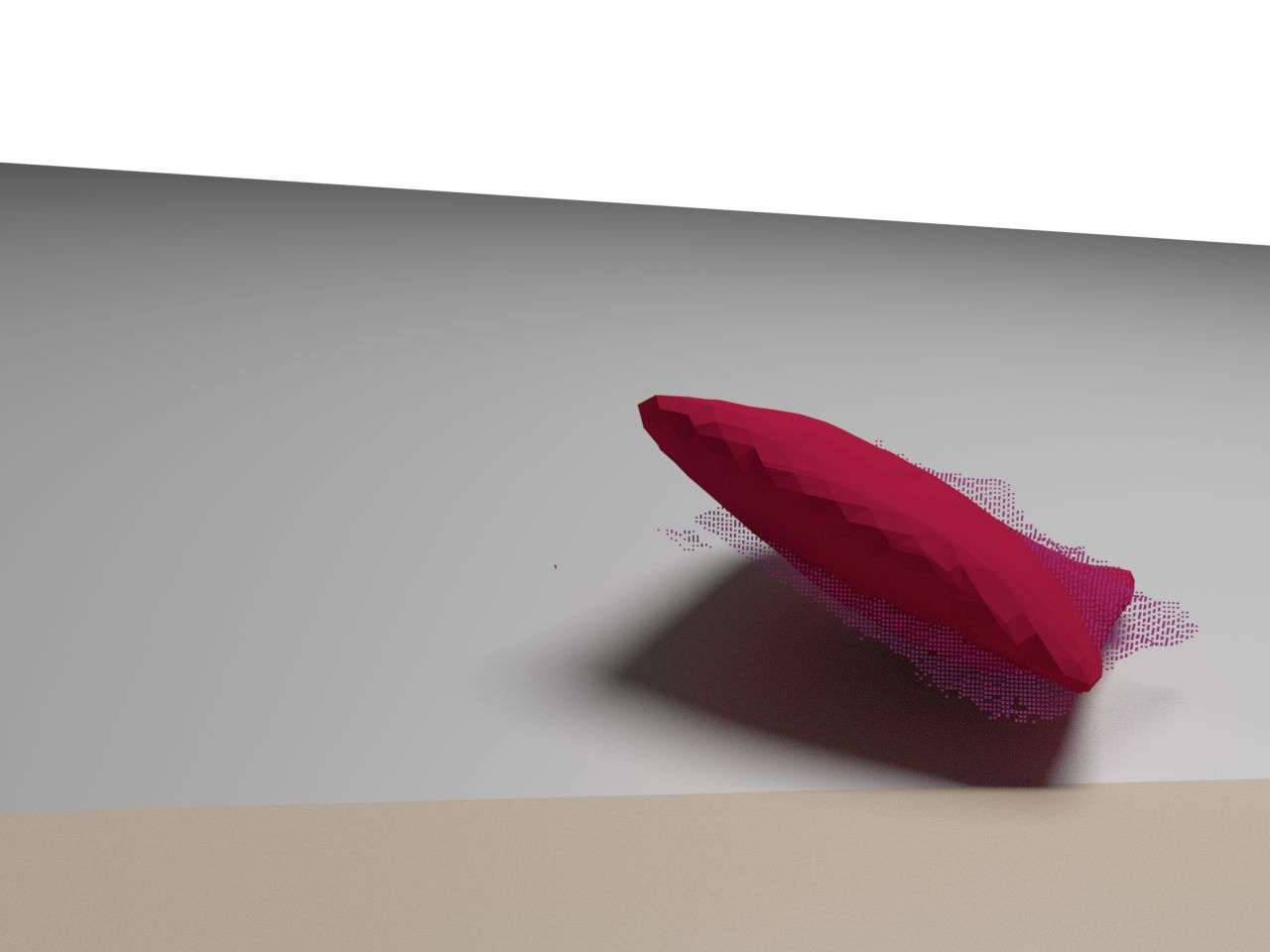}%
	%\caption{\sebi{Reconstructed solution}}
	\phantomsubcaption\label{fig:Pillow:Reconstruction}
\end{subfigure}
\caption{From top to bottom: Observed colors, observed depths, initial configuration, reconstructed model (purple dots indicate observations). Each row shows a sequence of steps over time. }
\label{fig:PillowColors}%
\end{figure*}

%% file: plots/RealWorld/TeddyPillowTableSmall.tex
\begin{tabular}{r|lll}
Testcase & Teddy & Pillow-Flat & Pillow-Ramp \\ \hline
Run & 14 & 8 & 7 \\
Initial Cost & 143.9 & 249.9 & 589.3 \\
Recon. Cost & 8.461 & 21.2 & 59.816 \\ \hline 
\sebi{camera framerate}  & 60Hz    & 60Hz    & 60 Hz   \\
\sebi{object size}       & 0.33x0.22x0.18m & 0.46x0.46x0.15m & 0.46x0.46x0.15m \\
\sebi{object mass}       & 0.256kg & 0.340kg & 0.340kg \\ \hline
Gravity & \sebi{$11.9\frac{m}{s^2}$} & \sebi{$8.03\frac{m}{s^2}$} & \sebi{$12.584\frac{m}{s^2}$} \\ %& -0.872 & -0.710 \\
Young's Modulus & \sebi{590Pa} & \sebi{151Pa} & \sebi{3430Pa} \\ %9.20 & 209.421 \\
Mass Damping & 0.240 & 0.078 & 0.068 \\
Stiffness Damping & 0.027 & 0.015 & 0.044 \\
Ground Height & - & - & 0.127\sebi{m}\\ 
Ground Theta & - & -  & \sebi{$21.9^\circ$} \\ %1.188 \\
Ground Phi & - & - & \sebi{$4.8^\circ$} \\%1.480\\ 
\end{tabular}

%% file: Appendix.tex
%====================================================================================
%====================================================================================
\section{Physics Simulation}\label{app:PhysicalModel}

In the following, we detail the physical model and governing equations underlying our work, and we describe the particular discretization scheme used.
%his section we present the linear elasticity theorem as introduced by \cite{ciarlet1988three} and show the strong and weak form of the PDE.

\paragraph{Strong Form}%\label{app:PhysicalModel:StrongForm}
Let the reference configuration be given in $\Omega^r$. The displacement at time $t$ is given by $u:\Omega^r \times t \rightarrow \R^3$.
Then the linear \textit{Green strain tensor} is given by 
\begin{equation}
	E(u) := \frac{1}{2}\left(\nabla u + (\nabla u)^T \right) \in \R^{3 \times 3}
\label{eq:LinearStrain}
\end{equation}
and the \textit{Piola-Kirchoff stress tensor}
\begin{equation}
	P(u) := 2 \mu E(u) + \lambda \tr(E(u)) \I
\label{eq:StressTensor2}
\end{equation}
with the \textit{Lam\'e coefficients} $\mu$ and $\lambda$ derived from the Young's modulus $k$ and the Poisson ratio $\rho$.
The dynamic elasticity problem in strong form is then defined as
\begin{subequations}
\label{eq:StrongPDEdynamic_long}
\small
\begin{alignat}{4}
	m\ddot{u}-\divergence P(u) &=& f_B &\text{\ \ in  } \Omega^r \times \R_0^{+} \label{eq:StrongPDEdynamic:a} \\
	u &=& u_D &\text{\ \ on  } \Gamma^r_D \times \R_0^{+}  \\
	P(u) \cdot \mathbf{n} &=& f_S &\text{\ \ in  } \Gamma^r_N \times \R_0^{+} \\
	u &=& u^0 &\text{\ \ in  } \Omega^r \times \{0\} \\
	\dot{u} &=& \dot{u}^0 &\text{\ \ in  } \Omega^r \times \{0\} \ , 
\end{alignat}
\end{subequations}
with the mass $m$, Dirichlet boundaries $\Gamma^r_D$ and Neumann boundaries $\Gamma^r_N$.

\paragraph{Weak Form}%\label{app:PhysicalModel:WeakForm}
To obtain the weak form, let $V := H^1(\bar{\Omega}^r \rightarrow \R^d)$ be the space of test and trial functions.
Because Neumann and Dirichlet boundaries are enforced weakly, the space of test and trial functions coincide.
Starting from the right hand side of Eq.~\eqref{eq:StrongPDEdynamic:a}, the generalized divergence theorem yields:
\begin{smalleq}
\begin{aligned}
	\int_\Omega f_B \cdot v \dx &= \int_\Omega -\divergence P(u) \cdot v \dx \\
	 %& = \int_\Omega P(u) : \nabla v \dx - \int_{\partial\Omega} P(u) \cdot \mathbf{n} \cdot v \ds + \int_{\Omega}{m \ddot{u} v \dx}\\
	 & = \int_\Omega \sum_{j=1}^d{ \left( \mu \left( \nabla u_j + \frac{\partial}{\partial x_j}u \right) +\lambda \sum_{i=1}^d{\frac{\partial u_i}{\partial u_j}} \I_j \right) \cdot \nabla v_j} \dx \\
	 & \phantom{=} - \int_{\partial\Omega} \sum_{j=1}^d{ ( (\mu\left(\nabla u_j + \frac{\partial}{\partial x_j}u\right) } \\
	 & \phantom{=} + \lambda\sum_{i=1}^d{\frac{\partial u_i}{\partial u_j}} \I_j ) \cdot \mathbf{n} ) v_j  \ds + \int_{\Omega}{m \ddot{u} v \dx}.
\end{aligned}
\label{eq:PDEWeakForm}
\end{smalleq}

\subsection{Hexahedral Finite Element Discretization}\label{app:PhysicalModel:FEM}

\begin{figure}[htb]
\begin{tabular}{p{0.45\linewidth} p{0.452\linewidth}}
	\begin{tikzpicture}[x=0.6cm,y=0.6cm,z=0.25cm,scale=2,baseline=(current bounding box.center)]
	  % Labels
		\node[circle,fill=black!75!white,label=below:{1}] (v0) at (0,0,0) {};
		\node[circle,fill=black!75!white,label=below:{2}] (v1) at (1,0,0) {};
		\node[circle,fill=black!75!white,label=left:{3}] (v2) at (0,1,0) {};
		\node[circle,fill=black!75!white,label=below:{4}] (v3) at (1,1,0) {};
		\node[circle,fill=black!75!white] (v4) at  (0,0,1) {};
		\node[circle,fill=black!75!white,label=right:{6}] (v5) at (1,0,1) {};
		\node[circle,fill=black!75!white,label=above:{7}] (v6) at (0,1,1) {};
		\node[circle,fill=black!75!white,label=above:{8}] (v7) at (1,1,1) {};
		\node (x) at (1.6,0,0) {$x$};
		\node (y) at (0,1.6,0) {$y$};
		\node (z) at (1,0,2) {$z$};
		% Box
		\draw (0,0,0) -- (1,0,0) -- (1,1,0) -- (0,1,0) -- cycle;
		\draw (0,0,1) -- (1,0,1) -- (1,1,1) -- (0,1,1) -- cycle;
		\draw (0,0,0) -- (0,0,1);
		\draw (1,0,0) -- (1,0,1);
		\draw (1,1,0) -- (1,1,1);
		\draw (0,1,0) -- (0,1,1);
		\draw[->] (1,0,0) -- (x);
		\draw[->] (0,1,0) -- (y);
		\draw[->] (1,0,1) -- (z);
		% Object
		\fill[fill=blue,opacity=0.3] (0,0,0) -- (1,0,0) -- (1,0.3,0) -- (0,0.2,0) -- cycle;
		\fill[fill=blue,opacity=0.3] (1,0,0) -- (1,0,1) -- (1,0.5,1) -- (1,0.3,0) -- cycle;
		\filldraw[draw=blue!50!black, fill=blue!75!white,line join=bevel] (0,0.2,0) -- (1,0.3,0) -- (1,0.5,1) -- (0,0.4,1) -- cycle;
		% Box
		\draw (0,0,0) -- (1,0,0) -- (1,1,0) -- (0,1,0) -- cycle;
		\draw (1,1,0) -- (1,1,1);
		\draw (0,1,0) -- (0,1,1);
	\end{tikzpicture}
	&
	{\hspace{-1cm} \small$\!\begin{aligned}
		N_1(\mathbf{x}) &:= \left(1-\frac{x}{h}\right)\left(1-\frac{y}{h}\right)\left(1-\frac{z}{h}\right) \\
		N_2(\mathbf{x}) &:= \frac{x}{h}\left(1-\frac{y}{h}\right)\left(1-\frac{z}{h}\right) \\
		& \vdots \\
		N_8(\mathbf{x}) &:= \frac{x}{h} \frac{y}{h} \frac{z}{h} \\
	\end{aligned}$}
\end{tabular}
\caption{Trilinear hexahedral element and embedded surface}
\Description{The standard basis functions on the hexahedral cell and the corner labeling.}
\label{fig:BasisFunctions}
\end{figure}
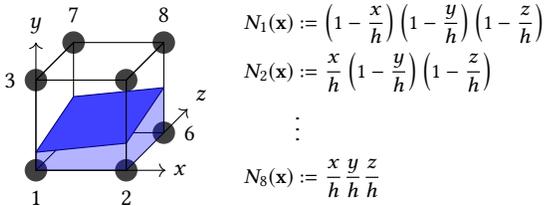

Trilinear shape functions are assigned to the finite hexahedral elements (see Fig.~\ref{fig:BasisFunctions}).

Thus, inside a cell,
any function $f$ can be approximated by trilinear interpolation of its values at the eight corners $\mathbf{v}_i$ as
\begin{equation}
	f(\mathbf{x}) \approx \sum_{i=1}^8{f(\mathbf{v}_i) N_i(\mathbf{x})} .
\label{eq:PartialIntegration3DAssumption}
\end{equation}

\paragraph{Partially Filled Cells}
We further embed the object boundary into the simulation grid, and consider cells that are partly filled with material.
For cells of size $[0,1]^3$, and the object given as signed distance function $\phi$,
we define the part of a cell that is contained in the object as
\begin{equation}
	\Omega^e := \{\mathbf{x} \in [0,1]^3 : \phi(\mathbf{x})\leq 1\} .
\label{eq:CellContents}
\end{equation}
For a trilinear interpolation within cells, integrals of arbitrary functions over cells can be approximated as
\begin{smalleq}
	\int_{\Omega^e} f(x) \dx \approx \int_{\Omega^e}{\sum_{i=1}^8{f(\mathbf{v}_i) N_i(x)} \dx} = \sum_{i=1}^8{f(\mathbf{v}_i) \underbrace{\int_{\Omega^e}{N_i(x) \dx}}_{=:w_v(e,i)}}
\label{eq:PartialIntegration3DVolume}
\end{smalleq}
where the integrals of the basis functions $w_v$ are precomputed and stored per cell.
As cell vertices do not necessarily lie on the object surface,
we use Nitsche's method
\cite{J.BenkM.UlbrichM.Mehl.2012, Juntunen.2009}
to incorporate Dirichlet boundaries.

%------------------------------------------------------------------------------------
\paragraph{Discrete Equations}%\label{sec:DiscreteEquations}
With the basis functions from Eq.~\eqref{eq:PartialIntegration3DVolume} we arrive at the following per-cell expressions:
\begin{align}
	M^e &= \int_{\Omega^e} \Phi^e(x)^T m \Phi^e(x) \dx \in \R^{24 \times 24} \label{eq:DiscreteM}\\
	K^e &= \int_{\Omega^e} B^e(x)^T C B^e(x) \dx \in \R^{24 \times 24} \label{eq:DiscreteK}\\
	f^e &= \int_{\Gamma_N^e} \Phi^e(s)^T \mathbf{f}_S(s) \ds \in \R^{24} \label{eq:Discretef}
\end{align}
Here, $\Phi^e(x) \in \R^{3 \times 24}$ and $B^e(x) \in \R^{6 \times 24}$, respectively, store for each coordinate the values of the basis functions $N_i$ and the derivatives. $C \in \R^{6 \times 6}$ is the regular material matrix for the chosen Lam\'e coefficients.

%====================================================================================
%====================================================================================
\paragraph{Corotation}%\label{app:PhysicalModel:Corotation}
To handle large rotations, we utilize the corotation formulation \cite{georgii2008CorotatedFE,Dick:2011:CUDAFEM,ChristianDick.2012,EftychiosD.Sifakis.2012,MichaelHauth.2003}.
First, the rotational part $R^e$ of the deformation of cell $e$ is extracted.
Let the average deformation gradient $F^e$ be computed as \cite{georgii2008CorotatedFE}
\begin{equation}
	F^e = \I_3 + \frac{1}{4h}\sum_{i=1}^8{\mathbf{u}_{s(e,i)} \begin{pmatrix} (-1)^i \\ (-1)^{\left\lceil i/2\right\rceil} \\ (-1)1^{\left\lceil i/4\right\rceil}\end{pmatrix}^T}.
\label{eq:CorotationJacobiGrid3D}
\end{equation}
The rotational component $R^e$ is then given by the polar decomposition $F^e = R^e S^e$ and can be computed with iterative 
procedures~\cite{MichaelHauth.2003,KenShoemake.}. 
If the polar decomposition should also handle flipped elements and inversions correctly, a more robust, but also more expensive 
Analytic Polar Decomposition~\cite{KKB18} would be needed (not used in this work).
Second, given $R^e$, the per-element term $K^e \mathbf{\underline{u}^e}$ is replaced by 
\begin{equation}
	T^e K^e ( (R^e)^T (\mathbf{\underline{x}^e} + \mathbf{\underline{u}^e}) - \mathbf{\underline{x}^e}).
\label{eq:CorotationalFormulation}
\end{equation}

%====================================================================================
%====================================================================================

%====================================================================================
%====================================================================================
\subsection{Blocked Expressions for Matrix Assembly}\label{app:Blocks}
One advantage of our chosen discretization is that the
per-element stiffness matrix $K^e$ (Eq.~\eqref{eq:DiscreteM}) has a $3\times3$ block structure
that is highly amenable to optimized implementations.
Each block describes the interactions between the coordinates of the two corresponding cell nodes.
Modulo index variations, the computation of these 3x3 blocks is identical for all 64 blocks. Furthermore,
the corotational strain formulation and Nitsche Dirichlet boundaries can be incorporated into the
block-wise decomposition in a straight forward way.

The regular block structure facilitates storing the stiffness matrix $K$ in a blocked compressed-sparse-row (CSR) format.
During assembly, work groups can process a single cell in parallel. % Each work group consists of 64 GPU threads that each processes one 3x3 block.
Reductions within the cell, as needed, e.g., for the computation of corotations and gradients, can be performed efficiently with warp-reductions.
By including analytic simplifications of the basis function evaluations (and their derivatives), we arrive at a highly
GPU-friendly algorithm that yields excellent performance.
%
%A more detailed discussion of the blockwise expressions of the stiffness matrix assembly is given in
%the following section.

The evaluation of the per-element stiffness matrix $K^e$ is performed blockwise.
Each of the $8 \times 8$ blocks $K_{i,j}^e\in\R{3 \times 3}$ are computed in parallel.
Let $\frac{\partial N_i(v_c)}{\partial \mathbf{x}_i}$ be the derivatives of the basis functions at the eight cell corner. This $8 \times 8 \times 3$ table only contains the entries $-\frac{1}{h}, 0$ or $\frac{1}{h}$. It is stored in constant memory on the GPU and hence allows fast access via the cache.
The per-block expression for the stiffness matrix $K^e$, see Eq.~\eqref{eq:DiscreteK} for the definition, is given by:
\begin{tinyeq}
\begin{aligned}
	K_{i,j}(x) =& 
	\begin{bmatrix}
		(2\mu+\lambda)\frac{\partial N_i(x)}{\partial x_1}\frac{\partial N_j(x)}{\partial x_1} + \mu\left(\frac{\partial N_i(x)}{\partial x_2}\frac{\partial N_j(x)}{\partial x_2} + \frac{\partial N_i(x)}{\partial x_3}\frac{\partial N_j(x)}{\partial x_3}\right) \\
		\mu\frac{\partial N_i(x)}{\partial x_1}\frac{\partial N_j(x)}{\partial x_2} + \lambda\frac{\partial N_i(x)}{\partial x_2}\frac{\partial N_j(x)}{\partial x_1} \\
		\mu\frac{\partial N_i(x)}{\partial x_1}\frac{\partial N_j(x)}{\partial x_3} + \lambda\frac{\partial N_i(x)}{\partial x_3}\frac{\partial N_j(x)}{\partial x_1}
	\end{bmatrix},\\
	&
	\begin{bmatrix}
		\mu\frac{\partial N_i(x)}{\partial x_2}\frac{\partial N_j(x)}{\partial x_1} + \lambda\frac{\partial N_i(x)}{\partial x_1}\frac{\partial N_j(x)}{\partial x_2} \\
		(2\mu+\lambda)\frac{\partial N_i(x)}{\partial x_2}\frac{\partial N_j(x)}{\partial x_2} + \mu\left(\frac{\partial N_i(x)}{\partial x_1}\frac{\partial N_j(x)}{\partial x_1} + \frac{\partial N_i(x)}{\partial x_3}\frac{\partial N_j(x)}{\partial x_3}\right) \\
		\mu\frac{\partial N_i(x)}{\partial x_2}\frac{\partial N_j(x)}{\partial x_3} + \lambda\frac{\partial N_i(x)}{\partial x_3}\frac{\partial N_j(x)}{\partial x_2}
	\end{bmatrix},\\
	&
	\begin{bmatrix}
		\mu\frac{\partial N_i(x)}{\partial x_3}\frac{\partial N_j(x)}{\partial x_1} + \lambda\frac{\partial N_i(x)}{\partial x_1}\frac{\partial N_j(x)}{\partial x_3}\\
		\mu\frac{\partial N_i(x)}{\partial x_3}\frac{\partial N_j(x)}{\partial x_2} + \lambda\frac{\partial N_i(x)}{\partial x_2}\frac{\partial N_j(x)}{\partial x_3}\\
		(2\mu+\lambda)\frac{\partial N_i(x)}{\partial x_3}\frac{\partial N_j(x)}{\partial x_3} + \mu\left(\frac{\partial N_i(x)}{\partial x_1}\frac{\partial N_j(x)}{\partial x_1} + \frac{\partial N_i(x)}{\partial x_2}\frac{\partial N_j(x)}{\partial x_2}\right)
	\end{bmatrix} .
\end{aligned}
\label{eq:StiffnessPart3D}
\end{tinyeq}

%====================================================================================
%====================================================================================

\subsection{Nitsche Dirichlet Boundaries}\label{app:Blocks:KDe}
To incorporate Dirichlet boundaries with Nitsche's method, the weak form from Eq.~\ref{eq:DiscreteM} is extended using productive zeros $u-u_0$:
\begin{smalleq}
	-\int_{\Gamma^r_N} P(u) \cdot \mathbf{n} \cdot v \ds - \int_{\Gamma^r_N} P(v) \cdot \mathbf{n} \cdot (u-u_0) \ds - \eta \int_{\Gamma^r_N} (u-u_0) \cdot v \ds .
\label{eq:DirichletNitsche}
\end{smalleq}
The first term makes the resulting linear system symmetric. The second term enforces the Dirichlet boundaries. The third term acts as a regularizer and the parameter $\eta$ has to be chosen as $\eta \geq ch^{-1}$ with $h$ being the grid size and $c$ a sufficient large constant. In our experiments, we chose $10^8$ for stable results.

These boundary conditions can also be formulated in am efficient, blocked matrix form.
First, the definition of $P(u)$ (Eq.~\eqref{eq:StressTensor2}) is used, leading to the following terms that are added to the weak form (Eq.~\eqref{eq:PDEWeakForm}):
\begin{tinyeq}
\begin{aligned}
	-& \underbrace{\int_{\Gamma_D^e}{\sum_{j=1}^3{\left( \mu\left(\nabla u_j \cdot n + \frac{\partial u}{\partial x_j} \cdot n\right) + \lambda n_j \sum_{i=1}^3\frac{\partial u_i}{\partial x_i} \right) v_j} \ds}}_{\text{(\rom{1})}} \\
	-&\underbrace{\int_{\Gamma_D^e}{\sum_{j=1}^3{\left( \mu\left(\nabla v_j \cdot n + \frac{\partial v}{\partial x_j} \cdot n\right) + \lambda n_j \sum_{i=1}^3\frac{\partial v_i}{\partial x_i} \right) u_j} \ds}}_{\text{(\rom{1}')}} \\
	-&\underbrace{\eta \int_{\Gamma_D^e}{u \cdot v \ds}}_{\text{(\rom{2})}}
\end{aligned}
\label{eq:Nitsche3D1}
\end{tinyeq}
Expressing $u$ and $v$ using the basis functions and their derivatives, combined in $\Phi^e$ and $B^e$, we obtain
\begin{equation}
	K^e \minuseq\int_{\Gamma_D^e}{\Phi^e(x)^T D^e B^e(x) \ds}
\label{eq:Nitsche3D2}
\end{equation}
for (\ref{eq:Nitsche3D1}.\rom{1}) with $D^e$ given by
\begin{tinyeq}
	D^e := 
	\begin{pmatrix}
		2\mu n_1 + \lambda n_1	& \lambda n_1							& \lambda n_1						 \\
		\lambda n_2							& 2\mu n_2 + \lambda n_2	& \lambda n_2						 \\
		\lambda n_3							& \lambda n_3							& 2\mu n_3 + \lambda n_3 
	\end{pmatrix} .
\label{eq:Nitsche3D3}
\end{tinyeq}
Here, the equation for (\ref{eq:Nitsche3D1}.\rom{1}') is the above equation, just transposed.
This integral is evaluated by computing the sum of the values at the eight corners weighted by $w_b(e,c)$.
The value of $\Phi_i$ evaluated at vertex $c$ has the special property that $\Phi_i=\I_3\,\mathds{1}_{i=c}$. 
This allows us to derive the following simplification and solution of Eq.~\eqref{eq:Nitsche3D2}, combining 
both (\ref{eq:Nitsche3D1}.\rom{1}) and (\ref{eq:Nitsche3D1}.\rom{1}'):
\begin{align}
	\text{KD}_{j,c} :=& 
	\begin{bmatrix}
		B_1 (\lambda n_1 + 2\mu n_1) + B_2 \mu n_2 + B_3 \mu n_3 \\
		B_2 \mu n_1 + B_1 \lambda n_2 \\
		B_3 \mu n_1 + B_1 \lambda n_3
	\end{bmatrix}, \nonumber \\
	&
	\begin{bmatrix}
		B_2 \lambda n_1 + B_1 \mu n_2 \\
		B_1 \mu n_1 + B_2 (\lambda n_2 + 2\mu n_2) + B_3 \mu n_3 \\
		B_3 \mu n_2 + B_1 \lambda n_3
	\end{bmatrix}, \nonumber \\
	&
	\begin{bmatrix}
		B_3 \lambda n_1 + B_1 \mu n_3 \\
		B_3 \lambda n_2 + B_2 \mu n_3 \\
		B_1 \mu n_1 + B_2 \mu n_2 + B_3 (\lambda n_3 + 2\mu n_3)
	\end{bmatrix} \in \R^{3 \times 3} \\
	& \text{with } B_i:=\frac{\partial N_j(v_c)}{\partial x_i} \nonumber \\
	K_{i,j} \minuseq & w_b(e,i) \text{KD}_{j,i} + w_b(e,j) \text{KD}_{i,j}^T  .
\label{eq:Nitsche3D4}
\end{align}
Part (\ref{eq:Nitsche3D1}.\rom{2}) even simplifies to the following expression:
\begin{equation}
	K_{i,j} \minuseq 1_{i=j} \eta w_b(e,i) \I_3 .
\label{eq:Nitsche3D5}
\end{equation}

%====================================================================================
%====================================================================================

\section{Additional Adjoint Code Example}\label{app:Adjoint}

The adjoint code of the damping parameters is presented here as an example.
For a variable $x$ of the forward step, $\hat{x}$ denotes the adjoint/gradient of that variable.
One simulation timestep is split as follows:
\begin{enumerate}
	\item Computation of stiffness matrix $K$.
	\item Computation of Rayleigh damping matrix $D = \alpha_1 M + \alpha_2 K$ where $\alpha_1$ and $\alpha_2$, respectively, are the damping on mass and stiffness.
	\item Newmark time integration Eq.~\eqref{eq:Newmark1}.
\end{enumerate}

In the adjoint code, these steps are performed in reverse order.
One adjoint simulation timestep is split as follows:
\begin{enumerate}
	\item Adjoint of Newmark time integration $\rightarrow$ $\hat{D}$
	\item Adjoint of Rayleigh damping:
		\begin{subequations}
		\begin{alignat}{2}
			\hat{M} &= \alpha_1 \hat{D} \ , \ 
			\hat{K}  = \alpha_2 \hat{D} \\
			\hat{\alpha}_1 &= \text{vec}(M) \bullet \hat{D} \ , \ 
			\hat{\alpha}_2  = \text{vec}(K) \bullet \hat{D} .
		\end{alignat}
		\end{subequations}
	\item Adjoint of stiffness matrix using $\hat{K}$ among others.
\end{enumerate}
The adjoint variables $\hat{\alpha}_1$ and $\hat{\alpha}_2$ are summed up for every timestep, giving rise to the final gradients for these parameters.

%====================================================================================
%====================================================================================

\section{Extended Stability Analysis}\label{app:Stability}

In the following we demonstrate
the robustness of our solver by comparing reconstructions to synthetic ground
truth values.

\begin{figure}[h]
    \centering
    \begin{subfigure}{0.24\linewidth}
    	%\includegraphics[trim=400 0 400 0,clip,width=\textwidth]{images/Torus/TorusMin_Rendering3.jpg}%
    	%\caption{Minimal initial Young's modulus}
    	%\label{fig:result:torus:min}
    	\begin{overpic}[trim=400 0 400 0,clip,width=\textwidth]{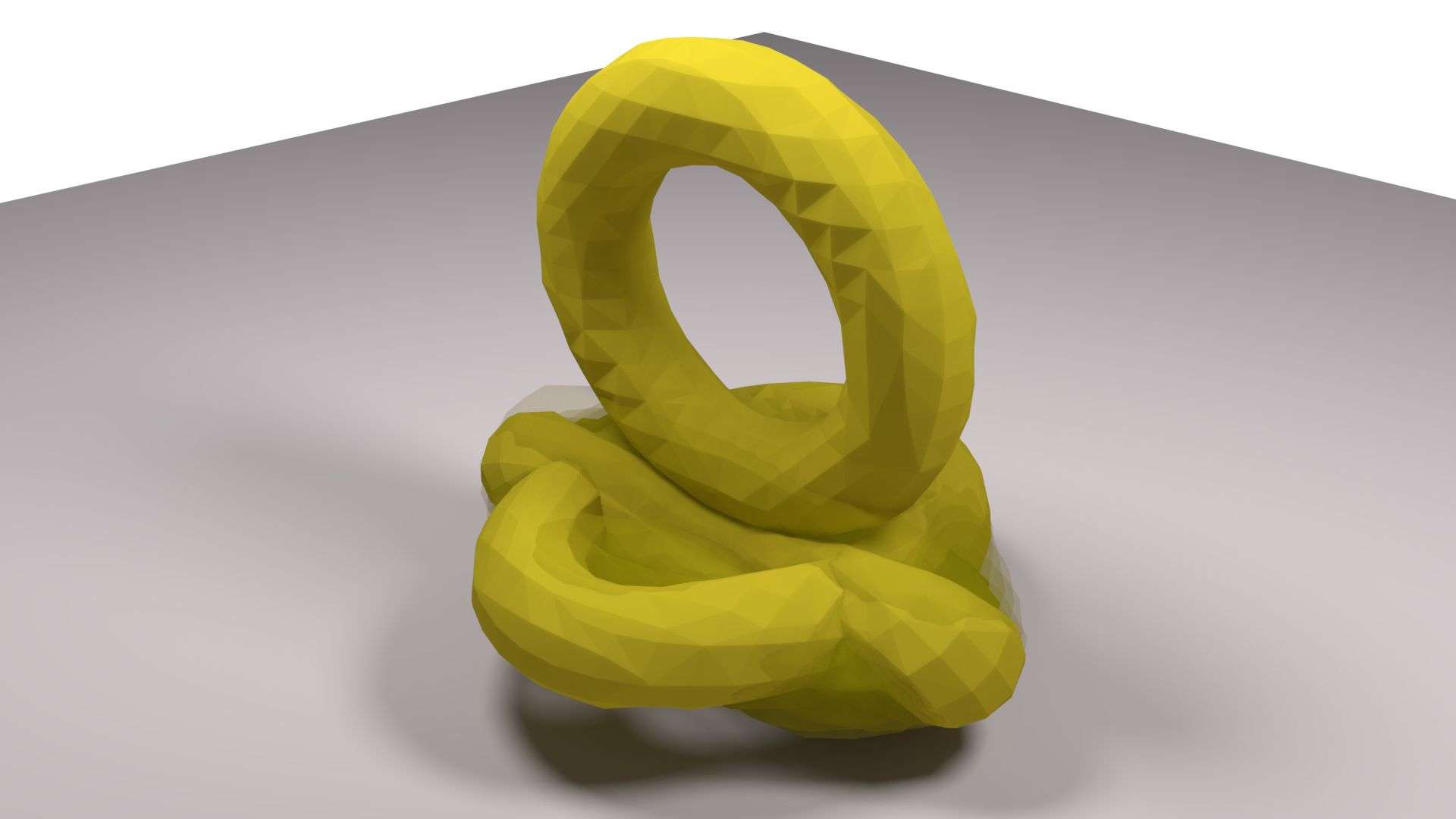}%
             \put(5,5){\textcolor{white}{(a)}}%
        \end{overpic}%
    \end{subfigure}
    \begin{subfigure}{0.24\linewidth}
    	%\includegraphics[trim=400 0 400 0,clip,width=\textwidth]{images/Torus/TorusMax_Rendering3.jpg}%
    	%\caption{Maximal initial Young's modulus}
    	%\label{fig:result:torus:max}
    	\begin{overpic}[trim=400 0 400 0,clip,width=\textwidth]{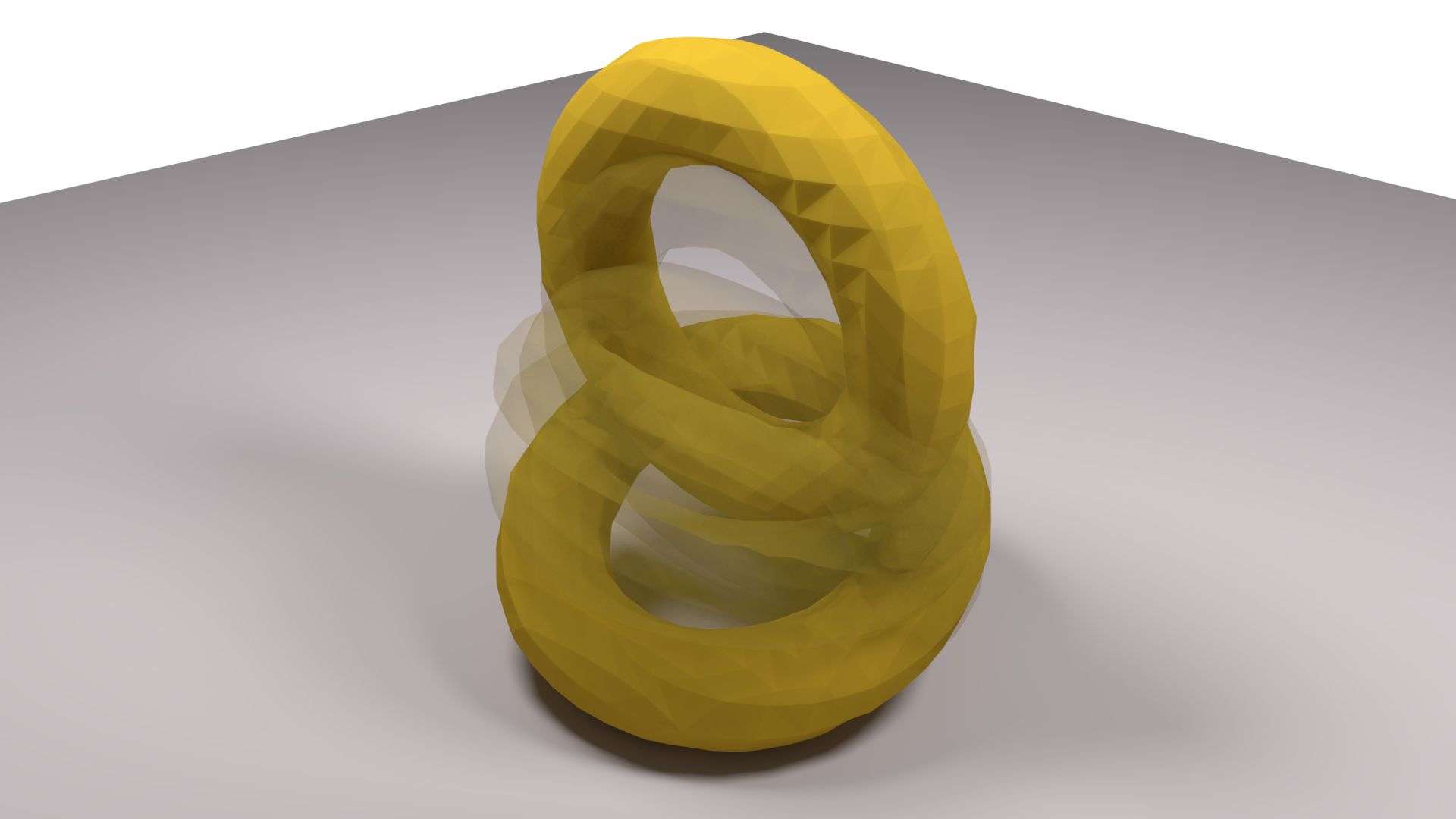}%
             \put(5,5){\textcolor{white}{(b)}}%
        \end{overpic}%
    \end{subfigure}
    \begin{subfigure}{0.24\linewidth}
    	%\includegraphics[trim=400 0 400 0,clip,width=\textwidth]{images/Torus/TorusRec_Rendering4.jpg}%
    	%\caption{Reconstruction}
    	%\label{fig:result:torus:rec}
    	\begin{overpic}[trim=400 0 400 0,clip,width=\textwidth]{images/Torus/TorusRec_Rendering3.jpg}%
             \put(5,5){\textcolor{white}{(c)}}%
        \end{overpic}%
    \end{subfigure}
    \begin{subfigure}{0.24\linewidth}
    	%\includegraphics[trim=400 0 400 0,clip,width=\textwidth]{images/Torus/TorusGT_Rendering3.jpg}%
    	%\caption{Ground truth Young's modulus}
    	%\label{fig:result:torus:gt}
    	\begin{overpic}[trim=400 0 400 0,clip,width=\textwidth]{images/Torus/TorusGT_Rendering3.jpg}%
             \put(5,5){\textcolor{white}{(d)}}%
        \end{overpic}%
    \end{subfigure}
    \caption{For different noise and camera settings, our optimizer was used to estimate the torus Young's modulus. From left to right, images show simulations using the minimal, maximal, reconstructed and ground truth values.
    Our reconstruction (c) closely matches the ground truth in (d).}
    %\caption{\sebi{Minimal, maximal, reconstructed and ground truth values of the Young's modulus when optimizing under different noise and camera settings.}}
    \label{fig:result:torus}
\end{figure}

%---------------------------------------------------------------------------------------------
%---------------------------------------------------------------------------------------------
\subsection{Gradient Stability for Varying Number of Timesteps and Noise}\label{app:Stability:TimeAndNoise}

%\sebi{Next, we analyze the influence of the number of timesteps and noise on the camera observations.}

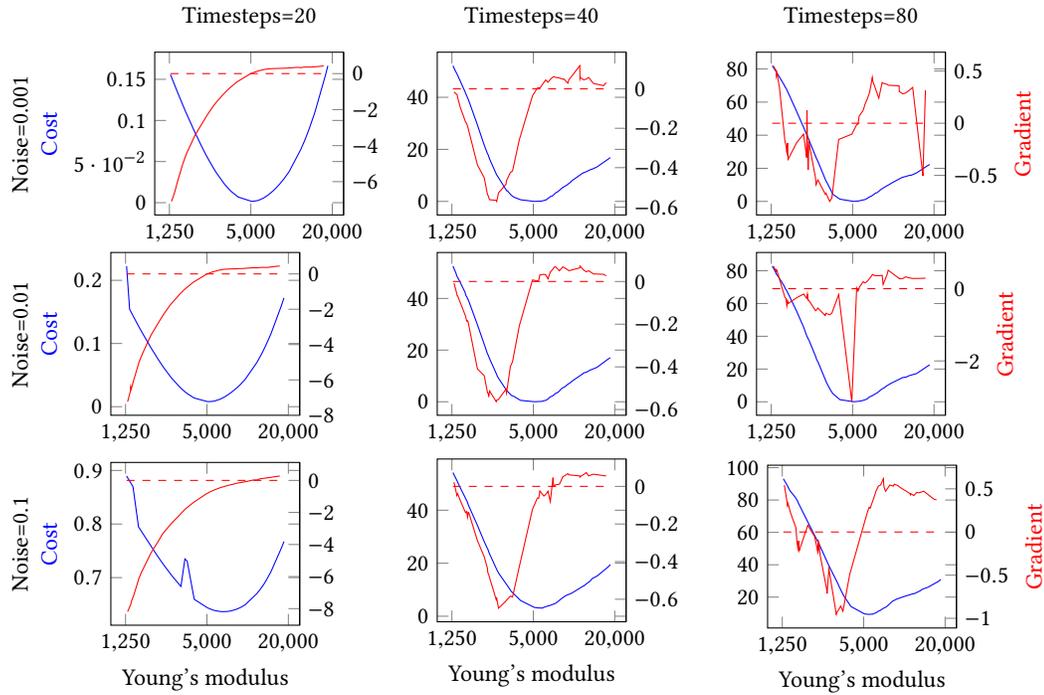
\begin{figure*}
\begin{subfigure}{0.2\textwidth}
	\begin{tikzpicture}
		\begin{axis}[
				title={Timesteps=20},
				xlabel={},
				ylabel style={align=center},
				ylabel={Noise=0.001\\\textcolor{blue}{Cost}},
				xmode=log,
				scale only axis,
                width=0.7\linewidth,
                xtick={1250,5000,20000},
                %ytick scale label code/.code={\pgfmathparse{int(-#1)}\textcolor{blue}{Cost$ \cdot 10^{\pgfmathresult}$}},
                %every y tick scale label/.style={at={(yticklabel cs:0.5)}, color=blue, anchor = south, rotate = 90},
                log ticks with fixed point,
				]
			\addplot[blue] table[col sep=tab] {plots/stability/TimeAndNoise/cost1.dat};
		\end{axis}
		\begin{axis}[
				xlabel={},
				xmode=log,
				title={},
				axis y line*=right,
				axis x line=none,
				scale only axis,
                width=0.7\linewidth,
                xtick={1250,5000,20000},
                ytick scale label code/.code={},
                every y tick scale label/.style={at={(yticklabel cs:0.5)}, color=red, anchor = south, rotate = 90},
                log ticks with fixed point,
				]
			\addplot[red] table[col sep=tab] {plots/stability/TimeAndNoise/grad1.dat};
			\addplot[mark=none, red, dashed, samples=2, domain=1250:20000] {0};
		\end{axis}
	\end{tikzpicture}
\end{subfigure}
\hspace{1.5cm}
\begin{subfigure}{0.2\textwidth}
	\begin{tikzpicture}
		\begin{axis}[
				title={Timesteps=40},
				xlabel={},
				xmode=log,
				scale only axis,
                width=0.7\linewidth,
                xtick={1250,5000,20000},
                ytick scale label code/.code={},
                every y tick scale label/.style={at={(yticklabel cs:0.5)}, color=blue, anchor = south, rotate = 90},
                log ticks with fixed point,
				]
			\addplot[blue] table[col sep=tab] {plots/stability/TimeAndNoise/cost4.dat};
		\end{axis}
		\begin{axis}[
				xlabel={},
				xmode=log,
				title={},
				axis y line*=right,
				axis x line=none,
				scale only axis,
                width=0.7\linewidth,
                xtick={1250,5000,20000},
                ytick scale label code/.code={},
                every y tick scale label/.style={at={(yticklabel cs:0.5)}, color=red, anchor = south, rotate = 90},
                log ticks with fixed point,
				]
			\addplot[red] table[col sep=tab] {plots/stability/TimeAndNoise/grad4.dat};
			\addplot[mark=none, red, dashed, samples=2, domain=1250:20000] {0};
		\end{axis}
	\end{tikzpicture}
\end{subfigure}
\hspace{0.5cm}
\begin{subfigure}{0.2\textwidth}
	\begin{tikzpicture}
		\begin{axis}[
				title={Timesteps=80},
				xlabel={},
				xmode=log,
				scale only axis,
                width=0.7\linewidth,
                xtick={1250,5000,20000},
                ytick scale label code/.code={},
                every y tick scale label/.style={at={(yticklabel cs:0.5)}, color=blue, anchor = south, rotate = 90},
                log ticks with fixed point,
				]
			\addplot[blue] table[col sep=tab] {plots/stability/TimeAndNoise/cost7.dat};
		\end{axis}
		\begin{axis}[
				xlabel={},
				xmode=log,
				ylabel={\textcolor{red}{Gradient}},
				title={},
				axis y line*=right,
				axis x line=none,
				scale only axis,
                width=0.7\linewidth,
                xtick={1250,5000,20000},
                ytick scale label code/.code={\pgfmathparse{int(-#1)}\textcolor{red}{Gradient$ \cdot 10^{\pgfmathresult}$}},
                every y tick scale label/.style={at={(yticklabel cs:0.5)}, color=red, anchor = south, rotate = 90},
                log ticks with fixed point,
				]
			\addplot[red] table[col sep=tab] {plots/stability/TimeAndNoise/grad7.dat};
			\addplot[mark=none, red, dashed, samples=2, domain=1250:20000] {0};
		\end{axis}
	\end{tikzpicture}
\end{subfigure}
\\
\begin{subfigure}{0.2\textwidth}
	\begin{tikzpicture}
		\begin{axis}[
				title={},
				xlabel={},
				ylabel style={align=center},
				ylabel={Noise=0.01\\\textcolor{blue}{Cost}},
				xmode=log,
				scale only axis,
                width=0.7\linewidth,
                xtick={1250,5000,20000},
                %ytick scale label code/.code={\pgfmathparse{int(-#1)}\textcolor{blue}{Cost$ \cdot 10^{\pgfmathresult}$}},
                %every y tick scale label/.style={at={(yticklabel cs:0.5)}, color=blue, anchor = south, rotate = 90},
                log ticks with fixed point,
				]
			\addplot[blue] table[col sep=tab] {plots/stability/TimeAndNoise/cost2.dat};
		\end{axis}
		\begin{axis}[
				xlabel={},
				xmode=log,
				title={},
				axis y line*=right,
				axis x line=none,
				scale only axis,
                width=0.7\linewidth,
                xtick={1250,5000,20000},
                ytick scale label code/.code={},
                every y tick scale label/.style={at={(yticklabel cs:0.5)}, color=red, anchor = south, rotate = 90},
                log ticks with fixed point,
				]
			\addplot[red] table[col sep=tab] {plots/stability/TimeAndNoise/grad2.dat};
			\addplot[mark=none, red, dashed, samples=2, domain=1250:20000] {0};
		\end{axis}
	\end{tikzpicture}
\end{subfigure}
\hspace{1.5cm}
\begin{subfigure}{0.2\textwidth}
	\begin{tikzpicture}
		\begin{axis}[
				title={},
				xlabel={},
				xmode=log,
				scale only axis,
                width=0.7\linewidth,
                xtick={1250,5000,20000},
                ytick scale label code/.code={},
                every y tick scale label/.style={at={(yticklabel cs:0.5)}, color=blue, anchor = south, rotate = 90},
                log ticks with fixed point,
				]
			\addplot[blue] table[col sep=tab] {plots/stability/TimeAndNoise/cost5.dat};
		\end{axis}
		\begin{axis}[
				xlabel={},
				xmode=log,
				title={},
				axis y line*=right,
				axis x line=none,
				scale only axis,
                width=0.7\linewidth,
                xtick={1250,5000,20000},
                ytick scale label code/.code={},
                every y tick scale label/.style={at={(yticklabel cs:0.5)}, color=red, anchor = south, rotate = 90},
                log ticks with fixed point,
				]
			\addplot[red] table[col sep=tab] {plots/stability/TimeAndNoise/grad5.dat};
			\addplot[mark=none, red, dashed, samples=2, domain=1250:20000] {0};
		\end{axis}
	\end{tikzpicture}
\end{subfigure}
\hspace{0.5cm}
\begin{subfigure}{0.2\textwidth}
	\begin{tikzpicture}
		\begin{axis}[
				title={},
				xlabel={},
				xmode=log,
				scale only axis,
                width=0.7\linewidth,
                xtick={1250,5000,20000},
                ytick scale label code/.code={},
                every y tick scale label/.style={at={(yticklabel cs:0.5)}, color=blue, anchor = south, rotate = 90},
                log ticks with fixed point,
				]
			\addplot[blue] table[col sep=tab] {plots/stability/TimeAndNoise/cost8.dat};
		\end{axis}
		\begin{axis}[
				xlabel={},
				xmode=log,
				ylabel={\textcolor{red}{Gradient}},
				title={},
				axis y line*=right,
				axis x line=none,
				scale only axis,
                width=0.7\linewidth,
                xtick={1250,5000,20000},
                ytick scale label code/.code={\pgfmathparse{int(-#1)}\textcolor{red}{Gradient$ \cdot 10^{\pgfmathresult}$}},
                every y tick scale label/.style={at={(yticklabel cs:0.5)}, color=red, anchor = south, rotate = 90},
                log ticks with fixed point,
				]
			\addplot[red] table[col sep=tab] {plots/stability/TimeAndNoise/grad8.dat};
			\addplot[mark=none, red, dashed, samples=2, domain=1250:20000] {0};
		\end{axis}
	\end{tikzpicture}
\end{subfigure}
\\
\begin{subfigure}{0.2\textwidth}
	\begin{tikzpicture}
		\begin{axis}[
				title={},
				xlabel={Young's modulus},
				ylabel style={align=center},
				ylabel={Noise=0.1\\\textcolor{blue}{Cost}},
				xmode=log,
				scale only axis,
                width=0.7\linewidth,
                xtick={1250,5000,20000},
                %ytick scale label code/.code={\pgfmathparse{int(-#1)}\textcolor{blue}{Cost$ \cdot 10^{\pgfmathresult}$}},
                %every y tick scale label/.style={at={(yticklabel cs:0.5)}, color=blue, anchor = south, rotate = 90},
                log ticks with fixed point,
				]
			\addplot[blue] table[col sep=tab] {plots/stability/TimeAndNoise/cost3.dat};
		\end{axis}
		\begin{axis}[
				xlabel={},
				xmode=log,
				title={},
				axis y line*=right,
				axis x line=none,
				scale only axis,
                width=0.7\linewidth,
                xtick={1250,5000,20000},
                ytick scale label code/.code={},
                every y tick scale label/.style={at={(yticklabel cs:0.5)}, color=red, anchor = south, rotate = 90},
                log ticks with fixed point,
				]
			\addplot[red] table[col sep=tab] {plots/stability/TimeAndNoise/grad3.dat};
			\addplot[mark=none, red, dashed, samples=2, domain=1250:20000] {0};
		\end{axis}
	\end{tikzpicture}
\end{subfigure}
\hspace{1.5cm}
\begin{subfigure}{0.2\textwidth}
	\begin{tikzpicture}
		\begin{axis}[
				title={},
				xlabel={Young's modulus},
				xmode=log,
				scale only axis,
                width=0.7\linewidth,
                xtick={1250,5000,20000},
                ytick scale label code/.code={},
                every y tick scale label/.style={at={(yticklabel cs:0.5)}, color=blue, anchor = south, rotate = 90},
                log ticks with fixed point,
				]
			\addplot[blue] table[col sep=tab] {plots/stability/TimeAndNoise/cost6.dat};
		\end{axis}
		\begin{axis}[
				xlabel={},
				xmode=log,
				title={},
				axis y line*=right,
				axis x line=none,
				scale only axis,
                width=0.7\linewidth,
                xtick={1250,5000,20000},
                ytick scale label code/.code={},
                every y tick scale label/.style={at={(yticklabel cs:0.5)}, color=red, anchor = south, rotate = 90},
                log ticks with fixed point,
				]
			\addplot[red] table[col sep=tab] {plots/stability/TimeAndNoise/grad6.dat};
			\addplot[mark=none, red, dashed, samples=2, domain=1250:20000] {0};
		\end{axis}
	\end{tikzpicture}
\end{subfigure}
\hspace{0.5cm}
\begin{subfigure}{0.2\textwidth}
	\begin{tikzpicture}
		\begin{axis}[
				title={},
				xlabel={Young's modulus},
				xmode=log,
				scale only axis,
                width=0.7\linewidth,
                xtick={1250,5000,20000},
                ytick scale label code/.code={},
                every y tick scale label/.style={at={(yticklabel cs:0.5)}, color=blue, anchor = south, rotate = 90},
                log ticks with fixed point,
				]
			\addplot[blue] table[col sep=tab] {plots/stability/TimeAndNoise/cost9.dat};
		\end{axis}
		\begin{axis}[
				xlabel={},
				xmode=log,
				ylabel={\textcolor{red}{Gradient}},
				title={},
				axis y line*=right,
				axis x line=none,
				scale only axis,
                width=0.7\linewidth,
                xtick={1250,5000,20000},
                ytick scale label code/.code={\pgfmathparse{int(-#1)}\textcolor{red}{Gradient$ \cdot 10^{\pgfmathresult}$}},
                every y tick scale label/.style={at={(yticklabel cs:0.5)}, color=red, anchor = south, rotate = 90},
                log ticks with fixed point,
				]
			\addplot[red] table[col sep=tab] {plots/stability/TimeAndNoise/grad9.dat};
			\addplot[mark=none, red, dashed, samples=2, domain=1250:20000] {0};
		\end{axis}
	\end{tikzpicture}
\end{subfigure}
\caption{Influence of number of timesteps and camera noise on the cost function and gradient estimates.
Test were performed on the torus data set with $\phi_{\text{max}}=5$.}
\label{fig:Stability:noiseTime}
\end{figure*}

We first investigate the robustness of the gradients computed
by our differentiable solver for the SSCs
when varying the number of steps over time, and when introducing noise.
The following tests are
performed with the torus data set shown in \autoref{fig:result:torus},
and show gradient evaluations when varying the Young's modulus estimate
around a ground truth value of 5000 along the x-axis. I.e. we expect the cost (shown in blue in the
following graphs) to have a minimum at 5000, while the gradient (shown in red in terms of it's magnitude) should be negative to the left of 5000, and positive on the right side.

The insights we gain from the plots in \autoref{fig:Stability:noiseTime} are twofold: First, the more timesteps are simulated, the more do numerical errors in the adjoint pass accumulate and the noisier the gradients become. Note, however, that the torus a rather difficult test case as it exhibits strong deformations.
Despite this effect, the gradients retain the correct sign, i.e., overall direction of the gradient, visible in \autoref{fig:Stability:noiseTime} from the fraction of the red curves above and below the dashed red line. The negative parts lie on the left side of the ground truth value of 5000, while positive gradients lie on the right side.
%The real world test cases don't have that large deformations, so that the adjoint pass becomes more stable and even more than 100 timesteps can be simulated without problems.
Second, the simulation becomes more stable with increasing noise magnitude. 
The increase in noise leads to a smoothing of the point assignments, reducing outliers, and hence smoothing the cost function.
%We believe that this is because larger noise values smooth over point mismatches in the cost function and tend to reduce the number of outliers.
Furthermore, note that the absolute value of the cost function is larger %in the optimum 
for higher noise values than for lower ones. This is because even in the optimal case, the observations can be quite far from the surface due to the noise.

%---------------------------------------------------------------------------------------------
%---------------------------------------------------------------------------------------------
\subsection{Finite difference based Gradient Estimation}\label{app:Stability:FiniteDifferences}

Next, we compare our proposed gradient estimation to a finite-difference based estimation. In particular, we shed light on the influence of finite-difference based gradient estimation using the hyper-parameter $\Delta x$ in the gradient approximation $f'(x) \approx \frac{f(x + \Delta x)-f(x)}{\Delta x}$ on reconstruction accuracy.

%The cost function looks quite smooth in the above plots, so one might think that %computing the gradients via finite differences might be a better approach than the %adjoint method which requires quite a lot of programming.}

%\sebi{From a conceptual perspective, finite differences already have two disadvantages:
%\begin{itemize}
%    \item Performance: finite differences require one extra forward pass for each %variable. The adjoint method requires only one backward pass which costs about 150\% of %a forward pass, independent of the number of parameters. Hence, the adjoint method is %already faster for only two parameters.
%    \item Hyper-parameters: finite differences require an additional parameter, the %delta $\Delta x$ on where to evaluate the gradient ($f'(x) \approx \frac{f(x + \Delta %x)-f(x)}{\Delta x}$).
%\end{itemize}

%This hyper-parameter $\Delta x$ has a strong influence of the robustness of the finite %difference method.}

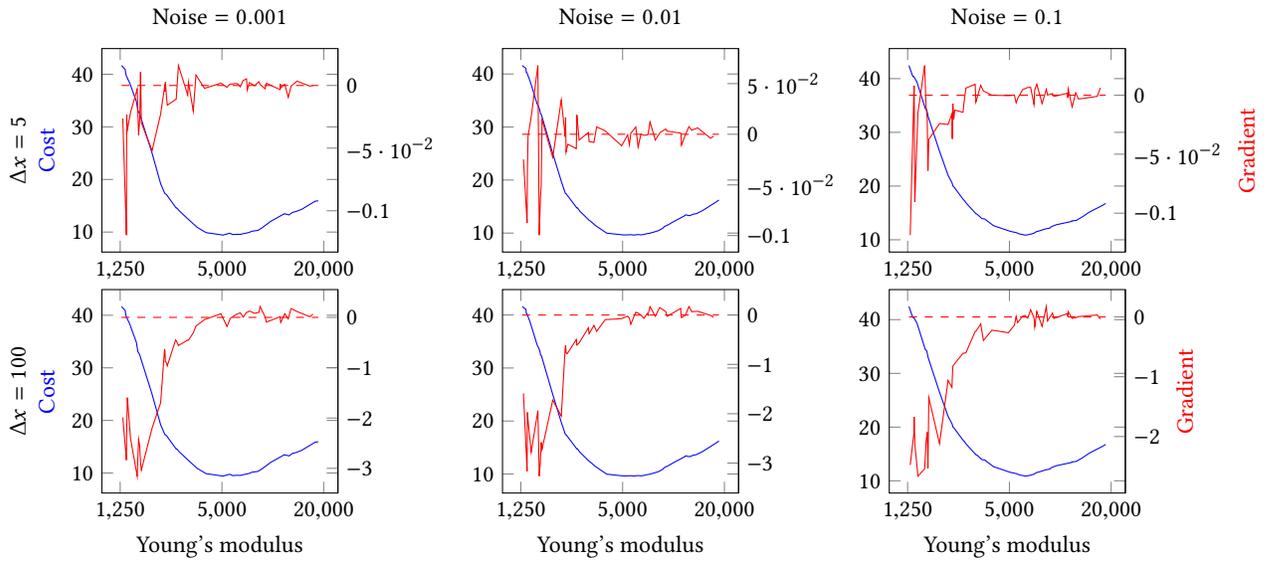
\begin{figure*}
\begin{subfigure}{0.25\textwidth}
	\begin{tikzpicture}
		\begin{axis}[
				title={$\text{Noise}=0.001$},
				xlabel={},
				ylabel style={align=center},
				ylabel={$\Delta x = 5$\\\textcolor{blue}{Cost}},
				xmode=log,
				scale only axis,
                width=0.7\linewidth,
                xtick={1250,5000,20000},
                log ticks with fixed point,
				]
			\addplot[blue] table[col sep=tab, x index=0,y index=1] {plots/stability/fd/TorusFD_t40_n1m_d5.dat};
		\end{axis}
		\begin{axis}[
				xlabel={},
				xmode=log,
				title={},
				axis y line*=right,
				axis x line=none,
				scale only axis,
                width=0.7\linewidth,
                xtick={1250,5000,20000},
                ytick scale label code/.code={},
                every y tick scale label/.style={at={(yticklabel cs:0.5)}, color=red, anchor = south, rotate = 90},
                log ticks with fixed point,
				]
			\addplot[red] table[col sep=tab, x index=0,y index=2] {plots/stability/fd/TorusFD_t40_n1m_d5.dat};
			\addplot[mark=none, red, dashed, samples=2, domain=1250:20000] {0};
		\end{axis}
	\end{tikzpicture}
\end{subfigure}
\hspace{1.5cm}
\begin{subfigure}{0.25\textwidth}
	\begin{tikzpicture}
		\begin{axis}[
				title={$\text{Noise}=0.01$},
				xlabel={},
				ylabel style={align=center},
				ylabel={},
				xmode=log,
				scale only axis,
                width=0.7\linewidth,
                xtick={1250,5000,20000},
                log ticks with fixed point,
				]
			\addplot[blue] table[col sep=tab, x index=0,y index=1] {plots/stability/fd/TorusFD_t40_n10m_d5.dat};
		\end{axis}
		\begin{axis}[
				xlabel={},
				xmode=log,
				title={},
				axis y line*=right,
				axis x line=none,
				scale only axis,
                width=0.7\linewidth,
                xtick={1250,5000,20000},
                ytick scale label code/.code={},
                every y tick scale label/.style={at={(yticklabel cs:0.5)}, color=red, anchor = south, rotate = 90},
                log ticks with fixed point,
				]
			\addplot[red] table[col sep=tab, x index=0,y index=2] {plots/stability/fd/TorusFD_t40_n10m_d5.dat};
			\addplot[mark=none, red, dashed, samples=2, domain=1250:20000] {0};
		\end{axis}
	\end{tikzpicture}
\end{subfigure}
\hspace{0.5cm}
\begin{subfigure}{0.25\textwidth}
	\begin{tikzpicture}
		\begin{axis}[
				title={$\text{Noise}=0.1$},
				xlabel={},
				xmode=log,
				scale only axis,
                width=0.7\linewidth,
                xtick={1250,5000,20000},
                ytick scale label code/.code={},
                every y tick scale label/.style={at={(yticklabel cs:0.5)}, color=blue, anchor = south, rotate = 90},
                log ticks with fixed point,
				]
			\addplot[blue] table[col sep=tab, x index=0,y index=1] {plots/stability/fd/TorusFD_t40_n100m_d5.dat};
		\end{axis}
		\begin{axis}[
				xlabel={},
				xmode=log,
				ylabel={\textcolor{red}{Gradient}},
				title={},
				axis y line*=right,
				axis x line=none,
				scale only axis,
                width=0.7\linewidth,
                xtick={1250,5000,20000},
                ytick scale label code/.code={\pgfmathparse{int(-#1)}\textcolor{red}{Gradient$ \cdot 10^{\pgfmathresult}$}},
                every y tick scale label/.style={at={(yticklabel cs:0.5)}, color=red, anchor = south, rotate = 90},
                log ticks with fixed point,
				]
			\addplot[red] table[col sep=tab, x index=0,y index=2] {plots/stability/fd/TorusFD_t40_n100m_d5.dat};
			\addplot[mark=none, red, dashed, samples=2, domain=1250:20000] {0};
		\end{axis}
	\end{tikzpicture}
\end{subfigure}
\\
\begin{subfigure}{0.25\textwidth}
	\begin{tikzpicture}
		\begin{axis}[
				title={},
				xlabel={Young's modulus},
				ylabel style={align=center},
				ylabel={$\Delta x = 100$\\\textcolor{blue}{Cost}},
				xmode=log,
				scale only axis,
                width=0.7\linewidth,
                xtick={1250,5000,20000},
                log ticks with fixed point,
				]
			\addplot[blue] table[col sep=tab, x index=0,y index=1] {plots/stability/fd/TorusFD_t40_n1m_d100.dat};
		\end{axis}
		\begin{axis}[
				xlabel={},
				xmode=log,
				title={},
				axis y line*=right,
				axis x line=none,
				scale only axis,
                width=0.7\linewidth,
                xtick={1250,5000,20000},
                ytick scale label code/.code={},
                every y tick scale label/.style={at={(yticklabel cs:0.5)}, color=red, anchor = south, rotate = 90},
                log ticks with fixed point,
				]
			\addplot[red] table[col sep=tab, x index=0,y index=2] {plots/stability/fd/TorusFD_t40_n1m_d100.dat};
			\addplot[mark=none, red, dashed, samples=2, domain=1250:20000] {0};
		\end{axis}
	\end{tikzpicture}
\end{subfigure}
\hspace{1.5cm}
\begin{subfigure}{0.25\textwidth}
	\begin{tikzpicture}
		\begin{axis}[
				title={},
				xlabel={Young's modulus},
				xmode=log,
				scale only axis,
                width=0.7\linewidth,
                xtick={1250,5000,20000},
                ytick scale label code/.code={},
                every y tick scale label/.style={at={(yticklabel cs:0.5)}, color=blue, anchor = south, rotate = 90},
                log ticks with fixed point,
				]
			\addplot[blue] table[col sep=tab, x index=0,y index=1] {plots/stability/fd/TorusFD_t40_n10m_d100.dat};
		\end{axis}
		\begin{axis}[
				xlabel={},
				xmode=log,
				title={},
				axis y line*=right,
				axis x line=none,
				scale only axis,
                width=0.7\linewidth,
                xtick={1250,5000,20000},
                ytick scale label code/.code={},
                every y tick scale label/.style={at={(yticklabel cs:0.5)}, color=red, anchor = south, rotate = 90},
                log ticks with fixed point,
				]
			\addplot[red] table[col sep=tab, x index=0,y index=2] {plots/stability/fd/TorusFD_t40_n10m_d100.dat};
			\addplot[mark=none, red, dashed, samples=2, domain=1250:20000] {0};
		\end{axis}
	\end{tikzpicture}
\end{subfigure}
\hspace{0.5cm}
\begin{subfigure}{0.25\textwidth}
	\begin{tikzpicture}
		\begin{axis}[
				title={},
				xlabel={Young's modulus},
				xmode=log,
				scale only axis,
                width=0.7\linewidth,
                xtick={1250,5000,20000},
                ytick scale label code/.code={},
                every y tick scale label/.style={at={(yticklabel cs:0.5)}, color=blue, anchor = south, rotate = 90},
                log ticks with fixed point,
				]
			\addplot[blue] table[col sep=tab, x index=0,y index=1] {plots/stability/fd/TorusFD_t40_n100m_d100.dat};
		\end{axis}
		\begin{axis}[
				xlabel={},
				xmode=log,
				ylabel={\textcolor{red}{Gradient}},
				title={},
				axis y line*=right,
				axis x line=none,
				scale only axis,
                width=0.7\linewidth,
                xtick={1250,5000,20000},
                ytick scale label code/.code={},
                log ticks with fixed point,
				]
			\addplot[red] table[col sep=tab, x index=0,y index=2] {plots/stability/fd/TorusFD_t40_n100m_d100.dat};
			\addplot[mark=none, red, dashed, samples=2, domain=1250:20000] {0};
		\end{axis}
	\end{tikzpicture}
\end{subfigure}
\caption{Comparison between gradient estimation using finite differences and the adjoint method. The test was performed on the torus test case with $\phi_{\text{max}}=5$ and 40 timesteps.} 
The jaggedness of the red curves, and especially the large number of sign flips visually indicate that the finite-difference approach is not suitable for optimizations.
\label{fig:Stability:FiniteDifferences}
\end{figure*}

\autoref{fig:Stability:FiniteDifferences} plots the cost function and corresponding gradient estimates for the torus test case when varying the Young's modulus. 
%, the same test that was used in \autoref{app:Stability:maxSdf} and %\autoref{app:Stability:TimeAndNoise}.
Especially for a small $\Delta x$ of 5 units, the resulting gradients exhibit strong noise and a large number of sign flips. This behavior is alleviated for $\Delta x=100$ in the negative part of the gradient, but remains critical on the positive side (above the ground truth value of 5000).
%When using finite differences, one observes tiny bumps in the cost function, which---for a small $\Delta x$ of 5 in units of the Young's modulus---lead to unstable gradients (first row in \autoref{fig:Stability:FiniteDifferences}).
%For a large $\Delta x$ of 100, the small bumps are mostly smoothed out, leading to consistently oriented gradients for Young's Moduli smaller than the ground truth. For larger Young's Moduli, where the gradient should be positive, there are still some outliers with negative gradients. 
In none of the full optimization runs performed in this work (as presented in the next section, \autoref{app:Stability:FullOptimization}) we were able to reach convergence with the finite difference approach. The adjoint method, in contrast, equipped with our proposed cost function produces numerically stable and smooth results. The corresponding case is shown in the middle column of Fig.~\ref{fig:Stability:FiniteDifferences}.

%---------------------------------------------------------------------------------------------
%---------------------------------------------------------------------------------------------
\subsection{Influence of Diffusion Distance}\label{app:Stability:maxSdf}
The diffusion distance $\phi_{\text{max}}$ (\autoref{sec:Cost}) specifies the size of the narrow band in the diffusion step, i.e., the maximal distance an observed point can have to the surface to be considered in the optimization. In the following, we investigate its influence on the gradient estimation.

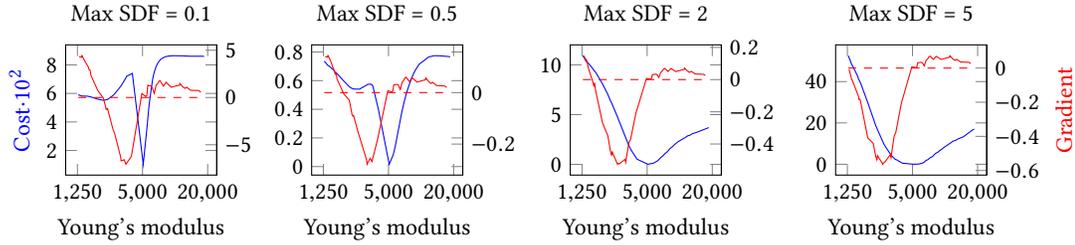
\begin{figure*}
\centering
\begin{subfigure}{0.16\textwidth}
	\begin{tikzpicture}
		\begin{axis}[
				title={Max SDF = 0.1},
				xlabel={Young's modulus},
				xmode=log,
				scale only axis,
                width=0.7\linewidth,
                xtick={1250,5000,20000},
                ytick scale label code/.code={\pgfmathparse{int(-#1)}\textcolor{blue}{Cost$ \cdot 10^{\pgfmathresult}$}},
                every y tick scale label/.style={at={(yticklabel cs:0.5)}, color=blue, anchor = south, rotate = 90},
                log ticks with fixed point,
				]
			\addplot[blue] table[col sep=tab] {plots/stability/maxSDF/cost1.dat};
		\end{axis}
		\begin{axis}[
				xlabel={},
				xmode=log,
				title={},
				axis y line*=right,
				axis x line=none,
				scale only axis,
                width=0.7\linewidth,
                xtick={1250,5000,20000},
                ytick scale label code/.code={},
                every y tick scale label/.style={at={(yticklabel cs:0.5)}, color=red, anchor = south, rotate = 90},
                log ticks with fixed point,
				]
			\addplot[red] table[col sep=tab] {plots/stability/maxSDF/grad1.dat};
			\addplot[mark=none, red, dashed, samples=2, domain=1250:20000] {0};
		\end{axis}
	\end{tikzpicture}
\end{subfigure}
\hspace{0.5cm}
\begin{subfigure}{0.16\textwidth}
	\begin{tikzpicture}
		\begin{axis}[
				title={Max SDF = 0.5},
				xlabel={Young's modulus},
				xmode=log,
				scale only axis,
                width=0.7\linewidth,
                xtick={1250,5000,20000},
                ytick scale label code/.code={},
                every y tick scale label/.style={at={(yticklabel cs:0.5)}, color=blue, anchor = south, rotate = 90},
                log ticks with fixed point,
				]
			\addplot[blue] table[col sep=tab] {plots/stability/maxSDF/cost3.dat};
		\end{axis}
		\begin{axis}[
				xlabel={},
				xmode=log,
				title={},
				axis y line*=right,
				axis x line=none,
				scale only axis,
                width=0.7\linewidth,
                xtick={1250,5000,20000},
                ytick scale label code/.code={},
                every y tick scale label/.style={at={(yticklabel cs:0.5)}, color=red, anchor = south, rotate = 90},
                log ticks with fixed point,
				]
			\addplot[red] table[col sep=tab] {plots/stability/maxSDF/grad3.dat};
			\addplot[mark=none, red, dashed, samples=2, domain=1250:20000] {0};
		\end{axis}
	\end{tikzpicture}
\end{subfigure}
\hspace{0.5cm}
\begin{subfigure}{0.16\textwidth}
	\begin{tikzpicture}
		\begin{axis}[
				title={Max SDF = 2},
				xlabel={Young's modulus},
				xmode=log,
				scale only axis,
                width=0.7\linewidth,
                xtick={1250,5000,20000},
                ytick scale label code/.code={},
                every y tick scale label/.style={at={(yticklabel cs:0.5)}, color=blue, anchor = south, rotate = 90},
                log ticks with fixed point,
				]
			\addplot[blue] table[col sep=tab] {plots/stability/maxSDF/cost5.dat};
		\end{axis}
		\begin{axis}[
				xlabel={},
				xmode=log,
				title={},
				axis y line*=right,
				axis x line=none,
				scale only axis,
                width=0.7\linewidth,
                xtick={1250,5000,20000},
                ytick scale label code/.code={},
                every y tick scale label/.style={at={(yticklabel cs:0.5)}, color=red, anchor = south, rotate = 90},
                log ticks with fixed point,
				]
			\addplot[red] table[col sep=tab] {plots/stability/maxSDF/grad5.dat};
			\addplot[mark=none, red, dashed, samples=2, domain=1250:20000] {0};
		\end{axis}
	\end{tikzpicture}
\end{subfigure}
\hspace{0.5cm}
\begin{subfigure}{0.16\textwidth}
	\begin{tikzpicture}
		\begin{axis}[
				title={Max SDF = 5},
				xlabel={Young's modulus},
				xmode=log,
				scale only axis,
                width=0.7\linewidth,
                xtick={1250,5000,20000},
                ytick scale label code/.code={},
                every y tick scale label/.style={at={(yticklabel cs:0.5)}, color=blue, anchor = south, rotate = 90},
                log ticks with fixed point,
				]
			\addplot[blue] table[col sep=tab] {plots/stability/maxSDF/cost6.dat};
		\end{axis}
		\begin{axis}[
				ylabel={\textcolor{red}{Gradient}},
				xmode=log,
				title={},
				axis y line*=right,
				axis x line=none,
				scale only axis,
                width=0.7\linewidth,
                xtick={1250,5000,20000},
                ytick scale label code/.code={\pgfmathparse{int(-#1)}\textcolor{red}{Gradient$ \cdot 10^{\pgfmathresult}$}},
                every y tick scale label/.style={at={(yticklabel cs:0.5)}, color=blue, anchor = south, rotate = 90},
                %log ticks with fixed point,
				]
			\addplot[red] table[col sep=tab] {plots/stability/maxSDF/grad6.dat};
			\addplot[mark=none, red, dashed, samples=2, domain=1250:20000] {0};
		\end{axis}
	\end{tikzpicture}
\end{subfigure}
%~\hspace{-0.2cm}
\caption{Influence of diffusion distance $\phi_{\text{max}}$ on cost function and gradient estimates for the torus data set (Young's modulus = 5000) with camera noise $0.01$ and 40 timesteps.}
\label{fig:Stability:maxSdf}
\end{figure*}

\autoref{fig:Stability:maxSdf} shows the effect of $\phi_{\text{max}}$ on the cost function and gradient estimates, as well as the reconstructed Young's modulus.
$\phi_{\text{max}}$ is given in terms of voxels. 
For small values below 1, only simulation points very close to the observation are used. This introduces additional local minima and leads to gradients with a wrong sign when the current simulation is far from the ground truth (first two plots, left side). Larger values of $\phi_{\text{max}}$ yield correct gradients for the full range of values, although the quality can deteriorate once values become too large and matching becomes ambiguous. In all of our experiments, we found a value of $\phi_{\text{max}}$ between 2 and 5 to produce stable and accurate results.

%---------------------------------------------------------------------------------------------
%---------------------------------------------------------------------------------------------
\subsection{Optimization Stability Analysis}\label{app:Stability:FullOptimization}

Next, we consider our full optimization, and compare
how the factors previously discussed for gradient estimation
%different optimizers fare across a larger number of perturbed optimization runs, in order to assess their relative performance.
influence our full algorithm.
We sample 20 random start values for the Young's modulus and let the optimization find the optimal value.
%\sebi{Next we repeat the above tests but run the whole optimization instead of only evaluating the gradient for different parameter values.}
%

%First, one can see that there seems to be a second local minimum with a higher cost at a Young's modulus of 1700 besides the ground truth of 5000. Almost all runs converge against one of those minima. With a noise setting of $0.1$, 14/20 ($\phi_{\text{max}}=1$) or 12/20 ($\phi_{\text{max}}=5$) runs converge against the ground truth. The best run for $\phi_{\text{max}}=5$ with a Young's modulus of $4888.87$ is also rendered in \autoref{fig:result:torus}c, there is no visual difference to the ground truth.}

\begin{figure*}
    \centering
    \includegraphics[width=\textwidth]{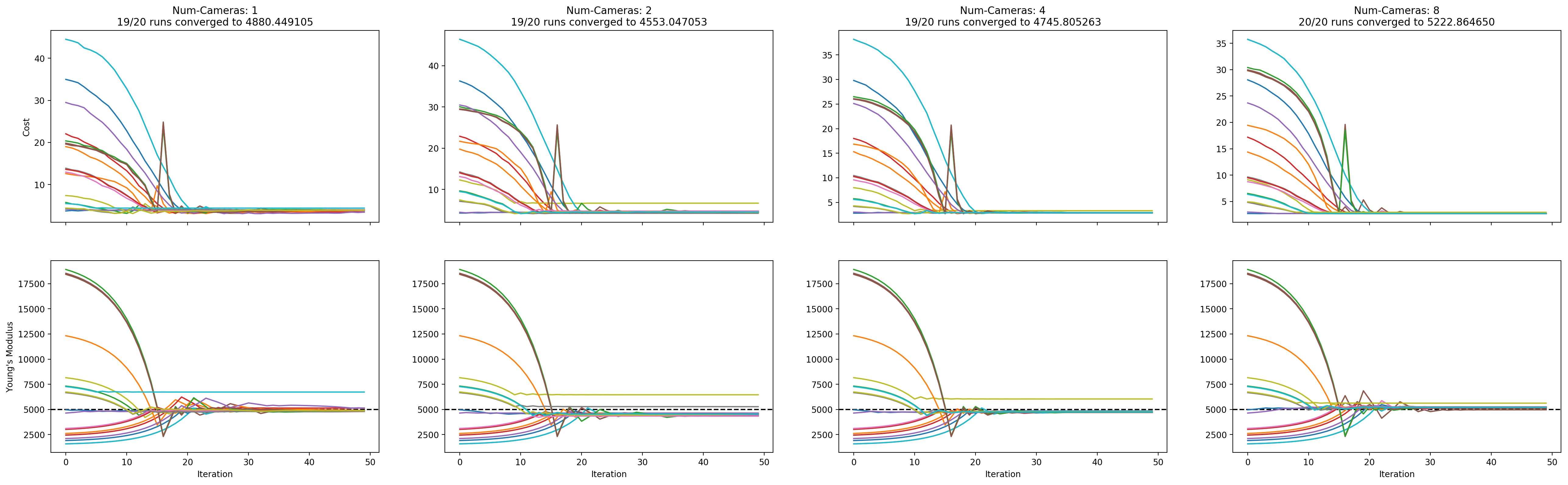}
    \caption{Influence of number of views on optimization convergence for the torus test case, using 40 timesteps, noise=0.1, maxSdf=5.0, and 20 runs with varying initial Young's modulus.
    The average relative error of the converged runs to the ground truth value is $2.39\%, 8.94\%, 5.08\%, 4.46\%$ for $1,2,4,8$ cameras, respectively.
    }
    \label{fig:Stability:NumCameras}
\end{figure*}
We first evaluate the influence of the number of different views on the optimization process. The camera locations are randomly sampled on the hemisphere above the ground and focus on the center of the torus.
As \autoref{fig:Stability:NumCameras} confirms, almost all runs ($95\%$ on average, i.e. 19 out of 20) converge and the number of cameras does not have an influence on stability, even a single camera provides enough observations, 
and multiple views do not negatively affect the reconstruction quality of our algorithm.
%and adding more points does not improve the results.}

%\sebi{All these tests have in common that the reconstructed value does not perfectly match the ground truth value, but all converging runs per test find the same value. This difference is because the depth images for the cameras are simulated differently (via marching cubes) than the observed points are used in the SSC cost function.}
Note that in many case, the Young's modulus estimates exhibit
slight deviations from the ground truth value of 5000.
This is caused by the fact that our data generation step relies on triangle surfaces generated with Marching Cubes. The SSC however, directly matches tri-linearly interpolated SDF values, which hence cannot be matched exactly in most cases.

\begin{figure*}
    \centering
    \includegraphics[width=\textwidth]{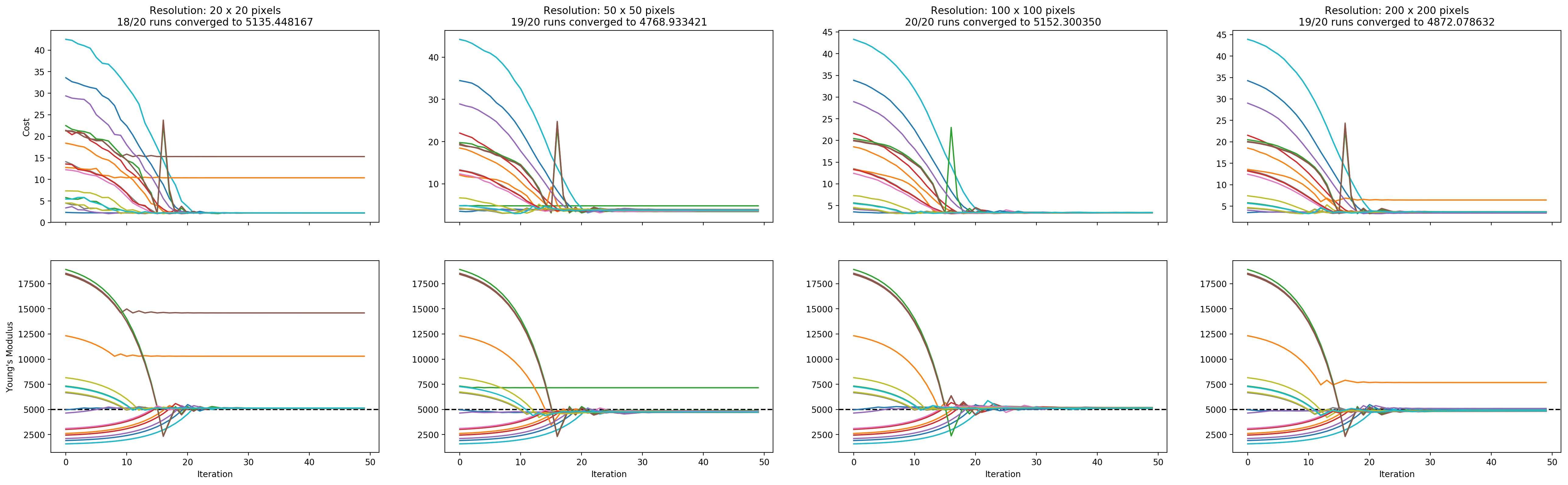}
    \caption{Influence of camera resolution on optimization convergence for the torus test case, using 40 timesteps, noise=0.1, $\phi_{\text{max}}=5.0$, 20 runs with varying initial Young's modulus.
    The average relative error of the converged runs to the ground truth value of 5000 is $2.71\%, 4.62\%, 3.05\%, 2.56\%$ for a resolution of $20^2, 50^2, 100^2, 200^2$, respectively.
    }
    \label{fig:Stability:Resolution}
\end{figure*}
The previous results are consistent with those we found when analyzing the influence of the camera resolution on optimization convergence (see  \autoref{fig:Stability:Resolution}). For one camera, increasing the resolution, and hence using more points in the cost function, does not improve the optimization.

\begin{figure*}
    \centering
    \includegraphics[width=\textwidth]{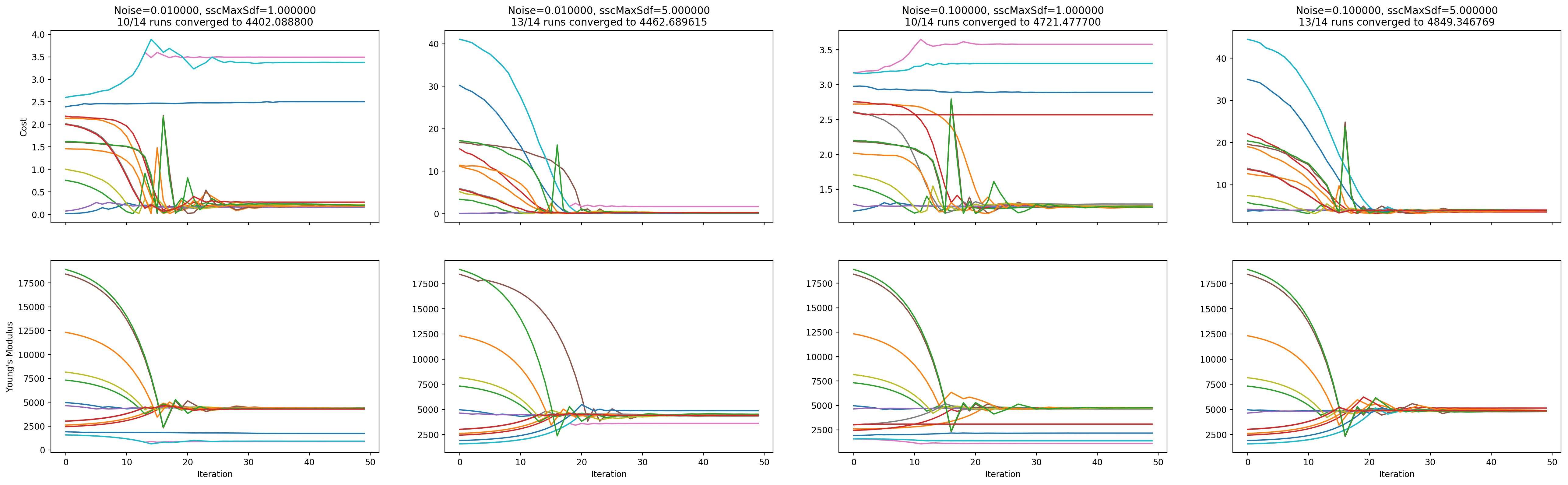}
    \caption{Influence of $\phi_{\text{max}}$ and camera noise on optimization convergence for the torus test case, using 40 timesteps and 14 runs with varying initial Young's modulus.
    The average relative error of the converged runs to the ground truth for the tests from left to right are: $11.96\%, 10.75\%, 5.57\%, 3.01\%$.
    }
    \label{fig:Stability:OptimNoiseSdf}
\end{figure*}
\begin{figure*}
    \centering
    \includegraphics[width=\textwidth]{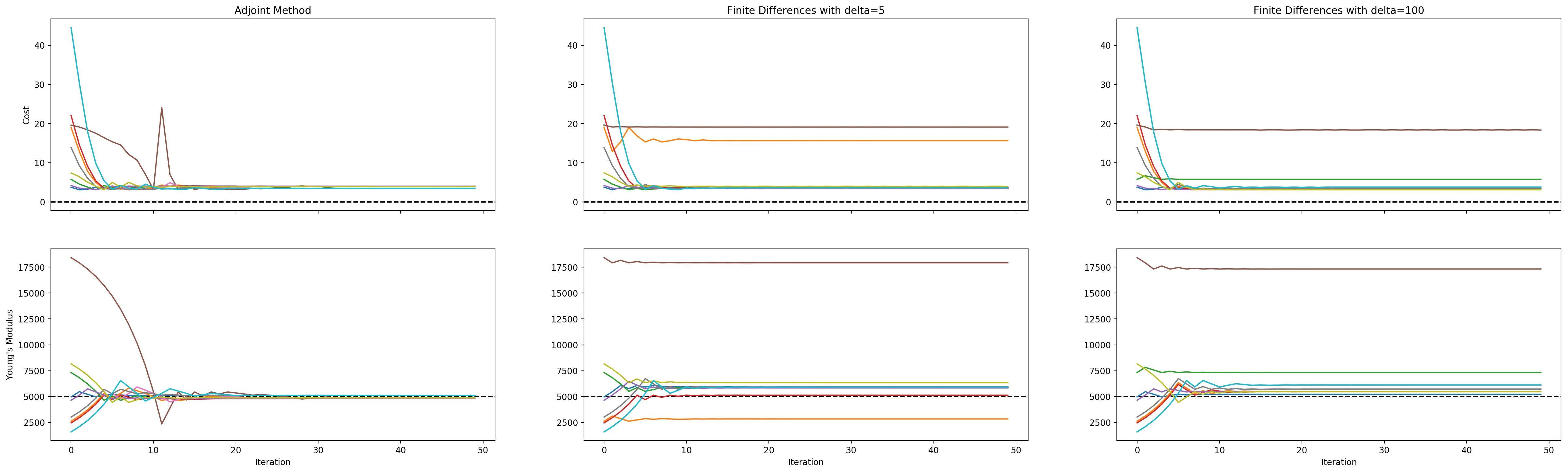}
    \caption{Comparison of the optimization when the gradients are computed with the adjoint method (left) or with finite differences with different values for $\Delta x$. %For a better comparison of the single runs, 
    For clarity, only seven runs are shown. Note that none of the runs with finite differences comes close to the ground truth solution.
    The average relative error of all runs to the ground truth value of 5000 is $1.41\%, 35.14\%, 39.33\%$ for the adjoint method and finite differences with $\Delta x=5, 100$, respectively. Especially, runs with initial values that are far away from the ground truth don't converge with the finite differences.}
    \label{fig:Stability:OptimFD}
\end{figure*}
We also vary the camera noise and diffusion distance, results for which are shown in \autoref{fig:Stability:OptimNoiseSdf}.
For a low value for $\phi_{\text{max}}$ of only one, only 10 of 14 runs converge, the other runs start with a very low Young's modulus and get stuck. For a value of $\phi_{\text{max}}=5$, 13 of 14 runs converge to the ground truth. This shows that the convergence rate decreases if the width of the narrow band is chosen too small.
In accordance to previous results from the gradient evaluations in \autoref{app:Stability:TimeAndNoise}, the amount of noise has little influence on the stability of the optimization. The results get slightly better with increased noise (more regularization).

Last, we compare the stability of the algorithm for gradients computed with the adjoint method and for gradients computed with the Finite Different Method and varying $\Delta x$, see \autoref{fig:Stability:OptimFD}.
None of the runs that use the Finite Different Method converge to the Ground Truth value for the Young's modulus, but rather converge to different local minima away from the ground truth value. This is in accordance to the gradient analysis in \autoref{app:Stability:FiniteDifferences} that show a very noisy behaviour for the gradients. In terms of average relative error, this manifests itself as an error of over $35\%$ for the runs with the finite difference method as compared to $1.41\%$ for the adjoint method
% average relative error of all runs to the ground truth value of 5000 is 1.09%, 2.5%, 19.98%

%\nils{Almost all runs ( $95\%$ on average, i.e. 19 out of 20) converge, and the number of cameras has no significant influence on the convergence rate. The resulting Young's modulus value typically slightly deviates from the ground truth, but each run converges to a stable estimate.
%because the observations were recorded in a slightly different way than they are used in the reconstruction. 
%In this case, the introduces a slight offsets (mentioned above) in terms of ground truth between Marching Cubes surfaces and SSC lead to these deviations. 
%}
%\commentNils{this case seem to not match with the varying cameras... there 19 from 20 converge to the correct YM?}
%consistent over the runs of the current setting. 

%---------------------------------------------------------------------------------------------
%---------------------------------------------------------------------------------------------
\subsection{Stability for Varying Boundary Conditions}\label{app:Stability:BoundaryConditions}

\begin{figure}
    \centering
    \begin{subfigure}{0.15\linewidth}
    	\includegraphics[trim=300 0 300 0,clip,width=\textwidth]{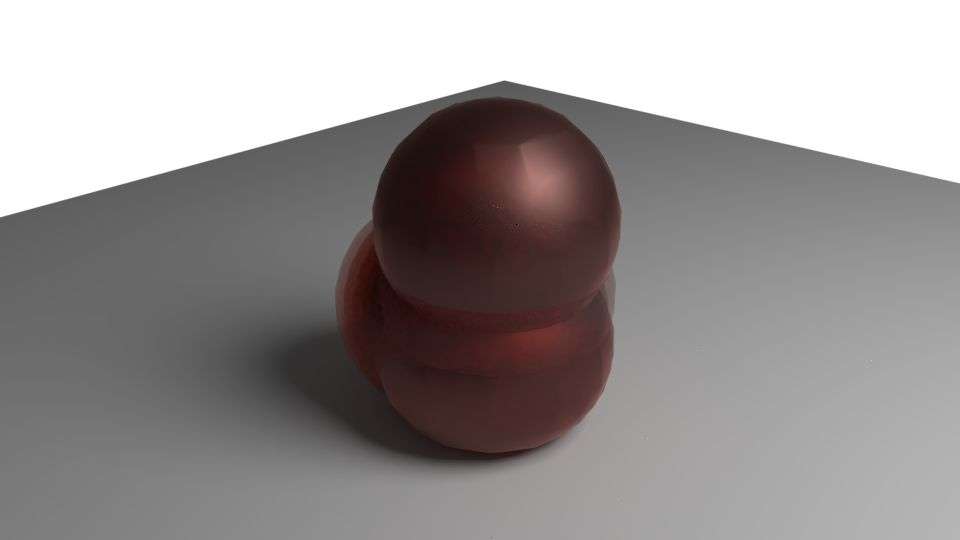}%
    \end{subfigure}
    \begin{subfigure}{0.15\linewidth}
    	\includegraphics[trim=300 0 300 0,clip,width=\textwidth]{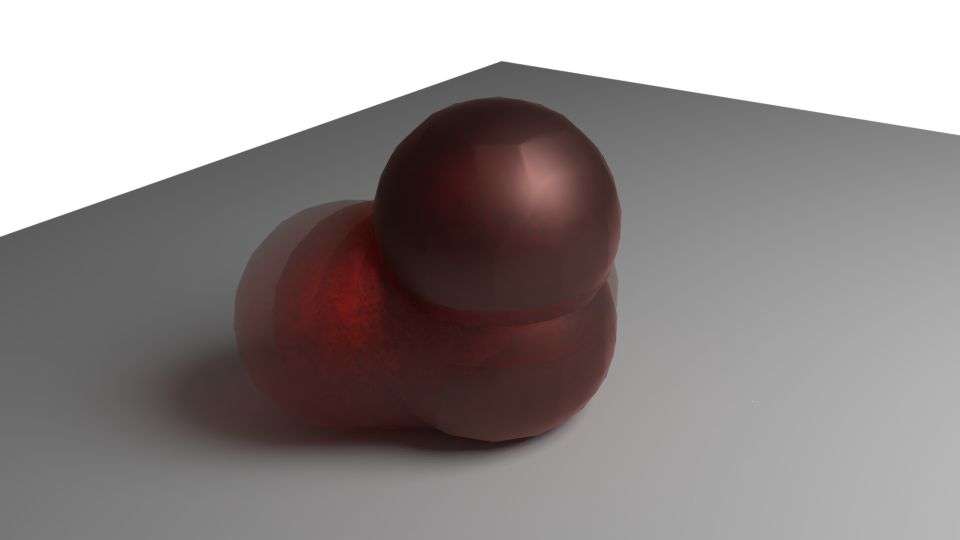}%
    \end{subfigure}
    \begin{subfigure}{0.15\linewidth}
    	\includegraphics[trim=300 0 300 0,clip,width=\textwidth]{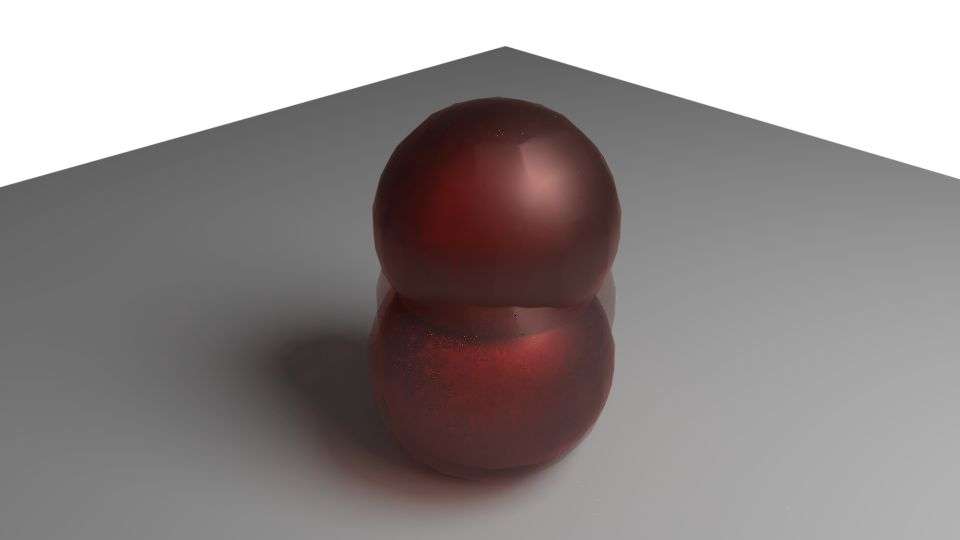}%
    \end{subfigure}
    \begin{subfigure}{0.15\linewidth}
    	\includegraphics[trim=300 0 300 0,clip,width=\textwidth]{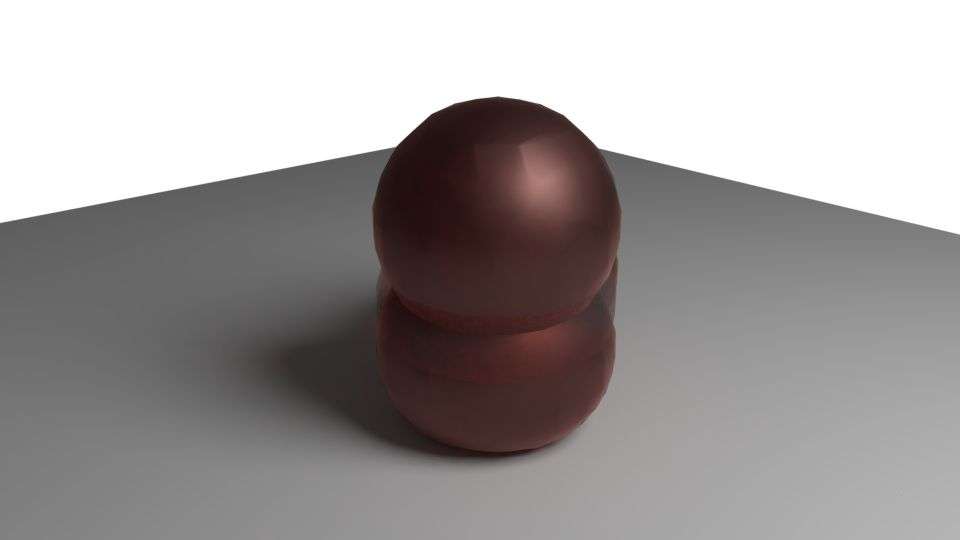}%
    \end{subfigure}
    \begin{subfigure}{0.15\linewidth}
    	\includegraphics[trim=300 0 300 0,clip,width=\textwidth]{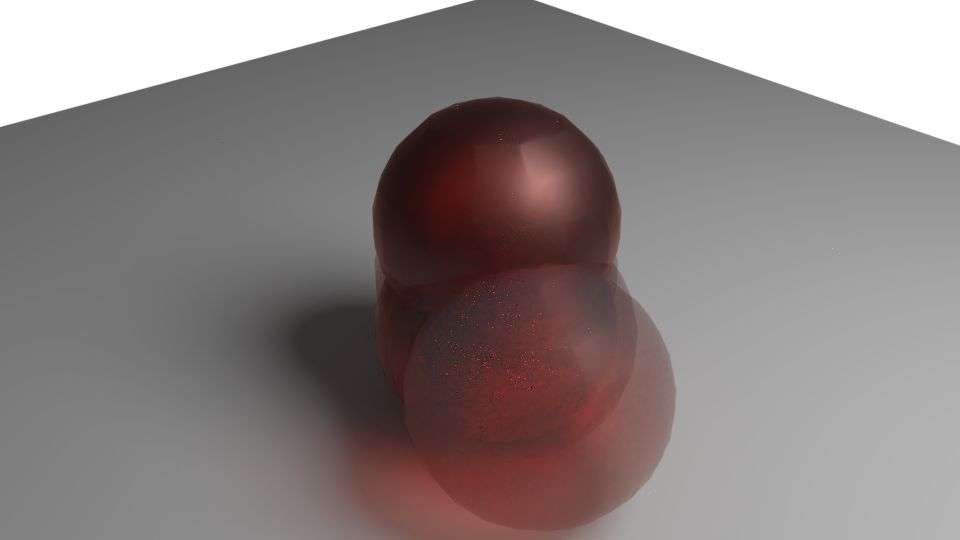}%
    \end{subfigure}
    \begin{subfigure}{0.15\linewidth}
    	\includegraphics[trim=300 0 300 0,clip,width=\textwidth]{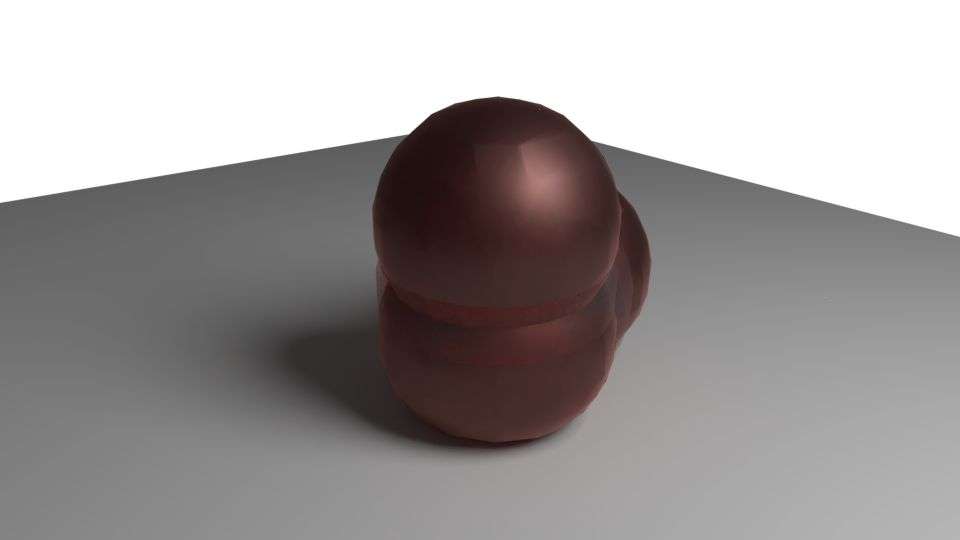}%
    \end{subfigure}
    \caption{The six ground orientations used to test the independence of the optimization from boundary conditions. The images show the substantially different collision behavior for each case.}
    \label{fig:result:ballOrientations}
\end{figure}

To analyze the dependency of the optimization process on the boundary conditions, we use a scene where a ball bounces on the ground plane. Six different settings are used, with randomly sampled orientation and position of the ground plane, and initial linear velocity of the ball, see \autoref{fig:result:ballOrientations}.
This leads to strongly differing behavior, i.e., the ball
rolling with very different speeds in different directions.
For each setting, 20 initial values for the Young's modulus between 0.1x and 10x the ground truth value of 2000 are randomly sampled. (Same 20 values for each of the six settings).
The simulation is performed over 20 timesteps (one bounce), recorded with one virtual camera of resolution 50x50 and a Gaussian noise with variance of $0.07$ voxels. The optimization is performed with a maxSDF value of 1 over 30 iterations.

\begin{figure*}
    \centering
    \includegraphics[width=\textwidth]{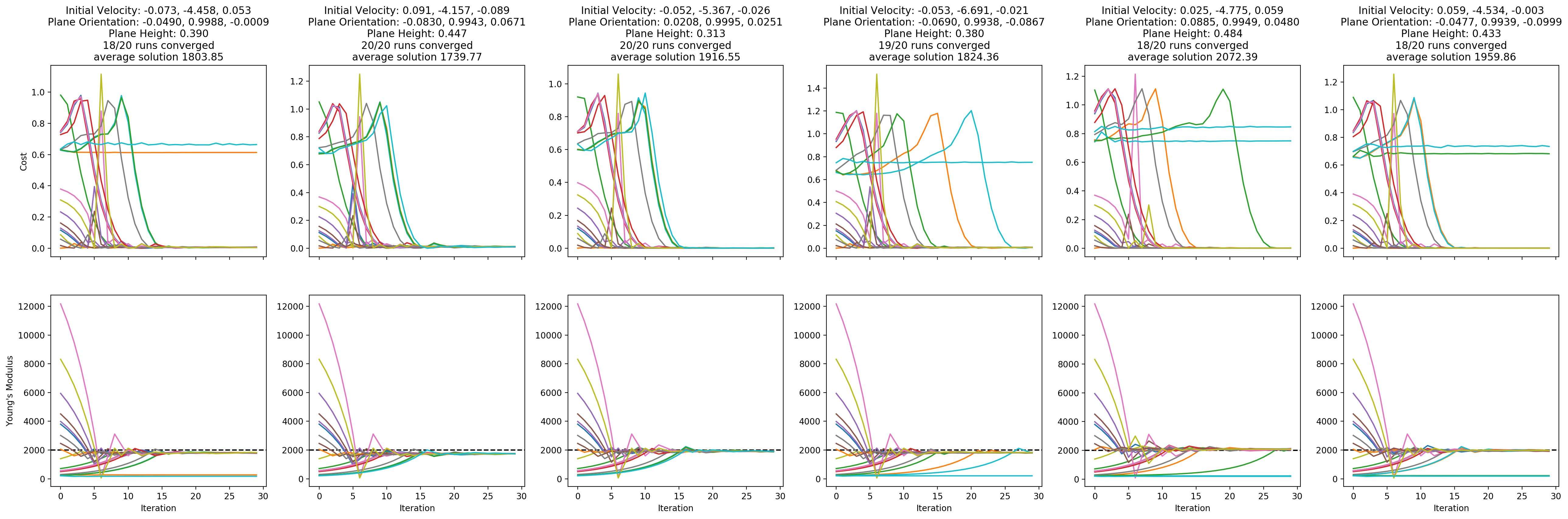}
    \caption{Same scene, a ball bounces on the floor, with different boundary conditions. Six random settings for the ground plane configuration and initial linear velocity.}
    \label{fig:Stability:Boundary}
\end{figure*}

The results are shown in \autoref{fig:Stability:Boundary}. 
Between 18 and 20 of the 20 initial values converge to the ground truth. %All non-converging runs start with very low Young's Moduli.
Interestingly, for Young's Moduli smaller than the ground truth, the cost function increases first before converging towards zero. This seems to indicate that the gradients from the adjoint method point into the right direction, even though this is not directly reflected in the cost function value. Despite this effect, these tests show 
that our method also robustly recovers the synthetic ground truth
under varying boundary conditionns.

%\sebi{The reconstructed value of the Young's modulus does not match the ground truth value of 2000 perfectly, it varies between $1739$ and $2072$. This might be due to small discrepancies between how the simulated camera observes the simulation and how those depth values are used in the reconstruction.}

%---------------------------------------------------------------------------------------------
%---------------------------------------------------------------------------------------------
\subsection{Recovering Multiple Parameters}\label{app:Stability:MultipleParameters}

\begin{figure*}
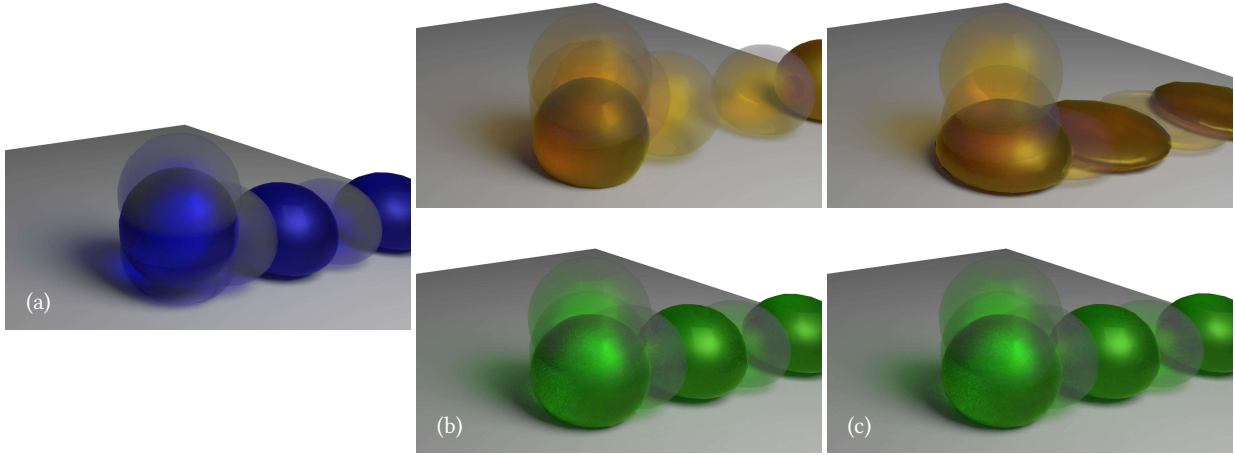

    \centering
    \begin{subfigure}[c]{0.3\linewidth}
    	\begin{overpic}[trim=300 100 0 0,clip,width=\textwidth]{images/BallMultipleVars/ball_gt}%
             \put(5,5){\textcolor{white}{(a)}}%
        \end{overpic}%
    \end{subfigure}
    \begin{subfigure}[c]{0.3\linewidth}
    	\includegraphics[trim=300 100 0 0,clip,width=\textwidth]{images/BallMultipleVars/ball_run0init}\\%
    	\begin{overpic}[trim=300 100 0 0,clip,width=\textwidth]{images/BallMultipleVars/ball_run0rec}%
             \put(5,5){\textcolor{white}{(b)}}%
        \end{overpic}%
    \end{subfigure}
    \begin{subfigure}[c]{0.3\linewidth}
    	\includegraphics[trim=300 100 0 0,clip,width=\textwidth]{images/BallMultipleVars/ball_run1init}\\%
    	\begin{overpic}[trim=300 100 0 0,clip,width=\textwidth]{images/BallMultipleVars/ball_run1rec}%
             \put(5,5){\textcolor{white}{(c)}}%
        \end{overpic}%
    \end{subfigure}
    \caption{Multi-parameter optimization for the bouncing ball test case. Ground truth (a) and the initial and reconstruction configuration for the two runs with the highest initial cost (b,c). Even though the reconstructed values are different from the ground truth, the output visually matches the ground truth.}
    \label{fig:Stability:BallMultipleVars:Images}
\end{figure*}

\begin{figure*}
    \centering
    \includegraphics[width=0.8\linewidth]{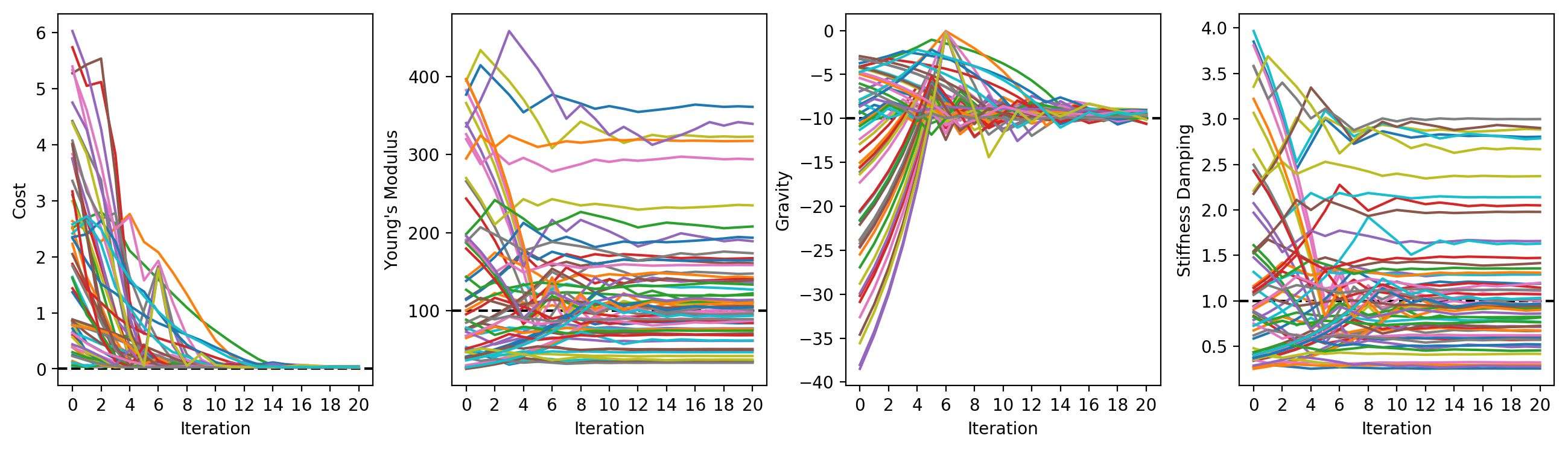}
    \caption{Bouncing ball, optimized for gravity, Young's modulus and stiffness damping. All 60 initial samples converge to a solution that has almost zero cost and visually same behaviour (see \autoref{fig:Stability:BallMultipleVars:Images}. The values for the Young's modulus and stiffness damping, however, strongly differ, showing their inter-dependency.}
    \label{fig:Stability:BallMultipleVars:Plots}
\end{figure*}

To analyze the stability of the optimization for multiple parameters,
we used the bouncing ball test case again and this time optimize for gravity, the Young's modulus and stiffness damping. We sampled 60 different initial configurations randomly and let the optimizer run for 20 iterations. The ground truth simulation is shown in \autoref{fig:Stability:BallMultipleVars:Images} on the left.
The convergence plots for all 60 runs are shown in \autoref{fig:Stability:BallMultipleVars:Plots}.
As one can see, all runs converge to a solution that has almost zero cost. The reconstructed value for the gravity is quite unique, but a strong inter-dependency between Young's modulus and stiffness damping is clearly seen. Despite differing values, the reconstructions very closely
match the ground truth.

%\sebi{
%This demonstrates the inter-dependency between the parameters, here %especially between the Young's modulus $k$ and the stiffness damping %$\alpha_2$ with ground truth values $k=100, \alpha_2=1.0$, respectively.
%For example, one run converges against $k=69$ and $\alpha_2=1.47$ with %initial cost of $3.17$ and final cost of $0.029$; %run_3_13
%another run converges against $k=137$ and $\alpha_2=0.715$ with initial cost %of $6.03$ and final cost of $0.026$. 
%}

%---------------------------------------------------------------------------------------------
%---------------------------------------------------------------------------------------------
\subsection{Different Optimizers}\label{app:Stability:Optimizers}

Throughout our work, we use the R-Prop optimizer for the reconstruction.
Here we compare it to a simple Gradient Descent optimizer and the L-BFGS optimizer.
While R-Prop only uses the sign of the gradient to determine the next search direction, Gradient Descent also uses the norm of the gradient with an additional adaptive step size, so we would expect faster convergence.
L-BFGS approximates the Hessian matrix and as a second-order method should converge even faster.

\begin{figure*}
    \centering
    \begin{subfigure}{0.23\linewidth}
    	\begin{overpic}[trim=450 0 450 0,clip,width=\textwidth]{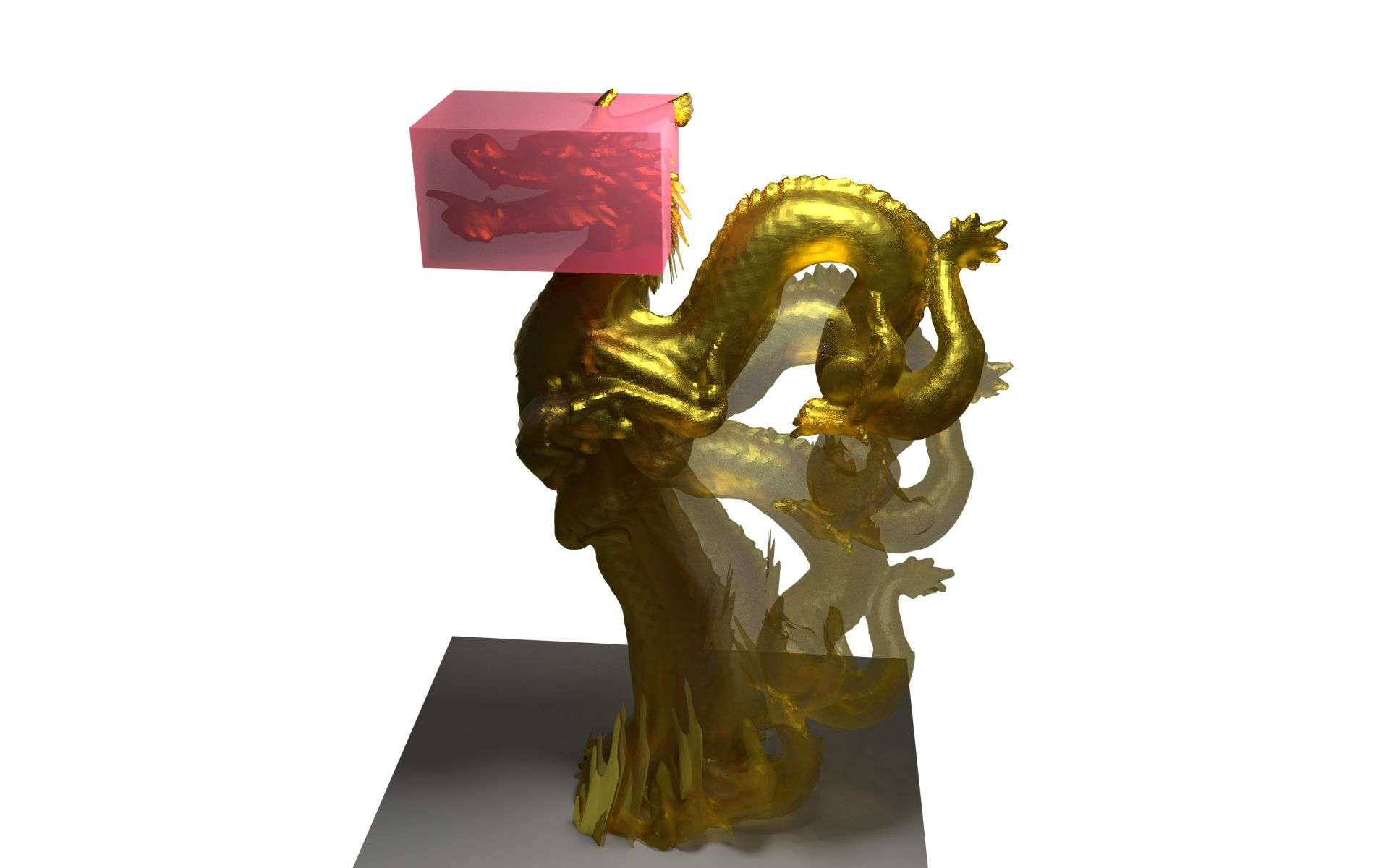}%
             \put(10,5){\textcolor{white}{(a)}}%
        \end{overpic}%
    	%\caption{Minimal initial value}
    	%\label{fig:result:dragon:min}
    \end{subfigure}
    \begin{subfigure}{0.23\linewidth}
    	\begin{overpic}[trim=450 0 450 0,clip,width=\textwidth]{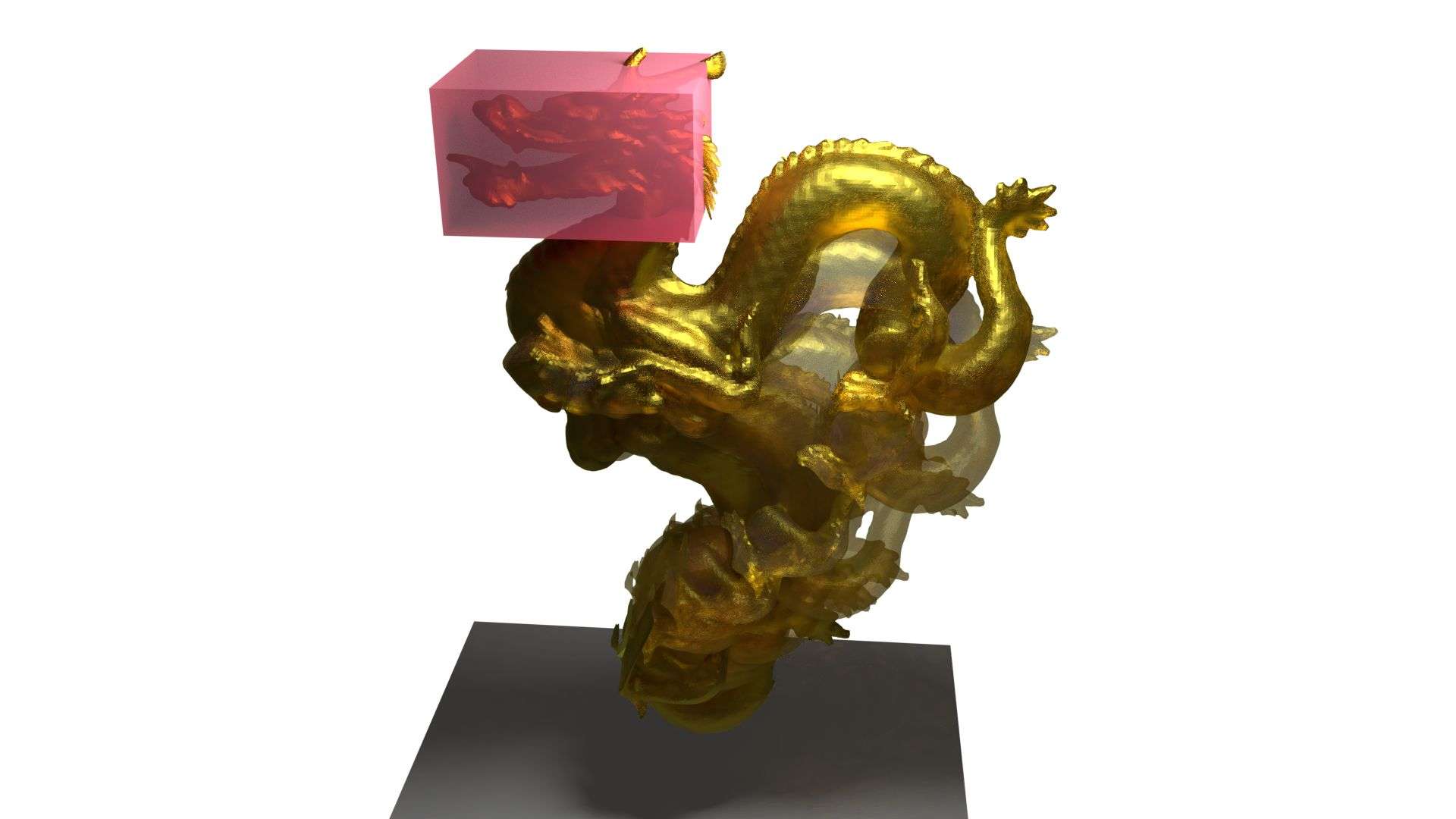}%
             \put(10,5){\textcolor{white}{(b)}}%
        \end{overpic}%
    	%\caption{Maximal initial value}
    	%\label{fig:result:dragon:max}
    \end{subfigure}
    \begin{subfigure}{0.23\linewidth}
    	\begin{overpic}[trim=450 0 450 0,clip,width=\textwidth]{images/Dragon/DragonRec_rendering2.jpg}%
             \put(10,5){\textcolor{white}{(c)}}%
        \end{overpic}%
    	%\caption{Reconstruction}
    	%\label{fig:result:dragon:rec}
    \end{subfigure}
    \begin{subfigure}{0.23\linewidth}
    	\begin{overpic}[trim=450 0 450 0,clip,width=\textwidth]{images/Dragon/DragonGT_rendering2.jpg}%
             \put(10,5){\textcolor{white}{(d)}}%
        \end{overpic}%
    	%\caption{Ground truth}
    	%\label{fig:result:dragon:gt}
    \end{subfigure}
    \caption{Optimization of the Young's modulus on the Dragon. From left to right: minimal and maximal initial value, reconstruction and ground truth.}
    \label{fig:result:dragon}
\end{figure*}

\begin{figure*}
\begin{tikzpicture}
\begin{groupplot}[
  group style={
    group size=3 by 2,
    group name=plots
  },
  width=0.3\textwidth,height=0.3\textwidth
]

% Dragon 1 - Cost
\nextgroupplot[
  legend entries={R-Prop, Gradient Descent, L-BFGS},
  legend to name=thelegend,
  title={a) Dragon 1},
  xlabel={Iteration},
  ylabel={Cost}]
\addplot+[mark=none, restrict x to domain=0:50] table[x index=0, y index=1] {plots/optimizer/dragon1000rprop.dat};
\addplot+[mark=none, restrict x to domain=0:50] table[x index=0, y index=1] {plots/optimizer/dragon1000gd.dat};
\addplot+[mark=none, restrict x to domain=0:50] table[x index=0, y index=1] {plots/optimizer/dragon1000lbfgs.dat};

% Dragon 2 - Cost
\nextgroupplot[title={b) Dragon 2}]
\addplot+[mark=none, restrict x to domain=0:50] table[x index=0, y index=1] {plots/optimizer/dragon7500rprop.dat};
\addplot+[mark=none, restrict x to domain=0:50] table[x index=0, y index=1] {plots/optimizer/dragon7500gd.dat};
\addplot+[mark=none, restrict x to domain=0:50] table[x index=0, y index=1] {plots/optimizer/dragon7500lbfgs.dat};

% Ball - Cost
\nextgroupplot  [title={c) Bouncing Ball}]
\addplot+[mark=none, restrict x to domain=0:20] table[x index=0, y index=1] {plots/optimizer/ball300rprop.dat};
\addplot+[mark=none, restrict x to domain=0:20] table[x index=0, y index=1] {plots/optimizer/ball300gd.dat};
\addplot+[mark=none, restrict x to domain=0:20] table[x index=0, y index=1] {plots/optimizer/ball300lbfgs.dat};

% Dragon 1 - Young's modulus
\nextgroupplot[
  title={},
  xlabel={Iteration},
  ylabel={Young's modulus},
  ymode=log,
  ytick={1000, 5000, 10000},
  log ticks with fixed point,
  x tick label style={/pgf/number format/1000 sep=\,}]
\addplot+[mark=none, restrict x to domain=0:50] table[x index=0, y index=2] {plots/optimizer/dragon1000rprop.dat};
\addplot+[mark=none, restrict x to domain=0:50] table[x index=0, y index=2] {plots/optimizer/dragon1000gd.dat};
\addplot+[mark=none, restrict x to domain=0:50] table[x index=0, y index=2] {plots/optimizer/dragon1000lbfgs.dat};
\addplot[mark=none, red, dashed, samples=2, domain=0:50] {5000};

% Dragon 2 - Young's modulus
\nextgroupplot[
  title={},
  xlabel={Iteration},
  ylabel={},
  ymode=log,
  ytick={3000, 5000, 7500},
  log ticks with fixed point,
  x tick label style={/pgf/number format/1000 sep=\,}]
\addplot+[mark=none, restrict x to domain=0:50] table[x index=0, y index=2] {plots/optimizer/dragon7500rprop.dat};
\addplot+[mark=none, restrict x to domain=0:50] table[x index=0, y index=2] {plots/optimizer/dragon7500gd.dat};
\addplot+[mark=none, restrict x to domain=0:50] table[x index=0, y index=2] {plots/optimizer/dragon7500lbfgs.dat};
\addplot[mark=none, red, dashed, samples=2, domain=0:50] {5000};

% Ball - Young's modulus
\nextgroupplot[
  title={},
  xlabel={Iteration},
  ylabel={},
  ymode=log]
\addplot+[mark=none, restrict x to domain=0:20] table[x index=0, y index=2] {plots/optimizer/ball300rprop.dat};
\addplot+[mark=none, restrict x to domain=0:20] table[x index=0, y index=2] {plots/optimizer/ball300gd.dat};
\addplot+[mark=none, restrict x to domain=0:20] table[x index=0, y index=2] {plots/optimizer/ball300lbfgs.dat};
\addplot[mark=none, red, dashed, samples=2, domain=0:20] {2000};

\end{groupplot}

% add legend
\node [below right,inner sep=0pt] at ([xshift=3mm]plots c3r1.north east) {\ref{thelegend}};

\end{tikzpicture}%
{\phantomsubcaption\label{fig:OptimizerComparison:Dragon1}}%
{\phantomsubcaption\label{fig:OptimizerComparison:Dragon2}}%
{\phantomsubcaption\label{fig:OptimizerComparison:Ball}}%
\caption{Comparison of the different optimizers for different scenes.}
\label{fig:OptimizerComparison}
\end{figure*}
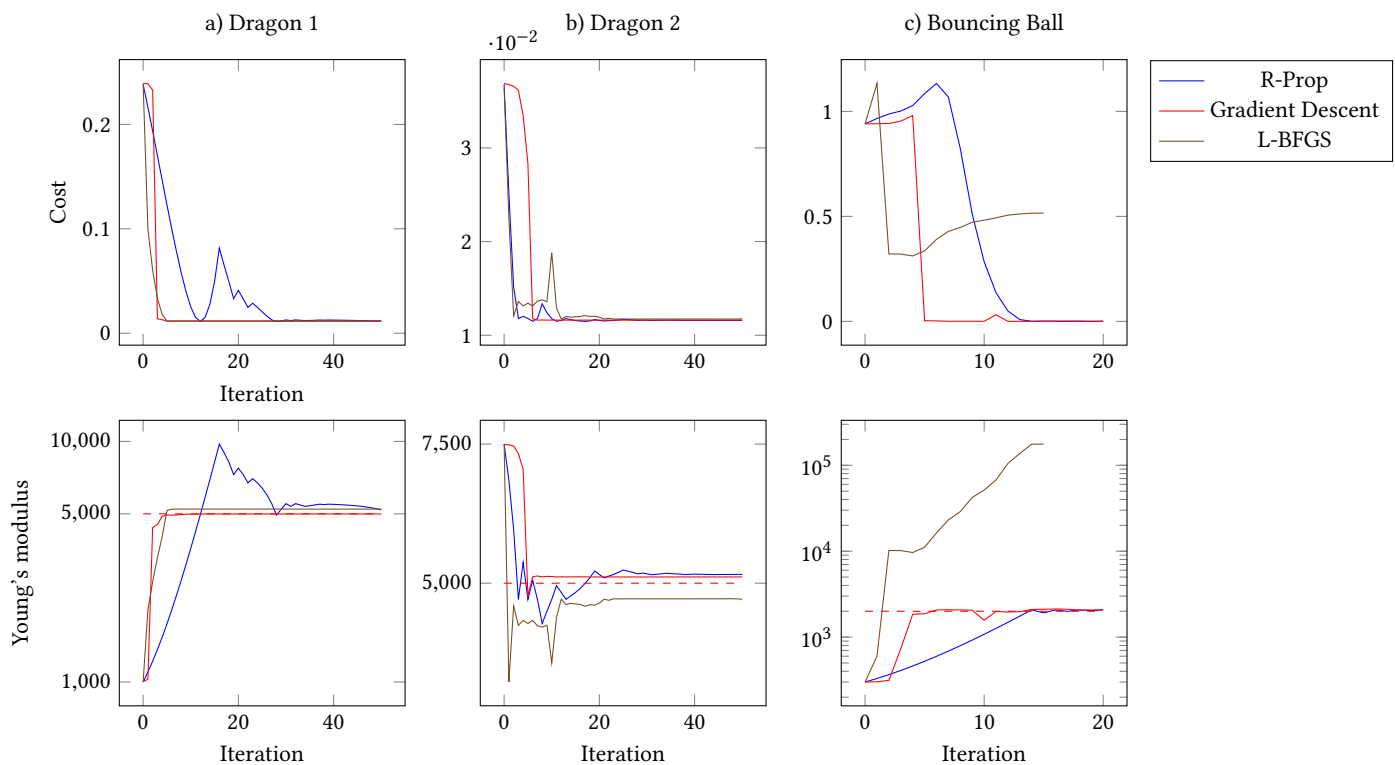

\begin{figure*}
    \centering
    \includegraphics[width=\textwidth]{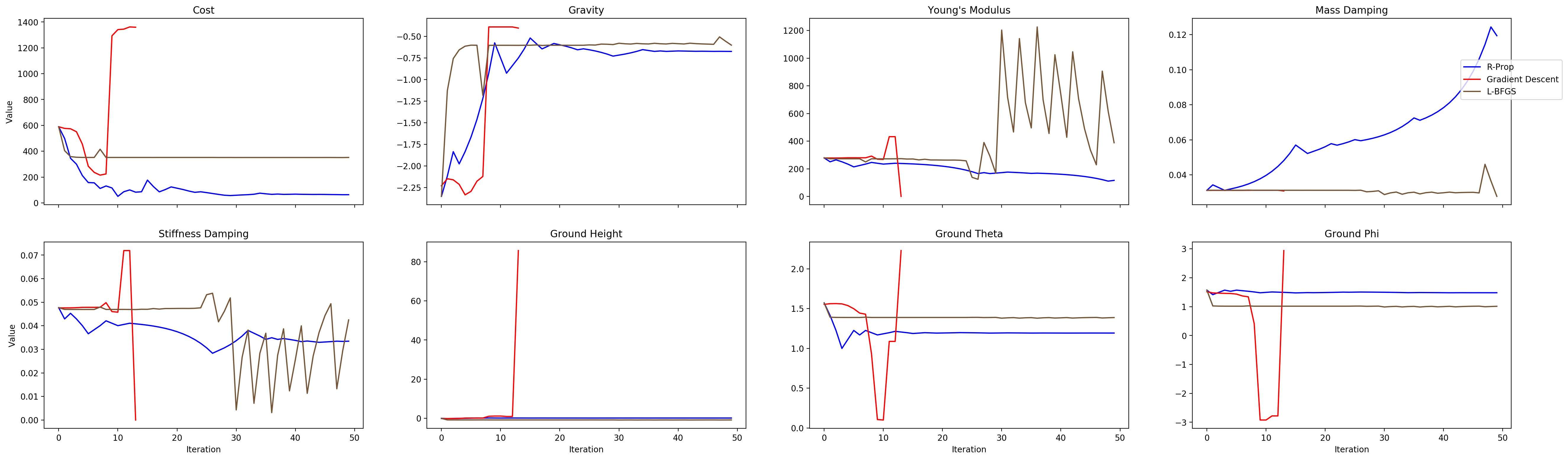}
    \caption{Comparison of the different optimizers for the Pillow-Ramp example}
    \label{fig:OptimizerComparison:Pillow}
\end{figure*}

We compared the three optimizers on three different test cases, the fixed Stanford dragon (see \autoref{fig:result:dragon}, the bouncing ball from \autoref{app:Stability:BoundaryConditions} and the pillow-ramp test case \autoref{fig:Pillow:Optimization}. 
To clearly see the behaviour of the different optimizers when started from the same initial configurations, we only used one or two runs per setting.
The results are shown in \autoref{fig:OptimizerComparison} and \autoref{fig:OptimizerComparison:Pillow}.

For the first two test cases, the Dragon and Bouncing Ball, both the R-Prop and the Gradient Descent algorithm converge to a solution (see \autoref{fig:result:dragon}c) that is indistinguishable from the ground truth.
Gradient Descent converges faster as expected.
%The L-BFGS algorithm in the form used in our implementation is more problematic. It 
The L-BFGS algorithm is often too aggressive in choosing the step size, which leads to the optimization getting stuck in sub-optimal local minima (see Figure~\ref{fig:OptimizerComparison:Dragon2}. Furthermore, it uses the values of the cost function itself to determine the step size, which can lead to instabilities if the cost function first increases when the solution gets better. This is the case for the bouncing ball test case when started from a low value for the Young's modulus, see \autoref{app:Stability:BoundaryConditions}.
These two sources of instability lead to a divergent optimization with the L-BFGS algorithm for the bouncing ball, see Figure~\ref{fig:OptimizerComparison:Ball}.

In the multi-parameters optimization, the Gradient Descent algorithm has severe problems. See e.g. \autoref{fig:OptimizerComparison:Pillow} for an example of the Pillow-Ramp test case where the Gradient Descent algorithm completely diverges and leads to a state where the simulation collapses. We believe that this is because Gradient Descent uses the same step size for all parameters. R-Prop and L-BFGS both use a different step size for each parameter. In the test case using the pillow, the L-BFGS starts to oscillate and gets stuck in a sub-optimal local minimum. R-Prop behaves much smoother and finds a good solution.
%
%\sebi{Based on these experiments, we chose R-Prop as the optimizer in all examples.}

%====================================================================================
%====================================================================================

\section{Real World Test Cases} \label{sec:Appendix:RealWorld}

\begin{figure*}%
\centering
\begin{subfigure}{\textwidth}
    \centering
	\includegraphics[width=0.24\textwidth]{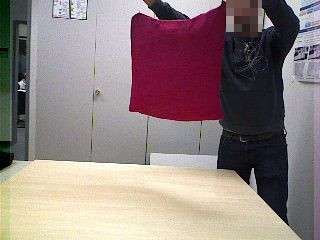}%
	\includegraphics[width=0.24\textwidth]{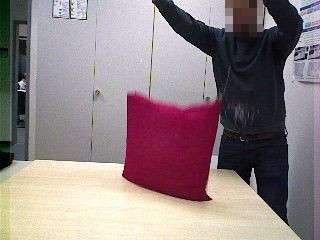}%
	\includegraphics[width=0.24\textwidth]{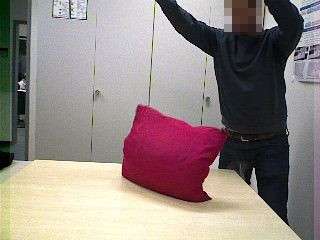}%
	\includegraphics[width=0.24\textwidth]{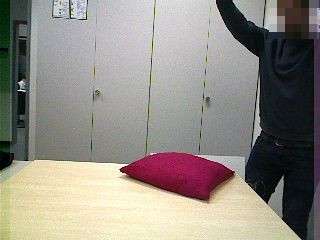}%
	\caption{Color Observation}
	\label{fig:PillowFlat:Color}
\end{subfigure}
\\
\begin{subfigure}{\textwidth}
    \centering
	\includegraphics[width=0.24\textwidth]{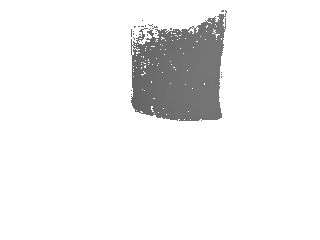}%
	\includegraphics[width=0.24\textwidth]{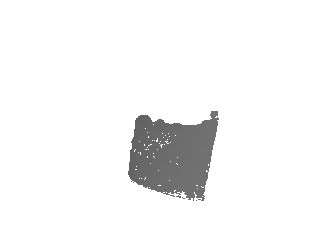}%
	\includegraphics[width=0.24\textwidth]{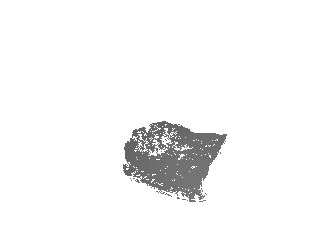}%
	\includegraphics[width=0.24\textwidth]{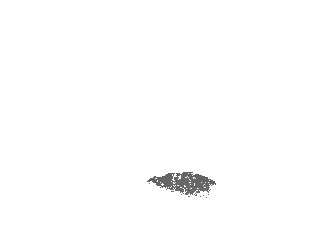}%
	\caption{Depth Observation}
	\label{fig:PillowFlat:Depth}
\end{subfigure}
\\
\begin{subfigure}{\textwidth}
    \centering
	\includegraphics[width=0.24\textwidth]{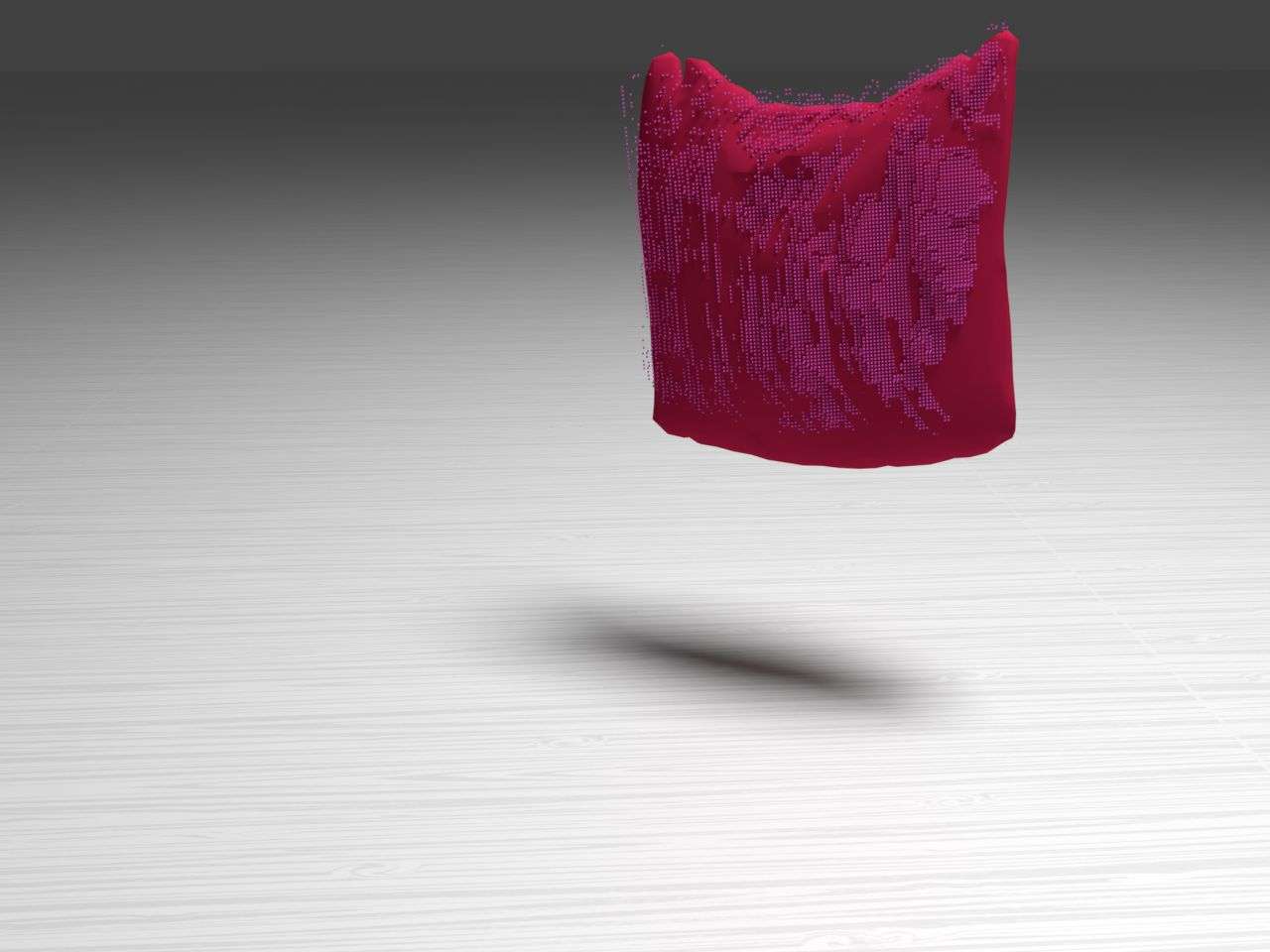}%
	\includegraphics[width=0.24\textwidth]{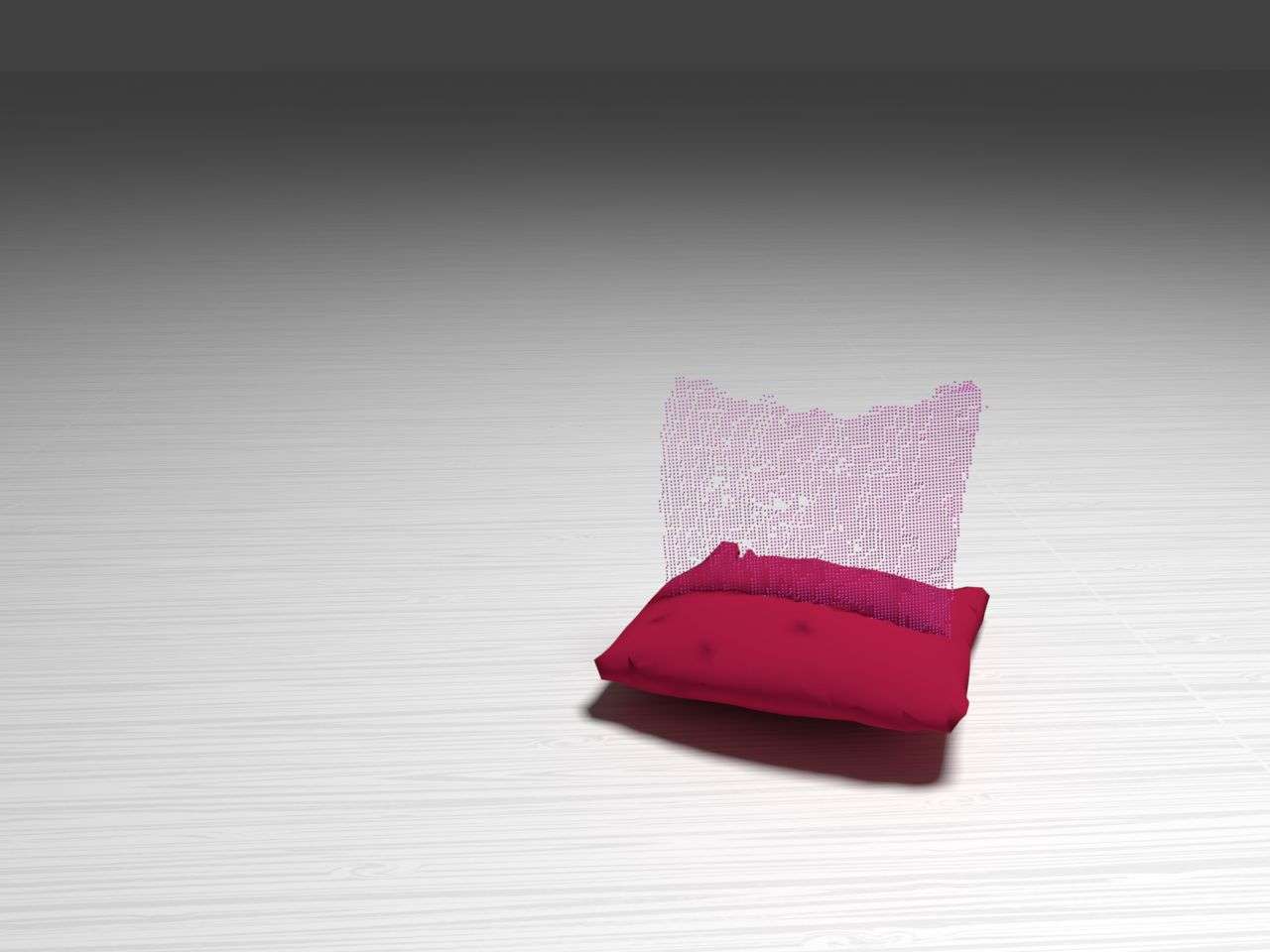}%
	\includegraphics[width=0.24\textwidth]{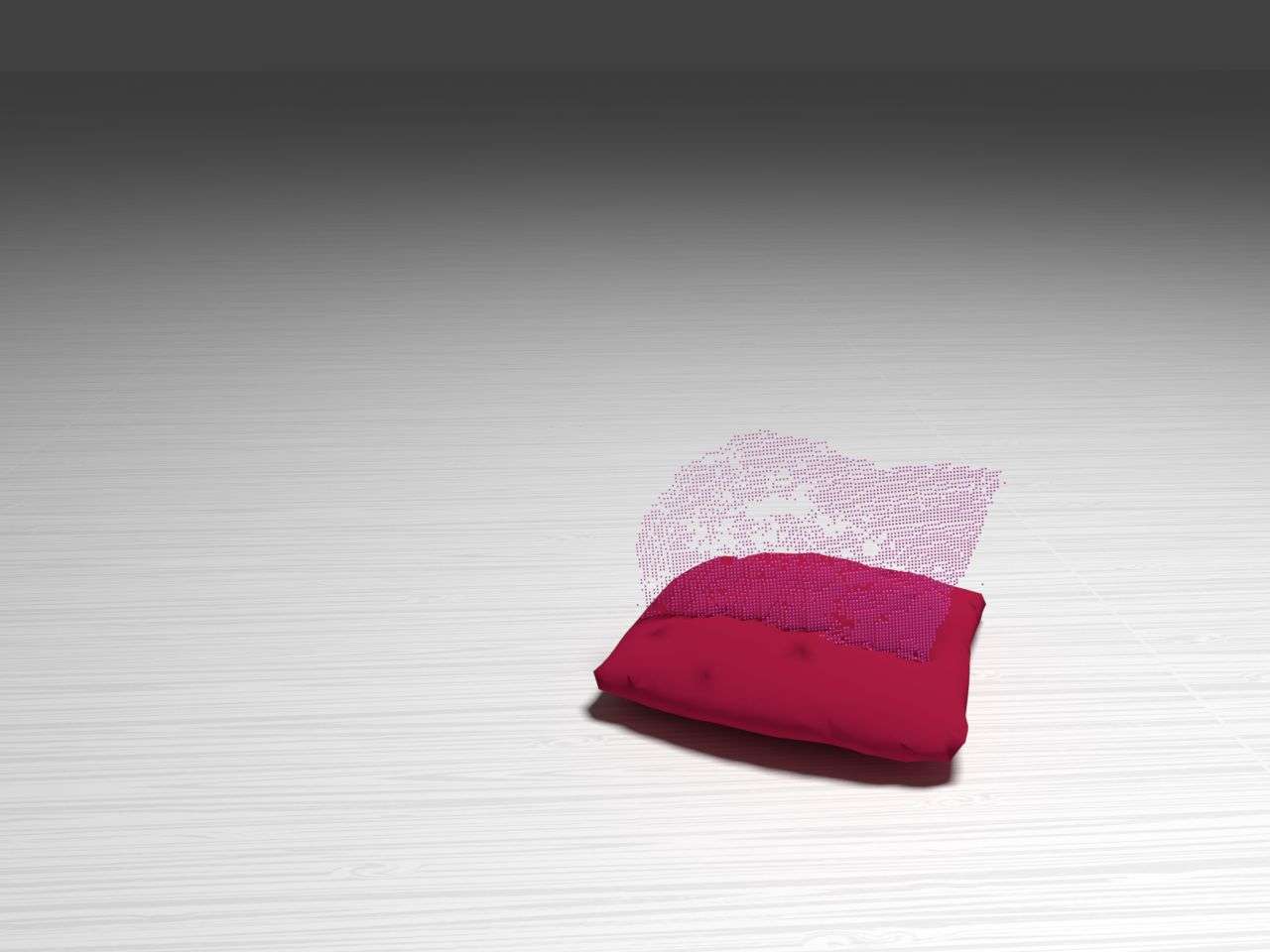}%
	\includegraphics[width=0.24\textwidth]{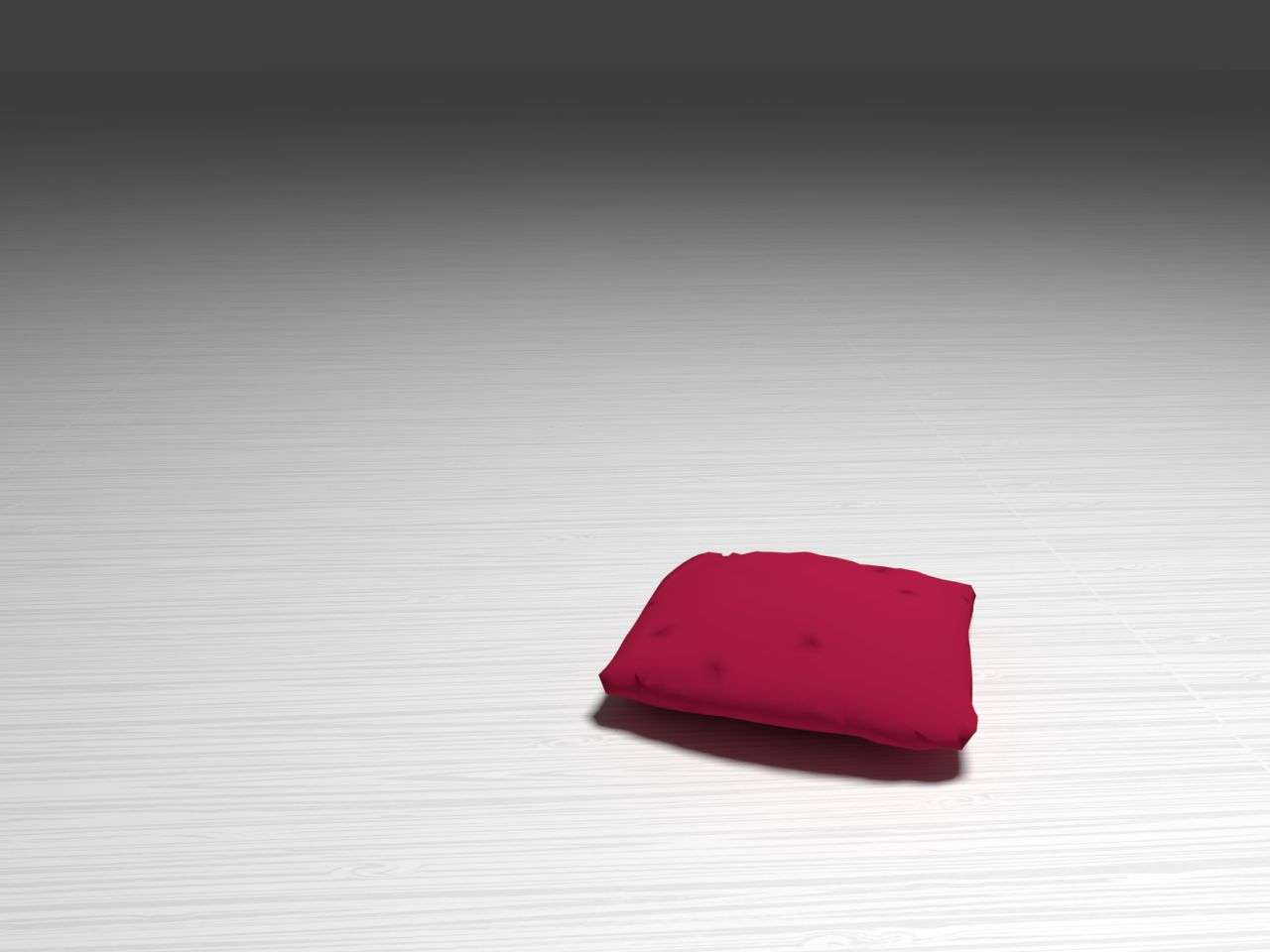}%
	\caption{Initial configuration for the optimization}
	\label{fig:PillowFlat:Initial}
\end{subfigure}
\\
\begin{subfigure}{\textwidth}
    \centering
	\includegraphics[width=0.24\textwidth]{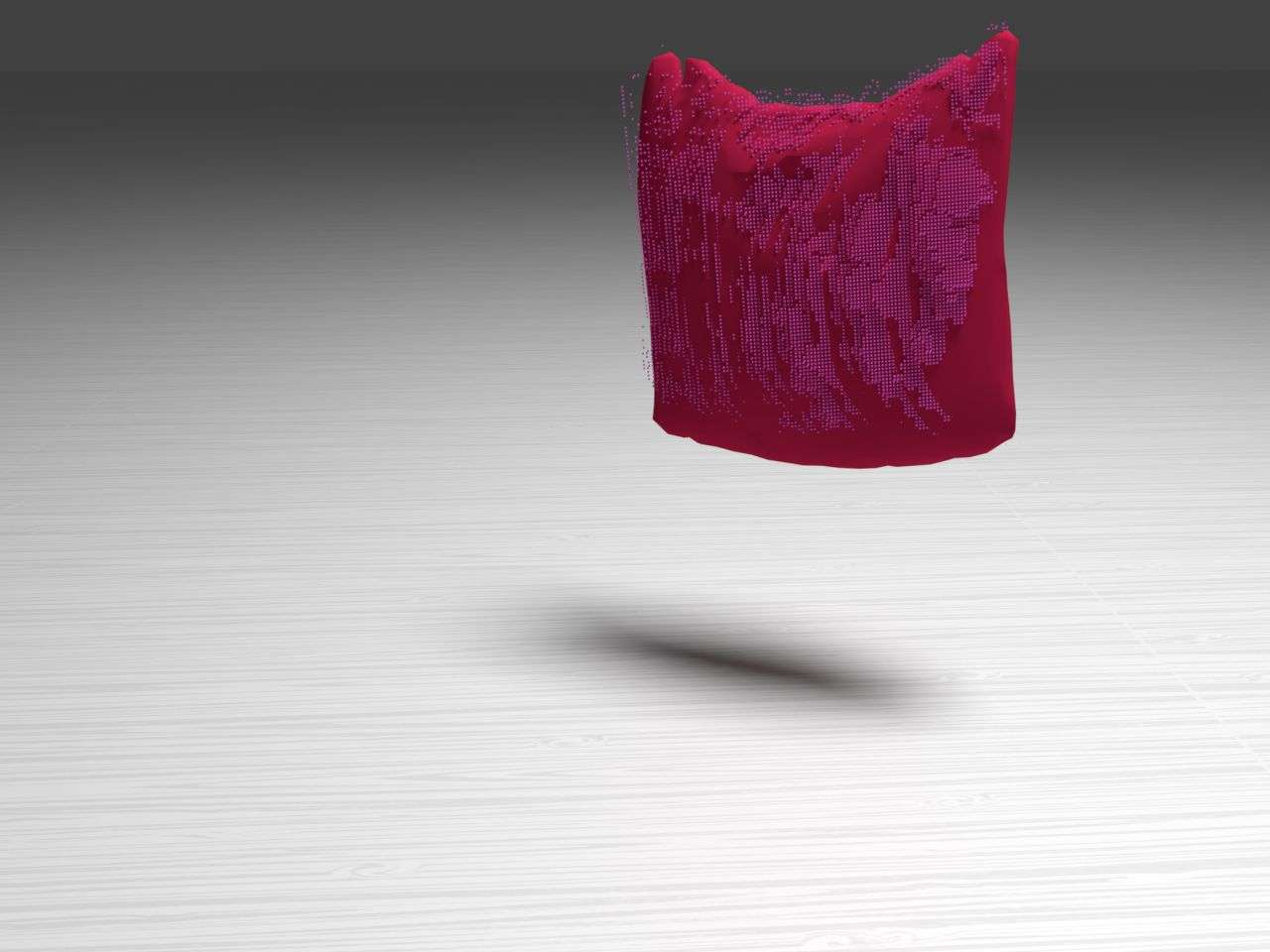}%
	\includegraphics[width=0.24\textwidth]{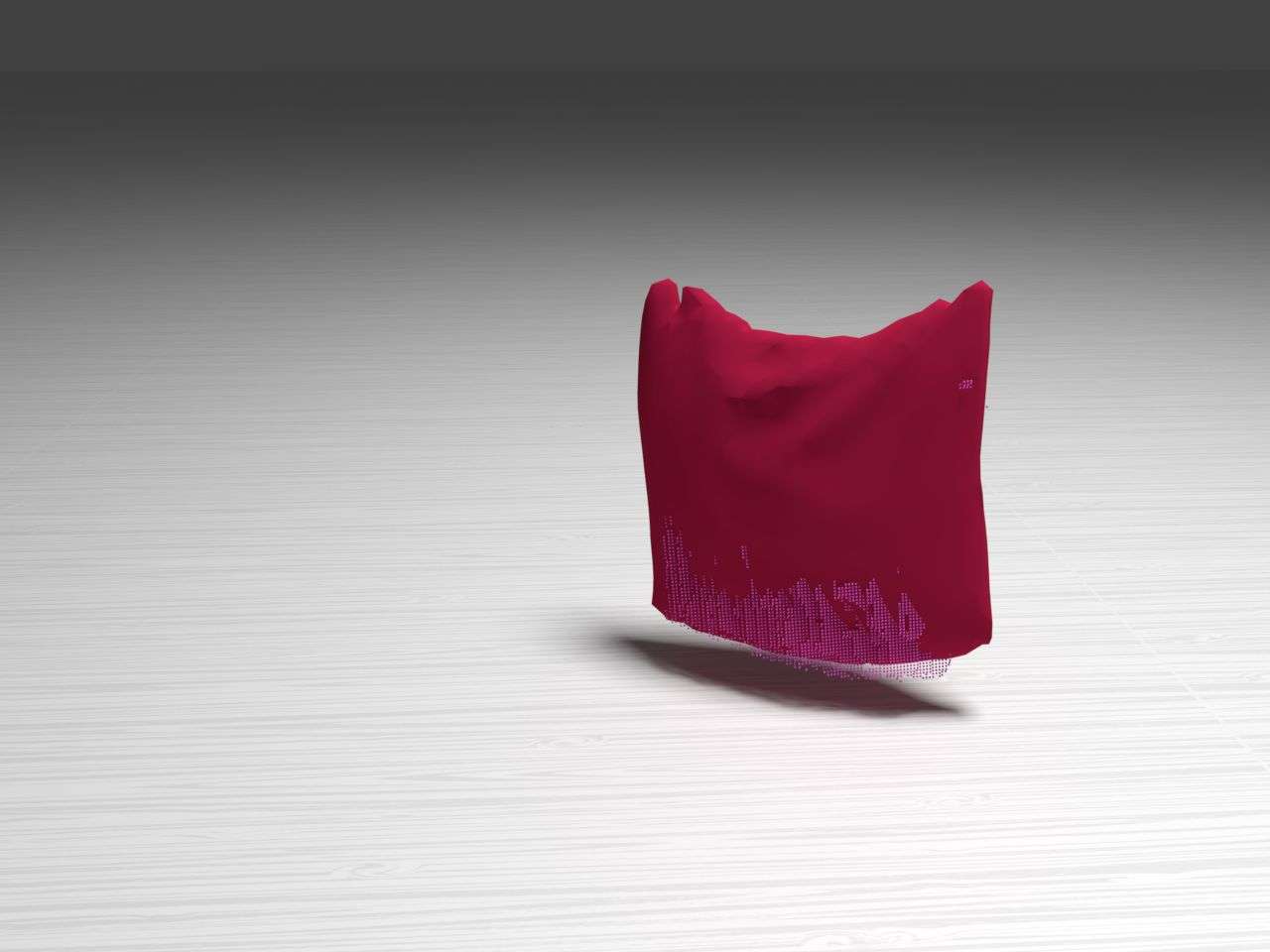}%
	\includegraphics[width=0.24\textwidth]{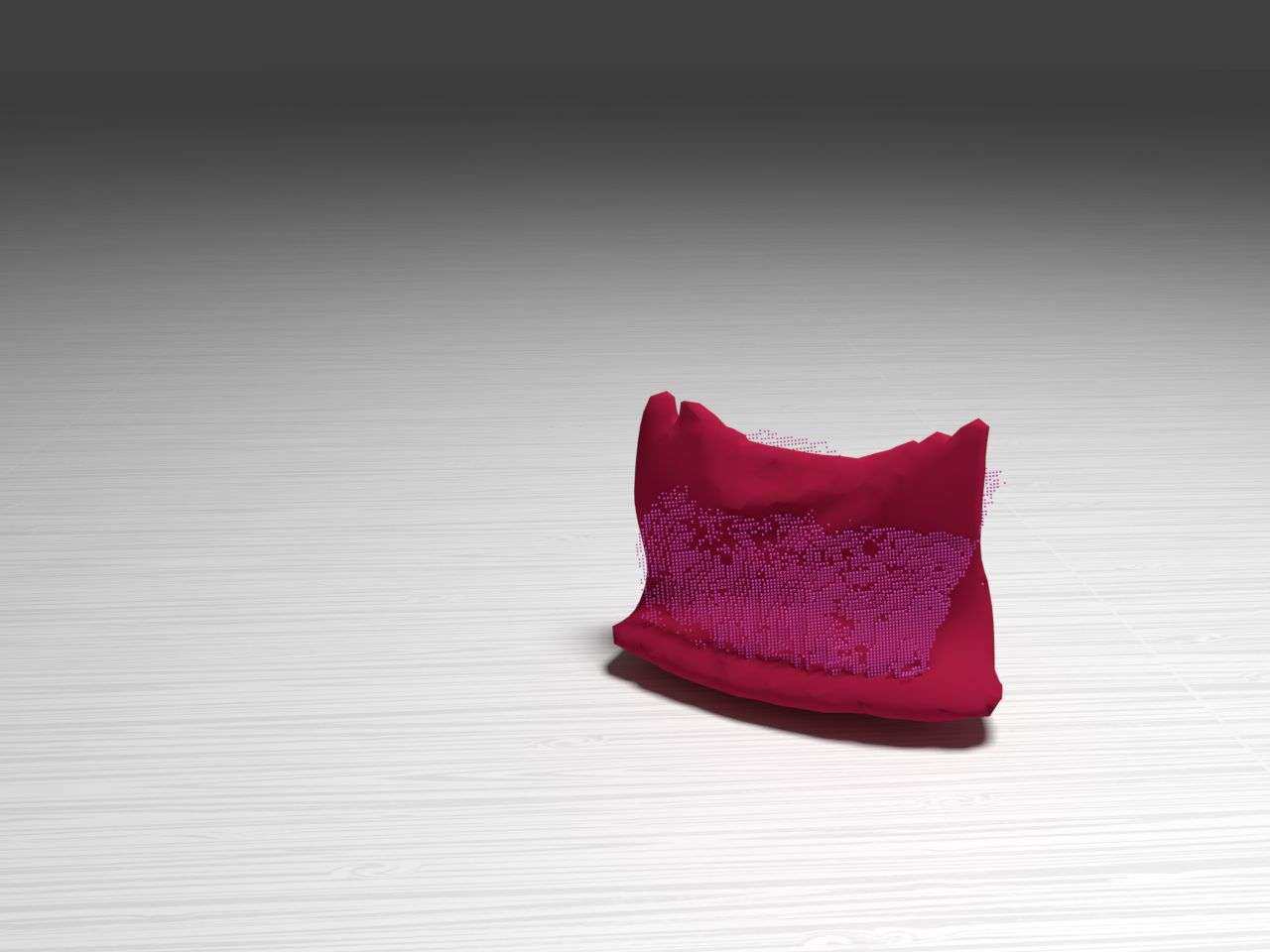}%
	\includegraphics[width=0.24\textwidth]{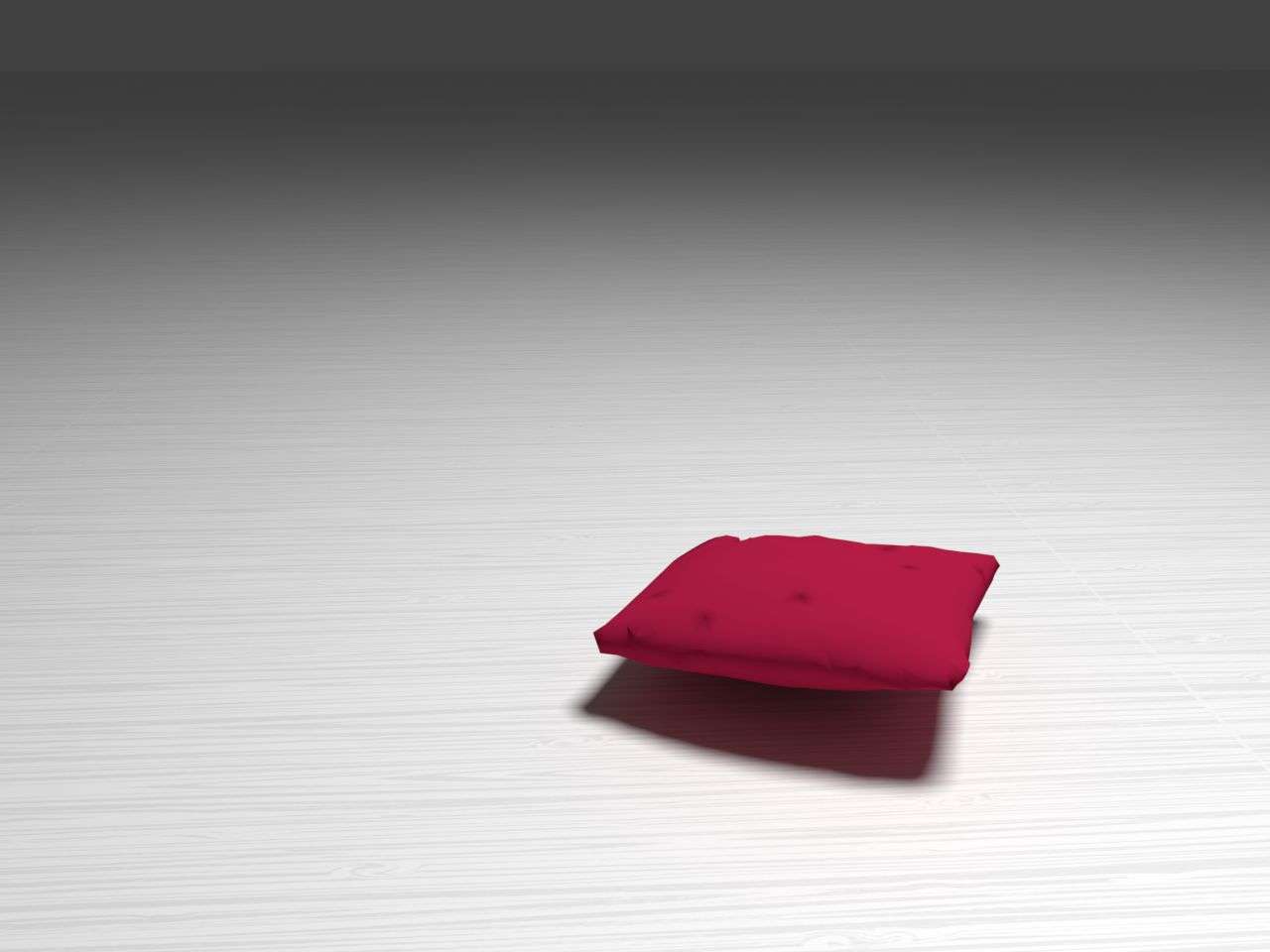}%
	\caption{Reconstructed solution}
	\label{fig:PillowFlat:Reconstruction}
\end{subfigure}
\\
\begin{subfigure}{0.9\textwidth}
    \includegraphics[width=\textwidth]{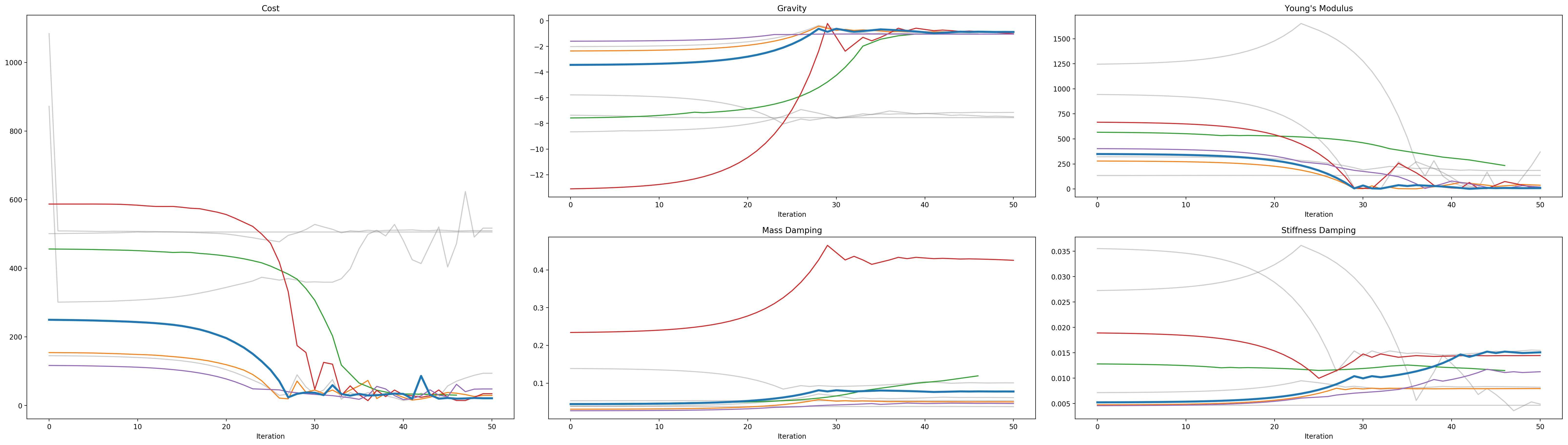}
    \caption{Optimization started from 10 different initial values. The best five runs are drawn in color, the very best run is drawn in thick lines and displayed in the renderings above.}
    \label{fig:PillowFlat:Optimization}
\end{subfigure}
\caption{Selected frames and plots of the optimization process for the Pillow-Flat test case. The rows of (a-d) each show a sequence of steps over time. }%
\Description{Selected frames from the pillow-flat test case: RGB-D observations from the real-world camera, initial configuration of the optimization and reconstructed solution.
	Plots of the optimization process of the pillow test case. Here, gravity, Young's modulus, mass- and stiffness damping are reconstructed.}
\label{fig:PillowFlatPlots}%
\end{figure*}

Here we analyze the convergence of our algorithm for real-world scenarios, as presented in \autoref{sec:realworld}. Reconstructed parameter values for the best five runs are given in \autoref{tab:RealWorld:Numbers:Teddy} for the Teddy,  \autoref{tab:RealWorld:Numbers:PillowRamp} for the Pillow-Ramp and  \autoref{tab:RealWorld:Numbers:PillowFlat} for the Pillow-Flat data set.
Furthermore, \autoref{fig:PillowFlatPlots} shows selected frames and the convergence plots for the Pillow-Flat example.

\begin{table*}
    \centering
    \input{plots/RealWorld/TeddyTable.tex}
    \caption{Reconstructed Parameter values from the teddy data set. Only the five best runs are shown.}
    \label{tab:RealWorld:Numbers:Teddy}
\end{table*}
\begin{table*}
    \centering
    \input{plots/RealWorld/PillowSlopeTable.tex}
    \caption{Reconstructed Parameter values from the Pillow-Ramp data set. Only the five best runs are shown.}
    \label{tab:RealWorld:Numbers:PillowRamp}
\end{table*}
\begin{table*}
    \centering
    \input{plots/RealWorld/PillowFlatTable.tex}
    \caption{Reconstructed Parameter values from the Pillow-Flat data set. Only the five best runs are shown.}
    \label{tab:RealWorld:Numbers:PillowFlat}
\end{table*}

%% file: plots/RealWorld/TeddyTable.tex
\begin{tabular}{r|lllll}
Run & 14 & 15 & 18 & 3 & 1\\ \hline 
Initial Cost & 143.924 & 109.387 & 134.841 & 136.685 & 136.551\\ \hline 
Recon. Cost & 8.461 & 8.506 & 8.693 & 8.747 & 8.810\\ \hline 
Gravity & -1.536 & -1.526 & -1.538 & -1.521 & -1.530\\ \hline 
Young's Modulus & 7.817 & 8.126 & 8.690 & 7.840 & 7.383\\ \hline 
Mass Damping & 0.240 & 0.086 & 0.080 & 0.122 & 0.102\\ \hline 
Stiffness Damping & 0.027 & 0.033 & 0.018 & 0.040 & 0.043\\ \hline 
\end{tabular}

%% file: plots/RealWorld/PillowSlopeTable.tex
\begin{tabular}{r|lllll}
Run & 7 & 12 & 4 & 18 & 9\\ \hline 
Initial Cost & 589.252 & 605.694 & 550.197 & 402.254 & 383.658\\ \hline 
Recon. Cost & 59.816 & 91.852 & 125.231 & 164.708 & 170.164\\ \hline 
Gravity & -0.710 & -1.430 & -1.681 & -0.805 & -1.573\\ \hline 
Young's Modulus & 209.421 & 142.042 & 914.497 & 839.310 & 1143.883\\ \hline 
Mass Damping & 0.068 & 0.064 & 0.071 & 0.176 & 0.044\\ \hline 
Stiffness Damping & 0.044 & 0.082 & 0.864 & 0.176 & 0.184\\ \hline 
Ground Height & 0.127 & 0.179 & 0.181 & -0.009 & 0.009\\ \hline 
Ground Theta & 1.188 & 1.229 & 1.034 & 1.288 & 1.222\\ \hline 
Ground Phi & 1.480 & 1.226 & 1.622 & 1.502 & 1.466\\ \hline 
\end{tabular}

%% file: plots/RealWorld/PillowFlatTable.tex
\begin{tabular}{r|lllll}
Run & 8 & 7 & 3 & 5 & 2\\ \hline 
Initial Cost & 249.871 & 154.268 & 456.128 & 587.173 & 116.719\\ \hline 
Recon. Cost & 21.126 & 29.303 & 30.233 & 34.395 & 48.241\\ \hline 
Gravity & -0.872 & -0.883 & -1.031 & -1.017 & -1.038\\ \hline 
Young's Modulus & 9.200 & 37.206 & 233.827 & 9.191 & 18.835\\ \hline 
Mass Damping & 0.078 & 0.051 & 0.119 & 0.425 & 0.046\\ \hline 
Stiffness Damping & 0.015 & 0.008 & 0.011 & 0.014 & 0.011\\ \hline 
\end{tabular}